\documentclass[Chicago]{fsuthesis}
\usepackage{epsfig,contigrefs}
\usepackage{color}
\include{finkdefs}
\title{TRANSVERSE RESUMMATION FOR DIRECT PHOTON PRODUCTION}
\author{Christopher Edmond Fink}
\uauthor{CHRISTOPHER EDMOND FINK}
\thesistype{dissertation}
\department{Department of Physics}

\headofdept{Kirby W.\ Kemper}
\deptheadtitle{Chairman}
\deanofschool{Donald J. Foss}
\degree{Doctor of Philosophy} \gradyear{2001} \dept{Physics}
\majorprof{Joseph F. Owens}
\majorprofdegree{Ph.D.}
\outcommmember{Jack Quine}
\commmembera{Howard Baer}
\commmemberc{Elbio Dagotto}
\commmemberb{Vasken Hagopian}

\semester{Spring} \college{Arts and Sciences}
\defensedate{March 23, 2001}

\begin{document}
\dedication{For my parents Robert and Catherine, my brother David, and his wife Lynn.}

\def\acknowledgementtext{
\hskip\parindent This dissertation could not have been achieved
without the support, patience, and guidance of Professor Joseph
(Jeff) Owens. I am deeply indebted for his close personal
interaction and concern. My sincere thanks to every member of the
High Energy Group at Florida State University for making this
doctoral study such an educational and enjoyable experience. I
would like also to extend my sincere appreciation to Professor
Howie Baer, Professor Vasken Hagopian, Professor Elbio Dagotto,
and Professor Jack Quine for reviewing this manuscript. Finally, a
special thanks to Professor Duncan Carlsmith at the University of
Wisconsin, Madison, without whose encouragement I might have
settled on a different line of work.}

\def\abstracttext{
Quantum Chromodynamics predicts radiative terms which, at low
transverse momentum $Q_T$ of an observed multi-particle system,
can be large enough to threaten the legitimacy of a fixed-order
perturbative treatment. We present a subtraction-style
calculation which sums these terms to all orders for a direct
photon plus jet final state, and show that the kinematic effect of
such a resummation is significant even after integration over the
jet. An improved matching algorithm is prescribed for the
transition to the high-$Q_T$ perturbative regime, and
re-examination of the nonperturbative regime indicates that
simple Gaussian parametrizations are to be favored, with
coefficients that depend not on the photon-jet mass $Q$, but on
the color structure of the initiated subprocess. Agreement with
data is significantly improved, and the method is expected to
have applicability in general for processes with steeply-falling
spectra.}
\dedicationtrue
\frontmatterformat
\makefsuabstract
\endfsuabstractpage
\titlepage
\maketitle
\makesignaturepage
\dedicationpage
\prefacesection{Acknowledgements}\acknowledgementtext
\tableofcontents
\listoftables
\listoffigures
\newpage
\abstractsection{ABSTRACT}
\maintext
\chapter{Introduction}
\label{Introduction} \pagenumbering{arabic}

\section{QCD as a Quantum Field Theory}

In particle physics we are continually searching for the most
fundamental description of matter and the forces which govern its
behavior. Those who wonder about the properties of macroscopic
objects may now look to molecular and atomic physics; those who
inquire about the structure of the atom may now be answered with
the details of electrons and nuclei. In each case a multitude of
types of matter has been replaced by a more fundamental set, the
variety of types being replaced by a variety of ways these types
can combine to form bound states. Concurrently, a unified
description has arisen for many forces once thought disparate.

The situation is no different for the atomic nucleus or its
inhabitants. In addition to radiation observed naturally, data on
nuclear structure came from experimental probes of the nucleus,
usually involving beams of particles known to be themselves
structureless. As more energy was made available, the same method
was used to probe individual nucleons (protons and neutrons). The
mid-20th century saw a proliferation of new strongly-interacting
particles (collectively known as {\it hadrons}) as the energy
barrier for their discovery was crossed, and again it became
natural to look for common properties (``good quantum numbers''),
clues to a symmetry on a deeper level.

In 1963, motivated by the observation that the known hadronic
families seemed to exhibit symmetries of a particular branch of
mathematics known as {\it Lie groups}, Gell-Mann ~\cite{Gell64}
and Zweig ~\cite{Zweig64} proposed a model of hadronic structure
based on a limited assortment of sub-nucleonic constituents
called {\it quarks}. Each quark came in a particular {\it
flavor}, had fractional electric charge, and had a corresponding
antimatter partner of opposite charge. Together, they were able
to be combined in ways that reproduced the known subnuclear
spectrum.

However, there was a spin-statistics problem. Systems of {\it
fermions} (particles of half-integer spin, like the quark) must
have wavefunctions which are totally antisymmetric under
interchange of any two particles, while total symmetry is
required of {\it boson} (integer-spin particle) collections.  The
quark model posited that all hadrons were quark (fermion)
combinations, those of half-integral spin (the {\it baryons},
such as the proton and neutron), being composed of three quarks,
and those of integral spin (i.e. {\it mesons}) composed of a quark
and antiquark. Unfortunately, some of these combinations required
symmetry in spin, space, and flavor, leading to a totally
symmetric wavefunction. If the above compositions of baryons and
mesons were correct, a new degree of freedom was required in which
to properly perform the antisymmetrization ~\cite{Han65}.

A new type of charge, posited by Greenberg ~\cite{Green64} and
which eventually became known as {\it color} ~\cite{Gell72}, thus
solved the statistics problem. Again, group theory was
instrumental (in this case a color $SU(3)$ fundamental
representation). With three colors, hadrons could be properly
antisymmetrized if each quark in a baryon was a different color,
but those in mesons were of the same color. The quark model seemed
to be rescued, despite the fact that no individual quarks, or any
other ``colored'' states, had been seen in the laboratory. The
lack of experimental evidence for anything but colorless states
could be postulated as a rule ({\it color confinement}), but
without an understanding of the forces present among quarks, such
rules would remain baseless.

Classical theory described interaction at a distance in terms of
fields, with forces that depended on the charges of the particles
involved and the distance between them. Modern quantum theory,
although maintaining use of the term ``field'', had replaced the
concept with the (macroscopically equivalent) exchange of quanta,
specific to the type of charge involved, which carried energy and
momentum across the intervening distance. One theory of this
type, Quantum Electrodynamics (QED), had already been successful
in remodeling the electromagnetic interaction between charged
particles (as well as all the known properties of light) in terms
of intermediate, long-range bosons called {\it photons}.

It had also long been known that symmetries of field theories
gave rise to conserved quantities, like charge and momentum. In
particular, conservation of electric charge in the coupled system
of photons and charged particles was shown to be a consequence of
the symmetry of the theory with respect to certain local {\it
gauge} transformations, performed simultaneously on all fields. As
the quarks had not only electric but color charge, both of which
were expected to be conserved, it was natural to look for gauge
bosons of the color force. However, unlike the electromagnetic
interaction, this force was known to become {\bf weaker} at
higher momentum transfer; that is, the individual quarks behave
more like free particles ~\cite{Bjork69,Pan68}. This property,
{\it asymptotic freedom}, was searched for in the known quantum
field theories, and found eventually to be associated exclusively
with those gauge theories in which the generators of the group
symmetry do not commute; i.e. are {\it non-Abelian}
~\cite{tHo72a,Gro73,Pol73,Zee73}.

Fritzsch and Gell-Mann proposed, then, that the group symmetry of
a non-Abelian gauge field be associated with color symmetry, and
Quantum Chromodynamics (QCD) was born ~\cite{FG72,FG73}. The gauge
bosons of this theory, called {\it gluons}, have eight members,
corresponding to the number of elements in the adjoint
representation of the $SU(3)$ color group. They couple to the
color charge of the quarks, binding them together within the
hadron. Being non-Abelian in nature, they also carry the color
charge themselves and thereby couple with one another (unlike the
photons of QED, which are Abelian). Not only does this explain the
strengthening of the strong force with distance, and consequently
why we do not see any free quarks or gluons (collectively known as
{\it partons}), but also has implications for the question of
color confinement. \footnote{The term ``parton'', coined by R.P.
Feynman ~\cite{Feyn69} before the advent of QCD (and in a purely
dynamical context), has since been identified with its basic
building blocks.} Indeed, Wilson ~\cite{Wilson74} has shown that,
for sufficiently strong coupling, the only finite-energy
asymptotic states of QCD are colorless. Any attempt to separate
one parton from another results only in the creation of a
quark/antiquark pair somewhere in the intervening distance, the
resulting colorless states propagating outward as everyday
hadrons. This is the process of ``fragmentation'' or {\it
hadronization}.

\section{The Parton Model}

Asymptotic freedom brings with it benefits and constraints. The
collision of two hadrons is an enormously complex problem, and
cannot be solved exactly, as the component partons can interact
in infinite combinations. First, as we are not able to directly
observe individual partons, we do not {\it apriori} know the
precise distribution of momentum among the quarks inside the
nucleon, nor can we predict from first principles the number of
each type represented at a given energy. Just as an electron is
at all times surrounded by its own cloud of photons and
electron-positron pairs produced by these photons, which tend to
alter the effective charge seen by a neighboring charge, so too
do the quarks constantly emit and reabsorb gluons, themselves
producing a cloud of quark-antiquark pairs. Increasing the
interaction energy only serves to probe deeper into the cloud,
thus changing the effective distribution ~\cite{KS74}. Thus we
find that the best description we can achieve of the nucleon at
high energy is, probabilistically, a set of continuous densities,
one for each flavor/antiflavor, plus a gluon density. These {\it
parton distribution functions} give the probability of finding a
certain type of parton (quark flavor or gluon) at a certain
momentum fraction and interaction energy within the nucleon.
Although prediction of these densities is ultimately a
non-perturbative endeavor, we {\bf can} perturbatively derive the
relationship between densities at different interaction energies.
Measurement of the densities at one energy thus provides
predictions for the rest.

Secondly, if the final state of interest contains particular
hadrons that are to come from partons involved in the scattering
process, instead of attempting to describe quantitatively the
mechanism by which this hadronization occurs, we can currently
account for it only phenomenologically, by introducing a {\it
fragmentation function} for each of these partons. In a manner
analogous to the parton distribution functions, these give the
probability that a certain hadron will be produced from the
fragmenting parton, with a specific fraction of that parent's
momentum. Both distribution functions and fragmentation functions
must ultimately be measured by experiment.

The good news of asymptotic freedom, however, is that at small
distances (and correspondingly high collision energies), the
strong force decreases and the partons behave more like free
particles ~\cite{PolPL77}. This allows us to {\it factorize} what
we can perturbatively describe (the interaction of free partons)
from what we can't (distributions and fragmentation). In the {\it
parton model} of high-energy hadronic collisions, the initial
reaction takes place over such a small time scale that it becomes
plausible to treat each hadron as a collection of free partons,
each traveling in the same direction as the parent hadron, and
only one of which takes place in any given ``hard scattering''.
That is, one may calculate basic subprocesses involving one
parton from each hadron, then sum incoherently over the
contributions of all partons via convolution with the parton
distribution functions. Similarly, the process of fragmentation is
assumed to take place over a time scale much longer than the hard
scattering, and can therefore also be treated independently. Thus
a description of hadronic collisions is necessarily built up from
both perturbative and non-perturbative pieces.

One of the conveniences the new quantum model allows is the
schematic representation of a subprocess as a {\it Feynman
diagram}, each fermion or boson being represented by a line. Line
intersections ({\it vertices}) can represent instances of
absorption, emission, or pair production, depending on the
direction of time. To each element of this picture a mathematical
expression is assigned, in such a way that one can easily derive
the correct expression for a subprocess: one simply draws its
diagram, starts at one end, and follows the lines and vertices,
writing down the corresponding factors in order. The result is
called the {\it amplitude} for the reaction, and its magnitude
squared is a measure of the probability that the incoming
particles will interact in precisely this way to form the
outgoing particles. Since it is the vertices at which the force
is felt, the overall strength of the reaction and its probability
for occurring become dependent upon the number of vertices in the
diagram.

Of course, for a given set of external lines, there is an
infinite number of diagrams that could be built. Each makes a
contribution to the total probability for the reaction, in
relation to its {\it order}, or exponent of the coupling
strength. It becomes clear that if we hope to make meaningful
predictions through this {\it perturbation theory}, this relation
must be inverse, {\it i.e.} the coupling constant (unit of charge)
used must be less than one. Otherwise, the diagrams with the
largest contribution would be the ones with the most vertices,
and we cannot analyze an infinite number of diagrams. With a small
coupling constant, we can usually stop at first or second order,
confident that the higher-order diagrams are of negligible
contribution (an important exception will be discussed in Section
\ref{sec:resneed}).

\section{The Perturbative Domain}
\label{sec:intropert}

Having a small coupling constant is not necessarily enough; if the
coefficients of successive orders grow, we may lose the
applicability of perturbation theory. Hence, while in many cases
we are justified in calculating only to leading order (LO), to
confirm the convergence of the series and provide corrections to
the leading-order result, we must often look at the
next-to-leading order (NLO) terms as well ~\cite{CPR81}. {\it
Loop} diagrams, having the same external legs but more internal
vertices, will of course contribute, as will the release of {\it
bremsstrahlung} radiation which inherently accompanies
acceleration of charged particles. Thus {\it inclusive} data is
considered (all events are kept which contain the desired final
state particles, regardless of the detection of additional
debris), and all contributing diagrams up to a given order are
calculated.

\begin{figure} \centering
\begin{fmffile}{cfsix1a}
\fmfframe(10,10)(10,10){
\begin{fmfgraph*}(60,60)
\fmfleft{i1,i2} \fmfright{o1,o2} \fmf{quark}{i1,v1,i2}
\fmf{photon,label=$\gamma$}{v1,v2} \fmf{fermion}{o1,v2,o2}
\fmflabel{$q$}{i1} \fmflabel{$\bar{q}$}{i2} \fmflabel{$\mu^-$}{o2}
\fmflabel{$\mu^+$}{o1} \fmflabel{$e$}{v1} \fmflabel{$e$}{v2}
\end{fmfgraph*}}
\end{fmffile}
\caption{Lowest Order (Born) Drell-Yan diagram, ${|{\cal M}|}^2
\sim {[e^2]}^2 \sim \alpha^2$.} \label{fig:six1a}
\end{figure}
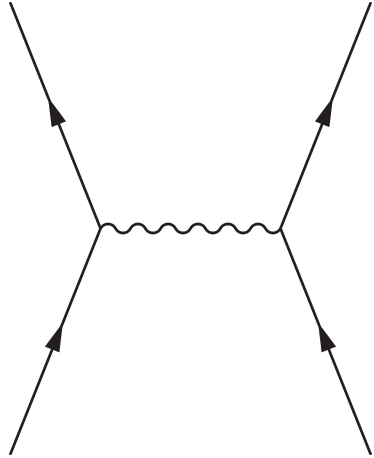

\begin{figure} \centering
\begin{fmffile}{cfsix1c}
\fmfframe(10,10)(10,10){
\begin{fmfgraph*}(60,60)
\fmfleft{i1,i2} \fmfright{o1,o2} \fmfbottom{b1}
\fmf{quark,tension=2}{i1,v1} \fmf{quark,tension=2}{v1,v2}
\fmf{quark}{v2,i2} \fmf{photon,label=$\gamma$}{v2,v3}
\fmf{fermion}{o1,v3,o2} \fmflabel{$q$}{i1}
\fmflabel{$\bar{q}$}{i2} \fmflabel{$\mu^-$}{o2}
\fmflabel{$\mu^+$}{o1} \fmflabel{$e$}{v2} \fmflabel{$e$}{v3}
\fmffreeze \fmf{gluon}{v1,b1} \fmfv{l=$g$,l.a=150,l.d=.05w}{v1}
\end{fmfgraph*}}
\end{fmffile}
\caption{Gluon Bremsstrahlung diagram, ${|{\cal M}|}^2 \sim
{[e^2g]}^2 \sim \alpha^2 \alpha_s$.} \label{fig:six1c}
\end{figure}
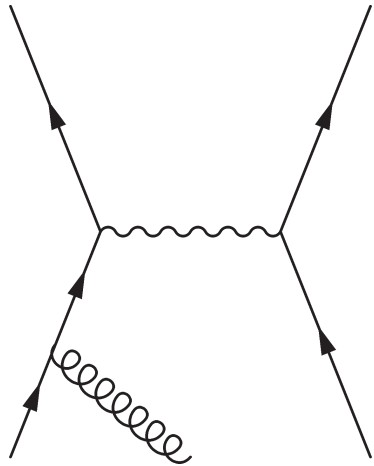

\begin{figure} \centering
\begin{fmffile}{cfsix1d}
\fmfframe(10,10)(10,10){
\begin{fmfgraph*}(60,60)
\fmfleft{i1,i2} \fmfright{o1,o2,o3} \fmf{quark}{i2,v2}
\fmf{quark,tension=0.5}{v2,v1} \fmf{quark}{v1,o1}
\fmf{gluon}{i1,v1} \fmf{photon,label=$\gamma$}{v2,v3}
\fmf{fermion}{o3,v3,o2} \fmflabel{$q$}{i2} \fmflabel{$q$}{o1}
\fmflabel{$e$}{v3} \fmflabel{$\mu^-$}{o2} \fmflabel{$\mu^+$}{o3}
\fmflabel{$e$}{v2} \fmflabel{$g$}{v1}
\end{fmfgraph*}}
\end{fmffile}
\caption{Compton diagram, ${|{\cal M}|}^2 \sim {[e^2g]}^2 \sim
\alpha^2 \alpha_s$.} \label{fig:six1d}
\end{figure}
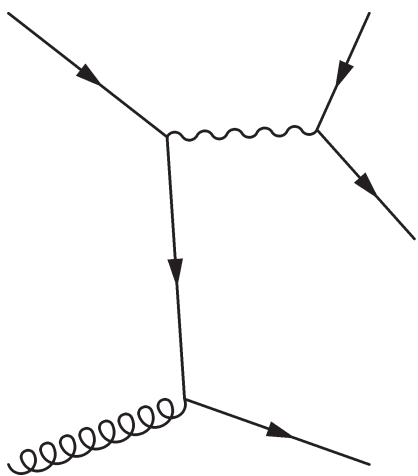

\begin{figure} \centering
\begin{fmffile}{cfsix1b}
\fmfframe(10,10)(10,10){
\begin{fmfgraph*}(60,60)
\fmfleft{i1,i2} \fmfright{o1,o2} \fmf{quark}{i1,v1,v2,v3,i2}
\fmf{photon,label=$\gamma$}{v2,v4} \fmf{fermion}{o1,v4,o2}
\fmflabel{$q$}{i1} \fmflabel{$\bar{q}$}{i2} \fmflabel{$\mu^-$}{o2}
\fmflabel{$\mu^+$}{o1} \fmflabel{$e$}{v2} \fmflabel{$e$}{v4}
\fmffreeze \fmf{gluon,left}{v1,v3} \fmfv{l=$g$,l.a=60,l.d=.1w}{v3}
\fmfv{l=$g$,l.a=-60,l.d=.1w}{v1}
\end{fmfgraph*}}
\end{fmffile}
\caption{Virtual Drell-Yan diagram, ${|{\cal M}|}^2 \sim
{[e^2g^2]}^2 \sim \alpha^2 \alpha_s^2$.} \label{fig:six1b}
\end{figure}
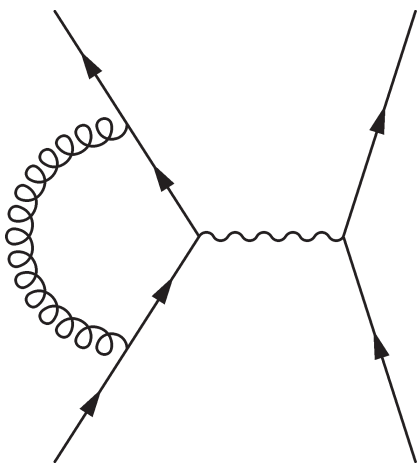

As an example, we'll show the leading order (LO) and
next-to-leading order (NLO) contributions to the production of a
muon pair in hadronic collisions. At lowest order, there is but
one contributing subprocess, proposed initially by Drell and Yan
~\cite{DY70,DY71}. A quark from one hadron annihilates with an
antiquark from the other to produce a vector boson (hereinafter
chosen to be a photon), which subsequently decays to a
muon-antimuon pair. The expansion so far consists of one term of
second order in the electromagnetic coupling $\alpha$, ${\rm
zero}^{th}$ order in the strong coupling $\alpha_s$ (see fig.
\ref{fig:six1a}). Order $\alpha^2 \alpha_s$ subprocesses include
gluon bremsstrahlung from one of the incoming quarks (fig.
\ref{fig:six1c}), and quark-gluon scattering (fig.
\ref{fig:six1d}), which produces an additional quark in the final
state. The exchange of a gluon between incoming quarks is a
contribution of order $\alpha^2 \alpha_s^2$ (fig.
\ref{fig:six1b}); its interference with the Born term, however,
leads to an $\alpha_s^1$ quantity. Photon bremsstrahlung or
exchange is also a possibility, but these order $\alpha^3$
processes are suppressed by the small size of $\alpha/\alpha_s$.
\footnote{In the diagrams, solid lines denote quarks, wavy lines
photons, and curly lines gluons. The coupling constants $e$
(electromagnetic) and $g$ (strong) are more commonly traded for
the quantities $\alpha \equiv e^2 / 4 \pi$, $\alpha_s \equiv g^2
/ 4 \pi$.}


\bigskip


\bigskip


\bigskip


As a rule, bremsstrahlung and loop terms contain various
mathematical singularities as masses and momenta are taken to
certain limits, but these divergences can all be removed
consistently in the full theory, as they must to avoid the
absurdity of an infinite cross section
~\cite{SW77,Kin62,Lee64,BN37}. {\it Infrared} singularities arise
from both bremsstrahlung and loop diagrams, in the limit of zero
radiative energy, and cancel if these contributions are both
included. Loop diagrams also contain {\it ultraviolet}
divergences at infinite loop momentum, and these are factored
into a redefinition, or {\it renormalization} of the
(unobservable) bare masses and couplings. The remaining class of
divergence arises in the limit that bremsstrahlung radiation
becomes parallel with the emitting particle. These {\it
collinear} singularities are factored into a redefinition of the
parton distribution functions (or {\it PDF}s, the probabilities
that the colliding hadrons contain the relevant partons). In both
renormalization and collinear factorization, the singularities
(or {\it poles}) are accompanied at order $\alpha_s^n$ by a power
series of logarithms involving an energy characteristic of the
subprocess, {\it e.g.} the invariant mass $Q$ of the muon-pair.
In the {\it leading log approximation}, the largest of these logs
at each order is summed over all orders and factored into the
couplings and PDFs also, making them energy-dependent.
\footnote{These energies are historically called {\it scales}
(the collinear in particular being the {\it factorization} scale,
which we'll denote as $M_f$). One of the assumptions of the
early, or ``na\"{\i}ve'', parton model was that the parton
distributions should ``scale with'' the momentum of the parent
hadron; i.e. should depend only on the momentum {\bf fraction}
involved. As we are here taking our picture of the hadron as a
hierarchy (or cloud) of partons which we sample at an effective
resolution $1/M_f$, and building it into a redefinition of the
parton distributions, this is no longer a valid assumption,
though we transfer the terminology.} As it has been shown that
(for each parton species) the divergent logs arise in the same
way for all processes with a large scale
~\cite{PolNP77,Ellis79,APV78}, once the new coupling and
distributions are calculated, they are transportable in this
approximation to other reactions. A lot of work is saved in the
rough calculation of a new process, as only the Born subprocess
need be derived, and the most dangerous higher-order terms
accounted for by convoluting the Born term with the new coupling
and parton distributions ~\cite{AP77,APP78}. Further improvement
comes, of course, with the inclusion of subleading higher-order
terms, but as we shall see, even this may not be sufficient for
processes which also involve a small scale.

\section{The Need for Resummation}
\label{sec:resneed}

During the late 1970's, the leading log approximation was applied
to a number of processes ~\cite{CTEQ95}, although we will focus
here on muon pair production. In addition to single-scale
observations such as the invariant mass $Q$ of muon pairs
~\cite{APV78,Sachpl78,AEM78,AEM79,KP79}, physicists looked at
correlations between the muons, such as the transverse momentum
imbalance, that is, the {\bf total} $Q_T$ of the pair (here and
throughout the text, we'll use ``$Q_T$'' to denote the transverse
momentum of a system of two or more particles, and ``$k_T$'' or
``$p_T$'' for single particles). To a good first approximation, at
relativistic speeds along a single beam line, the constituents of
hadronic interactions present no transverse momentum relative to
each other or the beam, and thus at Born level (just the
Drell-Yan subprocess), there should be no imbalance, and we get a
delta function at total $Q_T = 0$. \footnote{Indeed, the
``na\"{\i}ve" parton model takes this as a starting point; the
current ``effective constituents'' interpretation recognizes that
there are no natural cutoffs that would limit the transverse
momenta of partons relative to their parent hadrons, and hence
we're led to the idea of ``intrinsic'' parton $p_T$
~\cite{CG73,Par73}.} The available data ~\cite{Yoh78} indeed
showed the cross section to be largest at $Q_T=0$, falling off
gradually (although still steeply) with increasing $Q_T$. This
data being inclusive, the contribution at $Q_T \ne 0$ could be
due to ``recoil'' of the pair against bremsstrahlung from the
initial state partons, recoil against a final-state quark in the
quark-gluon ``Compton'' subprocess described above, or some
nonperturbative, ``intrinsic'' $p_T$ of the incoming partons (see
figure \ref{fig:intpt}) . The former having not yet been
calculated, it was unclear how much of a contribution was to be
ascribed to each of these effects, and the approach varied among
researchers.

\figboxh{intpt}{Intrinsic transverse momentum.}

A leading log calculation with simple Drell-Yan as the Born term
admitted only intrinsic $p_T$, and thus some attempted a
description solely through $p_T$-scaling of the parton
distribution functions ~\cite{Kog76,HS77}. This would ``smear''
the zero-$Q_T$ delta function into a more physical, Gaussian
shape, and it was hoped that the amount of intrinsic $p_T$ needed
would be small. This turned out not to be the case ~\cite{FFF78}.
When in fact the next-to-leading order bremsstrahlung and Compton
subprocesses were calculated, although the large-$Q_T$ region
seemed to be sufficiently accounted for, an unacceptable
low-$Q_T$ prediction remained. Some authors simply left their
results as valid only in the high-$Q_T$ region ~\cite{KR78,FM78}.
Others, notably Altarelli, Parisi, and Petronzio, were able to
devise a model of intrinsic $p_T$ which also brought the
low-$Q_T$ distribution under control ~\cite{APP78,KLR78}. The
average intrinsic $p_T$ required, however, was still a matter of
phenomenology, not derivation. In the same way that the
factorization and renormalization scale dependences of a
leading-log calculation are reduced when next-to-leading (and
higher) contributions are added, higher-order calculations were
needed which focused on transverse momentum effects.

It was recognized that the incoming quarks could radiate any
number of undetectable gluons, if the ``jets'' of hadrons thus
produced had energies less than the trigger threshold, and that
this additional, {\it soft} radiation could collectively
influence the transverse recoil of the observed final state. An
initial attempt to attack these contributions was undertaken in
1978 by Dokshitzer, D'Yakanov and Troyan (DDT) ~\cite{DDT78}, who
showed that in the soft limit, a pattern emerged as one went to
higher orders. Careful inclusion of $n$ gluons into the final
state shows that even after the cancellation of the infrared
divergence at $Q_T=0$, there remains a series of logarithms at
order $n$ in the strong coupling of the form \footnote{The ``+''
designation can here be thought of as a simple reminder that the
pole at $Q_T=0$ is removed. Later, in Section \ref{sec:factorize},
we'll introduce ``plus-distributions'' more formally. Neither
these logs nor the leading order delta function at $Q_T=0$ are
sufficiently physical representations of an actual
$Q_T$-distribution; their combination into such a representation
is the goal of {\it resummation}.}

\begin{equation}
{1 \over {Q_T}^2} \sum_{m=0}^{2n-1} {\bigl[ \ln(Q^2/{Q_T}^2)
\bigr]}_+^m \;. \label{logs}
\end{equation}

At first glance, more gluons seems to make the problem worse. At
high $Q_T$ (close to the mass $Q$), these contributions are
small. At low $Q_T$, however, we have a two scale problem, the
contributions of which may easily overwhelm the smallness of
${\alpha_s}^n$. The convergence of a perturbative expansion is
thus thrown into doubt. Furthermore, as was shown by Altarelli,
Ellis, and Martinelli (1979) ~\cite{AEM79}, this is the case no
matter how you define the parton distribution functions.

That a pattern existed, however, with at least the largest
($m=2n-1$) logs sharing a common coefficient across orders,
allowed DDT to {\it resum} this tier of logarithms over all
orders, yielding a finite result at $Q_T=0$. Their method, flawed
by a lack of regard for the conservation of momentum among the
emitted gluons, was improved upon by Parisi and Petronzio in 1979
~\cite{PP79,CG80}. The new procedure used the momentum-conserving
delta function to Fourier transform the cross section into impact
parameter space, perform the sum there, and transform back to
momentum space. The amount of intrinsic $p_T$ needed to reproduce
the data was drastically reduced, though the extremely low-$Q_T$
region still required some such nonperturbative help.

A later analysis (1981) by Collins and Soper ~\cite{CS81} for the
reverse Drell-Yan process uncovered a systematic way to include
subleading logs, thus improving the perturbatively-tractable
region, and provided a model for non-perturbative effects,
allowing the prediction to extend to even smaller transverse
momenta. In 1984, Altarelli, Ellis, Greco, and Martinelli
~\cite{AEGM84}, along with Collins, Soper and Sterman
~\cite{CSS85}, reapplied this formalism to the general case of
intermediate vector boson production in hadronic collisions, and
it became evident that reduction of the required intrinsic $p_T$
was linked in large part to systematic addition of further tiers
of subleading logs. This reflected the fact that ultimately both
intrinsic and recoil contributions derive from parton
interactions, whether internal or external to the unobservable
``dissociation'' of the hadron.

\section{Goals and Outline of this Work}

Although the bulk of resummation work has been devoted to
Drell-Yan and other intermediate vector boson processes, the
methods have recently been shown beneficial in the prediction of
other processes with steeply-falling transverse momentum spectra
~\cite{BOO92}. These include double jet ({\it dijet}) production
~\cite{CGS79} and direct double photon production
~\cite{CFG95,BBMY98}.

The study of hadronic photon production is important for a number
of reasons. Single photon production, due to its leading-order
dependence on the gluon content of the incoming hadrons, serves
as a good test of our knowledge of these densities. By comparing
the results of single and double photon production, the strong
coupling may be evaluated. Finally, in the ongoing search for the
Higgs, theory predicts a two photon decay, making precise
calculation of the direct reactions necessary for elimination of
background events.

However, the focus of most of the literature has been on observed
systems of two or more particles; less well-known is the effect
of resummed soft radiation on single-particle $p_T$ spectra. This
is not surprising, as the resummed logarithms arise naturally
only when asking questions about the properties of multi-particle
systems. Any transverse ``kick'' from unobserved radiation,
though, should distribute itself among the observed particles,
and it is only natural to ask about this distribution as well. In
this dissertation we apply the current resummation techniques
intermediately to a direct photon plus jet final state, and show
that the kinematic effect on the photon $p_T$ is significant even
after integration over the jet.

This requires that we keep track of more variables, and, unlike
many previous calculations, refrain from integrating over them
analytically. This is beneficial from an experimental standpoint
as well. In order to extract useful signals in practice, it is
sometimes necessary to make kinematic cuts on the detected
particles. These cuts may not be expressible analytically at the
parton level, where the integrations need to be done, as there is
usually a complicated mapping among the associated variables. A
multitude of convoluted holes may appear in the partonic phase
space, thus making of little versatility formulae in which the
relevant variables have already been integrated over a
continuous, predetermined range. In addition, jet definitions,
detector acceptances, and kinematic cuts are often changed,
necessitating full recalculation of analytic predictions.

We will handle experimental cuts by performing the required
integrals via a Monte Carlo approach. By taking advantage of the
method's ability to calculate Jacobians automatically, changes in
applied cuts are easily accomodated, and added benefits include
the ability to calculate any number of observables simultaneously
~\cite{HH64,Stroud71}. Precedence has been set for the
application of Monte Carlo to photon physics, as Bailey, Owens,
and Ohnemus ~\cite{BOO92,BOO90} performed such studies in the
early 1990's, incorporating a synthesis of NLO analytic and Monte
Carlo methods to predict the photon $p_T$ distribution. Inclusion
of soft gluon effects is the focus of the current work, with the
expectation that the resultant method will have applicability to
other single-particle predictions.

In Chapter \ref{ch:QCDpert}, the fundamentals of QCD perturbation
theory are presented, starting with the development of the
Lagrangian and the Feynman rules for calculating subprocess
matrix elements. We discuss the singularities that arise in
different types of diagrams, and the manner in which they
disappear in the full theory via the process of renormalization.
Finally, we describe the QCD hard scattering formalism, which
combines the perturbatively calculable matrix elements with the
experimentally determined parton distributions to arrive at
predictions for observable quantities.

Chapters \ref{ch:PPResum}, \ref{ch:CSS}, and \ref{ch:ktspace}
together present the evolution of current techniques for the
resummation of those large residual logarithms which arise
naturally in perturbation theory, but are problematic for its
convergence. Chapter \ref{ch:PPResum} motivates a statement of the
problem and offers the first attempts at a solution. Chapter
\ref{ch:CSS} describes the efforts of Collins, Soper, and Sterman
to derive a more comprehensive picture, and solution via a
transform to impact-parameter space. Chapter \ref{ch:ktspace}
reviews both the manner in which this formalism may be
analytically simplified and its connection with the idea of
intrinsic parton transverse momentum.

Using the diagrams and corresponding matrix elements of Appendix
\ref{ap:Diagrams}, in Chapter \ref{ch:photon} we begin the
process of calculating the pieces of a NLO direct photon (plus
jet) cross section. Contributions from three-body final states
are approximated in the singular regions of phase space, and
these approximations are both subtracted from the original (to
provide one finite piece) and added to the virtual two-body terms
(which are singular but of the opposite sign) to provide another
finite piece. Along the way, we perform a number of analytical
integrations, collected for reference in Appendix
\ref{ap:integrals}. Casting these finite pieces in a form
congruent to that of the NLO expansion of our resummation scheme,
we then simply read off the coefficients of the original resummed
form. Detailed calculational results are presented in Appendix
\ref{ap:Contribs}. Finally, we present a method for matching the
resummed and perturbative results at the boundary between their
regions of applicability.

Crucial to the versatility of this calculation is the use of
Monte Carlo techniques for performing remaining integrals, and a
short treatise on these techniques is included in Appendix
\ref{ap:MonteCarlo}. With the resulting FORTRAN program
(available upon request), we produce physical predictions for the
single direct photon $p_T$ spectrum at a variety of energies;
these results, and general conclusions, are presented in Chapter
\ref{ch:Results}. Observations on the form of the nonperturbative
parametrization are also discussed there.
\chapter{QCD Perturbation Theory}
\label{ch:QCDpert}

\section{The QCD Lagrangian}

We begin by constructing the QCD Lagrangian, from which the
equations of motion and scattering matrix follow. The Einstein
summation convention will be used throughout, in which repeated
indices are to be summed over.

The quarks will be represented by Fermi-Dirac fields $\psi_{f,i}
(x)$ of flavor $f$, color $i$, and mass $m_f$. Such fields obey
the Dirac equation, and so for a system of free quarks we would
have but

\begin{equation}
{\cal L}_q = {\bar{\psi}}_{f,i}  \bigl[ i \gamma^\mu
\partial_\mu - m_f \bigr] \psi^{f,i}  \;. \label{Lag1}
\end{equation}

Here the flavor index $f$ runs over the number of flavors $N_f$,
while the color index $i$ (and later, $j$) runs over the number
of colors $N_c$. Once we impose local $SU (N_c)$ gauge
invariance, the existence of gauge boson fields (the gluons)
$A_\mu^a (x)$ becomes necessary, as we must replace the gradient
$\partial_\mu$ with the covariant derivative $D_\mu =
\partial_\mu \delta^{ij} + ig{(T_a)}^{ij}A_\mu^a$. There are
$N_c^2-1$ such fields, corresponding to the number of elements in
the adjoint representation of the symmetry group, and we'll use
the indices $\{a,b,c\}$ to distinguish them. The generators $T_a$
form a basis for the Lie algebra of the $SU (N_c)$ symmetry
group, and obey the commutation relations

\begin{eqnarray}
\bigl[ T_a,T_b \bigr] &= if_{abc} T_c \; \label{Lag2a} \\ \bigl\{
T_a,T_b \bigr\} &= {\delta_{ab} \over 3} + d_{abc} T_c \;.
\label{Lag2b}
\end{eqnarray}

\noindent The $f_{abc}$ are totally antisymmetric, and are called
the {\it structure constants} of the group. The $d_{abc}$ are
totally symmetric.

Now that we have gluons, we need to include a kinetic term for
these fields which preserves local $SU(N_c)$ invariance. With
field strength tensors given by

\begin{equation}
F_{\mu\nu}^a = \partial_\mu A_\nu^a - \partial_\nu A_\mu^a + g
f_{abc} A_\mu^b A_\nu^c \;, \label{Lag3}
\end{equation}

\noindent a suitable construction is simply

\begin{equation}
{\cal L}_g = -{1\over 4} F_a^{\mu\nu} F_{\mu\nu}^a \;.
\label{Lag4}
\end{equation}

\noindent In contrast to the photons of QED, an Abelian theory,
the gluons of QCD couple to each other, as indicated by the third
term in equation \ref{Lag3}. As in QED, however, the boson fields
as currently written have extra, unphysical degrees of freedom. We
can get rid of these by imposing a {\it gauge-fixing} condition,
and there are many options, not all of which preserve
relativistic covariance. The {\it Lorentz} condition,
$\partial_\mu A_a^\mu =0$, does however, and leads to the
Lagrangian term

\begin{equation}
{\cal L}_{fix} = - {1\over {2\zeta}} {\biggl( \partial_\mu A_a^\mu
\biggr) }^2 \;. \label{Lag5}
\end{equation}

Use of a covariant gauge is not, unfortunately, a sufficient
constraint. The longitudinal polarization of the gluon fields
remains, and if not dealt with can destroy the unitarity of
scattering amplitudes. Fortunately, as was shown by Faddeev in
1967, one can couple a set of scalar {\it ghost} fields $\phi_a$
to the gluons in such a way as to cancel the unwanted degrees of
freedom. As there are $N_c^2 -1$ such degrees of freedom (one per
gluon field), there must be the same number of ghost fields,
although the coupling is not purely color-matched, as the
required addition to the Lagrangian shows:

\begin{equation}
{\cal L}_{ghost} = -\partial_\mu {\bar{\phi}}^a \bigl(
\partial^\mu \delta_{ab} + ig {(T^c)}_{ab} A_c^\mu \bigr) \phi^b
\;. \label{Lag6}
\end{equation}

\noindent The final QCD Lagrangian in a Lorentz gauge is thus
given by the sum

\begin{equation}
{\cal L}_{QCD} = {\cal L}_q + {\cal L}_g + {\cal L}_{fix}+{\cal
L}_{ghost} \;. \label{Lag7}
\end{equation}

\section{Feynman Rules for QCD}
\label{sec:FeynRule}

By {\it quantizing} the fields in the above Lagrangian, we can
derive quantum-mechanical amplitudes for particular field
interactions and propagation from one point to another in
spacetime. From these, physical predictions such as cross
sections and decay rates can be found. Quantization is not a
unique procedure, but all methods lead to the same physical
predictions. In the traditional canonical operator formalism, the
fields are identified with quantum operators which obey certain
commutation relations. The Green functions from which a
particular amplitude are built are calculated as vacuum
expectation values of products of these operators at particular
spacetime points. In the functional-integral approach, the Green
functions are obtained by integrating the product of the fields
over all their possible intermediate forms, with a
Lagrangian-dependent weight. A further method is that of
stochastic averaging, in which the fields are identified as
stochastic variables and the Green functions averages of field
products in equilibrium.

In momentum space, the Feynman diagrams for QCD and their
corresponding Green functions are shown in figure {fig:feynrule}.
Indices $i$ and $j$ run over the $N_c$ colors, indices $a$
through $d$ (as well as all $a_i$) will denote the $N_c ^2 -1$
members of the adjoint representation of color $SU(N_c)$, Greek
letters $\mu$ and $\nu$ (as well as all $\mu_i$) are 4-vector
indices, and momenta will be labeled by $p_i$ and $k_i$. Spin and
polarization indices are given as $\lambda$. In the denominator
of the propagator factors, it is customary to add a small
imaginary amount $i\epsilon$ in order to satisfy causal boundary
conditions, but this will be ignored for clarity. There is some
freedom in assigning factors of $i$ and minus signs with these
diagrams; the particular convention used here is that of Muta
~\cite{Muta87}.

\BFIG \hoffset=-0.1\linewidth \tiny
$$
\begin{array}{ccccccc}
{\begin{minipage}[b][19mm][c]{.1\linewidth}$ {{\rm outgoing} \atop
{\rm fermion}} $\end{minipage}} &
\begin{fmffile}{cfextof}
\fmfframe(2,4)(2,4){
\begin{fmfgraph*}(15,15)
\fmfleft{i1} \fmfright{o1} \fmfblob{5}{i1}
\fmflabel{$\lambda$}{o1} \fmf{quark,label=$p$}{i1,o1}
\end{fmfgraph*}}
\end{fmffile}
 & {\begin{minipage}[b][19mm][c]{.1\linewidth}${\bar{u}}_\lambda (p) $\end{minipage}} &\phantom{xx}&
{\begin{minipage}[b][19mm][c]{.1\linewidth}${{\rm outgoing} \atop
{\rm antifermion}} $\end{minipage}} &
\begin{fmffile}{cfextoaf}
\fmfframe(2,4)(2,4){
\begin{fmfgraph*}(15,15)
\fmfleft{i1} \fmfright{o1} \fmfblob{5}{i1}
\fmflabel{$\lambda$}{o1} \fmf{quark,label=$p$}{o1,i1}
\end{fmfgraph*}}
\end{fmffile}
 &{\begin{minipage}[b][19mm][c]{.1\linewidth}$ v_\lambda (p) $\end{minipage}}\\
{\begin{minipage}[b][19mm][c]{.1\linewidth}${{\rm incoming} \atop
{\rm fermion}} $\end{minipage}} &
\begin{fmffile}{cfextif}
\fmfframe(2,4)(2,4){
\begin{fmfgraph*}(15,15)
\fmfleft{i1} \fmfright{o1} \fmfblob{5}{o1}
\fmflabel{$\lambda$}{i1} \fmf{quark,label=$p$}{i1,o1}
\end{fmfgraph*}}
\end{fmffile}
 & {\begin{minipage}[b][19mm][c]{.1\linewidth}$u_\lambda (p) $\end{minipage}}&\phantom{xx}&
{\begin{minipage}[b][19mm][c]{.1\linewidth}${{\rm incoming} \atop
{\rm antifermion}} $\end{minipage}} &
\begin{fmffile}{cfextiaf}
\fmfframe(2,4)(2,4){
\begin{fmfgraph*}(15,15)
\fmfleft{i1} \fmfright{o1} \fmfblob{5}{o1}
\fmflabel{$\lambda$}{i1} \fmf{quark,label=$p$}{o1,i1}
\end{fmfgraph*}}
\end{fmffile}
 & {\begin{minipage}[b][19mm][c]{.1\linewidth}${\bar{v}}_\lambda (p) $\end{minipage}} \\
{\begin{minipage}[b][19mm][c]{.1\linewidth}${{\rm outgoing} \atop
{\rm boson}} $\end{minipage}} &
\begin{fmffile}{cfextob}
\fmfframe(2,4)(2,4){
\begin{fmfgraph*}(15,15)
\fmfleft{i1} \fmfright{o1} \fmfblob{5}{i1}
\fmflabel{$\lambda$}{o1} \fmf{gluon,label=$k$}{i1,o1}
\end{fmfgraph*}}
\end{fmffile}
 & {\begin{minipage}[b][19mm][c]{.1\linewidth}$ \epsilon_\lambda^{*\mu} (k) $\end{minipage}}&\phantom{xx}&
{\begin{minipage}[b][19mm][c]{.1\linewidth}${{\rm incoming} \atop
{\rm boson}} $\end{minipage}} &
\begin{fmffile}{cfextib}
\fmfframe(2,4)(2,4){
\begin{fmfgraph*}(15,15)
\fmfleft{i1} \fmfright{o1} \fmfblob{5}{o1}
\fmflabel{$\lambda$}{i1} \fmf{gluon,label=$k$}{i1,o1}
\end{fmfgraph*}}
\end{fmffile}
 & {\begin{minipage}[b][19mm][c]{.1\linewidth}$ \epsilon_\lambda^\mu (k) $\end{minipage}}\\
{\begin{minipage}[b][19mm][c]{.1\linewidth}${{\rm gluon} \atop
{\rm propagator}} $\end{minipage}} &
\begin{fmffile}{cfprpglu}
\fmfframe(2,4)(2,4){
\begin{fmfgraph*}(15,15)
\fmfleft{i1} \fmfright{o1} \fmf{gluon,label=$k$}{i1,o1}
\fmflabel{$a\mu$}{i1} \fmflabel{$b\nu$}{o1} \fmfdot{i1,o1}
\end{fmfgraph*}}
\end{fmffile}
 & {\begin{minipage}[b][19mm][c]{.1\linewidth}${\delta_{ab}\over k^2} d_{\mu \nu} (k) $\end{minipage}}&\phantom{xx}&
{\begin{minipage}[b][19mm][c]{.1\linewidth}${{\rm ghost} \atop
{\rm propagator}} $\end{minipage}} &
\begin{fmffile}{cfprpgho}
\fmfframe(2,4)(2,4){
\begin{fmfgraph*}(15,15)
\fmfleft{i1} \fmfright{o1} \fmf{ghost,label=$k$}{o1,i1}
\fmflabel{$a$}{i1} \fmflabel{$b$}{o1}
 \fmfdot{i1,o1}
\end{fmfgraph*}}
\end{fmffile}
 & {\begin{minipage}[b][19mm][c]{.1\linewidth}$-{\delta_{ab}\over k^2} $\end{minipage}}\\
{\begin{minipage}[b][19mm][c]{.1\linewidth}${{\rm quark} \atop
{\rm propagator}} $\end{minipage}} &
\begin{fmffile}{cfprpqua}
\fmfframe(2,4)(2,4){
\begin{fmfgraph*}(15,15)
\fmfleft{i1} \fmfright{o1} \fmf{quark,label=$p$}{o1,i1}
\fmflabel{$i$}{i1} \fmflabel{$j$}{o1} \fmfdot{i1,o1}
\end{fmfgraph*}}
\end{fmffile}
 & {\begin{minipage}[b][19mm][c]{.1\linewidth}${\delta_{ij}\over {m-\notp}} $\end{minipage}}&\phantom{xx}&
{\begin{minipage}[b][19mm][c]{.1\linewidth}${{\rm 3-gluon} \atop
{\rm vertex}} $\end{minipage}} &
\begin{fmffile}{cfglue3}
\fmfframe(2,4)(2,4){
\begin{fmfgraph*}(15,15)
\fmfleft{i1}  \fmfright{o1} \fmftop{t1}
\fmf{gluon,label=$k_2$}{i1,v1} \fmf{gluon,label=$k_3$}{o1,v1}
\fmf{gluon,label=$k_1$,tension=1/2}{t1,v1}  \fmflabel{$a_2
\mu_2$}{i1} \fmflabel{$a_3 \mu_3$}{o1}  \fmflabel{$a_1
\mu_1$}{t1} \fmfdot{v1}
\end{fmfgraph*}}
\end{fmffile}
 & {\begin{minipage}[b][19mm][c]{.1\linewidth}${{-igf^{a_1a_2a_3}}\atop {V_{\mu_1 \mu_2 \mu_3} (k_1,k_2,k_3)}} $\end{minipage}}\\
{\begin{minipage}[b][19mm][c]{.1\linewidth}${{\rm 4-gluon} \atop
{\rm vertex}} $\end{minipage}} &
\begin{fmffile}{cfglue4}
\fmfframe(2,4)(2,4){
\begin{fmfgraph*}(15,15)
\fmfleft{i1} \fmflabel{$a_2 \mu_2$}{i1} \fmfright{o1}
\fmflabel{$a_4 \mu_4$}{o1} \fmftop{t1} \fmflabel{$a_1 \mu_1$}{t1}
\fmfbottom{b1} \fmflabel{$a_3 \mu_3$}{b1} \fmf{gluon}{i1,v1}
\fmf{gluon}{t1,v1} \fmf{gluon}{o1,v1} \fmf{gluon}{b1,v1}
\fmfdot{v1}
\end{fmfgraph*}}
\end{fmffile}
 & {\begin{minipage}[b][19mm][c]{.1\linewidth}$-g^2 W_{\mu_1 \mu_2 \mu_3 \mu_4}^{a_1 a_2 a_3 a_4} $\end{minipage}}&\phantom{xx}&
{\begin{minipage}[b][19mm][c]{.1\linewidth}${{\rm gluon-ghost}
\atop {\rm vertex}} $\end{minipage}}&
\begin{fmffile}{cfglugho}
\fmfframe(2,4)(2,4){
\begin{fmfgraph*}(15,15)
\fmfleft{i1} \fmfright{o1} \fmftop{t1}
\fmf{ghost,label=$k$}{v1,i1} \fmf{gluon,tension=0.5}{t1,v1}
\fmf{ghost}{o1,v1} \fmflabel{$c$}{o1} \fmflabel{$b$}{i1}
\fmflabel{$a \mu$}{t1} \fmfdot{v1}
\end{fmfgraph*}}
\end{fmffile}
 & {\begin{minipage}[b][19mm][c]{.1\linewidth}$-igf^{abc}k_\mu $\end{minipage}}\\
{\begin{minipage}[b][19mm][c]{.1\linewidth}${{\rm gluon-quark}
\atop {\rm vertex}} $\end{minipage}} &
\begin{fmffile}{cfgluqua}
\fmfframe(2,4)(2,4){
\begin{fmfgraph*}(15,15)
\fmfleft{i1}  \fmfright{o1} \fmftop{t1} \fmf{quark}{v1,i1}
\fmf{gluon,tension=0.5}{t1,v1} \fmf{quark}{o1,v1}
\fmflabel{$i$}{i1} \fmflabel{$j$}{o1} \fmflabel{$a \mu$}{t1}
\fmfdot{v1}
\end{fmfgraph*}}
\end{fmffile}
 & {\begin{minipage}[b][19mm][c]{.1\linewidth}$g\gamma_\mu T_{ij}^a $\end{minipage}}
&\phantom{xx}&\phantom{xx} & \phantom{xx} & \phantom{xx} \\
\end{array}
$$
\normalsize \hoffset=0.0\linewidth \caption{Feynman Rules for
QCD.} \label{fig:feynrule} \EFIG

The functions $d_{\mu\nu}$, $V_{\mu_1 \mu_2 \mu_3}$, and
$W_{\mu_1 \mu_2 \mu_3 \mu_4}^{a_1 a_2 a_3 a_4}$ are given below:

\BQA d_{\mu\nu} (k) &\equiv& g_{\mu\nu} - (1-\alpha) {{k_\mu
k_\nu}\over k^2} \; \cr V_{\mu_1 \mu_2 \mu_3} (k_1,k_2,k_3)
&\equiv& {(k_1-k_2)}_{\mu_3} g_{\mu_1 \mu_2} + {(k_2-k_3)}_{\mu_1}
g_{\mu_2 \mu_3} + {(k_3-k_1)}_{\mu_2} g_{\mu_3 \mu_1} \; \cr
W_{\mu_1 \mu_2 \mu_3 \mu_4}^{a_1 a_2 a_3 a_4} &\equiv& (f^{13,24}
- f^{14,32}) g_{\mu_1 \mu_2} g_{\mu_3 \mu_4} + (f^{12,34} -
f^{14,23}) g_{\mu_1 \mu_3} g_{\mu_2 \mu_4} \; \cr &+& (f^{13,42} -
f^{12,34}) g_{\mu_1 \mu_4} g_{\mu_3 \mu_2} \;, \EQA

\noindent in which $f^{ij,kl}$ denotes the combination

\BQN f^{ij,kl} \equiv f^{a_i a_j a} f^{a_k a_l a} \; \EQN

\noindent and the Minkowski metric (in $D$ dimensions) is taken to
be

\BQN g_{\mu \nu} \equiv \delta_{\mu 0} \delta_{\nu 0} -
\sum_{i=1}^{D-1} \delta_{\mu i} \delta_{\nu i} \;. \EQN

\noindent There are additional requirements when loops arise in
diagrams. First of all, a ghost loop must be included whenever a
closed fermion loop appears. Secondly, for all loops, the loop
momentum $k$ must be integrated over:

$$ -i \int {{d^Dk}\over {(2\pi)}^D} \;,$$

\noindent with fermion and ghost loops carrying an extra minus
sign. Lastly, additional symmetry factors must be included for the
types of diagrams shown in figure \ref{fig:symfac}.

\BFIG \hoffset=-0.1\linewidth \tiny
$$
\begin{array}{ccccc}
\begin{fmffile}{cfsymm1}
\fmfframe(2,4)(2,4){
\begin{fmfgraph*}(30,30) \fmfset{curly_len}{2mm}
\fmfleft{i} \fmfright{o} \fmf{gluon}{i,v1} \fmf{gluon}{v2,o}
\fmf{gluon,left,tension=0.5}{v1,v2,v1}
\end{fmfgraph*}}
\end{fmffile}
& {\begin{minipage}[b][34mm][c]{.1\linewidth}${1\over {2!}}
$\end{minipage}} &\phantom{xx}&
\begin{fmffile}{cfsymm2}
\fmfframe(2,4)(2,4){
\begin{fmfgraph*}(30,30)  \fmfset{curly_len}{2mm}
\fmfleft{i} \fmfright{o} \fmf{gluon}{i,v} \fmf{gluon}{v,o}
\fmf{gluon,right,tension=0.5}{v,v} \fmfdot{v}
\end{fmfgraph*}}
\end{fmffile}
&{\begin{minipage}[b][34mm][c]{.1\linewidth}$ {1\over {2!}}$\end{minipage}}\\
\begin{fmffile}{cfsymm3}
\fmfframe(2,4)(2,4){
\begin{fmfgraph*}(30,30) \fmfset{curly_len}{2mm}
\fmfleft{i} \fmfright{o} \fmf{gluon}{i,v1,v2,o} \fmfdot{v1}
\fmfdot{v2} \fmffreeze \fmf{gluon,left,tension=0}{v1,v2,v1}
\end{fmfgraph*}}
\end{fmffile}
& {\begin{minipage}[b][34mm][c]{.1\linewidth}${1\over {3!}}
$\end{minipage}} &\phantom{xx}&
\begin{fmffile}{cfsymm4}
\fmfframe(2,4)(2,4){
\begin{fmfgraph*}(30,30) \fmfset{curly_len}{2mm}
\fmfleft{i} \fmfright{o1,o2} \fmf{gluon}{i,v1} \fmfdot{v1}
\fmfdot{v2} \fmf{gluon,left,tension=0.5}{v1,v2,v1}
\fmf{gluon}{o1,v2,o2}
\end{fmfgraph*}}
\end{fmffile}
&{\begin{minipage}[b][34mm][c]{.1\linewidth}$ {1\over {2!}}$\end{minipage}}\\
\end{array}
$$
\normalsize \hoffset=0.0\linewidth \caption{Symmetry factors.}
\label{fig:symfac} \EFIG

\section{Renormalization}
\label{sec:Renorm}

\subsection{Motivation}

In the calculation of the amplitude corresponding to a particular
Feynman diagram containing internal ({\it virtual}) particles,
there will necessarily be unobserved degrees of freedom which
must be integrated over. For example, a diagram with two external
legs and one loop will have two characteristic momenta, one
($p^\mu$) for the external legs and one ($k^\nu$) for an internal
leg (or {\it propagator}) of the loop, the other loop propagator
momentum being fixed by momentum conservation. The loop momentum
$k^\nu$, being unobserved, must be integrated over, but by the
uncertainty principle, it can be arbitrarily large as long as the
``size'' of the loop is correspondingly small. Thus the potential
arises for the integral to diverge. Each propagator will
contribute a power of $1$ or $2$ to the denominator of the
integrand, thus lessening the degree of divergence, but if the
dimension of the numerator is greater than that of the
denominator, we will still have a problem. By the same token, if
the rest mass of the loop particle is zero, there will be an
infrared divergence at the low end of the integral, $k^\nu k_\nu
= 0$.

As noted in Section \ref{sec:intropert}, there are divergences
even in purely tree-level graphs, and {\it regularization} is the
general term used to describe the process by which divergences are
parametrized and the amplitudes written as explicit functions of
these parameters. Being arbitrary, such parameters must
ultimately disappear in the prediction of any physical
observable, but in the mathematical calculation thereof, we can
temporarily excuse their presence. One main class of
regularization scheme is the {\it cutoff} method, in which
ultraviolet integrals are stopped at some high scale or infrared
divergences are avoided by introducing small masses. However,
such techniques tend to destroy either the Lorentz invariance or
the internal symmetries of the Lagrangian, and so are rarely
used. More common is the method of {\it dimensional
regularization}, in which integration over what would be a
4-dimensional momentum variable is analytically continued to a
space of arbitrary dimension $D$, usually parametrized as
$D=4-2\epsilon$. The integral is now formally finite for
$\epsilon > 0$, and the amplitude can be expressed as a function
of $\epsilon$. In the end, ultraviolet poles in the
regularization parameter $\epsilon$ must either cancel or be
absorbed into redefinitions of the observable masses and charges.
This latter procedure, regardless of the means of regularization,
is given the name {\it renormalization}.

The Lagrangian of a theory will be a function of the fields and
``bare'' parameters, such as masses $m_0$ and couplings $e_0$.
Derivation of the Feynman rules from such a Lagrangian will
result in a set of ``bare'' Green functions, the simple
propagators and vertices of the theory. Usually, the Lagrangian
is written as a sum of terms, each of which corresponds to one of
these functions. More complicated diagrams are built up from
these, and can be grouped into two classes -- those which have the
same external legs as the bare diagrams and those that do not. Of
the ones that do, some are single-particle-irreducible
(simply-connected diagrams which have external propagators
removed and which cannot be split into two by cutting one
internal propagator). We denote these by the term ``SPI'', and
for vertices, this is all we have in this sub-class. For
propagators there will be other diagrams, but these can be shown
to be just combinations of the SPI diagrams. We'll now define
``full'' propagators and vertices as sums of all diagrams with the
same external legs as their corresponding bare diagrams. For
vertices, this will then be the sum of all SPI vertex diagrams,
while for propagators, we have to first sum the SPI propagators,
and then link these sums in a geometric chain (see figure
\ref{fig:SPIchain}) ~\cite{PS95}. We now have a set of full Green
functions, each in one-to-one correspondence with its bare
counterpart. The remaining class of diagrams, those which don't
have the same external legs as the bare versions, were still
constructed from them, and so can be calculated now via the full
Green functions, expanded to the appropriate order of the
coupling.

\BFIG

\begin{eqnarray}
-i\Sigma (p,m_0) &\equiv& \parbox{20mm}{
\begin{fmffile}{cfspi1} \begin{fmfchar*}(20,15)
 \fmfleft{l} \fmfright{r} \fmf{quark,tension=1/3}{r,v,l} \fmffreeze
\fmfv{d.sh=circle,d.f=empty,d.si=.4w,l=SPI,l.d=-.15w}{v}
\end{fmfchar*} \end{fmffile} } \; \cr
&=&
\parbox{20mm}{\begin{fmffile}{cfspi2} \begin{fmfchar*}(20,15)
 \fmfleft{l} \fmfright{r} \fmf{plain}{r,v1,v2,l} \fmffreeze
 \fmf{photon,left,tension=0}{v1,v2}
\end{fmfchar*} \end{fmffile}}  + \quad
\parbox{20mm}{\begin{fmffile}{cfspi3} \begin{fmfchar*}(20,15)
 \fmfleft{l} \fmfright{r} \fmf{plain}{r,v1,v2,v3,v4,l} \fmffreeze
 \fmf{photon,left,tension=0}{v1,v3} \fmf{photon,right,tension=0}{v2,v4}
\end{fmfchar*} \end{fmffile}} + \quad
\parbox{20mm}{\begin{fmffile}{cfspi4} \begin{fmfchar*}(20,15)
 \fmfleft{l} \fmfright{r} \fmftop{t}
 \fmf{plain,tension=1/4}{r,v1,v2,l} \fmffreeze
 \fmf{photon,right,tension=0}{v1,t,v2} \fmfv{d.sh=circle,d.f=empty,d.si=.4w}{t}
\end{fmfchar*} \end{fmffile}}  + \quad ... \; \nonumber \end{eqnarray}

\begin{eqnarray}
-i\bar{\Sigma} (p,m) &\equiv&
\parbox{20mm}{\begin{fmffile}{cfspi5} \begin{fmfchar*}(20,15)
 \fmfleft{l} \fmfright{r} \fmf{quark}{r,v,l}
\fmfv{d.sh=circle,d.f=shaded,d.si=.4w}{v}
\end{fmfchar*} \end{fmffile} }\; \cr
&=&
\parbox{20mm}{\begin{fmffile}{cfspi6} \begin{fmfchar*}(20,15)
 \fmfleft{l} \fmfright{r} \fmf{quark}{r,l}
\end{fmfchar*} \end{fmffile}}  + \quad
\parbox{20mm}{\begin{fmffile}{cfspi7} \begin{fmfchar*}(20,15)
 \fmfleft{l} \fmfright{r} \fmf{quark,tension=1/3}{r,v,l}
\fmfv{d.sh=circle,d.f=empty,d.si=.4w,l=SPI,l.d=-.15w}{v}
\end{fmfchar*} \end{fmffile} } + \quad
\parbox{20mm}{\begin{fmffile}{cfspi8} \begin{fmfchar*}(20,20)
 \fmfleft{l} \fmfright{r} \fmf{quark,tension=1/4}{r,v1,v2,v3,l}
\fmfv{d.sh=circle,d.f=empty,d.si=.4w,l=SPI,l.d=-.2w}{v1}
\fmfv{d.sh=circle,d.f=empty,d.si=.4w,l=SPI,l.d=-.15w}{v3}
\end{fmfchar*} \end{fmffile} } + \quad ... \; \cr
&=& {i\over {\notp -m_0}} + {i\over {\notp -m_0}} {\Bigl(
{{\Sigma (\notp)}\over {\notp - m_0}} \Bigr)} + {i\over {\notp
-m_0}} {\Bigl( {{\Sigma (\notp)}\over {\notp - m_0}} \Bigr)}^2 +
\quad ... \; \cr &=& {i\over {\notp -m_0-\Sigma (\notp)}} \;
\nonumber
\end{eqnarray}

\caption[Building the full propagator.] {Building the full
propagator.} \label{fig:SPIchain} \EFIG

In general, these Green functions will be UV-divergent, as they
contain ``clouds'' of virtual activity. However, in a
renormalizable theory, to each simple propagator or vertex there
corresponds a characteristic quantity, potentially observable in
the ``full'' case, unobservable in the ``bare''. For propagators
this is the mass; for vertices it is the coupling. The
correspondence between the full and bare Green functions gives us
a means of dealing with these divergences, as well as a physical
picture of how higher orders of virtual activity ``shift'' the
masses and couplings from their bare values. The full Green
functions emerge as divergent functions of the bare parameters,
but if we define {\bf observable} quantities such that the bare
parameters are themselves divergent functions of the observables,
then the Green functions can be rendered as finite functions of
the observables.

\subsection{Canonical Procedure}

The exact process of renormalization depends on the theory under
consideration, but (without getting into these details) one can
describe, in a general way, the steps involved. We use one mass,
coupling, and regulator for simplicity.

1. Compute the full functions $\Gamma_i$ to a given order of the
bare coupling $e_0$, using a regulator $\Lambda$ when necessary.
One then has the functions $\Gamma_i (e_0, m_0, \Lambda,
\{p_j\})$ to order $N$, and these are divergent as the regulator
approaches some limit $\Lambda \rightarrow \Lambda_{lim}$. Here
$\{p_j\}$ are the momenta of external legs.

2. Impose {\it renormalization conditions} to define the form of
the renormalized Green functions at a particular scale $\mu =
f(\{p_j\})$. These forms should contain a dependence on the
renormalized couplings and masses $\{e,m\}$ and should be finite
as $\Lambda \rightarrow \Lambda_{lim}$. Suitable forms are
usually suggested by the bare Green functions themselves, {\it
e.g.} the full propagator should be able to be brought into the
same form as the bare propagator, the new, observable mass (now a
complicated function of the bare parameters) taking the place of
the bare mass (see figure \ref{fig:SPIchain}). Imposition of such
constraints on the Green functions $\Gamma_i$ results in
relations $e(e_0,m_0,\Lambda)$ and $m(e_0,m_0,\Lambda)$ between
the renormalized and bare quantities.

3. Solve for the bare parameters $\{e_0,m_0\}$ and substitute
into the functions $\Gamma_i$. The result is $\Gamma_i
(e,m,\Lambda)$, which will be finite as $\Lambda \rightarrow
\Lambda_{lim}$.

We'll call this the {\bf canonical} renormalization procedure,
and it will always work in a renormalizable theory, yet is
cumbersome at high orders $N$. An equivalent but better organized
approach to the bookkeeping is provided by the {\bf counterterm}
method, which anticipates the shift in parameter space by building
it into the Lagrangian at the start.

\subsection{Counterterm Method}

1. Rescale the bare parameters (including the field strengths
$\psi_0$) by introducing new, temporary unknowns $Z_i$, which
relate them to their observable counterparts; i.e. $\psi=Z_\psi
\psi_0$, $e=Z_e e_0$, etc.

2. Trade in these multiplicative unknowns $Z_i$ for additive ones
$\delta_i$ in such a way that, upon substitution of these
expressions for the bare parameters into the Lagrangian, one
reproduces the original Lagrangian (now in terms of the
renormalized observables), plus additional {\it counterterms}.
We'll refer to these pieces as ${\cal L}_{CG}$ and ${\cal
L}_{CT}$, respectively. The counterterms will have corresponding
Feynman rules of their own.

3. Specify the renormalization conditions, again as desired
relations of form between the unrenormalized and renormalized
Green functions at scale $\mu$.

4. Calculate the full functions using the rearranged Lagrangian.
To a given order $N$, one will find the usual divergences from
${\cal L}_{CG}$, plus new divergences from the counterterms.
However, we can now get these to cancel, as the set of
unobservable shifts $\delta_i$ are free parameters. By adjusting
these to maintain the renormalization conditions (which by
construction define nonsingular amplitudes), we can consistently
absorb the infinities of the theory into the bare parameters,
leaving Green functions that are finite and dependent upon only
the renormalized charges and masses.

The counterterms in ${\cal L}_{CT}$, with suitable
renormalization conditions, are commonly said to ``subtract off"
the divergences of the ``original'' Lagrangian ${\cal L}_{CG}$,
but there remains an ambiguity in this process, as one can always
subtract an extra, arbitrary, finite amount as well. The size of
this piece is, in general, related to the particular
renormalization conditions, and so together these two choices
define a {\it renormalization scheme}. We will discuss the most
commonly used schemes in a later section.

As an example, we take the renormalization of QED. The bare
Lagrangian is as follows (we denote bare quantities by the
subscript ``$0$''):

\begin{equation}
{\cal L} = -{1\over 4} {(F_0^{\mu \nu})}^2 + \bar{\psi_0} (i
\notpartial -m_0) \psi -e_0 \bar{\psi_0}\gamma_\mu \psi_0 A_0^\mu
\;. \label{REN1}
\end{equation}

\noindent We now impose the following sets of rescalings and
variable changes:

\begin{eqnarray}
\psi_0 &=& Z_2^{1/2} \psi \cr A_0^\mu &=& Z_3^{1/2} A^\mu \cr e_0
Z_2 Z_3^{1/2} &=& e Z_1 \; \label{REN2}
\end{eqnarray}

\begin{eqnarray}
\delta_m &=& Z_2 m_0 - m \cr \delta_1 &=& Z_1 -1 \cr \delta_2 &=&
Z_2 -1 \cr \delta_3 &=& Z_3 -1 \;, \label{REN3}
\end{eqnarray}

\noindent and thus transform the Lagrangian into a sum of two
pieces, one of which resembles the original Lagrangian, the other
provides the counterterms. The bare quantities and intermediate
scaling factors disappear:

\begin{eqnarray}
{\cal L} &=& {\cal L}_{CG} + {\cal L}_{CT} \cr {\cal L}_{CG} &=&
-{1\over 4} {(F^{\mu \nu})}^2 + \bar{\psi} (i \notpartial -m)
\psi -e \bar{\psi}\gamma_\mu \psi A^\mu \cr {\cal L}_{CT} &=&
-{1\over 4} \delta_3 {(F^{\mu \nu})}^2 + \bar{\psi} (i \delta_2
\notpartial - \delta_m) \psi -e \delta_1 \bar{\psi}\gamma_\mu \psi
A^\mu \;. \label{REN4}
\end{eqnarray}

\noindent For renormalization conditions, we use the following:

\begin{eqnarray}
\Sigma (\notp = m) &=& 0 \; \cr {{d\Sigma (\notp)}\over {d\notp}}
{\biggr|}_{\notp = m} &=& 0 \; \cr \Pi (q^2 = 0) &=& 0 \; \cr
\Gamma^\mu (\bar{p}=p) &=& \gamma^\mu \;, \label{REN5}
\end{eqnarray}

\noindent where the amplitudes $\Gamma^\mu$, $\Sigma$, and
$\Pi^{\mu \nu} \equiv (g^{\mu \nu} q^2 - q^\mu q^\nu) \Pi$ denote
the amputated full vertex, single-particle-irreducible (SPI)
electron propagator element, and SPI photon propagator element,
respectively. When calculated using the rearranged Lagrangian,
which includes counterterms, each will, in general, be a function
of the quantities $\{ e, m, \delta_1, \delta_2, \delta_3,
\delta_m, \notp, q \}$, and the regulator. Enforcing the above
conditions then determines $\{ \delta_1, \delta_2, \delta_3,
\delta_m \}$.

As we have said, the precise conditions used are a matter of
convention; the crucial point is that we define the full Green
functions (and thus the renormalized masses and couplings) to (1)
be finite, and (2) to take a certain form at some {\it
renormalization scale} $\mu$ or set of scales. In the QED example
above, we used a {\it momentum subtraction scheme}, in which the
scale used depends on the Green function being renormalized. We
could, however, have chosen a single scale, common to all
amplitudes, {\it i.e.} $p^2=-\mu^2$ and $q^2=-\mu^2$. The
counterterms then subtract only the divergent poles ({\it Minimal
Subtraction}, or ``$MS$"), or these plus standard constants {\it
Modified Minimal Subtraction}, ``$\overline{MS}$").  In this
dissertation we will use the {\bf $\overline{MS}$} convention.

Either way, we are left with renormalized Green's functions $G$
and couplings $\lambda$ which, though finite, seem to depend upon
a new parameter ($\mu$) not present in the original, bare versions
($G_0$, $\lambda_0$). We know that these must be related by the
(now calculable) rescalings $Z_i$; for example, there must be a
calculable function $Z_G$ of the scaling factors $Z_i$ such that

\begin{equation}
G (\mu, \lambda) = Z_G(\mu, \lambda) G_0 (\lambda) \;.
\label{REN6}
\end{equation}

\noindent However, we have calculated this relation to a fixed
order $N$ only. How can we generalize this to account for not
only the poles but the associated scale dependent pieces $\ln^N
\mu/p$ found at even higher orders? Specifically, if $\mu$ is
taken large enough, how do we justify a perturbative expansion,
unless the coupling $\sim \lambda^{2N}$ becomes correspondingly
small? There must be a {\it Renormalization Group Equation} (RGE)
which expresses the fact that a change in scale $\mu$ needs to be
compensated in order to keep the bare Green's function (and the
bare coupling) invariant. From equation \ref{REN6}:

\BQN \mu {d\over {d\mu}} {G \over Z_G} = 0 \;, \EQN

\noindent or

\BQN \Bigl[\mu {\partial \over {\partial \mu}} + \beta {\partial
\over {\partial \lambda}} + \gamma \Bigr] G (\mu, \lambda) = 0
\;, \label{REN7} \EQN

\noindent with

\begin{eqnarray}
\beta (\lambda) &\equiv& \mu {{d\lambda}\over {d\mu}} \; \label{REN8a} \\
\gamma (\lambda) &\equiv& -{\mu \over Z_G} {{dZ_G}\over {d\mu}}
\;. \label{REN8b}
\end{eqnarray}

Equation \ref{REN7} is known as the Callan-Symanzik equation
~\cite{Cal70,Sym70}. The functions $\beta$ and $\gamma$ are
unitless and must depend on the coupling only. Yet they describe
the compensating shifts in coupling and field strengths
(respectively) for a change in scale $\mu$. So these strengths
must themselves be functions of $\mu$, and equations \ref{REN8a},
\ref{REN8b} lead us to the perturbative form of this dependence.
Here we simplified to a theory involving one field and one
coupling; in general, there will be one $\gamma$ for each field
and one $\beta$ for each coupling.

\subsection{Application to QCD}

Now that we've described the basics of renormalization, let's
list the relations appropriate to QCD. As expected, they are
somewhat more complicated. We'll use a superscript ``$0$'' to
denote the bare quantities.

\begin{eqnarray}
g &=&  \mu^{-\epsilon} {g^0 \over \sqrt{Z_\alpha}} \; \cr m_j &=&
{m_j^0\over Z_m} \; \cr \zeta &=& {\zeta^0 \over Z_\zeta} \; \cr
 \psi_{q,i}(x) &=&  {\mu_r^\epsilon \over \sqrt{Z_{2F}}}
\psi_{q,i}^0 (x) \; \cr A_\mu^a (x) &=&  {\mu_r^\epsilon \over
\sqrt{Z_{3YM}}} A_\mu^{a0} (x) \; \cr
 \phi_a (x) &=&  {\mu_r^\epsilon \over \sqrt{\tilde{Z}_3}}
\phi_a^0 (x) \; \cr {(g \bar{\psi} A \psi)}(x) &=&
{\mu_r^{2\epsilon} \over Z_{1F}}
 {(g \bar{\psi} A \psi)}^0 (x) \;. \label{REN9}
\end{eqnarray}

\noindent The {\it running coupling} $\alpha_s \equiv g^2/{4\pi}$
is a solution of the Callan-Symanzik equation \ref{REN8a}, and
comes out (to two loops) as

\begin{equation}
\alpha_s (\mu) = {{2\pi}\over {b \ln (\mu^2/\Lambda^2)}} \Bigl[ 1
- {{2c}\over b} {{\ln [\ln (\mu^2/\Lambda^2)]}\over {\ln
(\mu^2/\Lambda^2)}} + O[\ln^{-2} (\mu^2/\Lambda^2)] \Bigr] \;,
\label{REN10}
\end{equation}

\noindent in which

\begin{eqnarray}
b &\equiv& {{11 {\rm N_C} -2 {\rm N_F}}\over 6} \; \cr c &\equiv&
{1\over {8b}} \biggl[ {34 \over 3} {\rm N_C}^2 - {13\over 3} {\rm
N_C} {\rm N_F} + {{\rm N_F}\over {\rm N_C}} \biggr] \;.
\label{REN11}
\end{eqnarray}

\noindent As expected, this exhibits the property of asymptotic
freedom -- at large $\mu$, the coupling $\alpha_s$ decreases.

\section{Factorization}
\label{sec:factorize}

With a renormalized theory, one can begin to calculate physical
quantities and compare these predictions with experimental
measurements. For a particular set of initial-state (incoming)
reactants $\{a,b\}$ and final-state products $\{c,d\}$ ({\it
e.g.}), one uses the Feynman rules to calculate an amplitude
${\cal M}_i$ for each topologically unique way this reaction can
occur. Quantum mechanics then dictates that the probability for
the reaction is proportional to the squared sum of these
amplitudes:

\BQN \sum_i {|{\cal M}_i|}^2 = \sum_i {\cal M}_i^* {\cal M}_i \;.
\EQN

In addition, the squared amplitude refers to (and is a function
of) particular values of the particle momenta $\{p_a^\mu, p_b^\mu,
p_c^\mu, p_d^\mu \}$. That is, the reaction rate per unit time
into volume element $d^3p_c d^3p_d$ is given by

$$ \sum_i {|{\cal M}_i|}^2 {{d^3p_c}\over {{(2\pi)}^3
2p_c^0}} {{d^3p_d}\over {{(2\pi)}^3 2p_d^0}} {(2\pi)}^4 \delta^4
(p_a^\mu + p_b^\mu - p_c^\mu - p_d^\mu) $$

\noindent and one must integrate over this {\it phase space} in
order to obtain the full reaction rate. Note that a delta
function is included to conserve momentum.

In the case of hadronic interactions, the initial state particles
will be partons, one from each hadron. By the collinear geometry
of the interaction, the parton momenta are known up to fractions
$x_a$ and $x_b$ of the incoming hadron momenta, and the {\it cross
section} (an experimentally measurable quantity) for two partons,
of flavors $a$ and $b$, with these momenta, to produce the given
final state, is found by dividing the above reaction rate by a
flux factor $2w(\hat{s},m_a,m_b)$, in which $\hat{s} \equiv
{(p_a^\mu+p_b^\mu)}^2$ and

\BQN w(x,y,z) \equiv \sqrt{x^2 + y^2 + z^2 -2xy-2xz-2yz} \;. \EQN

For all but the heaviest quarks, we can take the parton masses to
be zero (relative to the hadronic center of mass energy $S$), and
$w(\hat{s},m_a,m_b) \rightarrow \hat{s} = x_ax_bS$. The {\it
partonic} cross section for a particular $2\rightarrow 2$ particle
subprocess is thus given by the product

\BQN d\hat{\sigma} = {{\sum_i {|{\cal M}_i|}^2}\over {2\hat{s}}}
d^4 \Gamma_2 \;, \EQN

\noindent in which the 2-body phase space factor is

\BQN d^4 \Gamma_2 = {{d^3p_c}\over {{(2\pi)}^3 2p_c^0}}
{{d^3p_d}\over {{(2\pi)}^3 2p_d^0}} {(2\pi)}^4 \delta^4 (p_a^\mu
+ p_b^\mu - p_c^\mu - p_d^\mu) \;. \EQN

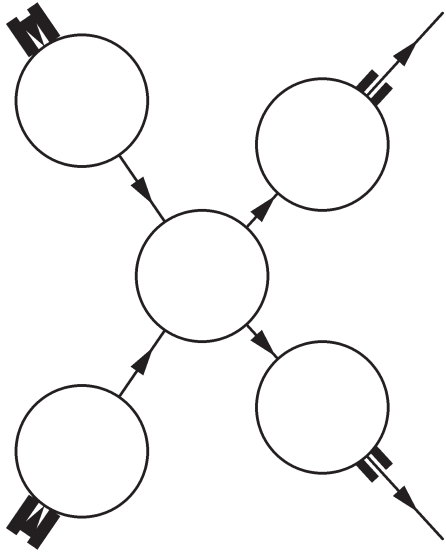
\begin{figure} \centering
\begin{fmffile}{cffactit} \fmfframe(5,5)(5,5){
\begin{fmfchar*}(70,70)   \fmfleft{i1,i2}
\fmfright{o1,o2} \fmf{fermion}{i1,v1} \fmf{fermion}{i2,v2}
\fmf{fermion}{v3,o3} \fmf{fermion}{v4,o4} \fmf{phantom}{o3,o1}
\fmf{phantom}{o4,o2}
\fmf{fermion,label=$x_bP_B$,l.s=left,tension=1/2}{v1,c}
\fmf{fermion,label=$x_aP_A$,l.s=right,tension=1/2}{v2,c}
\fmf{fermion,label=$p_c$,l.s=right,tension=1/2}{c,v4}
\fmf{fermion,label=$p_d$,l.s=left,tension=1/2}{c,v3}
\fmflabel{$P_A$}{i2} \fmflabel{$P_B$}{i1}
\fmflabel{$P_C=z_cp_c$}{o2} \fmflabel{$P_D=z_dp_d$}{o1}
\fmfv{d.sh=circle,d.f=empty,d.si=.25w,label=$d\hat{\sigma}$,l.a=45,l.d=.00w}{c}
\fmfv{d.sh=circle,d.f=empty,d.si=.25w,label=$q_{a/A}(x_a)$,l.a=135,l.d=.0w}{v2}
\fmfv{d.sh=circle,d.f=empty,d.si=.25w,label=$q_{b/B}(x_b)$,l.a=-135,l.d=.0w}{v1}
\fmfv{d.sh=circle,d.f=empty,d.si=.25w,label=$D_{C/c}(z_c)$,l.a=45,l.d=.0w}{v4}
\fmfv{d.sh=circle,d.f=empty,d.si=.25w,label=$D_{D/d}(z_d)$,l.a=-45,l.d=.0w}{v3}
\fmffreeze \fmf{dbl_plain_arrow,width=40,tension=0}{i1,v1}
\fmf{dbl_plain_arrow,width=40,tension=0}{i2,v2}
\fmf{dbl_plain,width=30,tension=0}{v3,o3}
\fmf{dbl_plain,width=30,tension=0}{v4,o4} \fmf{fermion}{v3,o1}
\fmf{fermion}{v4,o2}
\end{fmfchar*}}
\end{fmffile}
\caption[Factorization of the hadronic interaction.]
{Factorization of the hadronic interaction.} \label{fig:factor}
\end{figure}

The partonic cross section is perturbatively calculable to any
order $N$ if one is willing to calculate the required amplitudes.
However, it is not yet useful in practice. First, the above cross
section pertains to partons of definite flavor, color, spin, and
momentum fraction. In any real hadronic interaction, a multitude
of partonic flavors will contribute, at all momenta and with all
possible color and spin values. Spin and color averages are
abbreviated by a line over the squared amplitude sum, e.g.
$\bar{\sum}$. However, we do not {\it apriori} know the momentum
or flavor distribution of partons inside hadrons, nor do we have a
deterministic model for the hadronization of final state partons.
We are forced to find a way of {\it factorizing} the reaction
into component steps, some of which are dealt with
perturbatively, the rest requiring effective parametrizations and
experimental input (see figure \ref{fig:factor}).

In the {\it parton model} of high-energy hadronic collisions, the
initial reaction takes place over such a small time scale that it
becomes plausible to treat each hadron as a collection of free
partons, each travelling in the same direction as the parent
hadron, and only one of which takes place in any given ``hard
scattering''. That is, one may calculate basic subprocesses
involving one parton from each hadron, then sum incoherently over
the contributions of all partons. Similarly, the process of
hadronization is assumed to take place over a time scale much
longer than the hard scattering, and can therefore also be
treated independently.

Thus, we can assign to each initial state hadron a set of parton
distribution functions $q_{i/I}$ \footnote{Parton distributions
are variously denoted in the literature by $q_f$ or $G_f$ (where
$f$ is the flavor), or $f_q$ (where $q$ is the flavor).}, and, if
we are looking at particular hadrons in the final state, to each
final-state parton a given probability $D_{H/h}$ to fragment into
a state containing the desired hadron. The physically realizable
cross section for the reaction is then obtained by weighting the
subprocess cross section $d\hat{\sigma}$ with these distribution
and fragmentation functions, evaluated for particular flavors and
at particular fractions of the parent (or daughter) hadrons'
momenta, then summing over all flavors and momentum fractions:

\BQN d\sigma^{(AB \rightarrow CD)} = \sum_{a,b,c,d} \int_0^1 dx_a
\int_0^1 dx_b q_{a/A} (x_a) q_{b/B} (x_b) d{\hat{\sigma}}^{(ab
\rightarrow cd)} D_{c/C} (z_c) D_{d/D} (z_d) \;. \EQN

\noindent If suitable parametrizations of the distribution and
fragmentation functions are made at the outset, these quantities
may be fit to known cross section data and used in subsequent
calculations.

In the process of calculating $d\hat{\sigma}$ to higher orders,
two main classes of diagrams arise. First there are vertex and
propagator modifications, in which extra connections are made
between the legs of a diagram, without changing the number of
external legs. These produce {\it ultraviolet} (UV) and {\it
infrared} (IR) singularities (or ``poles'') at the upper and
lower limits of loop momentum, and in the previous section we
have seen that the UV poles may be regulated and absorbed into
the masses and couplings of the theory. The second class of
diagrams involve the emission of an extra particle into the final
state, and also give rise to IR singularities. Some of these
arise in gauge boson emission at zero boson momentum, and cancel
the IR poles from virtual contributions ~\cite{BN37}. As well,
there are {\it collinear} singularities at zero angle between the
daughter and parent particle.

The collinear singularities, arising as they do from the
inability to distinguish a bare parton from one accompanied by a
number of collinear partons, get moved into a redefinition of the
distribution and fragmentation functions, usually accompanied by
logarithms of a momentum scale characteristic of the subprocess.
The evolution of these quantities is now dependent upon the
change in momentum transfer at which they are sampled, a result
quantified for distribution functions by Altarelli and Parisi
~\cite{AP77}, and for the fragmentation functions by Owens and
Uematsu ~\cite{O78a}.

At order $N$ relative to the leading order, there can be as many
as  $N$ extra radiated partons, any or all of which may be
collinear. We thus find a power series of logarithmic divergences
at the NLO and higher orders. In the {\it leading log
approximation}, only the logarithm of highest power at each order
is retained. Iteration of the evolution equations effectively
sums the collinear logarithms to solve for the distribution and
fragmentation functions, with the original momentum of
parametrization as reference point. The leading log approximation
dictates that we can now simply use the modified coupling,
distribution, and fragmentation functions with the Born matrix
element. A number of calculations have been carried out in this
way ~\cite{CKR77,ORG78,CO80}.

To show how singularity cancellation and factorization works,
we'll look at a simple example, that of double photon production.
We won't be looking for particular hadrons in the final state, so
no fragmentation functions will enter into our calculation.

The ``Born' (leading order) subprocess cross section is calculated
using the diagrams shown in figure \ref{fig:qqbvv}. All
contributions have already been regulated using dimensional
regulation, with $\epsilon=(4-D)/2$. To order $\alpha^2$, we have

\begin{equation}
d\sigma_0 = {\sum_f \over {2S}} \int_0^1 {{dx_a}\over x_a}
\int_0^1 {{dx_b}\over x_b} H(x_a,x_b) d^D \Gamma_2 \bar{\Sigma}
{|{\cal M}|}^2_0 \;,
\end{equation}

\noindent where the luminosity function $H$ (which contains the
parton distributions $q_f$), the squared, summed and averaged
matrix element $\bar{\Sigma} {|M|}^2_0$, and the 2-body phase
space factor $d^D \Gamma_2$ are given as follows:

\begin{eqnarray}
H(x_a,x_b) &=&   q_{f/A} (x_a) q_{\bar{f}/B} (x_b)
 + (f \rightarrow \bar{f})  \; \\
\bar{\Sigma} {|{\cal M}|}^2_0 &=& {1\over 2} {1\over 3^2} {1\over
2^2} 8N_C {(4\pi)}^2 \mu^{4\epsilon} \alpha^2 Q_f^4 (1-\epsilon)
\Bigl[ (1-\epsilon) \biggl({\hat{t}\over \hat{u}} + {\hat{u}\over
\hat{t}} \biggr) -2\epsilon \Bigr] \; \\
d^D \Gamma_2 &=& {1\over {8\pi}} {\biggl( {{4\pi}\over \hat{s}}
\biggr)}^\epsilon {{[v(1-v)]}^{-\epsilon} \over {\Gamma
(1-\epsilon)}} dv \;.
\end{eqnarray}

\BFIG $$
\begin{array}{cc}

\begin{fmffile}{cf2gam01}
\fmfframe(10,10)(10,10){
\begin{fmfchar*}(20,20) \fmfset{curly_len}{2.0mm} \fmfleft{i1,i2}
\fmfright{o1,o2} \fmftop{t} \fmfbottom{b} \fmf{quark}{i2,t,b,i1}
\fmf{photon}{t,o2} \fmf{photon}{b,o1} \fmflabel{$p_1$}{i2}
\fmflabel{$p_2$}{i1} \fmflabel{$p_3$}{o2} \fmflabel{$p_4$}{o1}
\end{fmfchar*}
}\end{fmffile}

&

\begin{fmffile}{cf2gam02}
\fmfframe(10,10)(10,10){
\begin{fmfchar*}(20,20) \fmfset{curly_len}{2.0mm} \fmfleft{i1,i2}
\fmfright{o1,o2} \fmftop{t} \fmfbottom{b} \fmf{quark}{i1,t,b,i2}
\fmf{photon}{t,o2} \fmf{photon}{b,o1} \fmflabel{$p_1$}{i2}
\fmflabel{$p_2$}{i1} \fmflabel{$p_3$}{o2} \fmflabel{$p_4$}{o1}
\end{fmfchar*}
}\end{fmffile}

\\
\end{array} $$
\caption[LO $q\bar{q} \rightarrow \gamma \gamma$
diagrams.]{Leading order diagrams for $q\bar{q} \rightarrow
\gamma \gamma$.} \label{fig:qqbvv} \EFIG

\noindent Note here that the phase space factor has been
simplified by use of the momentum-conserving delta function, and
becomes dependent only upon a single variable $v={1\over
2}(1+\cos \theta_3)$, where $\theta_3$ is the angle $\vec{p}_3$
makes with $\vec{p}_1$ in the partonic center-of-mass frame. The
{\it Mandelstam} variables $\{\hat{s},\hat{t},\hat{u}\}$ are then
given by $\hat{s}=x_ax_bS$, $\hat{t}=-(1-v)\hat{s}$,
$\hat{u}=-v\hat{s}$.

In a leading log calculation, we would stop here, take $\epsilon
\rightarrow 0$ (as there are no divergences), and evaluate the
distribution functions at a factorization scale $M_f$
characteristic of the subprocess (perhaps the mass
$Q=\sqrt{\hat{s}}$ of the photon pair). If this leading-order
term had contained a dependence on the running strong coupling
$\alpha_s$, we would evaluate this, too, at a renormalization
scale $\mu$, usually taken to be equal to the factorization
scale. In fact, we will be adding higher-order terms, so the
strong coupling will show up in the final result.

At next-to-leading order, two things happen. First, we will have
contributions from tree diagrams in which a gluon is radiated
from one of the quark legs \footnote{We will also have
contributions in which a gluon initiates the process, and a quark
appears in the final state, but we'll ignore these for the sake
of this argument.} Second, we will have virtual diagrams in which
a gluon is radiated and reabsorbed, so that there are still only
two particles in the final state. Of course, emission and
reabsorption amounts to two QCD vertices, whereas the tree
diagrams only contribute one, so if we're working to NLO only, we
keep only the interference terms between the Born and virtual
amplitudes.

The tree diagrams for $q(p_1) \bar{q}(p_2) \rightarrow \gamma(p_3)
\gamma(p_4) g(p_5)$ are shown in figure \ref{fig:qqbvvg}, and
their contribution is as follows:

\begin{equation}
d\sigma_1 = {\sum_f \over {2S}} \int_0^1 {{dx_1}\over x_1}
\int_0^1 {{dx_2}\over x_2} H(x_1,x_2) d^D \Gamma_3 \bar{\Sigma}
{|M|}^2_1 \;,
\end{equation}

\noindent where here

\begin{eqnarray}
H(x_1,x_2) &=&   q_{f/A} (x_1) q_{\bar{f}/B} (x_2)
 + (f \rightarrow \bar{f}) \; \\
\bar{\Sigma} {|M|}^2_1 &=& {1\over 2} {1\over 3^2} {1\over 2^2}
16N_C {(4\pi)}^3 \mu^{6\epsilon} \alpha^2 \alpha_s Q_f^4 {{M_1
(p_3^\mu,p_5^\mu,x_1,x_2,\epsilon)}\over {(p_5^0 \sin\theta_5)}^2} \; \\
d^D \Gamma_3 &=& {{(2\pi)}^{4\epsilon-5}\over 4} {\Bigl[
2{(4\pi)}^{-\epsilon} {{\Gamma (1-\epsilon)}\over {\Gamma
(1-2\epsilon)}} \Bigr]}^2 \delta(p_4^2)\cr &\times&
{(p_3^0)}^{1-2\epsilon} \sin^{1-2\epsilon} \theta_3
\sin^{-2\epsilon} \phi_3 dp_3^0 d\theta_3 d\phi_3 \cr &\times&
{(p_5^0)}^{1-2\epsilon} \sin^{1-2\epsilon} \theta_5
\sin^{-2\epsilon} \phi_5 dp_5^0 d\theta_5 d\phi_5 \;.
\end{eqnarray}

\BFIG
$$
\begin{array}{cc}

\begin{fmffile}{cf2gam03}
\fmfframe(10,10)(10,10){
\begin{fmfchar*}(20,20) \fmfset{curly_len}{2.0mm} \fmfstraight \fmfleft{i1,i2}
\fmfright{o1,o2,o3} \fmftop{t} \fmfbottom{b}
\fmf{quark,tension=2}{i2,t,v,b,i1} \fmflabel{$p_1$}{i2}
\fmflabel{$p_2$}{i1} \fmflabel{$p_5$}{o3} \fmflabel{$p_4$}{o1}
\fmf{gluon}{t,o3} \fmf{photon}{b,o1} \fmffreeze
 \fmf{photon,tension=2}{v,o2}  \fmflabel{$p_3$}{o2}
\end{fmfchar*}
}\end{fmffile}

&

\begin{fmffile}{cf2gam04}
\fmfframe(10,10)(10,10){
\begin{fmfchar*}(20,20) \fmfset{curly_len}{2.0mm} \fmfstraight \fmfleft{i1,i2}
\fmfright{o1,o2,o3} \fmftop{t} \fmfbottom{b}
\fmf{quark,tension=2}{i1,vx,t,v,b,vx,i2} \fmflabel{$p_1$}{i2}
\fmflabel{$p_2$}{i1} \fmflabel{$p_5$}{o3} \fmflabel{$p_4$}{o1}
\fmf{gluon}{t,o3} \fmf{photon}{b,o1} \fmffreeze
 \fmf{photon,tension=2}{v,o2} \fmflabel{$p_3$}{o2}
\end{fmfchar*}
}\end{fmffile}

\\

\begin{fmffile}{cf2gam05}
\fmfframe(10,10)(10,10){
\begin{fmfchar*}(20,20) \fmfset{curly_len}{2.0mm} \fmfstraight \fmfleft{i1,i2}
\fmfright{o1,o2,o3} \fmftop{t} \fmfbottom{b}
\fmf{quark,tension=2}{i2,t,v,b,i1} \fmflabel{$p_1$}{i2}
\fmflabel{$p_2$}{i1} \fmflabel{$p_5$}{o1} \fmflabel{$p_3$}{o3}
\fmf{photon}{t,o3} \fmf{gluon}{b,o1} \fmffreeze
\fmf{photon,tension=2}{v,o2} \fmflabel{$p_4$}{o2}
\end{fmfchar*}
}\end{fmffile}

&

\begin{fmffile}{cf2gam06}
\fmfframe(10,10)(10,10){
\begin{fmfchar*}(20,20) \fmfset{curly_len}{2.0mm} \fmfstraight \fmfleft{i1,i2}
\fmfright{o1,o2,o3} \fmftop{t} \fmfbottom{b}
\fmf{quark,tension=2}{i1,vx,t,v,b,vx,i2} \fmflabel{$p_1$}{i2}
\fmflabel{$p_2$}{i1} \fmflabel{$p_5$}{o1} \fmflabel{$p_3$}{o3}
\fmf{photon}{t,o3} \fmf{gluon}{b,o1} \fmffreeze
\fmf{photon,tension=2}{v,o2} \fmflabel{$p_4$}{o2}
\end{fmfchar*}
}\end{fmffile}

\\

\begin{fmffile}{cf2gam07}
\fmfframe(10,10)(10,10){
\begin{fmfchar*}(20,20) \fmfset{curly_len}{2.0mm} \fmfstraight \fmfleft{i1,i2}
\fmfright{o1,o2,o3} \fmftop{t} \fmfbottom{b}
\fmf{quark,tension=2}{i2,t,v,b,i1} \fmflabel{$p_1$}{i2}
\fmflabel{$p_2$}{i1}  \fmflabel{$p_3$}{o3} \fmflabel{$p_4$}{o1}
\fmf{photon}{t,o3} \fmf{photon}{b,o1} \fmffreeze
\fmf{gluon,tension=2}{v,o2} \fmflabel{$p_5$}{o2}
\end{fmfchar*}
}\end{fmffile}

&

\begin{fmffile}{cf2gam08}
\fmfframe(10,10)(10,10){
\begin{fmfchar*}(20,20) \fmfset{curly_len}{2.0mm} \fmfstraight \fmfleft{i1,i2}
\fmfright{o1,o2,o3} \fmftop{t} \fmfbottom{b}
\fmf{quark,tension=2}{i1,vx,t,v,b,vx,i2} \fmflabel{$p_1$}{i2}
\fmflabel{$p_2$}{i1} \fmflabel{$p_3$}{o3} \fmflabel{$p_4$}{o1}
\fmf{photon}{b,o1} \fmf{photon}{t,o3} \fmffreeze
\fmf{gluon,tension=2}{v,o2} \fmflabel{$p_5$}{o2}
\end{fmfchar*}
}\end{fmffile}

\\

\end{array} $$
\caption[$q\bar{q} \rightarrow \gamma \gamma g$ diagrams.]
{Diagrams for $q\bar{q} \rightarrow \gamma \gamma g$.}
\label{fig:qqbvvg} \EFIG

The function $M_1 (p_3^\mu,p_5^\mu,x_1,x_2,\epsilon)$ is a
complicated one, and this makes integrating over the extra gluon
($p_5^\mu$) difficult. As we are here interested only in the
singularity structure, we will make an approximation. In the
limit that the gluon becomes collinear with an incoming quark
($\vec{p_5} || \vec{p_1}$, {\it e.g.}), its momentum, as well as
that of the daughter quark ($p_a^\mu$, which enters the
subprocess) are related to the parent in terms of a simple
fraction $z$. That is, $p_5^\mu = (1-z) p_1^\mu$ and $p_a^\mu =
zp_1^\mu$. The matrix element simplifies, the integrals become
easy, and the remaining phase space is that of a two-body final
state.

\begin{figure}[h]
\centering

\begin{fmffile}{cfta5col} \fmfframe(7,7)(7,7){
\begin{fmfgraph*}(30,30)
\fmfleft{i1} \fmfright{o2,o1} \fmfblob{30}{o2}
\fmflabel{$p_1^\mu$}{i1} \fmf{quark}{i1,v1} \fmf{quark}{v1,o2}
\fmflabel{$p_a^\mu = zp_1^\mu$}{o2} \fmflabel{$p_5^\mu = (1-z)
p_1^\mu$}{o1} \fmf{gluon}{v1,o1}
\end{fmfgraph*}}
\end{fmffile}

\bigskip
\caption{$p_1 \cdot p_5 =0$ collinear kinematics.}
\end{figure}
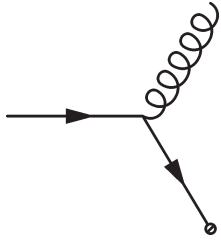

In short, the matrix element factorizes into a Born term
describing a $2 \rightarrow 2$ subprocess, and a {\it splitting
function} $\hat{P}_{qq} '$ describing the probability for a quark
to split into a collinear quark-gluon pair. This latter function
is independent of the particular subprocess, as long as it
contains an incoming quark. Of course, we get a pole from the
collinear singularity, and an important multiplicative factor.
The result, in $D=4-2\epsilon$ dimensions, is:

\begin{eqnarray}
H(x_a/z,x_b) &\rightarrow&   q_{f/A} (x_a/z)
q_{\bar{f}/B} (x_2) + (f \rightarrow \bar{f})  \; \\
\bar{\Sigma} {|M|}^2_1 &\rightarrow& \bar{\Sigma} {|M|}^2_0
 {{8 (4\pi) \alpha_s \mu^{2\epsilon} }\over {\hat{s} \sin^2 \theta_5}}
 {z\over {(1-z)}} \hat{P}_{qq} ' (z) \; \\
d^D \Gamma_3 &\rightarrow& d^D \Gamma_2 {{-1}\over \epsilon}
{\biggl( {{4\pi}\over \hat{s}} \biggr)}^\epsilon {{\Gamma
(1-\epsilon)}\over {\Gamma (1-2\epsilon)}} {\hat{s}\over
{(4\pi)}^2} {{dz}\over 4} {{(1-z)}^{1-2\epsilon} \over
z^{3-2\epsilon}}  \;,
\end{eqnarray}

\noindent which gives the cross section

\begin{eqnarray}
d\sigma_1 &=& {\sum_f \over {2S}} \int_0^1 {{dx_a}\over x_a}
\int_0^1 {{dx_b}\over x_b} d^D \Gamma_2 \bar{\Sigma} {|M|}^2_0
\cr &\times& {{-1}\over \epsilon} {\alpha_s \over {2\pi}}
{\biggl( {{4\pi \mu^2}\over \hat{s}} \biggr)}^\epsilon {{\Gamma
(1-\epsilon)}\over {\Gamma (1-2\epsilon)}} \int_{x_a}^1
{{dz}\over z} {\biggl[ {z\over {1-z}} \biggr]}^{2\epsilon}
H(x_a/z,x_b) \hat{P}_{qq} ' (z,\epsilon) \;.
\end{eqnarray}

\noindent Here $\hat{P}_{qq} ' (z,\epsilon)$ is the
aforementioned splitting function, pertinent to the process $q
\rightarrow q g$, where the quark on the right-hand side is
involved in the QCD subprocess. This function will always appear
in such a collinear limit, regardless of the subprocess, and has
the form

\begin{equation}
\hat{P}_{qq} ' (z,\epsilon) = C_F \biggl[ {{1+z^2}\over {1-z}}
-\epsilon (1-z) \biggr] \;.
\end{equation}

Unfortunately, there is also a soft pole here at $z=1$. We'd like
to separate this so we can focus on the purely collinear
contribution. We start by adding and subtracting, under the
$z$-integral, the $z \rightarrow 1$ limit of
$H(x_a/z,x_b)\hat{P}_{qq} ' (z,\epsilon)$. Remembering that all of
this multiplies the collinear ${{-1}\over \epsilon}$ pole, we have

\begin{eqnarray}
{{-1}\over \epsilon} \int_{x_a}^1 {{dz}\over z} {\biggl( {z\over
{1-z}} \biggr)}^\epsilon \Bigl[ H(x_a/z,x_b) \hat{P}_{qq} '
(z,\epsilon) -{{2C_F H(x_a,x_b)}\over {1-z}} \Bigr] \cr -
\int_{x_a}^1 {{dz}\over z} {\biggl( {z\over {1-z}}
\biggr)}^\epsilon {{2C_F H(x_a,x_b)}\over {1-z}} \;.
\label{softsub}
\end{eqnarray}

\noindent This allows us to express the first integrand (which is
finite as $z \rightarrow 1$) in terms of a  {\it
plus-distribution}, a handy notational device which is defined
upon convolution with a smooth function. We will have plenty of
use for these in subsequent chapters, as well. Here, the
plus-distribution ${(1-z)}_+$ is defined over a function $f(z)$ as

\BQN \int_0^1 dz {{f(z)}\over {(1-z)}_+} \equiv \int_0^1 dz
{{f(z)-f(1)}\over {1-z}} \;. \EQN

\noindent The first term in equation \ref{softsub} then works out
to

$${{-1}\over \epsilon} \int_{x_a}^1 {{dz}\over z}  H(x_a/z,x_b)
P_{qq}^+ (z) +{{C_F H(x_a,x_b)}\over \epsilon} \Bigl[ 2\ln
{{1-x_a}\over x_a}+{3\over 2} \Bigr]  + {\cal O}(1) $$

\noindent where $P_{qq}^+$ is the plus-distribution form of the
$\epsilon$-independent $qq$ splitting function:

\begin{equation}
P_{qq}^+ \equiv C_F \Bigl[ {{1+z^2}\over {(1-z)}_+} + {3\over 2}
\delta (1-z) \Bigr] \;;
\end{equation}

\noindent the second term becomes

$${{2C_F H(x_a,x_b)}\over \epsilon} \Bigl[ {1\over {2\epsilon}} -
\ln {{1-x_a}\over x_a}  + {\cal O}(\epsilon) \Bigr].$$

\noindent This operation splits $d\sigma_1$ into two
contributions, the first of which we'll call $d\sigma_{coll}$ and
the latter $d\sigma_{soft}$. The same terms arise, of course,
when we look at the limit in which the gluon is radiated from the
other incoming leg; in that case we'll be looking at terms
involving $H(x_a,x_b/z)$. When we put everything together, we'll
add in these contributions.

Meanwhile, the (renormalized) virtual contribution is:

\begin{equation}
d\sigma_{virt} = {\sum_f \over {2S}} \int_0^1 {{dx_a}\over x_a}
\int_0^1 {{dx_b}\over x_b}  H(x_a,x_b) d^D \Gamma_2 \bar{\Sigma}
{|M|}^2_v \;,
\end{equation}

\noindent where

\begin{equation}
\bar{\Sigma} {|M|}^2_v = \bar{\Sigma} {|M|}^2_0
 \Bigl[ -{{2C_F}\over \epsilon^2} - {{3C_F}\over \epsilon} + {\cal O}(1) \Bigr] {\alpha_s \over
{2\pi}} {\biggl( {{4\pi \mu^2}\over \hat{s}} \biggr)}^\epsilon
{{\Gamma (1-\epsilon)}\over {\Gamma (1-2\epsilon)}} \;.
\end{equation}

\noindent The virtual diagrams are shown in figure
\ref{fig:qqbvvirt}.

\BFIG
$$
\begin{array}{cc}

\begin{fmffile}{cf2gam09}
\fmfframe(10,7)(10,7){
\begin{fmfchar*}(20,20) \fmfset{curly_len}{2.0mm} \fmfstraight \fmfleft{i1,i2}
\fmfright{o1,o2} \fmftop{t} \fmfbottom{b}
\fmf{quark}{i2,t,v1,b,v2,i1} \fmflabel{$p_1$}{i2}
\fmflabel{$p_2$}{i1}  \fmflabel{$p_4$}{o1} \fmf{photon}{b,o1}
\fmf{photon}{t,o2}  \fmflabel{$p_3$}{o2} \fmffreeze
\fmf{gluon}{v1,v2}
\end{fmfchar*}
}\end{fmffile}

&

\begin{fmffile}{cf2gam10}
\fmfframe(10,7)(10,7){
\begin{fmfchar*}(20,20) \fmfset{curly_len}{2.0mm} \fmfstraight \fmfleft{i1,i2}
\fmfright{o1,o2} \fmftop{t} \fmfbottom{b}
\fmf{quark}{i2,t,v1,b,v2,i1} \fmflabel{$p_2$}{i2}
\fmflabel{$p_1$}{i1}  \fmflabel{$p_4$}{o1} \fmf{photon}{b,o1}
\fmf{photon}{t,o2}  \fmflabel{$p_3$}{o2} \fmffreeze
\fmf{gluon}{v1,v2}
\end{fmfchar*}
}\end{fmffile}

\\

\begin{fmffile}{cf2gam11}
\fmfframe(10,7)(10,7){
\begin{fmfchar*}(20,20) \fmfset{curly_len}{2.0mm} \fmfstraight \fmfleft{i1,i2}
\fmfright{o1,o2} \fmftop{t} \fmfbottom{b}
\fmf{quark}{i2,v1,t,b,v2,i1} \fmflabel{$p_2$}{i2}
\fmflabel{$p_1$}{i1}  \fmflabel{$p_4$}{o1} \fmf{photon}{b,o1}
\fmf{photon}{t,o2}  \fmflabel{$p_3$}{o2} \fmffreeze
\fmf{gluon}{v1,v2}
\end{fmfchar*}
}\end{fmffile}

&

\begin{fmffile}{cf2gam12}
\fmfframe(10,7)(10,7){
\begin{fmfchar*}(20,20) \fmfset{curly_len}{2.0mm} \fmfstraight \fmfleft{i1,i2}
\fmfright{o1,o2} \fmftop{t} \fmfbottom{b}
\fmf{quark}{i2,v1,t,b,v2,i1} \fmflabel{$p_1$}{i2}
\fmflabel{$p_2$}{i1}  \fmflabel{$p_4$}{o1} \fmf{photon}{b,o1}
\fmf{photon}{t,o2}  \fmflabel{$p_3$}{o2} \fmffreeze
\fmf{gluon}{v1,v2}
\end{fmfchar*}
}\end{fmffile}

\\

\begin{fmffile}{cf2gam13}
\fmfframe(10,7)(10,7){
\begin{fmfchar*}(20,20) \fmfset{curly_len}{2.0mm} \fmfstraight \fmfleft{i1,i2}
\fmfright{o1,o2} \fmftop{t} \fmfbottom{b}
\fmf{quark}{i2,v1,v2,t,b,i1} \fmflabel{$p_2$}{i2}
\fmflabel{$p_1$}{i1}  \fmflabel{$p_4$}{o1} \fmf{photon}{b,o1}
\fmf{photon}{t,o2}  \fmflabel{$p_3$}{o2} \fmffreeze
\fmf{gluon,right,tension=0}{v1,v2}
\end{fmfchar*}
}\end{fmffile}

&

\begin{fmffile}{cf2gam14}
\fmfframe(10,7)(10,7){
\begin{fmfchar*}(20,20) \fmfset{curly_len}{2.0mm} \fmfstraight \fmfleft{i1,i2}
\fmfright{o1,o2} \fmftop{t} \fmfbottom{b}
\fmf{quark}{i2,v1,v2,t,b,i1} \fmflabel{$p_1$}{i2}
\fmflabel{$p_2$}{i1}  \fmflabel{$p_4$}{o1} \fmf{photon}{b,o1}
\fmf{photon}{t,o2}  \fmflabel{$p_3$}{o2} \fmffreeze
\fmf{gluon,right,tension=0}{v1,v2}
\end{fmfchar*}
}\end{fmffile}

\\

\begin{fmffile}{cf2gam15}
\fmfframe(10,7)(10,7){
\begin{fmfchar*}(20,20) \fmfset{curly_len}{2.0mm} \fmfstraight \fmfleft{i1,i2}
\fmfright{o1,o2} \fmftop{t} \fmfbottom{b}
\fmf{quark}{i2,t,b,v1,v2,i1} \fmflabel{$p_2$}{i2}
\fmflabel{$p_1$}{i1}  \fmflabel{$p_4$}{o1} \fmf{photon}{b,o1}
\fmf{photon}{t,o2}  \fmflabel{$p_3$}{o2} \fmffreeze
\fmf{gluon,right,tension=0}{v1,v2}
\end{fmfchar*}
}\end{fmffile}

&

\begin{fmffile}{cf2gam16}
\fmfframe(10,7)(10,7){
\begin{fmfchar*}(20,20) \fmfset{curly_len}{2.0mm} \fmfstraight \fmfleft{i1,i2}
\fmfright{o1,o2} \fmftop{t} \fmfbottom{b}
\fmf{quark}{i2,t,b,v1,v2,i1} \fmflabel{$p_1$}{i2}
\fmflabel{$p_2$}{i1}  \fmflabel{$p_4$}{o1} \fmf{photon}{b,o1}
\fmf{photon}{t,o2}  \fmflabel{$p_3$}{o2} \fmffreeze
\fmf{gluon,right,tension=0}{v1,v2}
\end{fmfchar*}
}\end{fmffile}

\\

\begin{fmffile}{cf2gam17}
\fmfframe(10,7)(10,7){
\begin{fmfchar*}(20,20) \fmfset{curly_len}{2.0mm} \fmfstraight \fmfleft{i1,i2}
\fmfright{o1,o2} \fmftop{t} \fmfbottom{b}
\fmf{quark}{i2,t,v1,v2,b,i1} \fmflabel{$p_2$}{i2}
\fmflabel{$p_1$}{i1}  \fmflabel{$p_4$}{o1} \fmf{photon}{b,o1}
\fmf{photon}{t,o2}  \fmflabel{$p_3$}{o2} \fmffreeze
\fmf{gluon,right,tension=0}{v1,v2}
\end{fmfchar*}
}\end{fmffile}

&

\begin{fmffile}{cf2gam18}
\fmfframe(10,7)(10,7){
\begin{fmfchar*}(20,20) \fmfset{curly_len}{2.0mm} \fmfstraight \fmfleft{i1,i2}
\fmfright{o1,o2} \fmftop{t} \fmfbottom{b}
\fmf{quark}{i2,t,v1,v2,b,i1} \fmflabel{$p_1$}{i2}
\fmflabel{$p_2$}{i1}  \fmflabel{$p_4$}{o1} \fmf{photon}{b,o1}
\fmf{photon}{t,o2}  \fmflabel{$p_3$}{o2} \fmffreeze
\fmf{gluon,right,tension=0}{v1,v2}
\end{fmfchar*}
}\end{fmffile}

\\

\end{array} $$
\caption[$q\bar{q} \rightarrow \gamma \gamma$ virtual diagrams.]
{Virtual diagrams for $q\bar{q} \rightarrow \gamma \gamma$.}
\label{fig:qqbvvirt} \EFIG

In all, then, we have one LO and three NLO contributions, which
all add to give:

\begin{eqnarray}
d\sigma &=& {\sum_f \over {2S}} \int_0^1 {{dx_a}\over x_a}
\int_0^1 {{dx_b}\over x_b} d^D \Gamma_2 \bar{\Sigma} {|M|}^2_0
H(x_a,x_b) \; \cr &\times& \Biggl[ 1 + {1\over \epsilon}
{\alpha_s \over {2\pi}} {\biggl( {{4\pi \mu^2}\over \hat{s}}
\biggr)}^\epsilon {{\Gamma (1-\epsilon)}\over {\Gamma
(1-2\epsilon)}}  \bigl[ \Pi_{coll} + \Pi_{soft} + \Pi_{virt}
\bigr] \Biggr] \;, \label{Frag1}
\end{eqnarray}

\noindent where

\begin{eqnarray}
\Pi_{coll} &=&  -\int_{x_a}^1 {{dz}\over z} {{H(x_a/z,x_b)}\over
{H(x_a,x_b)}} P_{qq}^+ (z) - \int_{x_b}^1 {{dz}\over z}
{{H(x_a,x_b/z)}\over {H(x_a,x_b)}} P_{qq}^+ (z) \cr &+& 2C_F
\ln {{1-x_a}\over x_a}{{1-x_b}\over x_b} +3C_F  + {\cal O}(\epsilon) \; \\
\Pi_{soft} &=& 2C_F  \Bigl[ {1\over \epsilon} - \ln
{{1-x_a}\over x_a}{{1-x_b}\over x_b}  + {\cal O}(\epsilon) \Bigr] \; \\
\Pi_{virt} &=&  -{{2C_F}\over \epsilon} - 3C_F + {\cal
O}(\epsilon) \;.
\end{eqnarray}

One can see by inspection that, as predicted, all the poles cancel
except those associated with convolutions over splitting
functions. However, if we were to redefine our quark
distributions such that

\begin{equation}
q_f (x,M_f^2) \equiv q_f (x) - {1\over \epsilon} {\alpha_s \over
{2\pi}} {\biggl( {{4\pi \mu^2}\over M_f^2} \biggr)}^\epsilon
{{\Gamma (1-\epsilon)}\over {\Gamma (1-2\epsilon)}} \int_{x}^1
{{dz}\over z} q_f (x/z) P_{qq}^+ (z) \;, \label{pdfredef}
\end{equation}

\noindent then we could absorb the remaining divergence into the
distribution functions. The remaining NLO contributions would
come solely from the finite ${\cal O}(\epsilon)/\epsilon = {\cal
O}(1)$ remainders, we could take $\epsilon \rightarrow 0$
everywhere, and the distribution functions (as well as the
running coupling, to this order) would then be evaluated at a
factorization scale $M_f$, characteristic to the observed system
(in our case, $\hat{s}$).

\begin{equation}
d\sigma = {\sum_f \over {2S}} \int_0^1 {{dx_a}\over x_a} \int_0^1
{{dx_b}\over x_b} d^4 \Gamma_2 \bar{\Sigma} {|M|}^2_0
H(x_a,x_b,M_f) \Biggl[ 1 +  {{\alpha_s (M_f)} \over {2\pi}} {\cal
O}(1) \Biggr] \;.
\end{equation}

\noindent This reflects the interpretation noted above, that a
parton which participates in a subprocess is in fact a component
of (a component of a component of...) the parent hadron, and at
higher energies, we resolve more and more components. Of course,
we have here shown only the first term in such an expansion; at
higher orders (more splittings) we must absorb terms with more
poles, higher orders of the coupling, and more complicated
splitting factors.

Furthermore, we have shown only the $q \rightarrow {\boldmath q}
g$ splitting function (boldface type referring to the parton
participating in the subprocess). In general, there are splitting
functions for $q \rightarrow {\boldmath g} q$, $g \rightarrow
{\boldmath q} \bar{q}$, and $g \rightarrow {\boldmath g} g$.

The above redefinition is in fact a combination of two separate
reorganizations; the first absorbs the poles (and an arbitrary
constant), the second relates to energy dependence. Expanding the
order $\alpha_s$ factor above, we find

\begin{equation}
\Delta q_f (x) = {\alpha_s\over {2\pi}} \int_x^1 {{dz}\over z} q_f
(x/z) P_{qq}^+ (z) \Bigl[ {{-1}\over \epsilon} - \ln 4\pi +
\gamma_E + \ln {M_f^2\over \mu^2} - {\cal O}(1) \Bigr] \;,
\end{equation}

\noindent where we have included the ${{{\cal O}(\epsilon)}\over
\epsilon}  = {\cal O} (1)$ terms from \ref{Frag1}; that is, the
entire remainder. $\gamma_E$ is the Euler constant, $\simeq
0.577216$.

We can now absorb the pole and as much of the finite remainder as
we like into a new distribution function $\bar{q}_f (x)$. In the
{\bf DIS} scheme, everything is absorbed but the energy log:

\begin{equation}
\bar{q}_f (x) = q_f (x)+{\alpha_s\over {2\pi}} \int_x^1 {{dz}\over
z} q_f (x/z) P_{qq}^+ (z) \Bigl[ {{-1}\over \epsilon} - \ln 4\pi
+ \gamma_E - {\cal O}(1) \Bigr] \;.
\end{equation}

\noindent In the ${\bf \overline{MS}}$ scheme, only the factor
${{-1}\over \epsilon}-\ln 4\pi +\gamma_E$ is subtracted:

\begin{equation}
\bar{q}_f (x) = q_f (x)+{\alpha_s\over {2\pi}} \int_x^1 {{dz}\over
z} q_f (x/z) P_{qq}^+ (z) \Bigl[ {{-1}\over \epsilon} - \ln 4\pi
+ \gamma_E  \Bigr] \;.
\end{equation}

\noindent In this dissertation, we will be using the ${\bf
\overline{MS}}$ scheme exclusively.

Either way, the energy scale logarithm which remains is just as
ubiquitous as the constants $-\ln 4\pi + \gamma_E$ we have already
absorbed, so we continue with the further redefinition

\begin{equation}
q_f (x,M_f) \equiv \bar{q}_f + \Delta q_f (x,M_f) \;,
\end{equation}

\noindent where

\begin{equation}
\Delta q_f (x,M_f) = {\alpha_s\over {2\pi}} (\ln M_f^2 - \ln\mu^2)
\int_x^1 {{dz}\over z} q_f (x/z) P_{qq}^+ (z) \;.
\end{equation}

In the limit of infinitesimal change of scale, and taking into
account the other possible splitting functions, we are led to the
aforementioned {\it Altarelli-Parisi} equations:

\begin{eqnarray}
{{dq_f (x,M^2)}\over {d\ln M^2}} &=& {\alpha_s\over {2\pi}}
\int_x^1 {{dz}\over z} \Bigl[ q_f (x/z,M^2) P_{qq}^+ (z) + q_g
(x/z,M^2) P_{qg}^+ (z) \Bigr] \; \\
{{dq_g (x,M^2)}\over {d\ln M^2}} &=& {\alpha_s\over {2\pi}}
\int_x^1 {{dz}\over z} \Bigl[ q_f (x/z,M^2) P_{gq}^+ (z) + q_g
(x/z,M^2) P_{gg}^+ (z) \Bigr] \;.
\end{eqnarray}

\noindent For reference, Table \ref{Split} lists the terms
involved in first-order splitting functions; the conventions used
in this dissertation are:

\begin{table}
\caption{Splitting function pieces.} \vspace{.5cm}
\begin{center}
\thicklines
\begin {tabular} { lllll }
\hline \hline \thinlines
$ij$ & $\rho_{ij}^0$ & ${\hat{\rho}}_{ij}^+$ & $\rho_{ij}^\delta$ & $\rho_{ij}^1$ \\
\hline \thicklines
$qq$ & $C_F$ & ${{1+z^2}\over {(1-z)}_+}$ & ${3\over 2} \delta (1-z)$ & $-(1-z)$ \\
$qg$ & ${1\over {2(1-\epsilon)}}$ & $z^2+{(1-z)}^2$ & $0$ & $-1$ \\
$gq$ & $C_F$ & ${{1+{(1-z)}^2}\over z_+}$ & ${3\over 2} \delta (z)$ & $-z$ \\
$gg$ & $2N_C$ & ${z\over {(1-z)}_+}+{{1-z}\over z} +z(1-z)$ &
${{11-{2\over 3}N_f}\over 12} \delta (1-z)$ & $0$ \\
$\gamma q$ & $Q_f^2$ & ${{1+{(1-z)}^2}\over z_+}$ & ${3\over 2} \delta (z)$ & $-z$ \\
\hline \hline
\end{tabular}
\end{center}
\label{Split}
\end{table}

\begin{list}{}{}
\item 1. The most general version is ${P_{ij}^+} ' \equiv \rho_{ij}^0 \biggl[ {\hat{\rho}}_{ij}^+ +
\rho_{ij}^\delta + \epsilon \rho_{ij}^1 \biggr]$, where flavor $i$
is the daughter, $j$ the parent.
\item 2. Non-plus-distribution versions are obtained from the
above by simply removing the ``$+$"-signs.
\item 3. ``Primed" functions include the $\epsilon$-coefficient $\rho^1$, non-primed
functions do not.
\item 4. ``Hatted" functions do not include $\rho^\delta$.
\item 5. $P_{ij}^1 \equiv \rho_{ij}^0 \rho_{ij}^1$ includes only the
coefficient of $\epsilon$.
\end{list}

\noindent Examples (flavor labels suppressed):

\BQA \hat{P} &=& \rho^0 \hat{\rho} \cr P &=& \rho^0
(\hat{\rho}+\rho^\delta) \cr  \hat{P}^+ &=& \rho^0 \hat{\rho}^+
\cr P^+ &=& \rho^0 (\hat{\rho}^+ + \rho^\delta) \cr \hat{P}' &=&
\rho^0 (\hat{\rho}+\epsilon \rho^1) \cr P' &=& \rho^0 (\hat{\rho}
+ \rho^\delta + \epsilon \rho^1) \cr {\hat{P'}}^+ &=& \rho^0
(\hat{\rho}^+ + \epsilon \rho^1) \EQA

Thus we can derive the scale dependence of the parton
distributions, and take into account the leading dependence at
higher orders by iteratively solving the Altarelli-Parisi
equations. Just as in the case of renormalization, a full solution
requires that at some scale $M_0$ we {\bf measure} the
distributions by comparison with experiment. Use of the new
distributions in cross section calculations now requires only that
we calculate, but then drop (subtract) collinear poles (and
attendant finite pieces appropriate to the chosen factorization
scheme).

\chapter{Early Resummation Schemes}
\label{ch:PPResum}

\section{Origin of Terms in the Unresummed Distribution}

Here we will look at how the relevant logarithms arise for the
QED process $e^+ e^- \rightarrow \mu^+ \mu^-.$ We will then
describe the evolution of a viable resummation method, and apply
it finally to the more complicated QCD process $p p \rightarrow
\mu^+ \mu^- + X.$

As discussed in Section \ref{sec:resneed}, if we define $Q^2
\equiv {(p_{\mu^+}^\nu + p_{\mu^-}^\nu)}^2$ as the mass (squared)
of the observed muon pair, and $Q_T^2 = {(\vec{p}_{T\mu^+} +
\vec{p}_{T\mu^-})}^2$ the transverse momentum (squared) of the
pair, calculation of the leading-order contribution to a process
of this type results in a delta-function at $Q_T^2=0$. That is,
the $Q_T$-distribution is of the form

\begin{equation}
{1\over \sigma_0}{d\sigma \over {dQ^2 dQ_T^2}} = \delta({Q_T}^2)
\, \label{f2mu21} \end{equation}

\noindent where $\sigma_0 \equiv d\sigma / dQ^2$.

In the relativistic limit, adding a single radiated boson to one
of the incoming fermions modifies the matrix element thus:

\begin{figure} \centering
\begin{fmffile}{cfppr1} \fmfframe(5,5)(5,5){
\begin{fmfgraph*}(40,40)
\fmfleft{i1,i2} \fmfright{o1,o2} \fmfbottom{b1}
\fmf{fermion,tension=2,label=$u(p)$,l.s=left}{i1,v1}
\fmf{fermion,tension=2,label=$p'$,l.s=left}{v1,v2}
\fmf{fermion}{v2,i2}
\fmfv{d.sh=circle,d.f=empty,d.si=.2w,l=${\cal M}$,l.d=-.05w}{v2}
\fmf{fermion}{o1,v2,o2} \fmflabel{$e^-$}{i1} \fmflabel{$e^+$}{i2}
\fmflabel{$\mu^-$}{o2} \fmflabel{$\mu^+$}{o1} \fmffreeze
\fmf{photon}{v1,b1} \fmflabel{$\gamma$}{b1}
\end{fmfgraph*}}
\end{fmffile}
\caption{A single radiated photon in the initial state.}
\label{fig:ppr1}
\end{figure}
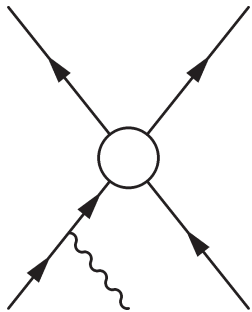

\begin{equation}
{\cal M}u(p) \rightarrow {-g {\cal M} \rlap /p' \rlap
/\epsilon(k) u(p) \over {p'}^2} = {-g {\cal M} (\rlap /p-\rlap
/k) \rlap /\epsilon(k) u(p) \over (p-k)^2} \;. \label{f2mu2mat}
\end{equation}

\noindent Here ${\cal M} u(p)$ is the Born matrix element, $p$ is
the incoming electron momentum, and $k$ is the momentum of the
emitted massless boson, which we take to be a photon (see figure
\ref{fig:ppr1}). For a high-mass muon pair, we may take the soft
radiation limit ($k$ small relative to $p$), and ignore terms
involving $k$ in the numerator of equation \ref{f2mu2mat}. This
leaves us with the {\it factorized} form

\begin{equation}
{\cal M'} = {g {\cal M} \rlap /p \rlap /\epsilon u(p) \over
2p\cdot k} \rightarrow {g {\cal M} {p\cdot \epsilon \over p\cdot
k} u(p)} \;. \label{f2mu26}
\end{equation}

\noindent By momentum conservation, $Q_T^2 =  k_T^2 \sim {(p
\cdot k)}^2$, and we find a {\it radiative correction\/} of order
$\alpha$ to the Born cross section:

\begin{equation}
{1\over \sigma_0}{d\sigma \over {dQ^2 dQ_T^2}} = \delta({Q_T}^2)
+ {\alpha \over 2 \pi Q_T^2} \bigl( \ln(Q^2/Q_T^2) + O(1) \bigr)
\;. \label{f2mu29}
\end{equation}

\noindent This cross section obviously diverges as $Q_T
\rightarrow 0$ (see figure \ref{fig:earlyres}, dot-dashed line),
a result of the soft and collinear divergences encountered when,
respectively, the photon is either emitted with zero energy or is
emitted in the same direction as the incident electron. In either
case, $p\cdot k =0$. In a full calculation, these divergences (or
{\it poles}), will be cancelled by the contribution of other
diagrams, and the remainder is usually described by means of a
{\it ``plus''-distribution}; that is, a distribution the value of
which at any part of the domain is meaningful only when
convoluted with another smooth function (see Section
\ref{sec:factorize} for one example).

In our case we have a divergent function $\nu(Q_T)$ given by

\BQN \nu(Q_T) =  {1 \over Q_T^2} \bigl( \ln(Q^2/Q_T^2) + O(1)
\bigr) \;. \EQN

\noindent We wish to separate off the singularities at $Q_T^2=0$,
which will be cancelled, and retain a function $\nu_+ (Q_T)$ which
is finite and integrable over the entire range of $Q_T^2$, {\it
i.e.} $0\leq Q_T^2 \leq Q^2$. That is, we want

\begin{equation}
\nu(Q_T^2) \rightarrow \nu_+ (Q_T^2) + {\rm singularities,} \;
\end{equation}

\noindent with the plus-distribution $\nu_+$ defined as follows:

\BQN \int_0^{Q^2} dQ_T^2 \nu_+ (Q_T) f(Q_T) \equiv  \int_0^{Q^2}
dQ_T^2 \nu (Q_T) \bigl[ f(Q_T) - f(0) \bigr] \;. \label{f2mu29b}
\EQN

\noindent At $Q_T^2=0$, $f(Q_T)-f(0)=0$ and the integral is
defined, even if the shape of $\nu_+(Q_T)$ is non-physical (like
the Born term delta function). We will thus rewrite equation
\ref{f2mu29} in the form

\BQA {1\over \sigma_0}{d\sigma \over {dQ^2 dQ_T^2}} &=&
\delta({Q_T}^2) + {\alpha \over 2 \pi} \nu_+(Q_T) \; \cr
\nu_+(Q_T) &\equiv& {\Bigl[{{\ln Q^2/Q_T^2}\over Q_T^2} \Bigr]}_+
+  {\Bigl[{{O(1)}\over Q_T^2} \Bigr]}_+ \;. \label{f2mu29c} \EQA

\begin{figure} \centering
\begin{fmffile}{cfppr2} \fmfframe(5,5)(5,5){
\begin{fmfgraph*}(40,40)
\fmfleft{i1} \fmfright{o1} \fmfbottom{b1,b2,b3,b4,b5}
\fmf{plain,tension=0.2}{i1,v1,v2,v3,v4,v5,o1}
\fmfv{d.sh=circle,d.f=empty,d.si=.2w,l=${\cal M}$,l.d=-.05w}{o1}
\fmflabel{$e^-$}{i1} \fmffreeze \fmflabel{$k_1$}{b1}
\fmf{photon}{v1,b1} \fmflabel{$k_2$}{b2} \fmf{photon}{v2,b2}
\fmflabel{$...$}{b3} \fmf{photon}{v3,b3} \fmflabel{$...$}{b4}
\fmf{photon}{v4,b4} \fmflabel{$k_n$}{b5} \fmf{photon}{v5,b5}
\end{fmfgraph*}}
\end{fmffile}
\caption{Multiple radiated photons in the initial state.}
\label{fig:ppr2}
\end{figure}
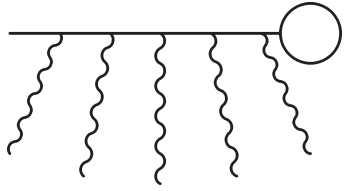

If we now consider multiple radiated photons (see figure
\ref{fig:ppr2}), we find that the matrix element---and thus the
cross section--- also factorizes in this soft approximation; that
is, for $n$-photon emission we obtain a matrix element of the form

\begin{equation}
{\cal M} {g^n \over n!} \prod_{i=1}^n {p\cdot \epsilon \over
p\cdot k_i} \;, \label{f2mu210}
\end{equation}

\noindent and a cross section

\begin{equation}
{1\over\sigma_0} {{d^n\sigma}\over {dQ^2 dQ_T^2}} = {{(\alpha /2
\pi)}^n \over n!} \prod_{i=1}^n \int_0^{Q_T^2} \nu_+({k_T}_i^2) \
d{k_T}_i^2 \ \delta^2 \Bigl( \vec {Q_T} - \sum_{i=1}^n {\vec
{k_T}}_i \Bigr) \;, \label{f2mu211}
\end{equation}

\noindent where $\nu_+ ({k_T}_i^2)$ is the first-order
contribution for a single photon (here ${\Bigl[{{\ln
Q^2/k_T^2}\over k_T^2} \Bigr]}_+ + {\Bigl[{{O(1)}\over k_T^2}
\Bigr]}_+ $), and the factor $n!$ arises from the
indistinguishability of the emitted photons.

\section{Development of a Resummation Scheme}

We have seen how the series of large logarithms (equation
\ref{logs}) arises at each order in $\alpha$. A first attempt at
an all-orders result would naturally be to sum over just the
largest log at each order. Our cross section contains a
momentum-conserving delta function, but given the presence of
$n!$, ignoring this function will allow us to directly form an
exponentiation in momentum space. We now have completely
independent photon emissions. If, as we have assumed so far, each
has a relatively small transverse momentum, limited by that of
the muon pair ($Q_T$), when we sum over all orders:

\begin{eqnarray}
{1\over \sigma_0} {d\sigma \over {dQ^2 dQ_T^2}} &=& {d\over
d{Q_T}^2} \sum_{n=1}^\infty {{(\alpha / 2\pi)}^n \over n!}
\prod_{i=1}^n \int_0^{{Q_T}^2} \nu_+ ({k_T}_i^2) d{k_T}_i^2  \;
\cr &=& {d\over d{Q_T}^2} \sum_{n=1}^\infty {{(\alpha / 2\pi)}^n
\over n!} \Bigl[ \int_0^{{Q_T}^2} \nu_+ ({k_T}^2) d{k_T}^2 {\Bigr]
}^n \; \cr &=& {d\over d{Q_T}^2} \Bigl[ e^{{\alpha \over 2\pi}
\int_0^{{Q_T}^2} \nu_+ d{k_T}^2} - 1 \Bigr]  \;, \label{f2mu212}
\end{eqnarray}

\noindent we find, for $ \nu_+ ({k_T}^2) = {\Bigl[
{\ln(\hat{s}/{k_T}^2) + O(1) \over {k_T}^2} \Bigr]}_+$, the result

\begin{eqnarray}
{1\over \sigma_0}{d\sigma \over {dQ^2 dQ_T^2}} &=& {d\over
d{Q_T}^2} \Bigl[ e^{-{\alpha \over 4\pi} \ln^2
(Q^2/Q_T^2)-{\alpha \over 2\pi} \ln (Q^2/Q_T^2)} \Bigr] \; \cr
&=& {\alpha \over 2\pi Q_T^2} \Bigl[ \ln{Q^2 \over Q_T^2} +O(1)
\Bigr] e^{-{\alpha \over 4\pi} \ln^2 (Q^2/Q_T^2)-{\alpha \over
2\pi} \ln (Q^2/Q_T^2)} \;, \label{f2mu213}
\end{eqnarray}

\noindent in which we have used the fact that the coefficient of
our plus-distribution is constant:

\BQA \int_0^{{Q_T}^2} \nu_+ ({k_T}^2) d{k_T}^2 &=& \int_0^{Q^2}
\nu_+ ({k_T}^2) d{k_T}^2 - \int_{Q_T^2}^{Q^2} \nu ({k_T}^2)
d{k_T}^2 \; \cr &\equiv& \int_0^{Q^2} \nu ({k_T}^2) (1-1) d{k_T}^2
- \int_{Q_T^2}^{Q^2} \nu ({k_T}^2) (1) d{k_T}^2 \; \cr &=& 0 -
{1\over 2} \ln^2 {Q^2\over Q_T^2} - O \bigl( \ln {Q^2\over Q_T^2}
\bigr) \;. \label{f2mu215} \EQA

This method was first attempted in 1978 by Dokshitzer, D'yakonov,
and Troyan (DDT) ~\cite{DDT78}. \footnote{Actually, they resummed
only the leading logs, $m=2n-1$. We include the $O(1)$ terms here
as well.} It gives, as we see in figure \ref{fig:earlyres} (dashed
line), an exponential {\it Sudakov suppression\/} of the $Q_T$
distribution as $Q_T^2 \rightarrow 0.$ This effect is a
consequence of the neglect of momentum-conservation among the
emitted bosons, effectively imposing the condition that all of
them have zero $k_T$ in order for a zero-$Q_T$ muon pair to be
produced. The exceedingly small phase space for n bosons to be
produced, all with near-zero $k_T$ at the same time, leads to the
suppression. We know, however, that this cannot be correct. It
should be possible for two or more bosons of non-negligible
$\vec{k}_T$ to be emitted and still maintain a zero total
$\vec{Q}_T$.

A subsequent analysis by Parisi and Petronzio ~\cite{PP79,CG80}
kept the factorized form for $d^n \sigma$, but also kept the full
momentum-conserving delta function and first-order contribution.
This time, the exponentiation develops automatically in
impact-parameter ($\vec{b}$) space. The delta function can be
rewritten as a Fourier transform in this space:

\begin{equation}
\delta^2 \bigl( \vec {Q_T} - \sum_i {\vec {k_T}}_i \bigr) =
{1\over {(2\pi)}^2} \int d^2b e^{-i {\vec b}\cdot (\vec {Q_T} -
\sum_i \vec {{k_T}_i} )} \;. \label{f2mu218}
\end{equation}

\noindent The cross section for n-boson emission then becomes a
Fourier transform and inverse-transform, with the exponentiation
sandwiched inbetween. To start,

\begin{eqnarray}
{1\over\sigma_0} {{d^n\sigma}\over {dQ^2 dQ_T^2}}
 &=& {1\over {(2\pi)}^2} {{(\alpha / 2\pi)}^n \over n!} \int d^2b e^{-i {\vec b}\cdot (\vec {Q_T} -
\sum_i \vec {k_T}_i )} \prod_{i=1}^n \int \nu_+ ({k_T}_i^2)
d^2{k_T}_i \; \cr &=& {1\over {(2\pi)}^2} \int d^2b \  {e^{-i \vec
b \cdot \vec {Q_T}} \over n!} \  \Bigl[ \int e^{i \vec b \cdot
\vec {k_T}} \  {\alpha\over 2 \pi} \   \nu_+ ({k_T}^2) d^2 k_T
{\Bigr]}^n \; \cr &=& {1\over {(2\pi)}^2} \int d^2b \   e^{-i
\vec b \cdot \vec {Q_T}} \ {{[{\alpha\over 2\pi} {\tilde\nu}_+
(b)]}^n \over n!} \;, \label{f2mu221}
\end{eqnarray}

\noindent where ${\tilde{\nu}}_+(b)$ is the transform of $\nu_+
(k_T)$ in impact parameter space.

The next step is clear. Summing over the contributions from all
numbers of photons, and including the $\delta$-function at
$Q_T=0$, Parisi and Petronzio arrived at

\begin{eqnarray}
{1\over\sigma_0} {{d\sigma}\over {dQ^2 dQ_T^2}} &=& {1\over
{(2\pi)}^2} \int d^2b e^{-i \vec b \cdot \vec {Q_T}} \Biggl[
\sum_{n=1}^\infty {1\over n!} \bigl[ {\alpha\over 2\pi}
{\tilde\nu}_+ (b) {\bigr] }^n +1 \Biggr] \cr &=& {1\over
{(2\pi)}^2} \int d^2b e^{-i \vec b \cdot \vec {Q_T}} \ e^{
{\alpha\over 2\pi} {\tilde\nu}_+ (b)} \;, \label{f2mu222}
\end{eqnarray}

\noindent where the last exponential on the right is sometimes
denoted by ${\tilde\sigma}.$ The prescription is, then, a Fourier
transform of an exponentiated transform of the next-to-leading
order (NLO) perturbative result. For our simple QED example, this
latter transform is

\BQA \tilde\nu_+ (b) &=& \int d^2k_T e^{i \vec b \cdot \vec {k_T}}
\Bigl[ {\ln (Q^2/k_T^2) + O(1) \over k_T^2} {\Bigr]}_+ \; \cr &=&
\int_0^{2\pi} d\phi \int_0^{Q^2} {dk_T^2 \over 2} (e^{i \vec b
\cdot \vec {k_T}} -1) \Bigl[ {\ln (Q^2/k_T^2) + O(1) \over k_T^2}
\Bigr] \;, \label{f2mu224} \EQA

\noindent where the upper limit on $k_T$ has been relaxed
relative to DDT and the plus-distribution definition \ref{f2mu29b}
again used. In the high-$b$ limit, this goes as $\tilde\nu(b)
\simeq - \ln^2 (1+b^2Q^2)$, giving

\begin{equation}
{1\over\sigma_0}{d\sigma \over {dQ^2 dQ_T^2}} {\Bigm|}_{{Q_T}^2=0}
\sim \int_0^\infty db^2 e^{-{\alpha\over 2\pi} \ln^2 (1+b^2Q^2)}
\simeq {e^{\pi / 2\alpha} \over Q^2} = {\rm constant.}
\label{f2mu225}
\end{equation}

\noindent The Sudakov suppression disappears, and one regains a
finite, non-zero cross section at $Q_T = 0$ (see figure
\ref{fig:earlyres}, solid line).

\BFIG
\centerline{\psfig{figure=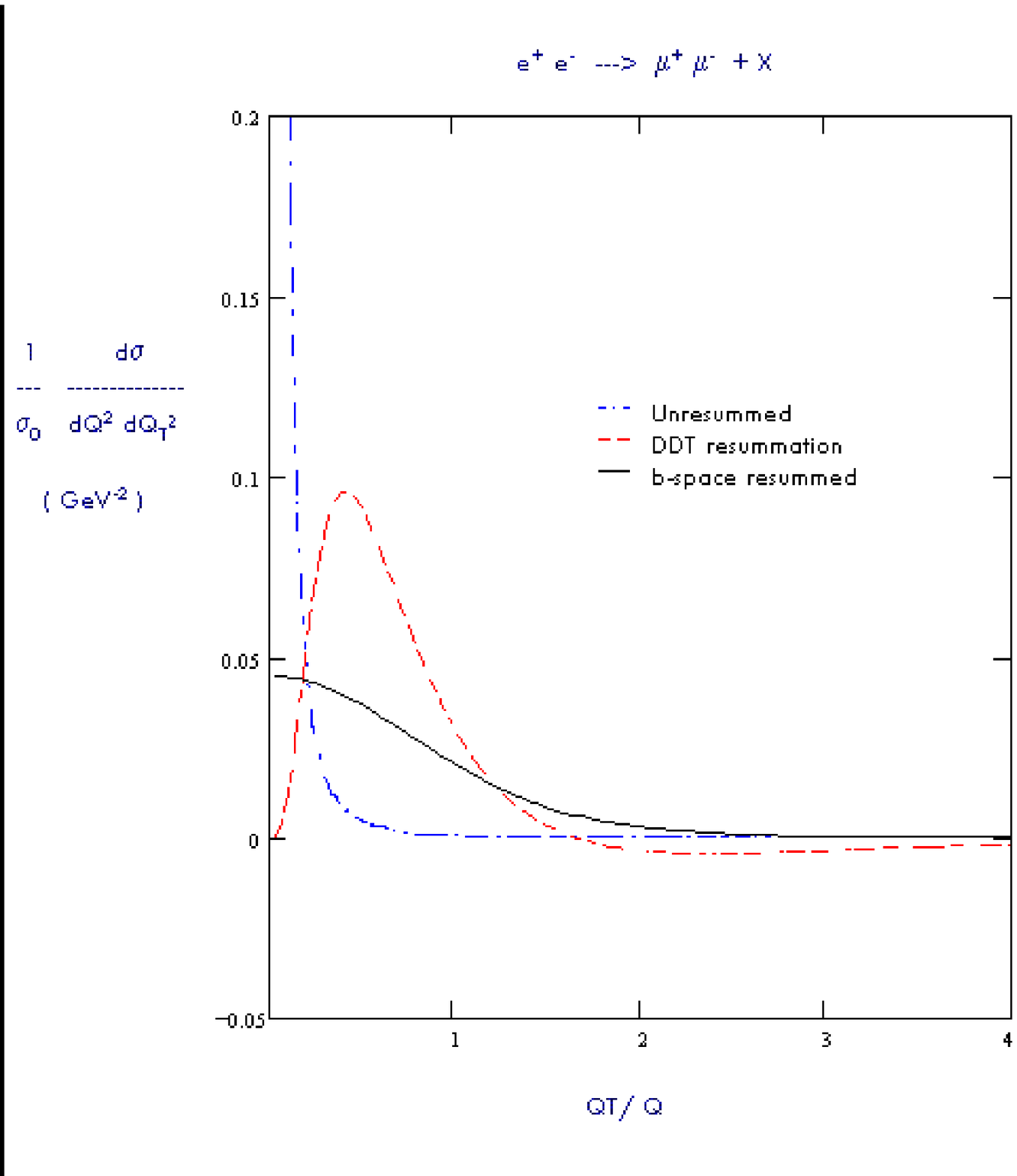,height=16.0cm,width=13.6cm}}
\caption[Resummation brings the low-$Q_T$ region under
control.]{Resummation brings the low-$Q_T$ region under control.}
\label{fig:earlyres} \EFIG

Resummation thus allows one to include the effects of multiple
soft boson radiation to all orders, while keeping the low-$Q_T$
region under control. The original perturbative prediction,
necessarily unphysical as a result of the delta-functions and
plus-distributions which dictated its bookkeeping, nevertheless
was integrable:

\BQN {1\over \sigma_0} {{d\sigma}\over {dQ^2}} = \int_0^{Q^2}
dQ_T^2 \Bigl\{ \delta(Q_T^2) + {\Bigl[{{\ln Q^2/Q_T^2}\over Q_T^2}
\Bigr]}_+ +  {\Bigl[{{O(1)}\over Q_T^2} \Bigr]}_+ \Bigr\} = 1 \;
\EQN

Resummation maintains the normalization while providing a more
physical model of the shape over the allowed $Q_T$ range.
Effectively, the delta function at $Q_T=0$ is smeared by the
combined recoil of multiple emissions, each of which may add or
subtract from the transverse momentum of the observed muon pair.
\footnote{The average transverse ``kick'' given to the system,
$<Q_T>$, will be a useful quantity to study in what follows.} It
is worth noting, though, that although the perturbative and DDT
results had an implied kinematic cutoff at $Q_T=Q$ (and in the
case of DDT resummation a similar restraint on the $k_T$ of each
emitted boson), no such requirement was built into this final
resummed result, and one must integrate over all $Q_T$ to recover
the correct normalization. In other words, while $3$-momentum
conservation was observed, energy conservation was not.

Imposition of such a constraint, if done smoothly, would tend to
reduce the smearing effect, but it is not the only effect yet to
be included. There remains the possibility (especially in QCD
applications) that the incoming reactants may have an
``intrinsic'' transverse momentum, a nonperturbative effect which
would again broaden the spectrum. Finally, there is the
expectation that, due to the approximations which allow the
above-described exponentiation to work, the correct shape of the
distribution at higher $Q_T$ (close to $Q$) would be better
described by the original perturbative result, as the logs in that
region do not overwhelm the expansion parameter. A proper
description of transverse momentum spectra will necessarily be a
delicate balance of perturbative, nonperturbative, and
resummation effects (see figure \ref{fig:pause}), and in the
chapters to come these will all be explored.

\figboxa{pause}{Three regions of interest.}

\section{Application to QCD}

Parisi and Petronzio applied their method to a QCD example, $pp
\rightarrow \mu^+ \mu^- + X,$ and compared the result with data.
Due to the differing structure of the theory, there are details
present here which do not apply to the QED case. Some of these
are irrelevant to their argument and will be ignored. Others
constitute new features of the procedure, or pitfalls that would
otherwise constrain its effectiveness, and these will be
discussed.

The structure of the resummed calculation is the same as that of
the previous section (equation \ref{f2mu222}), although here the
desired quantity is

\begin{eqnarray}
E {d^3\sigma \over d^3Q_T} {\Bigm|}_{y=0}^{\rm resummed} &=&
{1\over\pi}{d^3\sigma \over d{Q_T}^2 dQ dy} {\Bigm|}_{y=0}^{\rm
res} \; \cr  &\equiv& {\sigma_0 \over {(2\pi)}^2} \int d^2b \
e^{-i \vec b \cdot \vec {Q_T}} \ e^{\Delta(b)} \;,
\label{f2mu225d}
\end{eqnarray}

\noindent where the free variables are not only the mass $Q$ and
transverse momentum $Q_T$ of the muon pair, but also the rapidity
$y$ (although they look only at the $y=0$ region).
\footnote{$y={1\over 2} \ln (q^0+q^3)/(q^0-q^3)$, where
$q^\mu=p_{\mu^-}^\mu + p_{\mu^+}^\mu$.} The ``Sudakov'' exponent
$\Delta$ is still a transform to $b$-space, but as it contains the
effects of higher-order corrections which now depend not on the
QED coupling but the running QCD coupling $\alpha_s(k_T)$, the
latter cannot in general be pulled out of the integral. There is
instead

\BQN \Delta(b) = \int d^2k_T e^{i \vec b \cdot \vec {k_T}}
{\alpha_s ({k_T}^2) \over 2\pi} \nu_+ ({k_T}^2) \;, \EQN

\noindent in which $\nu_+$, like the Born cross section
$\sigma_0$, is defined by the unresummed perturbative result
after cancellation of the poles:

\begin{equation}
{1\over \pi}{d^3\sigma \over d{k_T}^2 dQ dy} {\Bigm|}_{y=0} =
\sigma_0 [\delta({k_T}^2) + {{\alpha_s ({k_T}^2)}\over {2\pi}}
\nu_+ ({k_T}^2)] \;. \label{f2mu229}
\end{equation}

The structure of the subprocess diagrams is the same as well, in
that to leading order there is but fermion-antifermion
annihilation, leading to an intermediate vector boson, which
subsequently decays to the muon pair. At next-to-leading order,
radiation from the initial state is included. The relevant
diagrams have already been shown in the Introduction, figures
\ref{fig:six1a}, \ref{fig:six1c}, \ref{fig:six1b}. \footnote{For
simplicity, the Compton process of figure \ref{fig:six1d}, in
which a quark from one proton interacts with a gluon from the
other, will not be considered here. The absence of a soft pole in
this case leads only to subleading terms which are either also
present in $q\bar{q}$ annihilation or which the current formalism
has no mechanism to express. These terms will be discussed later
in this chapter, and in Chapter \ref{ch:CSS}.}

Here, the fundamental interacting particles are the quarks which
make up the colliding protons, and so parametrizations of these
parton densities will necessarily be required. In the low-$Q_T$
approximation to which the resummation formalism applies, it can
be shown that these partons' momentum fractions depend solely
upon the rapidity of the muon pair and its mass relative to the
center-of-mass energy of the proton-proton system $\sqrt{S}$. In
this center-of-mass frame, the two momentum fractions are

\BQN x_{a \atop b} = {Q\over \sqrt{S}} e^{\pm y} \;, \EQN

\noindent and $\sigma_0$ will necessarily include sums over
contributing flavors of the form

$$ \sum_f Q_f^2 q_f^{\rm P} (x_a) {\bar{q}}_f^{\rm P} (x_b) \;, $$

\noindent in which $Q_f$ is the charge and $q_f$ the density of
flavor $f$. The sum extends over flavors and antiflavors, and
color and spin sums must, of course, also be taken into account.

The details had also been calculated by Altarelli, {\it et.al.}
(1978) ~\cite{APP78}, among others, and showed the expected
logarithms of $Q/Q_T$. The difference, relative to the rough model
of the previous section, is that in practice the coefficients of
these logarithms are particular to the subprocess under study. For
Drell-Yan production these are $2C_F$ and $-3C_F$, where
$C_F=4/3$ is a factor arising from the color structure involved.
Specifically, the details are:

\BQA \nu_+ (k_T^2) &=&  2C_F {\Bigl[{{\ln {Q^2/k_T^2}}\over k_T^2}
\Bigr]}_+ -3C_F {\Bigl[{1\over k_T^2} \Bigr]}_+ \; \cr \sigma_0
&=& {{16 \alpha^2} \over {9QS}} \sum_f Q_f^2 q_f^{\rm P} (x_a)
{\bar{q}}_f^{\rm P} (x_b) \;. \EQA

Before this calculation could be compared to data, a potential
problem in the transform to $b$-space needed to be
circumnavigated. This transform involves an integral over the
range $0\leq k_T^2 \leq Q^2$, but the running coupling $\alpha_s
(k_T^2)$ diverges at the small scale $\Lambda$:

\BQN \alpha_s (k_T^2) = {12\pi \over {(33-2N_f) \ln (k_T^2 /
\Lambda^2)}} \;. \EQN

\noindent This problem was avoided by replacing $k_T^2/\Lambda^2$
with $(k_T^2 + M^2)/\Lambda^2$, where $M$ is some scale greater
than $\Lambda$, but small enough that at large $k_T$ the
replacement doesn't affect the result.

Parisi and Petronzio then dealt with the possibility that the
annihilating partons might carry an intrinsic transverse
momentum. This would be a small effect, observable only at very
low $Q_T$ of the pair, and would tend to ``smear'' the resulting
low-$Q_T$ distribution further, as the additional momentum could
either add to or subtract from that provided by the radiative
effects. This is most easily incorporated in impact space by
tacking on a multiplicative factor to the $b$-space integrand:

\begin{equation} e^{\Delta(b)}  \rightarrow e^{\Delta(b)}
e^{-{1\over 4} b^2 <{{p_T}^2}_{\rm int}>} \;. \label{f2mu226}
\end{equation}

The Gaussian width is controlled by $<{{p_T}^2}_{\rm int}>$, the
average intrinsic $p_T^2$ assigned to the incoming partons. By
damping out the high-$b$ region, the effect of the
``high-frequency'' components is reduced and the momentum
distribution softened somewhat, as necessary. This is a simple
example of a ``nonperturbative parametrization" which was to be
more fully developed in later years, as the following chapter
will show.

Utilizing the scale-dependent distribution functions of the CTEQ
collaboration ~\cite{CTEQpp}, and with the small values
$<{{p_T}^2}_{\rm int}>$ = 1.2 GeV$^2$ and $M^2$ = 1.25 GeV$^2$,
one can achieve close agreement with the data of reference
~\cite{Yoh78} (see figure \ref{fig:six3}). Here $S = 750 {\rm
GeV}^2$, $Q^2 = 56 {\rm GeV}^2$, and a conversion factor ${\rm C}
\simeq 389 {\rm nbarn GeV}^2$ was applied to show the results in
units ${\rm nbGeV}^{-2}$. At higher center-of-mass energies,
nonperturbative effects are dwarfed by perturbative ones, the
results approach independence of the phenomenological parameters,
and the amount of intrinsic $p_T$ needed to reproduce the data is
considerably less.

\figboxf{six3}{$pp \rightarrow \mu^+ \mu^- X$ resummed
distribution.}

\clearpage

\section{Subleading Terms}
\label{sec:sublead}

One final note: not all next-to-leading order terms have been
included in this calculation. For both the gluon-radiative and
Compton contributions, there exist remnants of collinear pole
cancellations which go as ${({1\over Q_T^2})}_+$ times
convolutions of parton densities with splitting functions. These,
along with other finite terms, are not handled in the method of
Parisi and Petronzio, which is focused primarily on consolidating
the most singular terms.

In 1984, Altarelli, Ellis, Greco, and Martinelli (AEGM)
~\cite{AEGM84} revisited the $b$-space formalism with the goal of
including these terms. To start, they recast the calculation
within the framework of the {\it subtraction method}, which not
only performs the resummation operation upon low-$Q_T$
approximations of the NLO terms, but then subtracts these
approximations from the NLO terms to create additional finite
pieces without double-counting.

In addition to these, AEGM found a way to include the collinear
remnants mentioned above. Specifically, after transforming to
$b$-space, there are terms of the form

\BQA q_f(x,\mu^2) &+& {\alpha_s \over {2\pi}} \Biggl[ \ln
{Q_{Tmax}^2 \over \mu^2} + \int_0^{Q_{Tmax}^2} d^2Q_T e^{i\vec{b}
\cdot \vec{Q_T}} {\biggl( {1\over Q_T^2} \biggl)}_+ \Biggr] \bigl[
q_f(x,\mu^2) \circ P_{f \tilde{f}}(z) \bigr] \; \cr &+& {\alpha_s
\over {2\pi}} \bigl[ q_f(x,\mu^2) \circ C_f(z) \bigr] \;, \EQA

\noindent in which $q_f(x,\mu^2)$ is a parton density of flavor
$f$, evaluated at momentum fraction $x$ and scale $\mu$. $P_{f
\tilde{f}} (z)$ is an appropriate splitting function, and
$C_f(z)$ contains other finite terms. The symbol $[q_f \circ P]$
refers to a convolution of the parton density with the function
$P$ (similarly for $[q_f \circ C]$):

\BQN q_f(x,\mu^2) \circ P_{f \tilde{f}}(z) \equiv \sum_{\tilde{f}}
\int_x^1 {dz\over z} q_{\tilde{f}} \biggl( {x\over z},\mu^2
\biggr) P_{f \tilde{f}}(z) \;. \EQN

By defining a new factorization scale

\BQN \ln M_f^2 \equiv \ln Q_{Tmax}^2 + \int_0^{Q_{Tmax}^2} d^2Q_T
e^{i\vec{b} \cdot \vec{Q_T}} {\biggl( {1\over Q_T^2} \biggl)}_+
 \;, \EQN

\noindent they were able to absorb the remaining plus-distribution
terms into the parton densities via rescalings

\BQN q_f(x,M_f^2) = q_f(x,\mu^2) + {{\alpha_s(\mu^2)}\over
{2\pi}} \ln {M_f^2\over \mu^2} \sum_{\tilde{f}} \int_{x}^1
{dz\over z} q_{\tilde{f}}({x \over z} ,\mu^2) P_{f \tilde{f}} (z)
\;. \EQN

The remaining $[q_f \circ C]$ convolutions were left alone (and
not exponentiated), although they too were now evaluated at the
new scale $M_f^2$. As it turns out, $M_f \sim b_0/b$, which is the
value suggested almost concurrently by the work of Collins,
Soper, and Sterman (CSS), who used renormalization methods to
solidify the theoretical basis of the $b$-space formalism, and
extend it in a systematic way to include subleading tiers of
logarithms. We will study the CSS work in more detail in Chapter
\ref{ch:CSS}.

\chapter{The CSS Formalism}
\label{ch:CSS}

\section{Derivation}

In 1984, Collins, Soper, and Sterman (CSS) used renormalization
group methods to arrive at an improved resummation scheme, one
which not only preserves momentum conservation, but which can be
made increasingly more precise as additional terms are calculated
~\cite{CSS85}. Although derived for Drell-Yan processes, the
result is, in principle, generally applicable to all two-scale
processes.

It will be helpful here to recall a bit of the physical picture:
at leading order, two partons (one from each incoming hadron)
collide and produce a measured final state, of mass $Q$ and
transverse momentum $Q_T$. In the case of Drell-Yan production,
this final state is a pair of leptons. As the incoming partons are
assumed to be collinear with each other and the hadronic beam, if
no other particles are observed, there is nothing for the system
to recoil against, and one predicts that $Q_T$ will equal zero.
Similarly, by energy conservation, the mass of the system of
colliding partons (as measured in the same frame) must also have
been $Q$, each parton contributing a fraction of this energy.

In general, of course, other particles are generated from this
partonic collision, and one can go to higher orders in
perturbation theory to predict more accurately the
$Q_T$-dependence as measured in the lab. As noted in Chapter
\ref{ch:PPResum}, these predictions are plausible, without
modification, at high-$Q_T$, but in the limit $Q_T\ll Q$, the
biggest contributions come from cases in which the additional
particles are either too soft to individually detect or are
themselves nearly collinear with one or the other incoming parton.
Resummation is then required for a complete description.

Fortunately, in this limit the long-distance (low energy) and
short-distance (high energy) phenomena become roughly independent
of one another, and one can expect a factorization of the
cross-section as follows:

\begin{eqnarray}
\sigma_S &\sim&  \int d^2 k_{Ta} \int d^2 k_{Tb} \int d^2 k_{TS}
\delta^2 \bigl( \vec{Q_T} - \sum_i \vec{k_{Ti}} \bigr) \; \cr
&\times& J_a \Bigl( {p_a^0 \over \mu}, {k_{Ta} \over \mu}, g(\mu)
\Bigr) J_b \Bigl( {p_b^0 \over \mu}, {k_{Tb} \over \mu}, g(\mu)
\Bigr) \; \cr &\times& S \Bigl( {k_{TS} \over \mu}, g(\mu) \Bigr)
H \Bigl( {p_a^0 \over \mu}, {p_b^0 \over \mu}, {Q \over \mu},
g(\mu) \Bigr) \;. \label{CSS01}
\end{eqnarray}

Here high-energy quantities such as $Q$ and the parton energies
$p_{a,b}^0$ organize themselves within a hard scattering factor
$H$, while small, non-collinear transverse momenta $k_{TS}$
contribute to a soft function $S$. Associated with each incoming
parton (and the nearly collinear quanta $k_{Ti}$ surrounding it)
is a non-perturbative factor $J_i$, which remains dependent upon
both low and high-energy quantities ($k_{Ti}$ and $p_i^0$
respectively). For the transverse momenta shown, there is a
delta-function included which relates their vector sum to the
$\vec{Q}_T$ of the observed system. The hard quantities are
related to each other in a way: in the limit described, both
$p_a^0$ and $p_b^0$ are expressible as simple fractions of $Q$.
Thus the functions $J_{a,b}$ have an implicit dependence upon $Q$
which will become important in what follows. Each function depends
upon the QCD coupling strength $g(\mu)$ and its argument, the
renormalization scale $\mu$; the fact that the cross section
itself does not depend upon $\mu$ will also be of import.

Proof of such factorizations can be found, for example, in
references ~\cite{CS81,CSS89}; the particular choice of variables
which express the high and low-energy scales is different, but the
general argument is the same. The crucial point is that
factorization, coupled with invariance of the cross section under
certain transformations, leads to renormalization group
properties of the factored functions, and thus to exponentiation.
An all-orders resummation of the leading terms is the end result,
the corresponding leading coefficients determined by comparison
with the perturbatively-calculated result at fixed order.

Equation \ref{CSS01} above is in the form of a multiple
convolution; that is, only the matrix element, not the phase
space, has factorized. Significant simplification arises after a
Fourier transform to impact parameter space,

\BQN \int d^2Q_T e^{i\vec{b} \cdot \vec{Q_T}} \delta^2 (\vec{Q_T}
- \sum_i \vec{k}_{Ti} ) = e^{i \vec{b} \cdot \sum_i \vec{k}_{Ti}}
= \prod_i e^{i\vec{b} \cdot \vec{k}_{Ti}} \;, \label{CSS02} \EQN

\noindent in which the impact parameter is the two-dimensional
axial vector $\vec{b}$. Then, for each transverse momentum
$\vec{k}_{Ti}$, one obtains

\BQN \int d^2k_{Ti} e^{i\vec{b} \cdot \vec{k}_{Ti}} f(k_{Ti}) =
f(b) \;. \label{CSS03} \EQN

\noindent This produces the simple product

\begin{equation}
{\tilde{\sigma}}_S (b,Q) = {\tilde{J}}_a \Bigl( {p_a^0 \over
\mu}, b\mu, g(\mu) \Bigr) {\tilde{J}}_b \Bigl( {p_b^0 \over \mu},
b\mu, g(\mu) \Bigr) \tilde{S} \Bigl( b\mu, g(\mu) \Bigr)
\tilde{H} \Bigl( {p_a^0 \over \mu}, {p_b^0 \over \mu},{Q\over \mu}
, g(\mu) \Bigr) \;, \label{CSS04}
\end{equation}

\noindent where the tilde above each function is a reminder of the
transform that has taken place.

Taking the logarithmic derivative of both sides with respect to
the large scale $Q^2$, CSS arrived at

\BQN {{\partial\ln{\tilde{\sigma}}_S}\over {\partial\ln Q^2}} =
{{\partial\ln{\tilde{J}}_a}\over {\partial\ln Q^2}} +
{{\partial\ln{\tilde{J}}_b}\over {\partial\ln Q^2}} +
{{\partial\ln\tilde{H}}\over {\partial\ln Q^2}} \;. \label{CSS05}
\EQN

\noindent The logarithms of ${\tilde{J}}_a$ and ${\tilde{J}}_b$
now contain explicit dependences upon $Q$, and, as shown in
references ~\cite{CS81,CSS89,CLS96}, each derivative thereof
contributes to two separate quantities, one a function of $b\mu$
and $g(\mu)$ only, the other a function of $Q/\mu$ and $g(\mu)$
only:

\BQN {{\partial \ln {\tilde{J}}_a}\over {\partial \ln Q^2}} +
{{\partial \ln {\tilde{J}}_b}\over {\partial \ln Q^2}} = K \bigl(
b\mu,g(\mu) \bigr) + G_J \bigl( Q/ \mu, g(\mu) \bigr) \;.
\label{CSS06} \EQN

\noindent As $\tilde{H}$ doesn't depend on $b$, its derivative
contributes only to a function of $Q/\mu$ and $g(\mu)$:

\BQN {{\partial \ln \tilde{H}}\over {\partial \ln Q^2}} = G_H
\bigl( Q/ \mu, g(\mu) \bigr) \;. \EQN

\noindent Adding $G_J$ and $G_H$ to form a new function $G$, one
arrives at

\begin{equation}
{{\partial\ln{\tilde{\sigma}}_S}\over {\partial\ln Q^2}} = K
\bigl( b\mu,g(\mu) \bigr )+G \bigl( Q/\mu,g(\mu) \bigr)  \;.
\label{CSS07}
\end{equation}

CSS then made use of the fact that the cross section is
independent of renormalization scale $\mu$:

\BQN {\partial\over {\partial\ln Q^2}} {{\partial\ln
{\tilde{\sigma}}_S}\over {\partial\ln \mu}} = {\partial\over
{\partial\ln \mu}} {{\partial\ln {\tilde{\sigma}}_S}\over
{\partial\ln Q^2}} = {\partial\over {\partial\ln \mu}} \bigl[ K+G
\bigr] = 0 \;, \label{CSS08} \EQN

\noindent and so, since $K$ and $G$ share only a dependence on
$g$, it must be that there exists some function $\gamma$ such that

\begin{equation}
{{\partial K}\over {\partial\ln\mu}} = - {{\partial G}\over
{\partial\ln\mu}} \equiv -\lambda(g(\mu)) \;. \label{CSS09}
\end{equation}

\noindent All the $b$-dependence is in $K$, while all the
$Q$-dependence is in $G$, and both can be independently scaled,
since both satisfy their own evolution equations.

It will be instructive here to stop and see why this is
important. In a perturbative calculation, after dealing with the
soft and collinear poles, the cross section can be written as a
sum of two series in $\alpha_s / 2 \pi$:

\begin{equation}
\sigma = \sum_{N =0}^\infty {\bigl[ {\alpha_s(g(\mu)) \over
{2\pi}}\bigr]}^N \Bigl[ \sigma_S^{(N)} + \sigma_R^{(N)} \Bigr] \;,
\label{CSS09b}
\end{equation}

\noindent in which the coefficients are either integrably
divergent as $Q_T \rightarrow 0$ (here $\sigma_S^{(N)}$, the
``singular'' piece) or are zero in this limit and thus pose no
threat to the convergence of the expansion ($\sigma_R^{(N)}$, the
``regular'' piece). \footnote{This structure results from the use
of the subtraction method (see Section \ref{sec:sublead}).
Asymptotic limits of the fixed-order result (that is, the logs of
$Q/Q_T$) are subtracted to form $\sigma_R^{(N)}$, and then added
back in to form $\sigma_S^{(N)}$ after pole-cancellation.} The
former series ($\sigma_S^{(N)}$) corresponds to the cross section
we began with in equation \ref{CSS01}, and we know something about
its perturbative structure:

\begin{eqnarray}
\sigma_S^{(N)} &=& T_0^{(N)}(Q/\mu,g(\mu)) \delta({\vec Q_T}) \cr
&+& \sum_{m=0}^{\rm 2N-1} T^{(N,m)}(Q/\mu,g(\mu))
{\Bigl[{{\ln^m(Q^2/Q_T^2)} \over Q_T^2} \Bigr]}_+  \;.
\end{eqnarray}

\noindent Here the ``plus-distribution'' ${\bigl[ \quad \bigr]}_+$
denotes a regularization such that

\BQA \int_0^{Q^2} dQ_T^2 {\Bigl[{{\ln^m(Q^2/Q_T^2)} \over Q_T^2}
\Bigr]}_+ f(Q_T) &\equiv& \int_0^{Q^2} dQ_T^2
{\Bigl[{{\ln^m(Q^2/Q_T^2)} \over Q_T^2} \Bigr]} \Bigl[
f(Q_T)-f(0) \Bigr] \; \cr {\rm or} \; \cr \int_0^{p_T^2} dQ_T^2
{\Bigl[{{\ln^m(Q^2/Q_T^2)} \over Q_T^2} \Bigr]}_+ &=& -{1 \over
{m+1}} \ln^{m+1}(Q^2/p_T^2) \;. \EQA

CSS realized that, once the Fourier transform to $b$-space is
performed, as in equation \ref{CSS04}, the large logs of $Q/Q_T$
become large logs of $Qb$, or equivalently, logs of $Q/\mu$ and
$b\mu$:

\begin{equation}
\int d^2 Q_T e^{-i {\vec Q_T} \cdot {\vec b}}
{\Bigl[{{\ln^m(Q^2/Q_T^2)} \over Q_T^2} \Bigr]}_+ \sim
\sum_{n=0}^{m+1} \ln^n (Qb) \;.
\end{equation}

\noindent Since these large logs occur, the expansion doesn't
converge, and one can't even approximate $d{\tilde\sigma}_S$ well
by the sum of the leading log terms ($m=2N-1$).

Fortunately, given the separability of these logarithms (as shown
in equation \ref{CSS07}) and the independent renormalization
groups of $K$ and $G$ (equation \ref{CSS09}), CSS were able to do
something about this. \footnote{The derivation to be embarked
upon is rather involved, and the casual reader may wish to skip
to equation \ref{CSSpert}.} They scaled $\mu$ in $K$ up to order
$1/b$, and $\mu$ in $G$ to order $Q$. Both the large logs of
$b\mu$ in $K$ and the logs of $Q/\mu$ in $G$ then tended to $0$.
In practice, this need not be exact, and CSS allowed for variation
by using $c_1/b$ and $c_2Q$, where $c_1$ and $c_2$ are both of
order 1. The canonical choice for these constants is
$c_1=b_0\equiv 2e^{-\gamma_E}\simeq 1.123$, $c_2=1$. From
equation \ref{CSS09}:

\begin{eqnarray}
K(c_1,g(c_1/b))-K(b\mu,g(\mu)) &=& -\int_{\mu}^{c_1/b}
{{d\bar{\mu}} \over \bar{\mu}} \gamma(g(\bar{\mu})) \; \cr
G(1/c_2,g(c_2Q))-G(Q/\mu,g(\mu)) &=& \int_{\mu}^{c_2Q}
{{d\bar{\mu}} \over \bar{\mu}} \gamma(g(\bar{\mu})) \;,
\label{CSS10}
\end{eqnarray}

\noindent and thus

\begin{eqnarray}
[K(c_1,g(c_1/b))+G(1/c_2,g(c_2Q))] &-&
[K(b\mu,g(\mu))+G(Q/\mu,g(\mu))] \cr &=&
\int_{c_1^2/b^2}^{c_2^2Q^2} {d\bar{\mu}^2 \over \bar{\mu}^2}
{\gamma(g(\bar{\mu})) \over 2} \;. \label{CSS11}
\end{eqnarray}

\noindent Now, for any function $F(b,Q)$,

\begin{equation}
{dF(b,Q) \over {d(1/b^2)}} = {\partial F(b,Q) \over {\partial
\ln(1/b^2)}} {d\ln(1/b^2) \over {d(1/b^2)}} = {1 \over {(1/b^2)}}
{\partial F(b,Q) \over {\partial \ln(1/b^2)}} \;,
\end{equation}

\noindent so

\begin{eqnarray}
\int_{F(b,Q)}^{F(c_1/c_2Q,Q)} dF(\bar{b},Q) &=&
F(c_1/c_2Q,Q)-F(b,Q) \cr &=& \int_{1/b^2}^{c_2^2Q^2/c_1^2}
{d(1/{\bar b}^2) \over {(1/{\bar b}^2)}} {\partial F({\bar b},Q)
\over {\partial \ln(1/{\bar b}^2)}} \;.
\end{eqnarray}

\noindent If one takes $F(b,Q) \equiv
K(b\mu,g(\mu))+G(Q/\mu,g(\mu))$, it follows that

\begin{eqnarray}
K(b\mu)+G(Q/\mu) &=& K(c_1\mu/c_2Q,g(\mu))+G(Q/\mu,g(\mu)) \; \cr
&-& \int_{1/b^2}^{c_2^2Q^2/c_1^2} {d(1/{\bar b}^2) \over {(1/{\bar
b}^2)}} {\partial[K({\bar b}\mu)+G(Q/\mu)] \over {\partial
\ln(1/{\bar b}^2)}} \;, \label{CSS14}
\end{eqnarray}

\noindent where

\BQN {\partial[K({\bar b}\mu)+G(Q/\mu)] \over {\partial
\ln(1/{\bar b}^2)}} = {\partial[K(c_1,g(c_1/{\bar
b}))+G(1/c_2,g(c_2Q))] \over {\partial \ln(1/{\bar b}^2)}} +
{\gamma(g(c_1/{\bar b})) \over 2}\;, \EQN

\noindent by equation \ref{CSS11} and the theorem

\BQN {\partial \over {\partial x}} \int_a^x f(t) dt = f(x) \;,
\EQN

\noindent with $x \equiv c_1^2/{\bar b}^2$, $t \equiv
\bar{\mu}^2$, and $a \equiv c_2^2Q^2$. Meanwhile,

\BQN {{\partial[K(c_1,g(c_1/{\bar b}))+G(1/c_2,g(c_2Q))]} \over
{\partial \ln(1/{\bar b}^2)}} = \beta(g(1/\bar{b}^2)) {{\partial
K(c_1,g(1/\bar{b}^2))} \over {\partial g(1/\bar{b}^2)}} \; \EQN

\noindent for $\beta(g(x)) \equiv {{d g(x)} \over {d\ln x}}$, and

\BQN K(c_1\mu/c_2Q,g(\mu))+G(Q/\mu,g(\mu)) =
K(c_1,g(c_2Q))+G(1/c_2,g(c_2Q)) \;, \EQN

\noindent by equation \ref{CSS11} again, with $c_1/b \rightarrow
c_2Q$.

Performing the change of variables $\bar{\mu} \equiv c_1/{\bar
b}$ in the integral of equation \ref{CSS14}:

\begin{equation}
\int_{1/b^2}^{c_2^2Q^2/c_1^2} {d(1/{\bar b}^2) \over {(1/{\bar
b}^2)}} \rightarrow \int_{c_1^2/b^2}^{c_2^2Q^2}{d\bar{\mu}^2
\over \bar{\mu}^2} \;,
\end{equation}

\noindent CSS derived, in all,

\begin{equation}
K(b\mu,g(\mu))+G(Q/\mu,g(\mu)) =
-\int_{c_1^2/b^2}^{c_2^2Q^2}{d\bar{\mu}^2 \over \bar{\mu}^2}
A(c_1,g(\bar{\mu})) - B(c_1,c_2,g(c_2Q)) \;,
\end{equation}

\noindent where

\begin{eqnarray}
A(c_1,g(\bar{\mu})) &\equiv& {\gamma(g(\bar{\mu})) \over 2} + {1
\over 2} \beta(g(\bar{\mu})) {\partial K(c_1,g(\bar{\mu})) \over
{\partial g(\bar{\mu})}} \; \cr B(c_1,c_2,g(c_2Q)) &\equiv& -
[K(c_1,g(c_2Q))+G(1/c_2,g(c_2Q))] \;.
\end{eqnarray}

\noindent Thus, returning to equation \ref{CSS07},

\begin{equation}
{{\partial \ln \tilde{\sigma}_S(b,Q)} \over {\partial \ln Q^2}}  =
-  \int_{c_1^2/b^2}^{c_2^2Q^2} {d\bar{\mu}^2 \over \bar{\mu}^2}
A(c_1,g(\bar{\mu})) - B(c_1,c_2,g(c_2Q)) \;, \label{CSS19}
\end{equation}

\noindent where, since $K$,$G$,$\gamma$, and $\beta \partial K /
\partial g$ all have perturbative expansions in $\alpha_s$, so do
$A$ and $B$.

Treating equation \ref{CSS19} as an ordinary differential equation
in $c_2Q$ with a parameter $b$, CSS found the solution, and
generated the Sudakov exponent, as follows. Separating variables
and integrating both sides from $c_1/b$ to $c_2Q$ gives:

\begin{eqnarray}
\int_{c_1/b}^{c_2Q} d\ln {\tilde{\sigma}}_S (b,\bar{Q}) &=& \ln
{\tilde\sigma}_S(b,c_2Q) -\ln  {\tilde\sigma}_S(b,c_1/b) \equiv
-S(b,Q) \cr {\rm or} \cr {\tilde\sigma}_S (b,Q) &=&
{\tilde\sigma}_S(b,c_1/c_2b) e^{-S(b,Q)} \;, \label{CSS46}
\end{eqnarray}

\noindent where

\begin{equation}
S(b,Q) \equiv \int_{\ln(c_1/b)}^{\ln(c_2Q)} d\ln(c_2\bar{Q})
\Bigl[ \int_{c_1^2/b^2}^{c_2^2{\bar{Q}}^2} {d\bar{\mu}^2 \over
\bar{\mu}^2} A(\bar{\mu}) + B(c_2\bar{Q}) \Bigr] \;.
\end{equation}

\noindent This Sudakov exponent can be simplified as follows. The
first term,

\begin{equation} \int_{\ln(c_1/b)}^{\ln(c_2Q)} d\ln(c_2\bar{Q})
\int_{c_1^2/b^2}^{c_2^2{\bar{Q}}^2} {{d\bar{\mu}^2} \over
\bar{\mu}^2} A(\bar{\mu}) = 2 \int_{c_1/b}^{c_2Q} {{d\bar{\mu}}
\over \bar{\mu}} \int_{c_1/b}^{\bar{\mu}} {{d{\bar{\mu}'}} \over
{\bar{\mu}'}} A({\bar{\mu}'}) \;,
\end{equation}

\noindent followed by an integration by parts gives

\begin{eqnarray}
\int &=& 2  \ln\bar{\mu} \int_{c_1/b}^{\bar{\mu}} {d{\tilde{\mu}}
\over {\tilde{\mu}}} A({\tilde{\mu}})  \Bigr|_{c1/b}^{c_2Q} - 2
\int_{c1/b}^{c_2Q}{d\bar{\mu} \over \bar{\mu}} \ln\bar{\mu}
A(\bar{\mu})  \; \cr &=& 2 \ln{c_2Q} \int_{c1/b}^{c_2Q}
{d{\tilde{\mu}} \over {\tilde{\mu}}} A({\tilde{\mu}}) -
2\ln{c_1/b} \int_{c_1/b}^{c_1/b}{d{\tilde{\mu}} \over
{\tilde{\mu}}} A({\tilde{\mu}}) - 2\int_{c1/b}^{c_2Q} {d\bar{\mu}
\over \bar{\mu}} \ln\bar{\mu} A(\bar{\mu})  \cr &=&
\int_{c_1^2/b^2}^{c_2^2Q^2}{d\bar{\mu}^2 \over \bar{\mu}^2}
\ln\bigl( {c_2^2Q^2 \over \bar{\mu}^2}\bigr) A(\bar{\mu}) \;,
\end{eqnarray}

\noindent while

\begin{equation}
\int_{\ln(c_1/b)}^{\ln(c_2Q)} d\ln(c_2\bar{Q}) B(c_2\bar{Q}) = {1
\over 2}\int_{c_1^2/b^2}^{c_2^2Q^2}{d\bar{\mu}^2 \over
\bar{\mu}^2} B(\bar{\mu}) \;,
\end{equation}

\noindent and thus

\begin{equation}
S = \int_{c_1^2/b^2}^{c_2^2Q^2}{d\bar{\mu}^2 \over \bar{\mu}^2}
\Bigl[ \ln\bigl( {c_2^2Q^2 \over \bar{\mu}^2}\bigr) A(\bar{\mu})
+ B(\bar{\mu}) \Bigr] \;. \label{CSS52}
\end{equation}

Evaluating \ref{CSS04} with $Q=c_1/c_2b$ gives the coefficient of
the Sudakov exponential in equation \ref{CSS46}. Pulling out the
parton distributions from ${\tilde{J}}_{a,b}$ and rewriting the
rest as separable functions $C_{i/\tilde{i}}$, CSS obtained :

\begin{eqnarray}
{\tilde{\sigma}}_S(b,\mu) &=& \sum_{\tilde{a}} \int_{x_1^0}^1
{dx_1 \over x_1} f_{\tilde{a}/A} (x_1,\mu) C_{a/\tilde{a}}
(x_1^0/x_1,b,c_1/c_2,g(\mu)) \cr &\times& \sum_{\tilde{b}}
\int_{x_2^0}^1 {dx_2 \over x_2} f_{\tilde{b}/B} (x_2,\mu)
C_{b/\tilde{b}} (x_2^0/x_2,b,c_1/c_2,g(\mu)) \;, \label{CSS54}
\end{eqnarray}

\noindent for which they, in practice, set the scale $\mu=c_1/b$.

To regain a useful expression in $Q_T$-space, CSS performed the
inverse Fourier transform on equation \ref{CSS46}, added back in
the finite remainders $\sigma_R$ from equation \ref{CSS09b}, and
arrived at their final result for the perturbative region

\begin{equation}
\sigma = \Biggl\{ {1\over{2\pi}^2} \int d^2 \vec{b} e^{i\vec{b}
\cdot \vec{Q_T}} {\tilde{\sigma}}_S (b,c_1/c_2b) e^{-S(b,Q)}
\Biggr\} + \sigma_F \;, \label{CSSpert}
\end{equation}

\noindent in which

\begin{equation}
S = \int_{c_1^2/b^2}^{c_2^2Q^2}{d\bar{\mu}^2 \over \bar{\mu}^2}
\Bigl[ \ln\bigl( {c_2^2Q^2 \over \bar{\mu}^2}\bigr) A(\bar{\mu})
+ B(\bar{\mu}) \Bigr] \; \label{CSSsud}
\end{equation}

\BQN \sigma_F = \sum_{a,b} \int_{x_1^0}^1 {dx_1 \over x_1}
\int_{x_2^0}^1 {dx_2 \over x_2}  \sum_{\rm N=1}^\infty {\bigl[
{\alpha_s(\mu) \over 2\pi}\bigr]}^N \sigma_R^{(N)}
(a,b,Q_T,Q,\mu,{x_1^0 \over x_1},{x_2^0 \over x_2}) \;.
\label{CSSfin} \EQN

CSS resummation allows for the resummation of tiers of logarithms
two at a time. That is, given the first $N$ $\{A,B,C\}$
coefficients, one can resum the first $2N$ tiers of logarithms
over all orders. Implicitly, there occurs a reorganization of the
logs ($L \equiv \ln (Q/Q_T)$) such that successive orders grow by
$\alpha_s$ as opposed to $\alpha_s L^2$:

\begin{eqnarray}
\sigma &\sim& \alpha_s {\color{red}(L+1)} \cr &+& \alpha_s^2
\biggl[ {\color{red}(L^3+L^2)}+ {\color{green4}(L+1)} \biggr] \cr
&+& \alpha_s^3 \biggl[
{\color{red}(L^5+L^4)}+{\color{green4}(L^3+L^2)}+{\color{blue}(L+1)}
\biggr] \cr &+& \cdots \nonumber \end{eqnarray}

\begin{eqnarray} \rightarrow &\rightarrow& \alpha_s \biggl[
{\color{red}(L+1)} + \alpha_s {\color{red}(L^3+L^2)} + \alpha_s^2
{\color{red}(L^5+L^4)} + \cdots \biggr] \cr &+& \alpha_s^2
\biggl[ {\color{green4}(L+1)} + \alpha_s
{\color{green4}(L^3+L^2)} + \alpha_s^2 {\color{green4}(L^5+L^4)}
+ \cdots \biggr] \cr &+& \alpha_s^3 \biggl[ {\color{blue}(L+1)} +
\alpha_s {\color{blue}(L^3+L^2)} + \alpha_s^2
{\color{blue}(L^5+L^4)} + \cdots \biggr] \cr &+& \cdots \nonumber
\end{eqnarray}

To obtain these coefficients, one expands the resummed form to
the same order as one has calculated perturbatively, and compares
this expansion with the asymptotic approximation of the
perturbative result. Double-counting is avoided, as in the AEGM
result, by use of the subtraction method. Furthermore, the $C$
coefficients contain the unexponentiated subleading terms
discussed in Section \ref{sec:sublead}.

\section{Extension to the Nonperturbative Region.}
\label{sec:CSSNP}

Strictly, the above result is valid only at low $b$ ($b \ll
1/\Lambda_{QCD}$). At $b > \sim {1\over\Lambda_{QCD}}$, the
coefficients $A$ and $B$ in the Sudakov exponent become dependent
not only on $\alpha_s(\bar{\mu})$ but on the parton masses in the
form $m_f/\bar{\mu}$. A high enough impact parameter is reached
that perturbation theory is not a valid description, and
$\bar{\mu}$ is then allowed to go small enough that both
$g(\bar{\mu})$ and $m_f/\bar{\mu}$ blow up. As Parisi and
Petronzio have shown ~\cite{PP79}, ${\tilde{\sigma}}_S$ is
dominated by $b \simeq {1\over\Lambda_{QCD}}
{({Q\over\Lambda_{QCD}})}^{-0.41} \ll {1\over\Lambda_{QCD}}$ for
large $Q/Q_T$, but we don't with current technology obtain large
enough $Q$ to ignore the $b>{1\over\Lambda_{QCD}}$ region. At low
$b$, $c_1/b$ and hence $\bar{\mu}$ stays high enough that both
$m_f/\bar{\mu}$ and $g(\bar{\mu})$ are small, and one recovers the
perturbative results above.

It is clear that some arrangements must be made to parametrize the
effects of the nonperturbative region, while keeping the
perturbative formalism in use in its region of applicability. CSS
suggested the following: define a function

\begin{equation}
b^*(b) \equiv {b\over \sqrt{1+b^2/b_{max}^2}} \;,
\end{equation}

\noindent which goes no higher than $b_{max}$. Use this $b^*$ for
evaluations of the perturbative $d{\tilde{\sigma}}_S$, and rewrite
the full $d{\tilde{\sigma}}_S$ as

\begin{equation}
d{\tilde{\sigma}}_S (b) = d{\tilde{\sigma}}_S(b^*) \Bigl[
{{d{\tilde{\sigma}}_S(b)} \over {d{\tilde{\sigma}}_S(b^*)}} \Bigr]
\;,
\end{equation}

\noindent where the second factor is approximated by a
parametrization in $b$.

Perhaps the easiest way to predict the form of such a
parametrization is to return to equation \ref{CSS07} and directly
integrate, starting at some minimum scale $Q_0$ at which one would
deem a finite-order expansion in terms of $\alpha_s(Q_0)$
``sufficiently accurate'' ({\it e.g.} $Q_0=c_1/b_{max}$). Dividing
by the result taken at $b=b^*$ yields a ratio of the form
~\cite{CS82}

\BQN {{d{\tilde{\sigma}}_S(b)} \over {d{\tilde{\sigma}}_S(b^*)}}
= \exp \Bigl[ - h_K(b) \ln (Q^2/Q_0^2) - h_A (x_a^0,b) - h_B
(x_b^0,b) \bigr] \;. \EQN

We will hereafter refer to the exponent above as the
nonperturbative function $S_{NP}$.  Like the parton
distributions, the coefficient functions $h$ are intended to be
universal (generally applicable) and extractible from data. The
one constraint is that $exp(S_{NP}) \rightarrow 1$ as $b
\rightarrow b^* \rightarrow 0$.

To date, sufficient data have not been taken to study the flavor
dependence of the functions $h_{A,B}$, nor the dependence upon
$x_a^0$ or $x_b^0$ individually. However, attempts have been made
to fit simplified versions of the above form. Ladinsky and Yuan
(LY) ~\cite{LY94} in 1994 and Landry, Brock, Ladinsky, and Yuan
(LBLY) ~\cite{LBLY99} in 1999 used the three-parameter form shown
here:

\BQN S_{NP} = -g_2 b^2 \ln \Bigl( {Q\over {2Q_0}} \Bigr) - g_1b^2
- g_1g_3b \ln (100x_a^0x_b^0) \;. \EQN

A previous analysis by Davies, Webber, and Stirling (DWS)
~\cite{DWS85} in 1985 was made with a two-parameter form obtained
by setting $g_3=0$ in the above. LBLY also studied the
two-parameter form, and the results of all these efforts are
collected in Table \ref{NPparms} below. In Chapter
\ref{ch:Results}, we will provide evidence that a
single-parameter, $Q$-independent form may agree better with data.

\begin{table}[h]
\caption{Nonperturbative Parameters.} \vspace{.5cm}
\begin{center}
\thicklines
\begin {tabular} { lllll }
\hline \hline \thinlines
 &DWS&LBLY &LY &LBLY \\
\hline \thicklines
$g_1 (GeV^2)$& 0.40&0.24 & 0.11&0.15 \\
$g_2(GeV^2)$& 0.15&0.34 & 0.58 & 0.48 \\
$g_3(GeV^{-1})$& N/A & N/A & -1.5 & -0.58 \\
$Q_0(GeV)$& 2.0& 1.6  & 1.6 & 1.6  \\
$b_{max}(GeV^{-1})$& 0.5 & 0.5 & 0.5 & 0.5 \\
PDF&DO1&CTEQ3M&CTEQ2M&CTEQ3M\\
DATA&E288,R209&E288,R209,&E288,R209,&E288,R209,\\
    &         &CDF-Z,E605&CDF-Z     &CDF-Z,E605\\
\hline \hline
\end{tabular}
\end{center}
\label{NPparms}
\end{table}

In lieu of equation \ref{CSSpert} then, the final CSS result can
be written as:

\begin{equation}
d\sigma = \Biggl\{ {1\over{2\pi}^2} \int d^2 \vec{b} e^{i\vec{b}
\cdot \vec{Q_T}} d{\tilde{\sigma}}_S (b^*,c_1/c_2b^*)
e^{-S(b^*,Q)} e^{S_{NP}(b,Q)} \Biggr\} + d\sigma_F \;.
\label{CSSfull}
\end{equation}

\section{The Sudakov Exponent at very low $b$.}

Now, what about the very low-$b$ region, where $c_1/b > c_2Q$?
Here there are no large logs, but one does not want the Sudakov
exponent to change sign. In practice one may consider cutting off
the exponent, that is, simply taking $\exp[-S(b)] \rightarrow
\exp[0] =1$ for this region. But will this preserve the proper
normalization upon integration over $Q_T$? Altarelli, {\it et.
al.} (1984) ~\cite{AEGM84} calculated the exact first-order
result for the Sudakov form factor and arrived at

\BQN S(b,Q) = {\alpha_s\over {2\pi}} \int_0^{Q^2}
{{d{\bar{\mu}}^2}\over {\bar{\mu}}^2} \Bigl[A^{(1)} \ln \biggl(
{Q^2\over {\bar{\mu}}^2} \biggr) + B^{(1)} \Bigr]
\bigl(J_0(b\bar{\mu})-1 \bigr) \;, \EQN

\noindent which by inspection has the desired property of becoming
zero as $b \rightarrow 0$.

In the proposal of Ellis, {\it et. al.} (1997) ~\cite{ERV97}, the
CSS form for $S$ is maintained, but new scales $\lambda(b)$ and
$\mu(b)$, which never go above $c_2Q$, are used as lower limits:

\begin{equation}
S(b,Q) = \int_{\lambda^2(b)}^{c_2^2Q^2} {{d{\bar{\mu}}^2}\over
{\bar{\mu}}^2} \ln {Q^2\over {\bar{\mu}}^2} A(\alpha_s
(\bar{\mu})) + \int_{\mu^2(b)}^{c_2^2Q^2} {{d{\bar{\mu}}^2}\over
{\bar{\mu}}^2} B(\alpha_s (\bar{\mu})) \;.
\end{equation}

\noindent To agree with the Altarelli result, these scales must be
defined such that

\begin{eqnarray}
\int_\lambda^{c_2Q} {dx \over x} \ln {{c_2Q}\over x} &= {1\over
2} \ln^2 {{c_2Q}\over \lambda} &= \int_0^{c_2Q} {dx \over x} \ln
{{c_2Q}\over x} [1-J_0(bx)] \;, \cr \int_\mu^{c_2Q} {dx \over x}
&= \ln {{c_2Q}\over \mu} &= \int_0^{c_2Q} {dx \over x} [1-J_0(bx)]
\;,
\end{eqnarray}

\noindent and so must be

\BQA \mu(b) &=& Q \exp \Bigl\{ - \int_0^Q {dx\over x} \bigl(
1-J_0(bx) \bigr) \Bigr\} \; \cr \lambda(b) &=& Q \exp \Bigl\{ -
{\biggl[ \int_0^Q {dx\over x} \ln (Q/x) \bigl( 1-J_0(bx) \bigr)
\biggr]}^{1\over 2} \Bigr\} \;. \EQA

\noindent At large $b$, $\lambda$ and $\mu$ both $\rightarrow \sim
b_0/b$, in agreement with CSS.

\chapter{KT-Space Resummation}
\label{ch:ktspace}

As successful as the CSS formalism has been, it is not
necessarily easy to put into practice. Performing Fourier
transforms numerically is a slow process, requiring many
samplings of the parton distributions to achieve a prediction for
a single $Q_T$ value. The impact parameter integral formally
extends to $b=\infty$, but of course one must in practice halt
the integration at some $b_{max}$, and either drop the rest or
introduce an asymptotic expansion ~\cite{AK91}. Moreover, the
differing treatment of asymptotic pieces between resummed and
subtracted versions can affect their cancellation, leading to
unphysical behavior at large-$Q_T$. One must take pains to
smoothly yield to the ordinary perturbative result in this
region. It would be nice if one could perform the Fourier
transform analytically, or avoid it altogether.

A number of authors have undertaken this task, building on the
foundation laid by Dokshitzer, D'yakonov and Troyan (DDT)
~\cite{DDT78,DDTb78}. Their central result is essentially that of
equation \ref{f2mu213}, in a more general two-scale form:

\begin{equation}
{{d\sigma}\over {dQ^2 dQ_T^2 dy}} = {{d\hat{\sigma}}\over {dQ^2
dy}} \sum_{a,b} {d\over {dQ_T^2}} \Bigl[ f_a (x_a,Q_T) T_a
(Q^2,Q_T^2)  f_b (x_b,Q_T) T_b (Q^2,Q_T^2) \Bigr] \;, \label{DDT1}
\end{equation}

\noindent where the momentum fractions and Sudakov exponents are

\begin{eqnarray}
x_{a \atop b} &=& {Q\over \sqrt{S}} e^{\pm y} \; \\
T_{a \atop b} &=& \exp \Bigl[ - \int_{Q_T^2}^{Q^2}
{{d\lambda^2}\over \lambda^2} {{\alpha_s(\lambda^2)}\over {2\pi}}
\biggl( A_{a \atop b} \ln {Q^2\over \lambda^2} + B_{a \atop b}
\biggr) \Bigr] \;. \label{DDT2}
\end{eqnarray}

\section{Kimber, Martin, and Ryskin}

In 1999, Kimber, Martin, and Ryskin ~\cite{KMR99} revisited the
theory behind the DDT equation, and reinterpreted it in terms of
$k_T$-dependent parton distributions $f(x,k_T,Q)$,  e.g.:

\begin{equation}
f_a (x_a,k_a,Q) = {\partial \over {\partial \lambda^2}} \bigl[
f_a (x_a, \lambda) T_a (\lambda,Q) \bigr) {\biggr|}_{\lambda=k_a}
\;. \label{KMR1}
\end{equation}

That is, the perturbative terms encompassed in the Sudakov
exponential $T_a$ are combined with standard distributions $f_a
(x_a,\lambda)$ to form new, {\it unintegrated} parton
distributions, describing partons with transverse momentum
$\vec{k_a}$ at sampling energy $k_a \leq Q$.

The DDT form is then obtained by application of the {\it
strong-ordering condition} to the convolution over $f_a$ and
$f_b$. The $Q_T$-dependence being now described by a single
luminosity function ${\cal L}_{ab}$:

\begin{equation}
{{d\sigma}\over {dQ^2 dQ_T^2 dy}} = {{d\hat{\sigma}}\over {dQ^2
dy}} \sum_{a,b} {\cal L}_{ab} (x_a,x_b,Q_T,Q)\;, \label{KMR2}
\end{equation}

\noindent with

\begin{equation}
{\cal L}_{ab} = \int f_a (x_a,k_a,Q) f_b (x_b,k_b,Q) \delta^2
(\vec{k_a}+\vec{k_b}-\vec{Q_T}) {{d^2k_a}\over k_a^2}
{{d^2k_b}\over k_b^2} {1\over \pi} \;, \label{KMR3}
\end{equation}

\noindent application of strong-ordering allows only one of the
off-shell partons to contribute at a time. That is, either $k_a
\ll k_b \simeq Q_T$ or $k_b \ll k_a \simeq Q_T$. In the first
case, the delta function collapses to $\delta^2 (\vec{k_b} -
\vec{Q_T})$, while in the second case one gets $\delta^2
(\vec{k_a} - \vec{Q_T})$. Summing over both possibilities yields

\begin{equation}
{\cal L}_{ab} (x_a,x_b,Q_T,Q) = {d\over {d\lambda^2}} \Bigl[ f_a
(x_a,\lambda) T_a (\lambda, Q)  f_b (x_b,\lambda) T_b (\lambda,Q)
\Bigr] {\Bigr|}_{\lambda=Q_T} \;, \label{KMR4}
\end{equation}

\noindent which reproduces equation \ref{DDT1}.

Note that since $T_{a,b} (x_{a,b}, Q, Q) = 1$, it follows that
integrating the luminosity up to scale $Q$ recovers the standard
luminosity :

\begin{equation}
\int_0^{Q^2} dQ_T^2 {\cal L}_{ab} (x_a,x_b,Q_T,Q) = f_a (x_a,Q)
f_b (x_b, Q) \;, \label{KMR5}
\end{equation}

\noindent and in particular,

\begin{equation}
\int_0^{Q^2} {{dk_i^2}\over k_i^2} f_i (x_i,k_i,Q) = f_i(x_i,Q)
\;. \label{KMR6}
\end{equation}

\section{Ellis and Veseli}

In 1997, R.K. Ellis and S. Veseli (EV) came up with a similar
extension of the DDT result, while maintaining the improvements
of the CSS formalism ~\cite{EV97}. They did so by successfully
performing the CSS $b$-space Fourier transform analytically, for
the first three tiers of logarithms. Starting with the purely
perturbative CSS result

\begin{equation}
{{d\sigma}\over {dQ^2 dy dQ_T^2}} = \sum_{a,b}
{{\hat{\sigma_0}}\over 2} \int_0^\infty db b J_0 (Q_Tb) e^{-
S(b,Q)} {\acute{f}}_a (x_a, {b_0\over b}) {\acute{f}}_b
(x_b,{b_0\over b}) \;, \label{EV1}
\end{equation}

\noindent in which ${\acute{f}}_{a,b}$ are the convolutions over
finite, unresummed $C_{i/\tilde{i}}$ corrections

\begin{eqnarray}
{\acute{f}}_a (x_a, {b_0\over b}) &\equiv& \sum_{\tilde{a}}
\int_{x_a}^1 {dz \over z} f_{\tilde{a}/A} (x_a/z,b_0/b)
C_{a/\tilde{a}} (z,b_0/b) \cr {\acute{f}}_b (x_b,{b_0\over b})
&\equiv& \sum_{\tilde{b}} \int_{x_b}^1 {dz \over z}
f_{\tilde{b}/B} (x_b/z,b_0/b) C_{b/\tilde{b}} (z,b_0/b) \;,
\end{eqnarray}

\noindent EV defined a fixed scale ratio $t \equiv Q^2/S$ and
used the rapidity $y$ to write

\begin{eqnarray}
{{d\sigma}\over {dt dQ_T^2}} &=& \int_0^1 dx_a \int_0^1 dx_b
\delta (t-x_a x_b) \int dy \delta \bigl( y-{1\over 2} \ln {x_a
\over x_b} \bigr) {{d\sigma (t,y,Q_T)}\over {dt dy dQ_T^2}} \; \cr
&=& \int_0^1 dx_a \int_0^1 dx_b \delta (t-x_a x_b) {{d\sigma
(t,x_a/x_b,Q_T)}\over {dt dy dQ_T^2}} \;. \label{EV2}
\end{eqnarray}

\noindent Then, after substitution of equation \ref{EV1}, they
took the $N$th moment with respect to $t$:

\begin{eqnarray}
\Sigma (N) &\equiv& \int dt \; t^N {{d\sigma}\over {dt dQ_T^2}}
\; \cr &=& \sum_{a,b} {{\hat{\sigma_0}}\over 2} \int_0^\infty db
\quad b J_0 (Q_Tb) e^{- S(b,Q)} \; \cr &\cdot& \Bigl[ \int_0^1
dx_a x_a^N {\acute{f}}_a (x_a, {b_0\over b}) \Bigr] \Bigl[
\int_0^1 dx_b x_b^N {\acute{f}}_b (x_b, {b_0\over b}) \Bigr] \;
\cr &\equiv& \sum_{a,b} {{\hat{\sigma_0}}\over 2} \int_0^\infty
db \; b J_0 (Q_Tb) e^{-S(b,Q)} {\tilde{f}}_a (N, {b_0\over b})
{\tilde{f}}_b (N,{b_0\over b}) \;. \label{EV3}
\end{eqnarray}

\noindent The modified parton density moments ${\tilde{f}}_i$
evolve according to

\begin{eqnarray}
{{d{\tilde{f}}_i (N,\mu)}\over {d\ln \mu^2}} &=& {\gamma '}_N
{\tilde{f}}_i (N,\mu) \;, \quad \hbox{\rm or} \; \cr {\tilde{f}}_i
(N,b_0/b) &=& {\tilde{f}}_i (N,Q) \exp \Bigl[ -
\int_{b_0^2/b^2}^{Q^2} {{d\mu^2}\over \mu^2} {\gamma '}_N \Bigr]
\;, \label{EV4}
\end{eqnarray}

\noindent in which the ${\gamma '}_N$ are $\alpha_s
(\mu)$-dependent anomalous dimensions. After defining $x \equiv
Q_T b$, this allowed EV to reorganize the cross-section moment as

\begin{equation}
\Sigma (N) = {{\hat{\sigma_0}}\over 2} {{H_N (Q)}\over Q_T^2}
\int_0^\infty dx \; x J_0 (x) e^{{\cal U}_N (x,Q_T,Q)} \;,
\label{EV5}
\end{equation}

\noindent where $H_N (Q) \equiv \sum_{a,b} {\tilde{f}}_a (N,Q)
{\tilde{f}}_b (N,Q)$ and

\begin{eqnarray}
{\cal U}_N (x,Q_T,Q) &=& - \int_{b_0^2Q_T^2/x^2}^{Q^2}
{{d\mu^2}\over \mu^2} \Bigl[ A \ln {Q^2 \over \mu^2} + B +
2{\gamma '}_N \Bigr] \; \cr &\equiv& \sum_{n=1}^\infty
\sum_{m=0}^{n+1} {\biggl[ {{\alpha_s (Q^2)}\over {2\pi}}
\biggr]}^n \ln^m {{Q^2 x^2}\over {Q_T^2 b_0^2}} {_n D_m} \;.
\label{EV6}
\end{eqnarray}

That is, the new parameters ${_n D_m}$ are directly calculable
from the CSS parameters $\{ A_i, B_i \}$. With the relation

\begin{equation}
{d\over {dx}} \bigl[ x J_1 (x) \bigr] = x J_0 (x) \;, \label{EV7}
\end{equation}

\noindent equation \ref{EV5} can now be integrated by parts, with
the result

\begin{equation}
\Sigma (N) =  {{\hat{\sigma_0}}\over 2} {{H_N (Q)}\over Q_T^2}
\Biggl[ x J_1 (x) e^{{\cal U}_N} {\biggr|}_0^\infty -
\int_0^\infty dx \; x J_1 (x) {{de^{{\cal U}_N}}\over {dx}}
\Biggr] \;. \label{EV8}
\end{equation}

\noindent Due to the rapid damping of $\exp ({\cal U}_N)$ and
$J_1(x)$ as $x \rightarrow \infty$, one can ignore the boundary
term; and, using

\begin{equation}
{x\over {Q_T^2 dx}} = {x\over Q_T^2} {1\over {b dQ_T}} = {2\over
{dQ_T^2}} \;, \label{EV9}
\end{equation}

\noindent exhibit the similarity to the DDT formula:

\begin{equation}
\Sigma (N) = - {d\over {dQ_T^2}} \Bigl\{ \hat{\sigma_0} H_N (Q)
\int_0^\infty dx J_1 (x) e^{{\cal U}_N (x,Q_T,Q)} \Bigr\} \;.
\label{EV10}
\end{equation}

\noindent In fact, were it not for the logs of $x/b_0$ in ${\cal
U}_N$, one would have the standard Sudakov exponent, containing
only logs of $Q^2/Q_T^2$, which are, of course, the important
ones.

For this reason, EV split these off and defined a remainder
${\cal R}_N$. First, they defined an integrand ${\tilde{\cal
R}}_N$ via

\begin{eqnarray}
{\cal U}_N (x,Q_T,Q) &=& \exp \Bigl[ \sum_{n=1}^\infty
\sum_{m=0}^{n+1} {\biggl[ {\alpha_s \over {2\pi}} \biggr]}^n
\ln^m {{Q^2 x^2}\over {Q_T^2 b_0^2}} {_n D_m} \Bigr] \; \cr
&\equiv& \exp \Bigl[ \sum_{n=1}^\infty \sum_{m=0}^{n+1} {\biggl[
{\alpha_s \over {2\pi}} \biggr]}^n \ln^m {Q^2\over Q_T^2} {_n
D_m} \Bigr] + {\tilde{\cal R}}_N (x,Q_T,Q) \; \cr &=& {\cal U}_N
(b_0,Q_T,Q) + {\tilde{\cal R}}_N (x,Q_T,Q) \;, \label{EV11}
\end{eqnarray}

\noindent where

\begin{eqnarray}
{\tilde{\cal R}}_N (x,Q_T,Q) &\equiv& \exp \Bigl[
\sum_{n=1}^\infty \sum_{m=0}^{n+1} {\biggl[ {\alpha_s \over
{2\pi}} \biggr]}^n \ln^m {{Q^2 x^2}\over {Q_T^2 b_0^2}} {_n D_m}
\Bigr] \; \cr &-& \exp \Bigl[ \sum_{n=1}^\infty \sum_{m=0}^{n+1}
{\biggl[ {\alpha_s \over {2\pi}} \biggr]}^n \ln^m {Q^2\over
Q_T^2} {_n D_m} \Bigr] \;. \label{EV12}
\end{eqnarray}

\noindent Then, since ${\cal U}_N$ no longer depends on $x$, and
$\int_0^\infty dx J_1 (x) = 1$, equation \ref{EV10} becomes

\begin{equation}
\Sigma (N) = - {d\over {dQ_T^2}} \Bigl\{ \hat{\sigma_0} H_N (Q)
e^{{\cal U}_N (b_0,Q_T,Q)} + {\cal R}_N (Q_T,Q) \Bigr\} \;,
\label{EV13}
\end{equation}

\noindent where

\begin{equation}
{\cal R}_N (Q_T,Q) \equiv \hat{\sigma_0} H_N (Q) \int_0^\infty dx
J_1 (x) {\tilde{\cal R}}_N (x,Q_T,Q) \;. \label{EV14}
\end{equation}

Using equations \ref{EV4} and \ref{EV6}, the Sudakov exponent is
defined:

\begin{eqnarray}
H_N (Q) e^{{\cal U}_N (b_0,Q_T,Q)} &=& H_N (Q_T) e^{-S(b
\rightarrow b_0/Q_T,Q)} \; \cr &=& \sum_{a,b} \Bigl[ \int_0^1 dx_a
x_a^N {\acute{f}}_a (x_a, Q_T) \Bigr] \Bigl[ \int_0^1 dx_b x_b^N
{\acute{f}}_b (x_b, Q_T) \Bigr] e^{-S(b_0/Q_T,Q)} \;, \cr
&\phantom{\times}& \; \label{EV15}
\end{eqnarray}

\noindent at which point the inverse transformation back to
$Q$-space becomes easy. Comparison of equations \ref{EV15} and
\ref{EV3} leads to Ellis and Veseli's final result in the
perturbative regime

\begin{equation}
{{d\sigma}\over {dQ^2 dy dQ_T^2}} = \sum_{a,b} \hat{\sigma_0}
{d\over {dQ_T^2}} \Bigl[ {\acute{f}}_a (x_a, Q_T) {\acute{f}}_b
(x_b, Q_T) e^{-S(b_0/Q_T,Q)} \Bigr] + \Sigma^{-1} \Bigl( {{d{\cal
R}_N (Q_T,Q)}\over {dQ_T^2}} \Bigr) \;, \label{EV16}
\end{equation}

\noindent where $\Sigma^{-1}$ denotes the inverse transform $N
\rightarrow Q$, and the Sudakov exponent becomes the same as in
the $b$-space formalism (equation \ref{CSSsud}), except that the
lower limit $b_0^2/b^2$ is replaced by $Q_T^2$. By using known
integrals of $J_1 (x) \ln^m {x\over b_0}$, Ellis and Veseli were
able to prove that the remainder term ${\cal R}$ contributes no
more importantly than the NNNL series of logarithms, that is,
those corresponding to the CSS $B^{(2)} \alpha_s^2$ terms.

One still requires a prescription for dealing with the very
lowest $Q_T$ values, for which perturbation theory doesn't hold.
As in the CSS paper, this is handled by including a
non-perturbative function $F_{NP}$, this time of $Q_T$, while
sampling the Sudakov exponent and parton distributions at a scale
which never goes below a limiting value $Q_{Tlim}$. To maintain
proper normalization, while affecting only the small-$Q_T$
region, Ellis and Veseli chose the following parametrization:

\begin{eqnarray}
F_{NP} (Q_T) &=& 1-e^{-\tilde{a} Q_T^2} \;, \\
Q_{T*}^2 &=& Q_T^2 + Q_{Tlim}^2 \exp \biggl[ - {Q_T^2 \over
Q_{Tlim}^2} \biggr] \;. \label{EV17}
\end{eqnarray}

The resummed piece then becomes:
\begin{equation}
{{d\sigma_S}\over {dQ^2 dy dQ_T^2}} = \sum_{a,b} \hat{\sigma_0}
{d\over {dQ_T^2}} \Bigl[ {\acute{f}}_a (x_a, Q_T^*) {\acute{f}}_b
(x_b, Q_T^*) e^{-S(Q_T^*,Q)} F_{NP} (Q_T) \Bigr]  \;,
\label{EVfinal}
\end{equation}

\noindent where

\begin{equation}
S(Q_T,Q) = \int_{Q_T^2}^{Q^2}{d\bar{\mu}^2 \over \bar{\mu}^2}
\Bigl[ \ln\bigl( {Q^2 \over \bar{\mu}^2}\bigr) A(\bar{\mu}) +
B(\bar{\mu}) \Bigr] \;.
\end{equation}

This can be used with the same perturbative remainder $\sigma_F$
as in equation \ref{CSSfin}. Transforming to $b$-space is
avoided, drastically improving the speed of numerical
implementations, although the method remains susceptible to
pathologies at high-$Q_T$, especially at fixed-target energies,
as explained further in Section \ref{sec:matchme}.

\chapter{Direct Photon Production}
\label{ch:photon}

Here we begin the process of putting together a consistent,
resummed, next-to-leading order (NLO) description of single photon
hadroproduction. \footnote{All relevant diagrams and matrix
elements are collected in Appendix \ref{ap:Diagrams}.} However,
the resummation prescriptions described in Chapters
[\ref{ch:PPResum},\ref{ch:CSS},\ref{ch:ktspace}] all describe how
to reorganize logarithms of $Q^2/Q_T^2$, where $Q$ and $Q_T$ are
properties of the kinematics of a {\bf pair} of observed
particles. In predicting the cross section for inclusive direct
photon production, we are, in the end, concerned only with the
properties of the photon, so if we want to include the effects of
resummation we must, as an intermediate step, define a second
potentially observed particle for application of these
prescriptions. Fortunately, kinematics requires there to be a
second final-state particle at leading order anyway. To NLO, the
required Feynman diagrams involve up to two additional particles
in the final-state (besides the photon). At the subprocess level
these are either quarks or gluons; by the time they reach the
detector they have fragmented into hadronic {\it jets}, so named
because each has very little transverse spread about its original
partonic vector. That is, the hadrons into which each parton
fragments are concentrated within a narrow ``cone'' about the
parent parton's direction.

If one is ultimately unconcerned with precisely what kinds of
hadrons are produced (as in our case), then even if one could
identify each type within a particular jet, one would be ignoring
(summing over) this information, so in our calculation we need
only keep track of the kinematics of the jet itself (or,
equivalently, the parent parton); the fractional probabilities of
particular fragmentations all sum to one by definition.

That said, we are left with the following question: our NLO
calculation consists not just of $2 \rightarrow 2$-body leading
order (LO) contributions (for which assignment of the jet is
unambiguous), but also $2 \rightarrow 3$-body contributions
involving a photon and {\bf two} potential jets. To which do we
pair with the photon and apply the resummation formalism?

The answer is {\bf both}, in turn, but in the following way.
Experimentally, if the jets are far enough apart in direction as
to be distinguishable, we can define the ``pair'' to be the
photon plus the jet of greater transverse momentum. Given that
the experiment cannot tell which parton initiated that jet,
however, our calculation must consider both possibilities. And,
in the instance in which the jets are {\bf not} distinguishable,
our calculation must treat the resulting kinematics as that of a
{\bf single} jet. Each experiment has its own criteria for
distinguishability, the details of which are important only if
the final predictions include explicit jet characteristics. As
this will not be of concern to us, our calculation can consider
any nonzero separation distinguishable.

Thus, in addition to subprocess matrix elements ${\cal M}$ and
phase-space factors $d\Gamma$ (both of which are given in terms
of subprocess momenta $\{p_i^\mu\}$), the ingredients of our
calculation should also include a {\it jet definition}, that is,
a set of delta-functions the job of which is to define the jet
variables $p_j^\mu$ and pair variables $q^\mu \equiv p_\gamma^\mu
+ p_j^\mu$ in terms of these subprocess momenta.

The photon ($p_\gamma^\mu$), of course, can be unambiguously
associated with one of the outgoing subprocess momenta
$\{p_1^\mu,p_2^\mu,p_3^\mu\}$ (we'll choose $p_1^\mu$). However,
after application of momentum conservation, many of the variables
included in $\{p_1^\mu,p_2^\mu,p_3^\mu\}$ will not be free but
expressible only in terms of other variables in the set.
Depending on the contribution being calculated, the most
convenient remaining set for evaluation of that contribution may
not include $p_1^\mu$, and thus carrying around explicit
definitions for both the photon and jet would help keep track of
their properties in terms of the free subprocess variables. In
the end, integration over the phase space $d\Gamma$ then yields a
cross section in terms of the desired jet and/or pair
observables. For the sake of clarity we will suppress these
delta-functions in what follows.

The result of this chapter will be a collection of
singularity-free pieces which, when added, give an unresummed,
NLO cross section for $\gamma + jet$ production, but which are
also organized such that their assignment to CSS-style
resummation coefficients is self-evident. It is then a simple
matter to use these coefficients in a resummed calculation, and
integrate over the jet numerically. Indeed a FORTRAN program,
switchable between NLO and resummed output, has been written
concurrently with this dissertation, and is available upon
request.

The attempt is made throughout to keep as many free variables as
possible, consistent with the requirements of divergence
cancellation and infrared safety. This is required in order to
account for kinematic recoil of the photon-jet system against the
soft radiation.

A word on notation: the 4-momentum of particle $i$ will be
represented by $p_i^\mu$ throughout. The energy of such a particle
is $p_i^0$, while the magnitude and direction of 3-momentum are
$|\vec{p}_i|$ and $\hat{p}_i$, respectively. We use massless
particles exclusively, for which $|\vec{p}_i| = p_i^0$. The
transverse momentum, $|\vec{p}_i| \sin \theta_i$ will hereafter
be noted as simply $p_i$, while the rapidity is $y_i$. If we need
to refer to the {\it vector} transverse momentum of particle $i$,
we will denote this as $\vec{p}_{Ti}$. Dot-products of 4-vectors,
usually represented using Greek indices and the Einstein summation
convention, as in $p_i^\mu p_{j \mu}$, will for the sake of
clarity be shown as $p_i \cdot p_j$. Both describe the quantity
$p_i^0 p_j^0 - p_i^1 p_j^1 -p_i^2 p_j^2 - p_i^3 p_j^3$. The
following table, valid for massless particles, should help to
clarify our notation:

\bigskip

\BQA p_i^0 &= |\vec{p}_i| \phantom{\sin \theta_i \sin \phi_i} &=
p_i \cosh y_i \cr p_i^1 &= |\vec{p}_i| \sin \theta_i \cos \phi_i
&= p_i \cos \phi_i \cr p_i^2 &= |\vec{p}_i| \sin \theta_i \sin
\phi_i &= p_i \sin \phi_i \cr p_i^3 &= |\vec{p}_i| \cos \theta_i
\phantom{\sin \phi_i} &= p_i \sinh y_i \;. \EQA

\section{Two-Body Final States}
\label{sec:FS2}

The two-body phase space in arbitrary dimension $D=4-2\epsilon$ is

\BQN d^D\Gamma_2 = {{d^{D-1}p_1}\over {2p_1^0{(2\pi)}^{D-1}}}
{{d^{D-1}p_2}\over {2p_2^0{(2\pi)}^{D-1}}} {(2\pi)}^D \delta^D
\bigl( p_a^\mu + p_b^\mu - p_1^\mu - p_2^\mu \bigr) \;,
\label{JD04} \EQN

\noindent where the incoming parton vectors $\{p_a^\mu,p_b^\mu \}$
are, in the hadronic C.M., given in terms of the momentum
fractions $\{x_a,x_b\}$ as follows:

\BQA p_a^\mu &=& x_a {\sqrt{S}\over 2} (1,0,...,0,1) \cr p_b^\mu
&=& x_b {\sqrt{S}\over 2} (1,0,...,0,-1) \;. \label{JD05} \EQA

For a given matrix element ${\cal M}_i$, the corresponding 2-body
cross section is written as an integral over these momentum
fractions, weighted by the squared matrix element (suitably
summed and averaged over final and initial spins and colors), the
phase-space factor, and a luminosity function ${\cal L}$ which
gives the probabilities of finding incoming partons with those
momenta inside the parent hadrons. Thus the cross section for
subprocess $i$ is written:

\BQN d^D\sigma_i = \int_0^1 dx_a \int_0^1 dx_b d^D\Gamma_2 {\cal
L}_i (x_a,x_b) \overline{\sum} {|{\cal M}|}_i^2 \;, \label{JD06}
\EQN

\noindent in which the luminosity is a function of the relevant
parton densities $\{q_f^A, q_f^B \}$ inside hadrons $A$ and $B$:

\BQN {\cal L}_i (x_a,x_b) \equiv {{\sum_{f,\tilde{f}} q_f^A(x_a)
q_{\tilde{f}}^B(x_b)}\over {2x_ax_bS}}. \; \label{JD07} \EQN

The indices $f$ and $\tilde{f}$ each run over all flavors of
partons relevant to subprocess $i$ (here the gluon, as well as
all quarks and antiquarks, are considered ``flavors''). Thus,
{\it e.g.}, for the $qg \rightarrow \gamma q$ subprocess the sum
is simply $ \sum_f q_f^A (x_a) q_g^B (x_b)$, in which $f$ ranges
over quark flavors and antiflavors. We will sometimes refer to
this sum, for a given subprocess $i$, as $H_i(x_a,x_b)$.

It should be noted that charge factors, denoted in this work by
$Q_f$, are present at photon/quark vertices, and, as they depend
on the quark flavor $f$, need to be included in the sum. Each
gives the value of the relevant quark's electromagnetic charge in
units of the proton charge ($+e$). As they arise in the matrix
element, each $\overline{\sum} {|{\cal M}|}_i^2$ above should
properly have a subscript $f\tilde{f}$, and the summation symbol
$\sum_{f\tilde{f}}$ should be placed outside ${\cal L}$. For
clarity, we take these abbreviations to be understood in what
follows. Actually, we will no longer refer to the matrix element
as a whole; each can be written as the product of factors

\BQN \overline{\sum} {|{\cal M}|}_i^2 = \omega_{2\gamma} {\bigl[
K'T_{f\tilde{f}}'(v) \bigr]}_i \;, \label{defs02} \EQN

\noindent in which $\omega_{2\gamma} \equiv 2 {(4\pi)}^2 \alpha
\alpha_s C_F \mu^{4\epsilon}$ is independent of the particular
subprocess and $K'T_{f\tilde{f}}'(v)$ gives the angular
dependence as a function of $v \equiv e^{y_1}/(e^{y_1}+e^{y_2})$.
This, of course, {\bf is} dependent upon the particular
subprocess, and upon the quark charges involved. The ``primes''
denote $\epsilon$-dependence; when $\epsilon$ is taken to zero,
the primes will disappear.

We may now begin to simplify matters by integrating over some of
the unneeded variables in $d^D \Gamma_2$. Breaking down the
$p_1^\mu$ factor as

\BQA {{d^{D-1}p_1^\mu}\over {2p_1^0}}
\delta^D(p_a^\mu+p_b^\mu-p_1^\mu-p_2^\mu) &=& {dy_1\over 2}
\delta(p_1^+ - X^+) \delta(p_1^- - X^-) \; \cr &\times&
d^{D-2}{\vec p_{T1}} \delta^{D-2}({\vec p_{T1}}-{\vec X_T}) \cr
&=& {dy_1 \over S} \delta(x_a - \chi_a) \delta(x_b - \chi_b)
d^{D-2}{\vec p_{T1}} \delta^{D-2} ({\vec p_{T1}}-{\vec X_T}) \;,
\cr &\phantom{=}&  \label{JD08} \EQA

\noindent where

\BQA X^\mu &\equiv& p_a^\mu +p_b^\mu - p_2^\mu  \cr \chi_a &=&
{p_2 \over \sqrt{S}} \bigl(e^{y_1} + e^{y_2} \bigr) \cr \chi_b &=&
{p_2 \over \sqrt{S}} \bigl(e^{-y_1} + e^{-y_2} \bigr) \;, \EQA

\noindent we can integrate over $x_a$,$x_b$, and $\vec p_{T1}$,
which has the effect that $d^D\Gamma_2$ becomes

\BQN d^D\Gamma_2  = {{2\pi dy_1}\over S} {{d^{D-1}p_2}\over
{2p_2^0{(2\pi)}^{D-1}}}   \;. \label{JD09} \EQN

Since we'll require (at most) a transverse momentum, rapidity,
and azimuthal angle for each particle, we can further reduce our
variables as follows. Again, note that here and throughout this
dissertation, we will abbreviate single-particle transverse
momenta as $p_i \equiv |\vec{p_i}|\sin \theta_i$ :

\BQA {{d^{D-1}p_2}\over {2p_2^0 {(2\pi)}^{D-1}}} &=&
{{{|\vec{p_2}|}^{D-3}}\over {2 {(2\pi)}^{D-1}}} d|\vec{p_2}|
\sin^{D-3} \theta_2 d\theta_2 d^{D-3} \phi_2  \cr &=& {{p_2^{D-3}
dp_2 dy_2 d^{D-3} \phi_2}\over {2 {(2\pi)}^{D-1}}} \;,
\label{JD13} \EQA

\noindent Of course, we could stop here, and leave our result in
the following form:

\BQN d^D\sigma_i =  {\cal L} (x_a,x_b) \omega_{2\gamma} {\bigl[
K'T_{f\tilde{f}}'(v) \bigr]}_i d^D \bar{ \Gamma}_2  \;,
\label{JD15} \EQN

\noindent where

\BQN d^D \bar{\Gamma}_2  = {{dy_1 dp_2 dy_2 d^{D-3} \phi_2
p_2^{D-3}}\over {2S {(2\pi)}^{D-2}}}  \;. \label{JD16} \EQN

\noindent However, to use this result for prediction of photon
cross sections, it will be helpful to know what the associated
photon variables $\{p_\gamma,y_\gamma,\phi_\gamma \}$ are in
terms of the above. Since we are assigning the photon to
$p_1^\mu$ and the jet to $p_2^\mu$, we have the following:

\BQA p_\gamma = p_2 &p_j = p_2 \cr y_\gamma=y_1 &y_j=y_2 \cr
\phi_\gamma = \phi_2 + \pi &\phi_j = \phi_2 \;. \EQA

\noindent For a leading-order (LO) calculation, this is all that
would be necessary, and the reader could simply plug in the $ab
\rightarrow \gamma d$ matrix elements given in Appendix
\ref{ap:Diagrams}. \footnote{QCD Bremsstrahlung diagrams also
contribute to order $\alpha \alpha_s$; these will be discussed at
the end of the section.}

For a NLO calculation, however, in which the panorama of 2-body
contributions includes virtual and counterterm pieces, as well as
subtraction terms derived from 3-body asymptotic approximations,
all with poles in $\epsilon$, the constraint of infrared safety
forces us to choose some common set of free {\bf observable}
quantities, in order that addition of these pieces will result in
pole cancellation. We have four so far:
$\{y_\gamma,p_j,y_j,\phi_j \}$, but in order to fully define two
{\bf independent} massless 4-vectors (as our formalism requires),
we need six altogether. Specifically, in order to use the
resummation procedures described earlier in this work, we will
need the pair transverse momentum $Q_T$; in order to incorporate
smearing of the photon kinematics due to recoil, we will also
need an additional angle, which we take to be the axial direction
$\phi_q$ of the pair.

For the two-body contributions, of course, $Q_T$ is zero, and
$\phi_q$ is undefined. We will consequently tack on a factor

$$ dQ_T \delta (Q_T) {{d^{D-3} \phi_q}\over {\int d^{D-3} \phi_q}}
$$

\noindent for these pieces, to supply the additional two degrees
of freedom required.

Furthermore, there will be certain multiplicative factors common
to all pieces, which will be convenient to express explicitly
here. To that end, for our $ab \rightarrow \gamma d$
subprocesses, we take $D=4-2\epsilon$, and define the following
quantities:

\BQA d_\epsilon [\gamma j] &\equiv & dy_\gamma dp_j dy_j
d^{1-2\epsilon} \phi_j dQ_T d^{1-2\epsilon}\phi_q \cr \kappa_i
&\equiv & {{\alpha \alpha_s}\over S} {p_j \over {\int
d^{1-2\epsilon}\phi}} {\biggl( {p_j \over {2\pi \mu^2}}
\biggr)}^{-2\epsilon} 4C_F {[{\cal L}(x_a,x_b)
K'T_{f\tilde{f}}'(v)]}_i \;, \label{defs01} \EQA

\noindent through which one can show:

\BQA d\sigma_i^{\rm Born} &=& {\cal L} (x_a,x_b) \omega_{2\gamma}
{\bigl[ K'T_{f\tilde{f}}'(v) \bigr]}_i d^D \bar{ \Gamma}_2 \cr
&\rightarrow& d_\epsilon [\gamma j] \kappa_i \delta (Q_T) \;. \EQA

For a resummed result, the pair mass $Q$ also requires
definition; for these pieces it is $Q=\sqrt{x_ax_bS}=\sqrt{2p_j^2
\bigl( 1+\cosh(y_1-y_2) \bigr)}$, as can easily be verified via
the pair definition $q^\mu \equiv p_\gamma^\mu + p_j^\mu$.

Two-body Bremsstrahlung also contributes; here one final-state
parton in a purely QCD subprocess (such as $q\bar{q} \rightarrow
q\bar{q}$) fragments into hadronic material and an essentially
collinear photon. The probability for this to occur is thus most
conveniently described by {\it photon fragmentation functions}
$D_{\gamma/q}(z)$ and $D_{\gamma/g}(z)$,  each a function of the
fraction $z$ of the parent parton's momentum that is carried away
by the photon. The transverse momentum imbalance between the
photon and away-side jet in this case is not zero, but instead
given by $Q_T=p_\gamma-p_j=(1-z)p_j$.

There are 11 such subprocesses to consider (see Appendix
\ref{ap:Diagrams}), and the cross section for subprocess $i$, in
terms of the definitions \ref{defs01}, is:

\BQA d\sigma_i^{\rm Brem} &=& d_\epsilon [\gamma j] \kappa_i
{\alpha_s \over \alpha} \int_{z_{min}}^1 dz \delta \bigl( Q_T -
(1-z)p_j \bigr) D_{\gamma/c}(z) \; \cr &=& d_\epsilon [\gamma j]
\kappa_i {\alpha_s \over \alpha} {1\over p_j}  D_{\gamma/c}
\biggl( {{p_j-Q_T}\over p_j} \biggr) \theta \bigl(
(1-z_{min})p_j-Q_T \bigr) \;, \EQA

\noindent in which $D_{\gamma/c}(z)$ is the fragmentation
function relevant to parton $c$ in the subprocess ($ab
\rightarrow cd$). Here a lower bound $z_{min}$ is included for
the general case in which minimum photon energy constraints ({\it
isolation cuts}) are experimentally imposed  ~\cite{BOO92}.

One would expect such processes to be of order $\alpha
\alpha_s^2$; in fact, the $D_{\gamma/c}$ are of order
$\alpha/\alpha_s$, and so one obtains overall ${\cal O}(\alpha
\alpha_s)$ quantities. We use the leading-log fragmentation
functions of reference ~\cite{DO82}:

\BQA D_{\gamma/q}^{\rm LL}(z,\mu) &=& {{F(\mu)}\over z} \Biggl[
{{Q_q^2 (2.21-1.28z+1.29z^2)z^{0.049}}\over {1-1.63 \ln(1-z)}} +
0.0020{(1-z)}^{2.0} z^{-1.54} \Biggr] \cr  D_{\gamma/g}^{\rm
LL}(z,\mu) &=& {{F(\mu)}\over z} {{0.194}\over 8} {(1-z)}^{1.03}
z^{-0.97} \;, \EQA

\noindent where $F(z,\mu)=(\alpha/2\pi) \ln(\mu^2/\Lambda^2)$
with $\Lambda=0.2 {\rm GeV}$.

At NLO, we will require first-order expansions of the parton
distribution and fragmentation functions, in order to cancel
collinear singularities which arise in three-body final states
(see Section \ref{sec:factorize}, in particular equation
\ref{pdfredef}). In the $\overline{MS}$ scheme, with common
factorization scale $M_f$, these {\it counterterms} are as
follows:

\BQA  d\sigma_{a(i)}^{\rm CT} &=& d_\epsilon [\gamma j] \kappa_i
\beta_\epsilon \delta (Q_T) \sum_{\tilde a} \int_{x_a}^1 {dz\over
z} {H(x_a/z,x_b) \over H(x_a,x_b)} {{P_{a/{\tilde a}}^+ (z)}\over
\epsilon}  \; \\ d\sigma_{b(i)}^{\rm CT} &=& d_\epsilon [\gamma j]
\kappa_i \beta_\epsilon  \delta (Q_T) \sum_{\tilde b} \int_{x_b}^1
{dz\over z} {H(x_a,x_b/z) \over H(x_a,x_b)} {{P_{b/{\tilde b}}^+
(z)}\over \epsilon} \; \\ d\sigma_{1(i)}^{\rm CT} &=& d_\epsilon
[\gamma j] \kappa_i \beta_\epsilon \sum_q  {{P_{\gamma q}^+
(\tilde{z})}\over \epsilon}  \;, \EQA

\noindent where

\BQN \beta_\epsilon \equiv  {\alpha_s \over {2\pi}} {\biggl(
{{4\pi \mu^2}\over M_f^2} \biggr)}^\epsilon {{\Gamma
(1-\epsilon)}\over {\Gamma (1-2\epsilon)}} \;. \EQN

\noindent In addition, there will be virtual diagrams, which
contribute at NLO in the form

\BQN d\sigma_i^{\rm Virt} = d_\epsilon [\gamma j] \kappa_i
\beta_\epsilon \delta (Q_T) {\biggl[ {A_v^2\over \epsilon^2} +
{A_v^1 \over \epsilon} + A_v^0 +{B^v\over T_0} \biggr]}_i \;. \EQN

\noindent The virtual diagrams are shown in Appendix
\ref{ap:Diagrams}; their parameters $A_v^2, A_v^1, A_v^0, B^v$
are given in Appendix \ref{ap:Contribs}.

\section{Three-Body Final States}
\label{sec:FS3}

As a general guide we use the work of Ellis, Kunszt, and Soper
~\cite{EKS89,KS92}. These authors arrived at a general
subtraction algorithm for calculating infrared-safe quantities in
hadron-hadron collisions. The process is as follows: We begin
with the 3-body matrix elements for NLO single-photon production,
as listed in Appendix \ref{ap:Diagrams}. These will have
singularities in those regions of phase-space for which one of
the final-state particles become soft or collinear with another
particle. By the process of partial-fractioning, we can break
these matrix elements into smaller pieces, each of which has at
most two singular regions, one soft and one collinear. The
potential singularities manifest themselves as quantities in the
denominator which tend to zero in the singular regions of
phase-space, but as the numerators simplify in these regions
also, we can define approximate, simplified numerators which can
be added and subtracted in {\bf all} regions. Each
partial-fractioned 3-body piece pairs with a subtracted
approximate piece to form a function which is finite everywhere,
while the leftover (added) approximations are dealt with
separately. This is the essence of the {\it subtraction method},
as contrasted with the {\it Phase-Space-Slicing} (PSS) method, in
which the singular regions of phase space are cut out and their
effects added in later after simplification and analytical
integration. A NLO inclusive single photon calculation using the
PSS method, but without transverse-momentum resummation, was
performed by Baer, Ohnemus, and Owens (1990) ~\cite{BOO90}.

The leftover ``subtraction" pieces contain all the singularities
of the original 3-body piece, but are simpler to work with, as
many angular dependences fall away. What is required is that the
singularities be made {\bf explicit} so they may be cancelled
against virtual pieces or absorbed into distribution and
fragmentation functions. It should not matter, for a given
singular quantity, whether this end is achieved via integration
in $D$-dimensions or via extraction in terms of a pole and
accompanying plus-distribution. In the present calculation, we
use the former method for all but the required six free
variables, and handle poles in these variables with
plus-distributions.

Once the Born terms and virtual contributions are added, and the
poles taken care of, what remains is a completely finite NLO
result. From this, one may obtain parameters for use in a
resummed version of the calculation by expanding the resummed
form to NLO and comparing coefficients.

Crucial to the success of any such venture is the recognition
that one of the vectors we intend to observe, namely that of the
jet, is not uniquely defined in terms of the vectors we begin
with (call them $\{ p_1^\mu, p_2^\mu, p_3^\mu \}$). We may
declare that $p_1^\mu$ be assigned to the photon, but it is an
experimental reality that any jet we observe may correspond to
either {\bf or both} of the final-state partons emitted in our
3-body subprocesses ($p_2^\mu$ and $p_3^\mu$). Any attempt to
force a one-to-one correspondence will lead to uncancelled poles,
which is another way of saying that the result wouldn't be {\it
infrared-safe}. One must define the jet kinematics in such a way
that if the final-state partons are together within a certain
angular region, this is counted as a single jet (see figure
\ref{fig:jetdef}). One practical consequence of this is that
singularity cancellation can only be done after the free
variables have been reduced to those of the photon and jet, not
those of $\{ p_1^\mu, p_2^\mu, p_3^\mu \}$.

\figboxh{jetdef}{$\vec{p}_3$ inside jet cone.}

\subsection{Partial-fraction Three-Body matrix elements}
\label{subsec:PartFract}

Three-body phase space, in analogy with equation \ref{JD04}, is:

\BQA d^D\Gamma_3 &=& {{d^{D-1}p_1}\over {2p_1^0{(2\pi)}^{D-1}}}
{{d^{D-1}p_2}\over {2p_2^0{(2\pi)}^{D-1}}} {{d^{D-1}p_3}\over
{2p_3^0{(2\pi)}^{D-1}}} \cr &\times& {(2\pi)}^D \delta^D \bigl(
p_a^\mu + p_b^\mu - p_1^\mu - p_2^\mu - p_3^\mu \bigr) \;,
\label{JD17} \EQA

\noindent while the squared, summed, and averaged matrix element
(or partial-fractioned piece thereof) is denoted by
$\overline{\sum} {|{\cal M}|}^2$, and is a function of all five
momenta $\{p_a^\mu, p_b^\mu, p_1^\mu, p_2^\mu, p_3^\mu \}$.

As written, the final-state 4-vectors are allowed to range over
all space, and there are then up to 12 potential singularities:

\begin{list}{}{}
\item 1. $p_1^\mu$, $p_2^\mu$, $p_3^\mu$ soft,
\item 2. $p_1^\mu$, $p_2^\mu$, $p_3^\mu$ collinear with the beam
in either direction, and
\item 3. $p_1 \cdot p_2 =0$,  $p_1 \cdot p_3 =0$,  $p_2 \cdot p_3
=0$  collinear singularities.
\end{list}

If we assign $p_1^\mu$ to the photon, and through cuts make sure
it is never soft or collinear to the beam, this reduces us to 9
singularities. If we go further and declare $p_3^\mu$ to be that
parton with the lowest transverse momentum (written as simply
$p_3$), then the $p_2^\mu$ singularities disappear, providing we
symmetrize our matrix elements with respect to $p_2^\mu
\leftrightarrow p_3^\mu$. We will consequently define a
symmetrized squared matrix element (note the tilde instead of a
bar)

\BQN \widetilde{\sum} {|{\cal M}|}^2 \equiv \overline{\sum}
{|{\cal M}|}^2 \biggl( p_a^\mu, p_b^\mu, p_1^\mu, p_2^\mu,
p_3^\mu \biggr) + \overline{\sum} {|{\cal M}|}^2 \biggl( p_a^\mu,
p_b^\mu, p_1^\mu, p_3^\mu, p_2^\mu \biggr) \;, \label{JD18} \EQN

\noindent and introduce a theta function $\theta (p_2 - p_3)$.

We are left with 5 singular regions, corresponding to $p_3^\mu$
being either soft or collinear with one of the other 4 particles.
Since momentum conservation prohibits more than two of these from
occurring at once (for nonzero photon $p_T$), the divergent terms
with which we must deal will be at worst of the form

$$ {1\over {p_m \cdot p_3 p_n \cdot p_3}}. $$

\noindent This arrangement is singular in two different
``collinear regions''. To simplify to only one collinear
possibility, we may partial-fraction these terms as follows:

\BQN {1\over {p_m \cdot p_3 p_n \cdot p_3}} = {1\over {p_3 \cdot
(p_m + p_n)}} \Bigl[ {1\over {p_m \cdot p_3}} + {1\over {p_n \cdot
p_3}} \Bigr] \;. \label{PF3} \EQN

\noindent The factor ${[p_3 \cdot (p_m +p_n)]}^{-1}$ gives a
singularity only when $p_3^\mu$ is soft, but if this turns out to
be the case, then the simpler quantity $p_3$ will be zero as
well, and within an already partial-fractioned term, it will be
simpler to express the pole with $p_3$.

So, in all, we can split each (symmetrized, summed and averaged)
matrix element into four terms, each a product of an explicit
${[p_3 \; p_m \cdot p_3]}^{-1}$ singular factor and a general
function of $\{ p_1^\mu, p_2^\mu, p_3^\mu \}$:

\BQN \widetilde{\Sigma} {|{\cal M}|}^2_{ab \rightarrow 123} =
\sum_{m=a,b,1,2} {\Psi_m \over {p_3 \; p_m \cdot p_3}}  \;, \EQN

\noindent in which, by virtue of equation \ref{PF3}, $\Psi_m$
itself (and thus the cross section) can be written as a sum of
terms

\BQN \Psi_m \equiv \sum_{n=a,b,1,2} {\Psi_{mn} \over {p_3 \cdot
(p_m + p_n)}} \;. \label{PF5} \EQN

\noindent Refer to Appendix \ref{ap:Diagrams} for a list of the
3-body matrix elements which contribute. The $\Psi_{mn}$,
however, will not be listed.

\subsection{Reduce Three-body Phase Space.}
\label{subsec:PS3}

The structure of the cross section is analogous to the two-body
case:

\BQA d^D\sigma_3 &=& \int_0^1 dx_1 \int_0^1 dx_2 {\cal L}
(x_1,x_2) \widetilde{\sum} {|{\cal M}|}_{f \tilde{f}}^2
d^D\Gamma_3 \theta (p_2 - p_3) \;, \label{JD23} \EQA

\noindent with the phase-space factor

\BQN d^D\Gamma_3 = {{d^{D-1}p_1}\over {2p_1^0{(2\pi)}^{D-1}}}
{{d^{D-1}p_2}\over {2p_2^0{(2\pi)}^{D-1}}} {{d^{D-1}p_3}\over
{2p_3^0{(2\pi)}^{D-1}}} {(2\pi)}^D \delta^D \bigl( p_a^\mu +
p_b^\mu - p_1^\mu - p_2^\mu -p_3^\mu \bigr) \;,  \EQN

\noindent but after the manipulations given in equations
\ref{JD08} and \ref{JD13}, we now find

\BQA d^D\Gamma_3 &=& {{2\pi dy_1}\over S} {{d^{D-1}p_2}\over
{2p_2^0{(2\pi)}^{D-1}}} {{d^{D-1}p_3}\over {2p_3^0{(2\pi)}^{D-1}}}
\; \label{JD24a} \\ x_{1 \atop 2} &\equiv & {1\over\sqrt{S}} (p_1
e^{\pm y_1} + p_2 e^{\pm y_2} + p_3 e^{\pm y_3}) \;.
\label{JD24e} \EQA

\noindent in which

\BQA p_1 &\equiv& \sqrt{p_2^2 + p_3^2 + 2p_2p_3 \cos(\phi_2 -
\phi_3)} \cr \phi_1 &\equiv& \tan^{-1} \Biggl( {{p_2 \sin\phi_2 +
p_3 \sin\phi_3}\over {p_2 \cos\phi_2 + p_3 \cos\phi_3}} \Biggr)
+\pi \;. \label{JD31} \EQA

We must now arrange our cross-section ingredients in a form which
makes the function to be approximated explicit. We'll call this
function ${\cal F}_{mn}$, corresponding to the partial-fractioned
matrix element numerators defined in section
\ref{subsec:PartFract}. The phase space and singular factors will
not be a part of this function, while the luminosity factor and
$\Psi_{mn}$ will. By analogy with equations \ref{defs02} and
\ref{defs01}, we'll also find it convenient to extract the
overall factor $\omega_3 = 4{(4\pi)}^3 C_F \alpha \alpha_s^2
\mu^{6\epsilon}$ common to all our 3-body matrix elements, and
define the new quantities

\BQA d_\epsilon[123] &\equiv & dy_1 dp_2 dy_2
d\phi_2^{1-2\epsilon} dp_3 dy_3 d\phi_3^{1-2\epsilon} \;, \cr
\chi(\epsilon) &\equiv & {{\alpha\alpha_s^2\mu^{2\epsilon}} \over
{{(2\pi)}^2 S}} {\biggl( {p_2 \over {2\pi\mu}}
\biggr)}^{-2\epsilon} {\biggl( {p_3 \over {2\pi\mu}}
\biggr)}^{-2\epsilon} p_2 p_3 4\rm{C_F} \;, \label{defs5} \EQA

\noindent by which our partial-fractioned cross-section pieces
can then be expressed as

\BQN d^D\sigma_{mn} = {{d_\epsilon[123]\chi(\epsilon)}\over {p_3
\; p_m \cdot p_3}} {\cal F}_{mn} \theta (p_2 - p_3)  \;, \EQN

\noindent with

\BQN {\cal F}_{mn} ( p_1^\mu, p_2^\mu, p_3^\mu ) \equiv {\cal L}
(x_1,x_2) {\Psi_{mn} \over {\omega_3 \; p_3 \cdot (p_m + p_n)}}
\;. \EQN

\noindent Keep in mind, of course, that we will often refer to
$d\sigma_m$ and ${\cal F}_m$, which are, by equation \ref{PF5},
sums over $d\sigma_{mn}$ and ${\cal F}_{mn}$, respectively.

Strictly speaking, the above is used only for the $m=\{a,b,2\}$
contributions. The $m=1$ contribution contains poles in the limit
$\hat{p}_3 \rightarrow \hat{p}_1$, and we'd like to take
advantage of the math already done for the $\hat{p}_3 \rightarrow
\hat{p}_2$ case. Thus we begin by switching $1 \leftrightarrow 2$
in the phase space element $d\Gamma_3$ {\it only} (after
approximation, we will switch back in the subtracted piece, in
order to use the same Monte-Carlo event generator). Thus, in place
of equations \ref{defs5}, we wind up with

\BQA d_\epsilon [213] &\equiv& dy_2 dp_1 dy_1 d^{1-2\epsilon}
\phi_1 dp_3 dy_3 d^{1-2\epsilon} \phi_3 \; \cr
\tilde{\chi}(\epsilon) &\equiv&
{{\alpha\alpha_s^2\mu^{2\epsilon}} \over {{(2\pi)}^2 S}} {\biggl(
{p_1 \over {2\pi\mu}} \biggr)}^{-2\epsilon} {\biggl( {p_3 \over
{2\pi\mu}} \biggr)}^{-2\epsilon} p_1 p_3 4\rm{C_F} \;
\label{tilchi} \EQA

\noindent in which $x_1$ and $x_2$ retain their usual definitions,
but this time $p_1$ and $\phi_1$ are fixed, while $p_2^\mu$ is a
function of $\{ y_2, p_1^\mu, p_3^\mu \}$:

\BQA p_2 &\equiv& \sqrt{p_1^2 + p_3^2 +2p_1 p_3 \cos
(\phi_1-\phi_3)} \; \cr \phi_2 &\equiv& \tan^{-1} \Bigl[ {{p_1
\sin \phi_1 + p_3 \sin \phi_3}\over {p_1 \cos \phi_1 + p_3 \cos
\phi_3}} \Bigr] + \pi \;. \EQA

\subsection{Define and Subtract Approximations}
\label{subsec:Subtract}

We'll now define the additions and subtractions of approximate
versions of these functions ${\cal F}_m$. Each multiplies a
factor ${[p_3 \; p_m \cdot p_3]}^{-1}$ which becomes singular in
the regions in which $p_3^\mu$ either disappears altogether
($p_3^0 \rightarrow 0$, the {\bf soft} region) or becomes
collinear with particle $m$ (the {\bf collinear} region,
$\hat{p}_m \cdot \hat{p}_3 \rightarrow 1$). The essence of the
subtraction method is, as previously stated, to subtract off a
simpler asymptotic approximation which retains the same pole
structure. This ``{\bf subtracted} piece'' (or {\it asymptotic
piece}) pairs with the original three-body partial-fractioned
piece to form a function $d\sigma_m^{finite}$, finite everywhere
in phase space, and which goes to zero as the singular regions are
approached.

This process then, of course, requires us to add back in that
which was subtracted; so our singularities re-enter our
calculation, but now in a much simplified form. These ``{\bf
subtraction} pieces'' can now be analytically manipulated in order
to extract their poles explicitly, these being subsequently
cancelled by the addition of virtual and counterterm corrections.

In practice, since two types of poles exist in each original
partial-fractioned piece, it is usually convenient to allow for
two separate subtractions in order that each resulting subtraction
piece contain only one type of pole. There is a further reason: as
stated in Section \ref{sec:factorize}, we expect the soft limits
to reduce to known 2-body subprocess matrix elements, and the
collinear limits to show up as these times convolutions of
familiar splitting functions and luminosity factors. Thus for the
$m=\{a,b\}$ (initial-state) terms, it will be convenient to have
one version which fixes the direction of $p_3^\mu$ while taking
$p_3^0$ to zero (the soft approximation), and another which takes
$\hat{p}_3 \rightarrow \hat{p}_m$ (collinear). Then, to keep our
soft singularities in one place and avoid double counting, we
subtract the soft limit from our collinear piece.

The above is only a rough statement for the moment, however, as
we did not leave ourselves with $p_3^0$ and $\hat{p}_3$ (the
energy and direction) as free variables, but instead the set
$\{p_3,y_3,\phi_3\}$. We must be more precise.

First of all, we eventually intend to integrate over $y_3$, so we
must be clear about the limits of this integral. Allowing $y_3$
to go to infinity in either direction, without a corresponding
decrease in the magnitude $|\vec{p_3}|$, would at some point
violate momentum conservation, as the momentum fractions of the
incoming partons cannot exceed $1$. In other words, when we take
initial-state soft and collinear approximations (in which the
transverse momentum $p_3$ necessarily tends to zero), we can do
so only if these factors of $p_3$ stand alone; if $p_3$
multiplies some divergent function of $y_3$, we must take
additional care. So far, we have not explicitly included these
bounds, but we do so now in terms of restrictions on $y_3$.

From equations \ref{JD24e} and \ref{JD31}, we have the momentum
fractions

\BQA x_1 &= & {1\over\sqrt{S}} (p_1 e^{y_1} + p_2 e^{y_2} + p_3
e^{y_3}) \cr  x_2 &= & {1\over\sqrt{S}} (p_1 e^{-y_1} + p_2
e^{-y_2} + p_3 e^{-y_3}) \;, \cr {\rm where} \cr p_1 &\equiv&
\sqrt{p_2^2 + p_3^2 + 2p_2p_3 \cos(\phi_2 - \phi_3)}
\label{x12}\;. \EQA

\noindent Restricting $x_{1 \atop 2} \leq 1$ is thus equivalent to
the restrictions

\BQA y_3 &\leq &\ln {{\sqrt{S}-p_1 e^{y_1}-p_2 e^{y_2}}\over p_3}
\cr y_3 &\geq &-\ln {{\sqrt{S}-p_1 e^{-y_1}-p_2 e^{-y_2}}\over
p_3} \;. \EQA

\noindent However, since the factors of $p_3$ in $p_1$ do not
multiply exponentials of $y_3$, these can be dropped in the
low-$p_3$ limit, leaving

\BQA y_3 &\leq &\ln {{(1-x_a)\sqrt{S}}\over p_3} \cr y_3 &\geq
&-\ln {{(1-x_b)\sqrt{S}}\over p_3} \;, \EQA

\noindent in which $x_a, x_b$ are the usual two-body fractions

\BQN x_{a \atop b} \equiv {p_2\over \sqrt{S}} \bigl( e^{\pm y_1} +
e^{\pm y_2} \bigr) \label{xab} \;. \EQN

In practice, we can ignore one or both of these limits if the
relevant integrand is not singular as $y_3$ approaches infinity
from that direction. Our initial-state soft pieces must therefore
retain both limits, but our initial-state collinear pieces can
ignore one of the limits, and in the case of final-state pieces,
both limits are ignored.

So, although na\"{\i}vely expecting our soft limit to be
adequately represented by the conditions ``$p_3 \rightarrow 0$,
$y_3$ and $\phi_3$ fixed'' ignores the fact that $p_3^0 = p_3
\cosh y_3$ does not necessarily go to zero if $y_3$ is taken
large enough, now that we have finite limits on $y_3$, these
conditions make sense, and we will write our soft limits in the
form

$${\cal F}_m \bigl( p_3=0, y_3, \phi_3 \bigr) \theta_a \theta_b \;,$$

\noindent where $\theta_a$ and $\theta_b$ express the $y_3$
limits above:

\BQA \theta_a &\equiv & \theta (\ln [(1-x_a)\sqrt{S} /p_3] - y_3)
\;, \cr \theta_b &\equiv & \theta (y_3 + \ln [(1-x_b)\sqrt{S}
/p_3]) \label{thetas} \;. \EQA

Unfortunately, the same notation won't work for the initial-state
collinear terms. What we wish to express is that as $p_3^\mu$
becomes parallel with $p_a^\mu$, for example, $p_3 \cdot p_a \sim
p_3e^{-y_3}$ and $p_3$ alone go to zero, while $p_3 \cdot p_b
\sim p_3e^{y_3}$ does not. The descriptions we will use are:
\footnote{Moving to light-cone coordinates $p_3^\mu \equiv
\{p_3^+, p_3^-, \vec{p}_{T3}\}$, in which $p_3^+ \equiv (p_3^0 +
p_3^3)/\sqrt{2} = p_3e^{y_3}/\sqrt{2}$ and $p_3^- \equiv (p_3^0 -
p_3^3)/\sqrt{2} = p_3e^{-y_3}/\sqrt{2}$ would allow us to write
our collinear limits more cleanly, but at the expense of
additional notation. We will have to be content with the above
descriptions.}

\BQA {\cal F}_a \bigl( p_3e^{y_3}, p_3e^{-y3}=0, \vec{p}_{T3} =
\vec{0} \bigr) \theta_a \cr  {\cal F}_b \bigl( p_3e^{y_3}=0,
p_3e^{-y3}, \vec{p}_{T3} = \vec{0} \bigr) \theta_b \;, \EQA

\noindent the soft limits of which are

\BQA {\cal F}_a \bigl( p_3e^{y_3}=0, p_3e^{-y3}=0, \vec{p}_{T3} =
\vec{0} \bigr) \theta_a \cr  {\cal F}_b \bigl( p_3e^{y_3}=0,
p_3e^{-y3}=0, \vec{p}_{T3} = \vec{0} \bigr) \theta_b \;, \EQA

\noindent and will be called {\it soft-collinear} limits. These
will be subtracted from the collinear terms.

For $m=a$, then, we have in the end these pieces:

\BQN d^D\sigma_a^{soft} \equiv
{{d_\epsilon[123]\chi(\epsilon)}\over {p_3 p_a \cdot p_3}} \theta
(p_2 - p_3) \theta_a \theta_b {\cal F}_a \bigl( p_3=0, y_3,
\phi_3 \bigr) \; \EQN

\BQA d^D\sigma_a^{coll} &\equiv&
{{d_\epsilon[123]\chi(\epsilon)}\over {p_3 p_a \cdot p_3}} \theta
(p_2 - p_3) \theta_a  \cr &\times& \biggl[ {\cal F}_a \bigl(
p_3e^{y_3}, p_3e^{-y3}=0, \vec{p}_{T3} = \vec{0} \bigr) - {\cal
F}_a \bigl( p_3e^{y_3}=0, p_3e^{-y3}=0, \vec{p}_{T3} = \vec{0}
\bigr) \biggl] \cr &\phantom{\times}& \; \EQA

\BQA d^D\sigma_a^{finite} &\equiv&
{{d_\epsilon[123]\chi(\epsilon)}\over {p_3 p_a \cdot p_3}} \theta
(p_2 - p_3) \cr &\times& \biggl[ {\cal F}_a
(y_1,p_2,y_2,\phi_2,p_3,y_3,\phi_3) - {\cal F}_a \bigl(
p_3e^{y_3}, p_3e^{-y3}=0, \vec{p}_{T3} = \vec{0} \bigr)\theta_a
\cr &+& {\cal F}_a \bigl( p_3e^{y_3}=0, p_3e^{-y3}=0,
\vec{p}_{T3} = \vec{0} \bigr)\theta_a   - {\cal F}_a \bigl(
p_3=0, y_3, \phi_3 \bigr) \theta_a \theta_b \biggl] \cr
&\phantom{\times}& \;. \label{FINab} \EQA

Note that $d^D\sigma_a^{finite}$ is finite everywhere, so we can
in fact take $\epsilon \rightarrow 0$ in this term, thus
replacing $d_\epsilon[123] \chi (\epsilon)$ with the notations
$d_0[123] \chi_0$ and $D=4-2\epsilon \rightarrow 4$. Meanwhile,
$d^D\sigma_a^{coll}$ has a collinear singularity but no soft
ones, and $d^D\sigma_a^{soft}$ has both soft and collinear
divergences. The sum
$d\sigma^{soft}+d\sigma^{coll}+d\sigma^{finite} \sim {\cal
F}(y_1,p_2,y_2,\phi_2,p_3,y_3,\phi_3)$, which is what we started
with.

It will become evident in Section \ref{subsec:inty3} that
significant simplification of our $y_3$ integrals arises if we
take advantage of the further breakdown $d\sigma_a = d\sigma_{ab}
+ d\sigma_{a2}$ for the soft pieces ($\{mn\} = \{a1,b1,21\}$ do
not occur via the arguments of section \ref{subsec:PartFract}). In
addition to $d\sigma_a^{finite}$, then, we have just defined
three {\it subtraction pieces}, two soft (labeled $\{mn\} =
\{ab,a2\}$) and one collinear. Analogous terms are similarly
constructed for $m=b$.

The final-state pieces have their own organizational requirements,
but thankfully are simpler as regards notation. For $m=2$, we
subtract the soft-collinear piece from the soft term instead,
leaving it with no collinear poles.

\BQA d^D\sigma_2^{soft} &\equiv&
{{d_\epsilon[123]\chi(\epsilon)}\over {p_3 p_2 \cdot p_3}} \theta
(p_2 - p_3) \cr &\times& \biggl[ {\cal F}_2 (p_3=0, y_3, \phi_3)
- {\cal F}_2 (p_3=0, y_3=y_2, \phi_3=\phi_2) \biggr] \; \EQA

\BQN d^D\sigma_2^{coll} \equiv
{{d_\epsilon[123]\chi(\epsilon)}\over {p_3 p_2 \cdot p_3}} \theta
(p_2 - p_3) {\cal F}_2 (p_3, y_3=y_2, \phi_3=\phi_2)  \; \EQN

\BQA d^4\sigma_2^{finite} &\equiv& {{d_0[123]\chi_0}\over {p_3 p_2
\cdot p_3}}  \theta (p_2 - p_3) \cr &\times& \biggl[ {\cal F}_2
(y_1,p_2,y_2,\phi_2,p_3,y_3,\phi_3) - {\cal F}_2 (p_3, y_3=y_2,
\phi_3=\phi_2) \cr &+& {\cal F}_2 (p_3=0, y_3=y_2, \phi_3=\phi_2)
- {\cal F}_2 (p_3=0, y_3, \phi_3) \biggl]  \;. \label{FIN2} \EQA

\noindent Again, it will be useful to break the $m=2$ soft piece
into two partial-fractioned terms $\{mn\} = \{2a,2b\}$.

\bigskip

For $m=1$, we will worry about a collinear piece only, as it can
be shown that there are no soft poles:

\BQN d^D\sigma_1^{coll} \equiv {{d_\epsilon[213]
\tilde{\chi}(\epsilon)} \over {p_1 \cdot p_3}} \theta (p_2 - p_3)
{\cal F}_1 (p_3, y_3=y_1, \phi_3=\phi_1) \; \EQN

\BQA d^4\sigma_1^{finite} &\equiv& {{d_0[123] \chi_0} \over {p_1
\cdot p_3}} \theta (p_2 - p_3) \cr &\times& \biggl[ {\cal F}_1
(y_1,p_2,y_2,\phi_2,p_3,y_3,\phi_3) - {\cal F}_1 (p_3, y_3=y_1,
\phi_3=\phi_1) \biggl] \;. \label{FIN1} \EQA

\noindent Note that, as promised, we have switched back to
$p_1^\mu \leftrightarrow p_2^\mu$ in the phase space of
$d^D\sigma_1^{finite}$ in order to take advantage of one common
event generator in our Monte Carlo program.

In all, we have ten subtraction pieces; six correspond to soft
singularities and are labeled $\{mn\} = \{ab,ba,a2,b2,2a,2b\}$.
The other four correspond to the legs with which $p_3^\mu$ may be
collinear, and these are labeled $m = \{a,b,1,2\}$. The next
section is devoted to performing the approximations herein
defined.

\subsection{Perform Approximations}
\label{subsec:approx}

Since the ${\cal F}_m$, like the squared matrix elements from
which they are derived, are given in terms of invariant
dot-products $p_i \cdot p_j$, we must first see what these look
like in terms of our free variables
$\{y_1,p_2,y_2,\phi_2,p_3,y_3,\phi_3\}$.

\BQA 2p_a \cdot p_b &=& x_1 x_2 S \cr 2p_a \cdot p_1 &=& x_1
\sqrt{S} p_1 e^{-y_1} \cr 2p_a \cdot p_2 &=& x_1 \sqrt{S} p_2
e^{-y_2} \cr 2p_a \cdot p_3 &=& x_1 \sqrt{S} p_3 e^{-y_3} \cr
2p_b \cdot p_1 &=& x_2 \sqrt{S} p_1 e^{y_1} \cr 2p_b \cdot p_2
&=& x_2 \sqrt{S} p_2 e^{y_2} \cr 2p_b \cdot p_3 &=& x_2 \sqrt{S}
p_3 e^{y_3} \cr 2p_1 \cdot p_2 &=& 2p_1 p_2 \bigl[ \cosh (y_1
-y_2) - \cos (\phi_1 - \phi_2) \bigr] \cr 2p_1 \cdot p_3 &=& 2p_1
p_3 \bigl[ \cosh (y_1 -y_3) - \cos (\phi_1 - \phi_3) \bigr] \cr
2p_2 \cdot p_3 &=& 2p_2 p_3 \bigl[ \cosh (y_2 -y_3) - \cos
(\phi_2 - \phi_3) \bigr] \;, \EQA

\noindent in which

\BQA x_{1 \atop 2} &=& {1\over \sqrt{S}} \biggl( p_1e^{\pm y_1} +
p_2e^{\pm y_2} + p_3e^{\pm y_3} \biggr) \cr p_1 &=&
\sqrt{p_2^2+p_3^2+2p_2p_3 \cos (\phi_2 - \phi_3)} \cr \phi_1 &=&
\tan^{-1} {{p_2 \sin \phi_2 + p_3 \sin\phi_3}\over {p_2\cos\phi_2
+ p_3 \cos\phi_3}} \;. \EQA

\begin{table}[h]
\caption{Soft limits of invariants.} \vspace{.5cm}
\begin{center}
\thicklines
\begin {tabular} { llll }
\hline \hline \thinlines $2p_a \cdot p_b = \hat{s}$ & & &
\\ $2p_a \cdot p_1 = -\hat{t}$ & $2p_b \cdot p_1 =
-\hat{u}$ & & \\  $2p_a \cdot p_2 = -\hat{u}$ & $2p_b \cdot p_2 =
-\hat{t}$ & $2p_1 \cdot p_2 = \hat{s}$ &
\\ $2p_a \cdot p_3 = 0$ & $2p_b \cdot p_3 = 0$ & $2p_1 \cdot p_3 = 0$ & $2p_2
\cdot p_3 = 0$ \\
\hline \hline
\end{tabular}
\end{center}
\label{softtab}
\end{table}

In the soft limit, $p_3 \rightarrow 0$ while $y_3$ and $\phi_3$
remain fixed. We then obtain $p_1 \rightarrow p_2$ and
$\{x_1,x_2\} \rightarrow \{x_a,x_b\}$, where

\BQN x_{a \atop b} = {p_2\over \sqrt{S}} \bigl( e^{\pm y_1} +
e^{\pm y_2} \bigr) \;, \EQN

\noindent leading to the soft limits expressed in Table
\ref{softtab}, in which the $ab \rightarrow 12$ subprocess
invariants $\hat{s},\hat{t},\hat{u}$ are given (for massless
particles) as

\BQA \hat{s} &=& 2p_a \cdot p_b = 2p_1 \cdot p_2 \cr &=& x_ax_bS
= 2p_2^2 \bigl( 1+\cosh(y_1 -y_2) \bigr) \\ \hat{t} &=& -2p_a
\cdot p_1 = -2p_b \cdot p_2 \cr &=& -x_a \sqrt{S} p_2e^{-y_1} =
-p_2^2 \bigl(1+e^{y_2-y_1} \bigr) \\ \hat{u} &=& -2p_a \cdot p_2
= -2p_b \cdot p_1 \cr &=& -x_a \sqrt{S} p_2e^{-y_2} = -p_2^2
\bigl(1+e^{y_1-y_2} \bigr) \;. \EQA

\noindent In fact, for 2-body subprocesses, these invariants are
not independent. For massless participants, $\hat{t}$ and
$\hat{u}$ can be written in terms of $\hat{s}$ and a single
angular variable $v$:

\BQA \hat{t} &=& -(1-v) \hat{s} \cr \hat{u} &=& -v \hat{s} \cr v
&\equiv& {1\over2} (1+\cos\theta_1) = {e^{y_1}\over {e^{y_1} +
e^{y_2}}} \label{defv} \;. \EQA

\noindent In this way, we can show that, as expected, the soft
limits ${\cal F}_m (p_3=0, y_3, \phi_3)$ reduce to products of
2-body matrix elements $T_{f\tilde{f}}' (v)$ and luminosity
functions ${\cal L} (x_a, x_b)$. The denominators will retain
certain $y_3,\phi_3$ dependences after approximation, and we will
integrate over these in what follows.

As an example of how the {\bf collinear} limits are taken, we look
at the case in which a gluon is radiated from one of the incoming
legs of a subprocess, as in the diagram below. As the gluon (here
$p_3^\mu$) becomes collinear with the parent quark
$p_{\tilde{a}}^\mu$, its momentum, as well as that of the
daughter quark ($p_a^\mu$, which enters the subprocess) are
related to the parent in terms of a simple fraction $z$. That is,
$p_3^\mu = (1-z) p_{\tilde{a}}^\mu$ and $p_a^\mu =
zp_{\tilde{a}}^\mu$:

\begin{figure}[h]
\centering
\parbox{60mm}{
\begin{fmffile}{cft15col}
\begin{fmfgraph*}(40,40)
\fmfleft{i1} \fmfright{o2,o1} \fmfblob{30}{o2}
\fmflabel{$p_{\tilde{a}}^\mu$}{i1} \fmf{quark}{i1,v1}
\fmf{quark}{v1,o2} \fmflabel{$p_a^\mu = zp_{\tilde{a}}^\mu$}{o2}
\fmflabel{$p_3^\mu = (1-z) p_{\tilde{a}}^\mu$}{o1}
\fmf{gluon}{v1,o1}
\end{fmfgraph*}
\end{fmffile}
}
\bigskip
\caption{$p_{\tilde{a}} \cdot p_3 =0$ collinear kinematics.}
\end{figure}
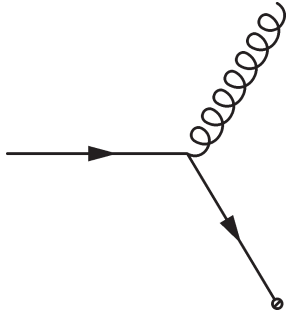

\bigskip
The functions to be approximated, ${\cal F}_m$, are in general
functions of the invariants $p_i \cdot p_j$, where $\{i,j\} =
\{\tilde{a}, b, 1,2,3\}$ for a three-body scattering
$p_{\tilde{a}}^\mu + p_b^\mu \rightarrow p_1^\mu + p_2^\mu
+p_3^\mu$. \footnote{Labeling of the partons involved in a
collinear splitting can be confusing. Having started with the
3-body labels $ab \rightarrow 123$, it would seem appropriate to
leave the parent label unchanged (e.g. $\boldmath a$), and add a
tilde to the daughter (e.g. $\boldmath \tilde{a}$). The
convention, however, is that the parent always carries the tilde.}
In the collinear limit of massless particles, however, these
invariants reduce to functions of $z$ and the subprocess
invariants (see Table \ref{t15coltab}).

\begin{table}[h]
\caption{$p_{\tilde{a}} \cdot p_3 =0$ collinear limits of
invariants.} \vspace{.5cm}
\begin{center}
\thicklines
\begin {tabular} { llll }
\hline \hline \thinlines $2p_{\tilde{a}} \cdot p_b = {\hat{s}
\over z}$ & & &  \\ $2p_{\tilde{a}} \cdot p_1 = -{\hat{t}\over
z}$ & $2p_b \cdot p_1 = -\hat{u}$ & & \\  $2p_{\tilde{a}} \cdot
p_2 = -{\hat{u}\over z}$ & $2p_b \cdot p_2 = -\hat{t}$ & $2p_1
\cdot p_2 = \hat{s}$ & \\ $2p_{\tilde{a}} \cdot p_3 = 0$ & $2p_b
\cdot p_3 = {{1-z}\over z}\hat{s}$ & $2p_1 \cdot p_3 =
-{{1-z}\over z}\hat{t}$ & $2p_2 \cdot
p_3 = -{{1-z}\over z}\hat{u}$ \\
\hline \hline
\end{tabular}
\end{center}
\label{t15coltab}
\end{table}

The fraction $z$ is not arbitrary; its value in terms of our free
variables is, as it should be, given by the $\{p_3e^{y_3},
p_3e^{-y_3} =0, \hat{p}_3 =0 \}$ limit of the ratio
$x_a/x_{\tilde{a}}$:

\BQN z_a \equiv {x_a\over x_{\tilde{a}}} = {x_a\over x_1}
\rightarrow {{x_a \sqrt{S}}\over {x_a \sqrt{S} + p_3e^{y_3}}} \;.
\label{defz} \EQN

\noindent One can easily verify that in the soft-collinear limit,
when $p_3e^{y_3}$ itself goes to zero, $z \rightarrow 1$ and the
invariants in Table \ref{t15coltab} reduce to those in Table
\ref{softtab}.

For the $m=1$ collinear contribution, $p_3^\mu$ becomes collinear
with $p_1^\mu$, and their sum balances $p_2^\mu$. A new momentum
fraction $\tilde{z}$ is thus defined via
$p_1=\tilde{z}(p_1+p_3)$, and the momenta of the incoming partons
becomes the $\{y_3,\phi_3\} \rightarrow \{y_1,\phi_1\}$ limit of
$x_{1\atop 2}$, namely

\BQN \tilde{x}_{a \atop b} = {{(p_1+p_3)}\over \sqrt{S}} \bigl(
e^{\pm y_1} + e^{\pm y_2} \bigr) \; \label{tilx} \EQN

\noindent (see equations \ref{JD24e}, \ref{tilchi}). For $m=2$,
we have the $\{y_3,\phi_3\} \rightarrow \{y_2,\phi_2\}$ limits

\BQA \bar{z} &\equiv & {p_2 \over {p_2+p_3}} \; \cr \bar{x}_{a
\atop b} &\equiv & {{(p_2+p_3)}\over \sqrt{S}} (e^{\pm
y_1}+e^{\pm y_2}) \label{barx} \;. \EQA

In this way, all collinear ${\cal F}_m$ can be shown to factorize
into two functions, a splitting function $\hat{P}_{ij}'$ which
depends on only the relevant momentum fraction $z$, and a
two-body matrix element $T_{f\tilde{f}}' (v)$ characteristic of
the subprocess but independent of the earlier splitting.

Adding up the contributions to each $d^D \sigma_m^{soft}$ and
$d^D \sigma_m^{coll}$ from all our 3-body matrix elements, and
grouping by 2-body subprocess $i$ (see Appendix \ref{ap:Diagrams}
for a list), we can finally write out our subtraction pieces as
follows. With the definitions (see equations \ref{defs5},
\ref{x12}, \ref{xab}, \ref{thetas}, \ref{defv}):

\BQA x_{a \atop b} &\equiv & {p_2 \over \sqrt{S}} (e^{\pm y_1} +
e^{\pm y_2}) \; \cr v &\equiv & {e^{y_1} \over {e^{y_1}+e^{y_2}}}
\; \cr x_{1 \atop 2} &\equiv & {1\over\sqrt{S}} (p_1 e^{\pm y_1}
+ p_2 e^{\pm y_2} + p_3 e^{\pm y_3}) \; \cr p_1 &\equiv &
\sqrt{p_2^2+p_3^2+2p_2p_3\cos(\phi_2-\phi_3)} \; \cr p_a^\mu
&\equiv & {x_a \over x_1} p_{\tilde a}^\mu = x_a {\sqrt{S} \over
2} \pmatrix{1&0&0&1} \; \cr p_b^\mu &\equiv & {x_b \over x_2}
p_{\tilde b}^\mu = x_b {\sqrt{S} \over 2} \pmatrix{1&0&0&-1} \;
\cr \theta_a &\equiv & \theta (\ln [(1-x_a)\sqrt{S} /p_3-y_3]) \;
\cr \theta_b &\equiv & \theta (y_3 + \ln [(1-x_b)\sqrt{S} /p_3])
\; \cr d_\epsilon[123] &\equiv & dy_1 dp_2 dy_2
d\phi_2^{1-2\epsilon} dp_3 dy_3 d\phi_3^{1-2\epsilon} \; \cr
\chi(\epsilon) &\equiv & {{\alpha\alpha_s^2\mu^{2\epsilon}} \over
{{(2\pi)}^2 S}} {\biggl( {p_2 \over {2\pi\mu}}
\biggr)}^{-2\epsilon} {\biggl( {p_3 \over {2\pi\mu}}
\biggr)}^{-2\epsilon} p_2 p_3 4C_F \;, \label{defs8} \EQA

\noindent our soft pieces become, for subprocess $i$:

\BQA d\sigma_{ab(i)}^{\rm soft} &=&
{{d_\epsilon[123]\chi(\epsilon)}\over{p_a \cdot p_3}} {\biggl[
\psi_{ab}^0 {\cal L} (x_a,x_b) K'T_{f\tilde{f}}'(v) \biggr]}_i
{{p_a \cdot
p_b}\over{p_a \cdot p_3 + p_b \cdot p_3}} \theta_a \theta_b \; \\
d\sigma_{ba(i)}^{\rm soft} &=&
{{d_\epsilon[123]\chi(\epsilon)}\over{p_b \cdot p_3}} {\biggl[
\psi_{ba}^0{\cal L} (x_a,x_b) K'T_{f\tilde{f}}'(v) \biggr]}_i
{{p_a \cdot
p_b}\over{p_a \cdot p_3 + p_b \cdot p_3}} \theta_a \theta_b \; \\
d\sigma_{a2(i)}^{\rm soft} &=&
{{d_\epsilon[123]\chi(\epsilon)}\over{p_a \cdot p_3}} {\biggl[
\psi_{a2}^0{\cal L} (x_a,x_b) K'T_{f\tilde{f}}'(v) \biggr]}_i
{{p_a \cdot
p_2}\over{p_a \cdot p_3 + p_2 \cdot p_3}} \theta_a \theta_b \; \\
d\sigma_{b2(i)}^{\rm soft} &=&
{{d_\epsilon[123]\chi(\epsilon)}\over{p_b \cdot p_3}} {\biggl[
\psi_{b2}^0{\cal L} (x_a,x_b) K'T_{f\tilde{f}}'(v) \biggr]}_i
{{p_b \cdot
p_2}\over{p_b \cdot p_3 + p_2 \cdot p_3}} \theta_a \theta_b \; \\
d\sigma_{2a(i)}^{\rm soft} &=&
{{d_\epsilon[123]\chi(\epsilon)}\over{p_2 \cdot p_3}} {\biggl[
\psi_{2a}^0{\cal L} (x_a,x_b) K'T_{f\tilde{f}}'(v) \biggr]}_i
\biggl[{{p_2 \cdot p_a}\over{p_a \cdot p_3 + p_2 \cdot p_3}} -
{p_2 \over p_3} \biggr] \; \\ d\sigma_{2b(i)}^{\rm soft} &=&
{{d_\epsilon[123]\chi(\epsilon)}\over{p_2 \cdot p_3}} {\biggl[
\psi_{2b}^0{\cal L} (x_a,x_b) K'T_{f\tilde{f}}'(v) \biggr]}_i
\biggl[{{p_2 \cdot p_b}\over{p_b \cdot p_3 + p_2 \cdot p_3}} -
{p_2 \over p_3} \biggr] \;, \EQA

\noindent in which the ${[K'T_{f\tilde{f}}'(v)]}_i$ are given in
Appendix \ref{ap:Diagrams} and

\BQN {[\psi_{mn}^0]}_i =  {[\psi_{nm}^0]}_i  =(2C_F - N_C)
\delta_{mq} \delta_{nq} + N_C (1-\delta_{mq} \delta_{nq}) \;. \EQN

\noindent As an example, for the subprocess $q\bar{q} \rightarrow
\gamma g$, $\psi_{ab}^0 = \psi_{ba}^0 = 2C_F-N_C$, as both $m=a$
and $n=b$ refer to quarks, while $\psi_{\{a2,2a,b2,2b\}}^0$ all
equal $N_C$ for this subprocess.

\bigskip

With the further definitions (see equations \ref{tilchi},
\ref{defz}, \ref{tilx}, \ref{barx})

\BQA z_a &\equiv & {{x_a \sqrt{S}} \over {x_a \sqrt{S} + p_3
e^{y_3}}} \; \cr z_b &\equiv & {{x_b \sqrt{S}} \over {x_b
\sqrt{S} + p_3 e^{-y_3}}} \; \cr \tilde{z} &\equiv& {p_1 \over
{p_1+p_3}} \; \cr \tilde{x}_{a \atop b} &\equiv&
{{(p_1+p_3)}\over \sqrt{S}} \bigl( e^{\pm y_1} + e^{\pm y_2}
\bigr) \; \cr d_\epsilon [213] &\equiv& dy_2 dp_1 dy_1
d^{1-2\epsilon} \phi_1 dp_3 dy_3 d^{1-2\epsilon} \phi_3 \; \cr
\tilde{\chi}(\epsilon) &\equiv&
{{\alpha\alpha_s^2\mu^{2\epsilon}} \over {{(2\pi)}^2 S}} {\biggl(
{p_1 \over {2\pi\mu}} \biggr)}^{-2\epsilon} {\biggl( {p_3 \over
{2\pi\mu}} \biggr)}^{-2\epsilon} p_1 p_3 4\rm{C_F} \; \cr \bar{z}
&\equiv & {p_2 \over {p_2+p_3}} \; \cr \bar{x}_{a \atop b}
&\equiv & {{(p_2+p_3)}\over \sqrt{S}} (e^{\pm y_1}+e^{\pm y_2})
\;, \EQA

\noindent we find the collinear contributions:

\BQA d\sigma_{a(i)}^{\rm coll} &=& {{d_\epsilon [123]
\chi(\epsilon)}\over {p_a \cdot p_3}} {[{\cal
L}(x_a,x_b)K'T_{f\tilde{f}}'(v)]}_i \theta_a \cr &\times&
\sum_{\tilde a} \biggl[ {{{\cal L}(x_a/z_a,x_b)}\over {{\cal
L}(x_a,x_b)}} \hat{P}_{a{\tilde a}}' (z_a) - {{2{\rm C}(a)
z_a} \over {1-z_a}} \delta_{g3} \biggr] \;  \\
d\sigma_{b(i)}^{\rm coll} &=& {{d_\epsilon [123]
\chi(\epsilon)}\over {p_b \cdot p_3}} {[{\cal
L}(x_a,x_b)K'T_{f\tilde{f}}'(v)]}_i \theta_b \cr &\times&
\sum_{\tilde b} \biggl[ {{{\cal L}(x_a,x_b/z_b)}\over {{\cal
L}(x_a,x_b)}} \hat{P}_{b{\tilde b}}' (z_b) - {{2{\rm C}(b)
z_b} \over {1-z_b}} \delta_{g3} \biggr] \;  \\
d\sigma_{1(i)}^{coll} &=& {{d_\epsilon[213] \tilde{\chi}
(\epsilon)}\over {p_1 \cdot p_3}} {[{\cal
L}(\tilde{x}_a,\tilde{x}_b)K'T_{f\tilde{f}}'(v)]}_i \sum_q
\hat{P'}_{\gamma q} (\tilde{z})  \; \\ d\sigma_{2(i)}^{\rm coll}
&=& {{d_\epsilon [123] \chi (\epsilon)}\over {p_2 \cdot p_3}}
{[{\cal L}(\bar{x}_a,\bar{x}_b)K'T_{f\tilde{f}}'(v)]}_i
\sum_{\tilde d} \hat{P}_{{\tilde d}d}' (\bar{z}) \;. \EQA

\noindent Here $\delta_{g3}=1$ if the radiated third particle is a
gluon; otherwise $\delta_{g3}=0$.

\subsection{Perform $y_3$-Integrals.}
\label{subsec:inty3}

We will now integrate analytically over the rapidity $y_3$. This
process will introduce further new notation, but as the
functional form of these integrals is shared by more than one
subtraction term, the net increase in clarity should render the
abbreviations forgivable.

To start, the notation we're currently using for our free,
unobservable phase-space differentials is {\bf
$d_\epsilon[123]$}. After integration over $y_3$ it may be
easiest to simply write

\BQN  d_\epsilon[123-y_3] \equiv  dy_1 dp_2 dy_2
d\phi_2^{1-2\epsilon} dp_3 d\phi_3^{1-2\epsilon} \;. \EQN

\noindent Next, we'll see that for our initial-state subtraction
terms, the limits on $y_3$ (as shown in the functions $\theta_a$
and $\theta_b$) will appear in the final result. These limits
will be written as

\BQN y_{3m}^{a,b} \equiv  \ln \bigl[ (1-x_{a,b} )\sqrt{S}/p_3
\bigr] \;, \EQN

\noindent and will provide us, eventually, with the logs of $Q_T$
we expect.

The integrals we'll need are performed in Appendix
\ref{ap:integrals}; the results are as follows:

\BQA I_{ab \atop ba}^s(x) &\equiv & \int_{-y_{3m}^b}^{y_{3m}^a}
dy_3 {e^{\pm y_3} \over {e^{\pm y_3}+xe^{\mp y_3}}} = \bigl(
y_{3m}^{a\atop b} -{1\over 2}\ln x \bigr) \pm {1\over 2} \ln
\biggl[ {{1+x^{\pm 1}e^{- 2y_{3m}^a}} \over {1+x^{\mp 1}e^{-
2y_{3m}^b}}} \biggr] \; \label{iabba}\\ I_{a2 \atop b2}^s(x)
&\equiv & \int_{-y_{3m}^b}^{y_{3m}^a} dy_3 {e^{\pm (y_3-y_2)}
\over {xe^{\mp (y_3-y_2)} + \cosh(y_3-y_2) -
\cos(\phi_3-\phi_2)}} \cr  &=& \pm \ln \biggl[ {{{(1+2x)}^{\pm
1}e^{-2(y_{3m}^a - y_2)} + 1} \over {{(1+2x)}^{\mp
1}e^{-2(y_{3m}^b + y_2)} + 1}} \biggr] \cr &+&
2(\pi-\alpha(x))\cot\alpha(x) -\ln(1+2x) +2(y_{3m}^{a\atop b} \mp
y_2) \; \label{ia2b2} \\ I_{2a \atop 2b}^s(x) &\equiv &
\int_{-\infty}^\infty dy_3 {1\over {\cosh(y_3 -
y_2)-\cos(\phi_3-\phi_2)}} \; \cr &\times& \biggl[ {x\over
{xe^{\mp y_3} + \cosh(y_3 - y_2)-\cos(\phi_3-\phi_2)}} \biggr]
\cr  &=& \ln(1+2x) + 2(\pi-(\phi_3-\phi_2))\cot(\phi_3-\phi_2) -
2(\pi-\alpha(x))\cot\alpha(x) \; \cr &\phantom{\times}& \label{i2a2b}\\
I_2^c &\equiv & \int_{-\infty}^\infty {dy_3 \over {\cosh(y_3 -
y_2)-\cos(\phi_3-\phi_2)}} = {{2(\pi-(\phi_3 - \phi_2))}\over
{\sin(\phi_3-\phi_2)}} \; \label{i2c} \\ I_{\tilde{a}}^c &\equiv
& \int_{-\infty}^{y_{3m}^a} dy_3 {p_3z_a\over {x_a \sqrt{S}
e^{-y_3}}} \biggl[ {{H(x_a/z_a,x_b)} \over H(x_a,x_b)}
\hat{P}_{a{\tilde a}}' (z_a) - {{2{\rm C}(a)} \over {1-z_a}}
\delta_{g3} \biggr] \cr &=& \int_{x_a}^1 {{dz_a}\over z_a}
\biggl[ {{H(x_a/z_a,x_b)} \over H(x_a,x_b)} \hat{P}_{a{\tilde
a}}' (z_a) - {{2{\rm C}(a)} \over {1-z_a}} \delta_{g3} \biggr] \;
\label{iac}\\ I_{\tilde{b}}^c &\equiv & \int_{-y_{3m}^b}^\infty
dy_3 {p_3z_b\over {x_b \sqrt{S} e^{y_3}}} \biggl[
{{H(x_a,x_b/z_b)} \over H(x_a,x_b)} \hat{P}_{b{\tilde b}}' (z_b)
- {{2{\rm C}(b)} \over {1-z_b}} \delta_{g3} \biggr] \cr &=&
\int_{x_b}^1 {{dz_b}\over z_b} \biggl[ {{H(x_a,x_b/z_b)} \over
H(x_a,x_b)} \hat{P}_{b{\tilde b}}' (z_b) - {{2{\rm C}(b)} \over
{1-z_b}} \delta_{g3} \biggr] \;, \label{ibc} \EQA

\noindent in which we've abbreviated

\BQA \cos\alpha(x) &\equiv&  {{\cos(\phi_3 - \phi_2)} \over
\sqrt{1+2x}} \cr H(x_1,x_2) &\equiv& 2x_1x_2S {\cal L}(x_1,x_2)\;.
\EQA

\noindent The subtraction terms then become:

\BQA d\sigma_{ab(i)}^{\rm soft} &=&
d_\epsilon[123-y_3]\chi(\epsilon) {\biggl[ \psi_{ab}^0 {\cal
L}(x_a,x_b)K'T_{f\tilde{f}}'(v) \biggr]}_i {2\over p_3^2} I_{ab}^s
\bigl({x_a \over x_b}\bigr) \; \\ d\sigma_{ba(i)}^{\rm soft} &=&
d_\epsilon[123-y_3]\chi(\epsilon) {\biggl[ \psi_{ba}^0 {\cal
L}(x_a,x_b)K'T_{f\tilde{f}}'(v) \biggr]}_i {2\over p_3^2} I_{ba}^s
\bigl({x_b \over x_a}\bigr) \; \\ d\sigma_{a2(i)}^{\rm soft} &=&
d_\epsilon[123-y_3]\chi(\epsilon) {\biggl[ \psi_{a2}^0 {\cal
L}(x_a,x_b)K'T_{f\tilde{f}}'(v) \biggr]}_i {1\over p_3^2}
I_{a2}^s \bigl(
{{{x_a\sqrt{S}}e^{-y_2}} \over {2p_2}} \bigr) \; \\
d\sigma_{b2(i)}^{\rm soft} &=& d_\epsilon[123-y_3]\chi(\epsilon)
{\biggl[ \psi_{b2}^0 {\cal L}(x_a,x_b)K'T_{f\tilde{f}}'(v)
\biggr]}_i {1\over p_3^2} I_{b2}^s \bigl( {{{x_b\sqrt{S}}e^{y_2}}
\over {2p_2}} \bigr) \; \\ d\sigma_{2a(i)}^{\rm soft} &=&
d_\epsilon[123-y_3]\chi(\epsilon) {\biggl[ \psi_{2a}^0 {\cal
L}(x_a,x_b)K'T_{f\tilde{f}}'(v) \biggr]}_i {1\over p_3^2} \biggl[
I_{2a}^s \bigl( {{{x_a\sqrt{S}}e^{-y_2}} \over {2p_2}}
\bigr) - I_2^c \biggr] \cr &\phantom{\times}& \; \\
d\sigma_{2b(i)}^{\rm soft} &=& d_\epsilon[123-y_3]\chi(\epsilon)
{\biggl[ \psi_{2b}^0 {\cal L}(x_a,x_b)K'T_{f\tilde{f}}'(v)
\biggr]}_i {1\over p_3^2} \biggl[ I_{2b}^s \bigl(
{{{x_b\sqrt{S}}e^{y_2}} \over {2p_2}} \bigr) - I_2^c \biggr]  \cr
&\phantom{\times}& \;
\\ d\sigma_{a(i)}^{\rm coll} &=& d_\epsilon [123-y_3]
\chi(\epsilon) {[{\cal L}(x_a,x_b)K'T_{f\tilde{f}}'(v)]}_i
{2\over p_3^2} \sum_{\tilde a} I_{\tilde{a}}^c \; \\
d\sigma_{b(i)}^{\rm coll} &=& d_\epsilon [123-y_3] \chi(\epsilon)
{[{\cal L}(x_a,x_b)K'T_{f\tilde{f}}'(v)]}_i {2\over
p_3^2} \sum_{\tilde b} I_{\tilde{b}}^c \; \\
d\sigma_{1(i)}^{\rm coll} &=& d_\epsilon [213-y_3] \tilde{\chi}
(\epsilon) {[{\cal
L}(\tilde{x}_a,\tilde{x}_b)K'T_{f\tilde{f}}'(v)]}_i {I_2^c \over
{p_1p_3}} \sum_q \hat{P}_{\gamma q}' (\tilde{z}) \;
\\ d\sigma_{2(i)}^{\rm coll} &=& d_\epsilon [123-y_3] \chi
(\epsilon) {[{\cal
L}(\bar{x}_a,\bar{x}_b)K'T_{f\tilde{f}}'(v)]}_i {I_2^c \over
{p_2p_3}} \sum_{\tilde d} \hat{P}_{{\tilde d}d}' (\bar{z}) \;.
\EQA

\subsection{Map to Photon-Jet space}
\label{subsec:jetmap}

It's now time to translate to a common set of observables
$\{y_\gamma,p_j,y_j,\phi_j,Q_T,\phi_q\}$ so that Born, virtual,
counterterm, and subtraction terms may all be added and poles
cancelled. As stated in Section \ref{sec:FS3}, the poles in
$\epsilon$ will not cancel unless this is done first.

For the initial-state contributions $m=\{a,b\}$, we simply have
$\{y_1,p_2,y_2,\phi_2 \} \rightarrow \{y_\gamma,p_j,y_j,\phi_j
\}$ as in the two-body case, but as we have a non-zero $p_3$
which contributes transverse momentum to the pair, we take $p_3
\rightarrow Q_T$ and $\phi_3 \rightarrow \phi_q - \pi$. As this
affects the photon transverse momentum, we also list the smeared
dependent variables

\BQA p_\gamma &=&
\sqrt{p_j^2+Q_T^2+2p_jQ_T\cos(\phi_j-\phi_q-\pi)} \cr \phi_\gamma
&=& \tan^{-1}\Biggl( {{p_j \sin\phi_j + Q_T \sin\phi_q}\over {p_j
\cos\phi_j + Q_T \cos\phi_q}} \Biggr) +\pi \;. \label{defs9} \EQA

The final-state soft pieces $\{mn\}=\{2a,2b\}$ will not be
resummed, so we could approximate the kinematics as well as the
weights, and simply set $Q_T=0$ here. Unfortunately, this would
require an analytic integration over $\phi_3$ (namely $\int
d^{1-2\epsilon} \phi_3 (\pi-\alpha) \cot \alpha$), and our lack
of a closed form for such an expression makes it difficult to
explicitly show the cancellation of poles. As these terms
describe radiation which, in general, falls outside the jet cone,
and thus contributes to the $Q_T$ of the photon-jet pair, it will
be more natural to treat these terms in the same way as the
$\{mn\}=\{a2,b2\}$ pieces above.

For the $m=2$ collinear term, we unquestionably have zero $Q_T$,
and due to the collinear configuration of particles 2 and 3, they
are counted as a single jet of transverse momentum $p_2+p_3$.
With the definition of $\bar{z}$ used earlier, we can take

\BQN  \int dp_2 dp_3 \rightarrow p_j dp_j d\bar{z}  \;, \EQN

\noindent while $\{y_1,y_2,\phi_2 \} \rightarrow
\{y_\gamma,y_j,\phi_j \}$ and $p_\gamma=p_j$,
$\phi_\gamma=\phi_j+\pi$. As we now have purely $2\rightarrow 2$
kinematics for this piece, $\phi_q$ is undefined, and so we (as
in the two-body case) include a factor $d^{1-2\epsilon}\phi_q /
\int d^{1-2\epsilon}\phi_q$. The integrations over $\phi_3$ and
$\bar{z}$ must be done analytically.

Finally, for the $m=1$ collinear term, we do have a nonzero
$Q_T=p_3$, and $p_j=p_2$ as in the initial-state pieces. However,
as we have switched $p_1^\mu \leftrightarrow p_2^\mu$ in the
phase space for calculational simplicity, $p_2$ is no longer a
free variable. Instead we have its approximation $p_j = p_1+p_3$.
Thus we write $p_1 (\equiv p_\gamma) \rightarrow p_j-p_3 =
p_j-Q_T $. The splitting functions will continue to be expressed
in terms of the momentum fraction $\tilde{z}$, where now
$\tilde{z}=(p_j-Q_T)/p_j$. As for the rest of the variables,
$\{y_1,y_2 \} \rightarrow \{y_\gamma,y_j \}$ and $\phi_1 (\equiv
\phi_\gamma)  \rightarrow \phi_j+\pi$, while $\phi_3 \rightarrow
\phi_q-\pi$ as in the initial-state terms.

After performing the above operations, we can split each $\chi$
and $\tilde{\chi}$ factor into three pieces. The first,
$\kappa_i$, will be common to all contributions (2-body and
subtraction), the second, $\beta_\epsilon$, common to all
higher-order contributions, and the third, $\chi_{\rm fac}$, will
be expanded through the balance of the relevant contribution.
With the definitions:

\BQA d_\epsilon [\gamma j] &\equiv & dy_\gamma dp_j dy_j
d^{1-2\epsilon} \phi_j dQ_T d^{1-2\epsilon}\phi_q \; \cr x_{a
\atop b} &\equiv & {p_j \over \sqrt{S}} (e^{\pm y_\gamma} +
e^{\pm y_j}) \; \cr \kappa_i &\equiv & {{\alpha \alpha_s}\over S}
{p_j \over {\int d^{1-2\epsilon}\phi}} {\biggl( {p_j \over {2\pi
\mu^2}} \biggr)}^{-2\epsilon} 4C_F{[{\cal L}(x_a,x_b)
K'T_{f\tilde{f}}'(v)]}_i \; \cr \beta_\epsilon &\equiv & {\alpha_s
\over {2\pi}} {\biggl( {{4\pi \mu^2}\over M_f^2} \biggr)}^\epsilon
{{\Gamma (1-\epsilon)}\over {\Gamma (1-2\epsilon)}} \; \cr
\chi_{\rm fac} &\equiv & {{\Gamma (1-2\epsilon)}\over {\Gamma^2
(1-\epsilon)}} {\biggl( {M_f^2 \over p_j^2} \biggr)}^\epsilon \;,
\EQA

\noindent in which $M_f$ is the factorization scale, we can now
write all our subtraction pieces in the form:

\BQA d\sigma_{(i)} &=& d_\epsilon [\gamma j] \kappa_i
\beta_\epsilon \Gamma_{(i)} \;, \EQA

\noindent where \BQA \Gamma_{ab(i)}^{\rm soft} &=& \chi_{\rm fac}
\psi_{ab(i)}^0 {2\over Q_T} {\biggl( {p_j^2 \over Q_T^2}
\biggr)}^\epsilon I_{ab}^s \bigl({x_a \over x_b}\bigr) \; \\
\Gamma_{ba(i)}^{\rm soft} &=& \chi_{\rm fac} \psi_{ba(i)}^0
{2\over Q_T} {\biggl( {p_j^2 \over Q_T^2}
\biggr)}^\epsilon I_{ba}^s \bigl({x_b \over x_a}\bigr) \; \\
\Gamma_{a2(i)}^{\rm soft} &=&  \chi_{\rm fac} \psi_{a2(i)}^0
{1\over Q_T} {\biggl( {p_j^2 \over Q_T^2} \biggr)}^\epsilon
I_{a2}^s \bigl( {1 \over {2(1-v)}} \bigr) \; \\
\Gamma_{b2(i)}^{\rm soft} &=& \chi_{\rm fac} \psi_{b2(i)}^0 {1
\over Q_T} {\biggl( {p_j^2 \over Q_T^2}
\biggr)}^\epsilon  I_{b2}^s \bigl( {1 \over {2v}} \bigr) \; \\
\Gamma_{2a(i)}^{\rm soft} &=& \chi_{\rm fac} \psi_{2a(i)}^0
{1\over Q_T} {\biggl( {p_j^2 \over Q_T^2} \biggr)}^\epsilon
 \biggl[ I_{2a}^s \bigl( {1 \over
{2(1-v)}} \bigr) - I_2^c \biggr] \; \\
\Gamma_{2b(i)}^{\rm soft} &=&  \chi_{\rm fac} \psi_{2b(i)}^0
{1\over Q_T} {\biggl( {p_j^2 \over Q_T^2} \biggr)}^\epsilon
\biggl[ I_{2b}^s \bigl( {1 \over {2v}} \bigr) - I_2^c
\biggr]  \; \\
\Gamma_{a(i)}^{\rm coll} &=& \chi_{\rm fac} {2\over Q_T} {\biggl(
{p_j^2 \over Q_T^2} \biggr)}^\epsilon \sum_{\tilde a} I_{\tilde{a}}^c \; \\
\Gamma_{b(i)}^{\rm coll} &=& \chi_{\rm fac}
 {2\over Q_T} {\biggl( {p_j^2 \over Q_T^2} \biggr)}^\epsilon
\sum_{\tilde b} I_{{\tilde b}}^c \; \\ \Gamma_{1(i)}^{\rm coll}
&=& \chi_{\rm fac} {I_2^c \over p_j}
{[\tilde{z}(1-\tilde{z})]}^{-2\epsilon} \sum_q \hat{P}_{\gamma
q}' (\tilde{z}) \; \\ \Gamma_{2(i)}^{\rm coll} &=& \chi_{\rm fac}
\delta (Q_T) {{\int d^{1-2\epsilon} \phi_3 I_2^c(\phi_3)}\over
{\int d^{1-2\epsilon} \phi_3}} \int d\bar{z}
{[\bar{z}(1-\bar{z})]}^{-2\epsilon} \sum_{\tilde d}
\hat{P}_{{\tilde d}d}' (\bar{z})  \;. \EQA

\subsection{Extract Poles in $\epsilon$}

Toward this end, we notice that for most contributions, the poles
reside purely at $Q_T=0$, and can be extracted with the help of:

\BQA \int_0^{p_j} {dQ_T \over Q_T} {\biggl( {p_j^2 \over Q_T^2}
\biggr)}^\epsilon f(Q_T) &=& \int_0^{p_j} dQ_T \biggl[
{\biggl({1\over Q_T} \biggr)}_+ - {{\delta(Q_T)} \over
{2\epsilon}} \biggr] f(Q_T) \;
\\ \int_0^{p_j} {dQ_T \over Q_T} {\biggl( {p_j^2 \over Q_T^2}
\biggr)}^\epsilon 2y_{3m}^i f(Q_T) &=& \int_0^{p_j} dQ_T \Biggl[
\ln\biggl( {{{(1-x_i)}^2 S} \over p_j^2} \biggr) \biggl[ {\biggl(
{1\over Q_T} \biggr) }_+ - {{\delta(Q_T)} \over {2\epsilon}}
\biggr] \cr &-& 2 {\biggl( {{\ln Q_T/p_j} \over Q_T} \biggr)}_+
+{{\delta(Q_T)} \over {2\epsilon^2}} \Biggr] f(Q_T) \; \EQA

\noindent whereas for the $m=2$ collinear piece, there are poles
at $\bar{z}=1$ and $\phi_3=\phi_j+\pi$. We then need:

\BQA  \bar{\cal Z}(q) &\equiv & \sum_{\tilde d} \int_{1\over 2}^1
d\bar{z} {[\bar{z}(1-\bar{z})]}^{-2\epsilon} {\hat{P}}_{{\tilde
d}q}' (\bar{z}) \cr &\simeq& {\rm C_F} \biggl[ -{1\over \epsilon}
- {3\over 2} + \epsilon \biggl( {{2 \pi^2}\over 3} - {13\over 2}
\biggr) \biggr] \; \\ \bar{\cal Z}(g) &\equiv & \sum_{\tilde d}
\int_{1\over 2}^1 d\bar{z} {[\bar{z}(1-\bar{z})]}^{-2\epsilon}
{\hat{P}}_{{\tilde d}g}' (\bar{z}) \cr &\simeq& {\rm N_C} \biggl[
-{1\over \epsilon} - {11\over 6} + \epsilon \biggl( {{2
\pi^2}\over 3} - {67\over 9} \biggr) \biggr] + {\rm N_f} \biggl[
{1\over 3} + {23\over 18}\epsilon \biggr] \; \\ {\cal Z}_2^c
&\equiv& \int d^{1-2\epsilon} \phi_3 I_2^c (\phi_3) \Bigl/ \int
d^{1-2\epsilon} \phi \cr &=& -{{16}^{-\epsilon}\over \epsilon}
{{\Gamma^4(1-\epsilon)}\over{\Gamma^2(1-2\epsilon)}} \;.  \EQA

Finally, the $m=1$ collinear piece has nonzero $Q_T$ (implicit in
our definition of $\tilde{z}$) and therefore dependence upon a
defined pair angle $\phi_q$. Unlike the $m=2$ collinear case,
then, integration over the function $I_2^c$ is not independent of
our free variables, and we must extract its $\phi_q=\phi_j+\pi$
pole in the sense of a distribution as follows:

\BQN \int d^{1-2\epsilon} \phi_q I_2^c (\phi_q) = \int
d^{1-2\epsilon} \phi_q \Biggl[ {{2(\pi-\Delta \phi)} \over {\sin
\Delta \phi}}\bigl[ 1-\cos \Delta \phi \bigr] - {1\over \epsilon}
\Biggr]  \;, \EQN

\noindent with $\Delta\phi \equiv {\rm MOD} (\| \phi_q-\phi_j-\pi
\| /\pi)$.

This brings us, finally, to:
\BQA \Gamma_{ab(i)}^{\rm soft} &=&  \psi_{ab(i)}^0 \Biggl[ \ln
{{{(1-x_a)}^2 S x_b} \over {x_a p_j^2}} \biggl[ {\bigl( {1\over
Q_T} \bigr)}_+ - {{\delta (Q_T)} \over {2\epsilon}} \biggr] \cr
&-& 2{\bigl( {{\ln Q_T / p_j} \over Q_T} \bigr)}_+ + {{\delta
(Q_T)}\over {2\epsilon^2}} + {1\over Q_T} {\biggl( {p_j^2 \over
Q_T^2} \biggr)}^\epsilon \ln \biggl[ {{1+{x_a\over
x_b}e^{-2y_{3m}^a}}\over {1+{x_b\over x_a}e^{-2y_{3m}^b}}}
\biggr] \Biggr] \chi_{\rm fac} \; \\ \Gamma_{ba(i)}^{\rm soft}
&=& \psi_{ba(i)}^0 \Biggl[ \ln {{{(1-x_b)}^2 S x_a} \over {x_b
p_j^2}} \biggl[ {\bigl( {1\over Q_T} \bigr)}_+ - {{\delta (Q_T)}
\over {2\epsilon}} \biggr] \cr &-& 2{\bigl( {{\ln Q_T / p_j}
\over Q_T} \bigr)}_+ + {{\delta (Q_T)}\over {2\epsilon^2}} -
{1\over Q_T} {\biggl( {p_j^2 \over Q_T^2} \biggr)}^\epsilon \ln
\biggl[ {{1+{x_a\over x_b}e^{-2y_{3m}^a}}\over {1+{x_b\over
x_a}e^{-2y_{3m}^b}}} \biggr] \Biggr] \chi_{\rm fac} \; \\
\Gamma_{a2(i)}^{\rm soft} &=& \psi_{a2(i)}^0 \Biggl[ \biggl[
2(\pi-\alpha_a)\cot\alpha_a-2y_j+ \ln {{{(1-x_a)}^2 S (1-v)}
\over {(2-v) p_j^2}} \biggr] \biggl[ {\bigl( {1\over Q_T}
\bigr)}_+ - {{\delta (Q_T)} \over {2\epsilon}} \biggr] \cr &-&
2{\bigl( {{\ln Q_T / p_j} \over Q_T} \bigr)}_+ + {{\delta
(Q_T)}\over {2\epsilon^2}} + {1\over Q_T} {\biggl( {p_j^2 \over
Q_T^2} \biggr)}^\epsilon \ln \biggl[ {{1+{{2-v}\over
{1-v}}e^{-2(y_{3m}^a-y_j)}}\over {1+{{1-v}\over
{2-v}}e^{-2(y_{3m}^b+y_j)}}} \biggr] \Biggr] \chi_{\rm fac}  \;
\\ \Gamma_{b2(i)}^{\rm soft} &=&  \psi_{b2(i)}^0 \Biggl[ \biggl[
2(\pi-\alpha_b)\cot\alpha_b+2y_j+ \ln {{{(1-x_b)}^2 S v} \over
{(1+v) p_j^2}} \biggr] \biggl[ {\bigl( {1\over Q_T} \bigr)}_+ -
{{\delta (Q_T)} \over {2\epsilon}} \biggr] \cr &-& 2{\bigl( {{\ln
Q_T / p_j} \over Q_T} \bigr)}_+ + {{\delta (Q_T)}\over
{2\epsilon^2}} - {1\over Q_T} {\biggl( {p_j^2 \over Q_T^2}
\biggr)}^\epsilon \ln \biggl[ {{1+{v\over
{1+v}}e^{-2(y_{3m}^a-y_j)}}\over {1+{{1+v}\over
v}e^{-2(y_{3m}^b+y_j)}}} \biggr] \Biggr] \chi_{\rm fac} \; \\
\Gamma_{2a(i)}^{\rm soft} &=& \psi_{2a(i)}^0 \biggl[ {\bigl(
{1\over Q_T} \bigr)}_+ - {{\delta (Q_T)} \over {2\epsilon}}
\biggr] \cr &\times& \biggl[ \ln {{2-v}\over {1-v}} -
2(\pi-\Delta\phi) {{(1-\cos\Delta\phi)}\over {\sin\Delta\phi}}
-2(\pi-\alpha_a) \cot\alpha_a \biggr] \chi_{\rm fac} \; \\
\Gamma_{2b(i)}^{\rm soft} &=& \psi_{2b(i)}^0 \biggl[ {\bigl(
{1\over Q_T} \bigr)}_+ - {{\delta (Q_T)} \over {2\epsilon}}
\biggr] \cr &\times& \biggl[ \ln {{1+v}\over v} -
2(\pi-\Delta\phi) {{(1-\cos\Delta\phi)}\over {\sin\Delta\phi}}
-2(\pi-\alpha_b) \cot\alpha_b  \biggr] \chi_{\rm fac} \; \\
\Gamma_{a(i)}^{\rm coll} &=& \sum_{\tilde a} I_{{\tilde a}}^c
\biggl[ {\bigl( {2\over Q_T} \bigr)}_+ - {{\delta (Q_T)} \over
\epsilon} \biggr] \chi_{\rm fac} \; \\ \Gamma_{b(i)}^{\rm coll}
&=& \sum_{\tilde b} I_{{\tilde b}}^c \biggl[ {\bigl( {2\over Q_T}
\bigr)}_+ - {{\delta (Q_T)} \over \epsilon} \biggr] \chi_{\rm fac}
\; \\ \Gamma_{1(i)}^{\rm coll} &=&
{[\tilde{z}(1-\tilde{z})]}^{-2\epsilon} \sum_q \hat{P}_{\gamma
q}' (\tilde{z}) \Biggl[ {{2(\pi-\Delta \phi)} \over {\sin \Delta
\phi}}\bigl[ 1-\cos \Delta \phi \bigr] - {1\over \epsilon}
\Biggr] {\chi_{\rm fac} \over p_j} \; \\ \Gamma_{2(i)}^{\rm coll}
&=& \delta (Q_T) \bar{\cal Z}(d) {\cal Z}_2^c \chi_{\rm fac} \;
\EQA

\noindent where the $d$ in $\bar{\cal Z}(d)$ is the type of
parton ($q$ or $g$) on leg $d$ (for subprocess $ab \rightarrow
\gamma d$), and

\BQA v &\equiv& {e^{y_\gamma}\over {e^{y_\gamma}+e^{y_j}}} \; \cr
\cos \alpha_a &\equiv& \cos \Delta \phi \sqrt{{1-v}\over {2-v}} \;
\cr \cos \alpha_b &\equiv& \cos \Delta \phi \sqrt{v\over {1+v}} \;
\cr y_j &=& \ln {{x_a (1-v)}\over {x_b v}} \; \cr \Delta\phi
&\equiv& {\rm MOD} (\| \phi_q-\phi_j-\pi \| /\pi) \;. \EQA

Note that in $\Gamma_{ab,ba,a2,b2}^{\rm soft}$, there exist
complicated logarithms of the form

$$\ln \biggl[ {{1+xe^{-2y_{3m}^a}}\over {1+{1\over x}e^{-2y_{3m}^b}}}
\biggr]\;.$$

\noindent Since these tend to zero as $Q_T \rightarrow 0$, we can
drop them for the sake of simplicity. This amounts to a
redefinition of our asymptotic form, and is valid as long as we
get rid of these terms in the finite piece also. This has been
done in the Monte Carlo program.

\section{Add Born, Virtual, and Counterterm Contributions.}

We may now add in the two-body pieces from Section \ref{sec:FS2}:

\BQA d\sigma_i^{\rm Born} &=& d_\epsilon [\gamma j] \kappa_i
\delta (Q_T) \; \\ d\sigma_i^{\rm Virt} &=& d_\epsilon [\gamma j]
\kappa_i \beta_\epsilon \Gamma_{{\rm Virt}(i)} \;
\\ d\sigma_{a(i)}^{\rm CT} &=& d_\epsilon [\gamma j] \kappa_i
\beta_\epsilon  \Gamma_{a(i)}^{\rm CT} \;
\\ d\sigma_{b(i)}^{\rm CT} &=& d_\epsilon [\gamma j] \kappa_i
\beta_\epsilon \Gamma_{b(i)}^{\rm CT} \; \\ d\sigma_{1(i)}^{\rm
CT} &=& d_\epsilon [\gamma j] \kappa_i \beta_\epsilon
\Gamma_{1(i)}^{\rm CT} \;, \EQA

\noindent where

\BQA \Gamma_{{\rm Virt}(i)} &=& \delta (Q_T)
{\biggl[ {A_v^2\over \epsilon^2} + {A_v^1 \over \epsilon} + A_v^0 +{B^v\over T_0} \biggr]}_i \; \\
\Gamma_{a(i)}^{\rm CT} &=& \delta (Q_T) \sum_{\tilde a}
\int_{x_a}^1 {dz\over z} {H(x_a/z,x_b) \over H(x_a,x_b)}
{{P_{a{\tilde a}}^+ (z)}\over \epsilon} \;
\\ \Gamma_{b(i)}^{\rm CT} &=& \delta (Q_T) \sum_{\tilde b}
\int_{x_b}^1 {dz\over z} {H(x_a,x_b/z) \over H(x_a,x_b)}
{{P_{b{\tilde b}}^+ (z)}\over \epsilon} \; \\
\Gamma_{1(i)}^{\rm CT} &=&  \sum_q  {{P_{\gamma q}^+
(\tilde{z})}\over {\epsilon p_j}} \;. \EQA

\noindent and the virtual parameters $A_v^2, A_v^1, A_v^0, B^v$
are given in Appendix \ref{ap:Contribs}.

\section{Cancel Poles and Take $\epsilon \rightarrow 0$}

For simplicity of notation, we allow the virtual and counterterm
pieces to share $\{mn\}$ designations of their own, and write the
sum

\BQA \sum_{mn} \Gamma_{mn(i)} &\equiv& \Gamma_{{\rm Virt}(i)} +
\Gamma_{ab(i)}^{\rm soft} + \Gamma_{ba(i)}^{\rm soft} +
\Gamma_{a2(i)}^{\rm soft} + \Gamma_{b2(i)}^{\rm soft} +
\Gamma_{2a(i)}^{\rm soft} + \Gamma_{2b(i)}^{\rm soft} \cr &+&
\Gamma_{a(i)}^{\rm coll} + \Gamma_{b(i)}^{\rm coll} +
\Gamma_{2(i)}^{\rm coll} + \Gamma_{1(i)}^{\rm coll} +
\Gamma_{a(i)}^{\rm CT} + \Gamma_{b(i)}^{\rm CT} +
\Gamma_{1(i)}^{\rm CT} \;, \EQA

\noindent which should now be finite. The full NLO result is then

\BQN d\sigma = d\sigma^{sing} + d\sigma^{finite} \;,
\label{singreg} \EQN

\noindent with $d\sigma^{finite}$ given by the finite corrections
of equations [\ref{FINab},  \ref{FIN2},  \ref{FIN1}], and

\BQN d\sigma^{sing} = \sum_i  d_\epsilon [\gamma j] \kappa_i
\Biggl[ \delta(Q_T)+ \beta_\epsilon \sum_{mn} \Gamma_{mn(i)}
\Biggr] \;. \label{sigsing} \EQN

After pole cancellation, of course, we can take $\epsilon
\rightarrow 0$, and

\BQA d_\epsilon[\gamma j] &\rightarrow& dy_\gamma dp_j dy_j d
\phi_j dQ_T d\phi_q \cr \kappa_i &\rightarrow & {{\alpha
\alpha_s}\over S} {p_j \over {2\pi}} 4C_F{[{\cal L}(x_a,x_b)
KT_{f\tilde{f}}(v)]}_i \cr \beta_\epsilon &\rightarrow& {\alpha_s
\over {2\pi}} \;. \EQA

\noindent With the exception of the $m=1$ Bremsstrahlung pieces,
which we will handle separately, each $\Gamma_{mn}$ can be
written in the form (subprocess label $i$ suppressed)

\begin{eqnarray}
\Gamma_{mn} &\equiv & \delta(Q_T) \biggl( {{_2 \Gamma_{mn}}\over
\epsilon^2} + {{_1\Gamma_{mn}}\over \epsilon} \biggr) +
\delta(Q_T) {_\delta C_{mn}} \cr &+& \delta(Q_T) \biggl[
\sum_{\tilde{a}} \int_{x_a}^1 {dz\over z} {{H(x_a/z,x_b)}\over
H(x_a,x_b)} {_{\tilde{a}} C_{mn}} (z) \cr &+& \sum_{\tilde{b}}
\int_{x_b}^1 {dz\over z} {{H(x_a,x_b/z)}\over H(x_a,x_b)}
{_{\tilde{b}} C_{mn}} (z) \biggr] \cr &-&4 A_{mn} {\biggl[ {{\ln
Q_T/Q}\over Q_T} \biggr]}_+ + 2B_{mn} {\biggl[ {1\over Q_T}
\biggr] }_+ \cr &+& {\biggl[ {2\over Q_T} \biggr]}_+ \biggl[ {_a
D_{mn}} \sum_{\tilde{a}} \int_{x_a}^1 {dz\over z}
{{H(x_a/z,x_b)}\over H(x_a,x_b)} P_{a\tilde{a}}^+ (z) \cr &+& {_b
D_{mn}} \sum_{\tilde{b}} \int_{x_b}^1 {dz\over z}
{{H(x_a,x_b/z)}\over H(x_a,x_b)} P_{b\tilde{b}}^+ (z) \biggr] \;,
\label{Gamform} \end{eqnarray}

\noindent the coefficients of which are given in Appendix
\ref{ap:Contribs}. Note (there) that the sum of all ${_2
\Gamma_{mn}}$ coefficients is zero, as required, as is the sum of
${_1 \Gamma_{mn}}$ coefficients. Also note that logs of $Q_T/p_j$
have been expanded into logs of $Q_T/Q$ plus logs of $Q/p_j$ in
order to anticipate comparison with the expansion of the resummed
form.

\bigskip

\section{Extract Resummation Parameters}
\label{sec:ABC}

Each resummation formalism described in the text has, as
parameters, a set of functions which depend on the particular
process under study. In the $b$-space formalism, these parameters
are $A$, $B$, $C_{a/\tilde{a}}$, and $C_{b/\tilde{b}}$, and we
have seen in Chapter \ref{ch:ktspace} that these same parameters
can be used for resummation in $k_T$-space.  Now that we have a
set of finite terms which together constitute the cross section
to a fixed order (here NLO), we have the material necessary to
calculate these parameters.

This can be done in a couple of ways, each of which involves an
expansion of the resummed form to the same order in $\alpha_s$ as
our perturbative result, followed by identification of $A$, $B$,
$C_{a/\tilde{a}}$, and $C_{b/\tilde{b}}$ in the expansion with
corresponding coefficients in that perturbative result.

However straightforward this procedure may sound, it is not
without its complications. Although the $b$-space and $k_T$-space
resummed expressions are formally finite at $Q_T =0$ (even
without the non-perturbative parametrizations), this convergence
is inextricably tied to the all-orders nature of the Sudakov
exponentiation. Attempts to expand either result directly, to a
fixed order, will lead to expressions which are unregulated at
zero $Q_T$.

For the $b$-space formalism, one way to resolve this is to ``meet
the expansion halfway" by comparing the Fourier transform of the
NLO result to an expansion of just the $b$-space integrand. Our
starting point is equation \ref{CSSpert}, minus the finite
correction:

\BQN d\sigma_S^{pert} = {1\over{2\pi}^2} \int d^2 \vec{b}
e^{i\vec{b} \cdot \vec{Q_T}} d{\tilde{\sigma}}^S (b) e^{-S(b,Q)}
\;. \EQN

The $C$ parameters live inside  $d{\tilde{\sigma}}^S$, and can be
expanded to NLO in $\alpha_s$. Since we seek a form in which the
partons are sampled at the same energy as in the NLO expression,
we must also expand the parton distributions $f_{a,b} (b_0/b)$
into $f_{a,b} (M_f)$ plus appropriate convolutions over splitting
functions $P^+$. The resulting expansion of $d{\tilde{\sigma}}^S$
takes the form:

\BQA d{\tilde{\sigma}}^S &\simeq& \sum_{a,b} \Biggl[ f_{a/A}
(x_a,M_f) f_{b/B} (x_b,M_f) \cr &+& {\alpha_s \over {2\pi}}
\sum_{\tilde{a}} \int_{x_a}^1 {{dz}\over z} f_{\tilde{a}/A}
(x_a/z,M_f) f_{b/B} (x_b,M_f) \biggl( C_{a/\tilde{a}}^{(1)} (z) -
2 \ln {{M_f b} \over b_0} P_{a\tilde{a}}^+ (z) \biggr) \cr &+&
{\alpha_s \over {2\pi}} \sum_{\tilde{b}} \int_{x_b}^1 {{dz}\over
z} f_{a/A} (x_a,M_f) f_{\tilde{b}/B} (x_b/z,M_f) \biggl(
C_{b/\tilde{b}}^{(1)} (z) - 2 \ln {{M_f b} \over b_0}
P_{b\tilde{b}}^+ (z) \biggr) \Biggr] \cr &\phantom{\times}& \;.
\EQA

The $A$ and $B$ coefficients live inside the Sudakov exponential
$S(b,Q)$, and can also be expanded to NLO in $\alpha_s$.
Performing the integral over $\mu$ in $S$, while keeping
$\alpha_s (\mu) \simeq \alpha_s (M_f)$ to first order, yields
logarithms of $Qb/b_0$:

\BQN e^{-S(b,Q)} \simeq 1-{\alpha_s \over {2\pi}} \biggl[ 2A^{(1)}
\ln^2 {{Qb}\over b_0} + 2B^{(1)} \ln {{Qb}\over b_0} \biggr] \;.
\EQN

Multiplying these two expressions and keeping only terms of order
$1$ or $\alpha_s$ gives an expansion which can be compared with
the Fourier transform of the NLO result. It is perhaps easiest to
go back a few steps in our NLO derivation to directly transform
the regulated $Q_T^{-1+2\epsilon} \ln^m Q_T/Q$ pieces, but the
transforms over plus-distributions can be done as well. Either
procedure yields the required logs of $Qb/b_0$, so that
identification of the resummation parameters can proceed. The
following integrals are useful in this regard:

\BQN T_+ (x) \equiv \int_0^x {{dt}\over t} \bigl[ J_0 (t) -1
\bigr] + \int_x^\infty {{dt}\over t}  J_0 (t) = \ln {b_0 \over x}
\; \EQN

\BQN T_+^{\ln} (x) \equiv \int_0^x {{dt}\over t} \bigl[ J_0 (t) -1
\bigr] \ln t + \int_x^\infty {{dt}\over t}  J_0 (t) \ln t =
{1\over 2} \ln {b_0 \over x} \Bigl[ 2\ln b_0 - \ln {b_0 \over x}
\Bigr] \; \EQN

Since our plus-distributions have been defined with the upper
limit $p_j$, the Fourier transforms we would need are $T_+ (bp_j)$
and $T_+^{\ln} (bp_j)$. These give logs of $bp_j/b_0$, which can
be expanded into the required logs of $bQ/b_0$, plus logs of
$p_j/Q$. Since any finite upper limit to our $Q_T$-integrals is
immaterial, the latter logs should cancel similar logs in the NLO
result (those created when the plus-distributions were
introduced).

That said, the actual method we will use to extract the
resummation parameters is even simpler, and it relies on the fact
that in the $k_T$-space formalism, the transform from $b$-space
back to transverse momentum space has already been done. We still
cannot directly compare a finite-order expansion, but we can
integrate both the resummed form and the NLO expression up to
some arbitrary limit, and {\it then} expand the former. Beginning
with equation \ref{EV16} (without the remainder term ${\cal R}$),
we integrate on $Q_T^2$ up to some $p_T^2$:

\BQA \int_0^{p_T^2} dQ_T^2 &\sum_{a,b}& \hat{\sigma_0} {d\over
{dQ_T^2}} \Bigl[ {\acute{f}}_a (x_a, Q_T) {\acute{f}}_b (x_b,
Q_T) e^{-S(Q_T,Q)} \Bigr]  = \cr &\sum_{a,b}& \hat{\sigma_0}
\Bigl[ {\acute{f}}_a (x_a, p_T) {\acute{f}}_b (x_b, p_T)
e^{-S(p_T,Q)} - {\acute{f}}_a (x_a, 0) {\acute{f}}_b (x_b, 0)
e^{-S(0,Q)} \Bigr]  \;. \cr &\phantom{\times}& \EQA

Since the Sudakov exponent goes like $S \sim \sum_m \ln^m Q/Q_T$,
taking $Q_T \rightarrow 0$ kills the second term above, and we're
left with only the first term, which, when expanded, gives

\BQA \phantom{x} &\int_0^{p_T^2}& dQ_T^2 d{\sigma}^S \simeq
\sum_{a,b}  \hat{\sigma_0} \Biggl[ f_{a/A} (x_a,M_f) f_{b/B}
(x_b,M_f) \cr &+& {{\alpha_s (M_f)} \over {2\pi}}
\sum_{\tilde{a}}  \int_{x_a}^1 {{dz}\over z} f_{\tilde{a}/A} ({x_a
\over z},M_f) f_{b/B} (x_b,M_f) \biggl(
C_{a/\tilde{a}}^{(1)}(z)-2\ln {M_f \over p_T} P_{a\tilde{a}}^+
(z) \biggr) \cr &+& {{\alpha_s (M_f)}\over {2\pi}}
\sum_{\tilde{b}} \int_{x_b}^1 {{dz}\over z} f_{a/A} (x_a,M_f)
f_{\tilde{b}/B} ({x_b \over z},M_f) \biggl(
C_{b/\tilde{b}}^{(1)}(z)-2 \ln {M_f \over p_T} P_{b\tilde{b}}^+
(z) \biggr) \cr &-& {{\alpha_s (M_f)}\over {2\pi}} \biggl[
2A^{(1)} \ln^2 {Q\over p_T} + 2B^{(1)} \ln {Q\over p_T} \biggr]
\Biggr] \;. \label{KTexpand} \EQA

\noindent To this we must compare the integral of the NLO
expression. As can be seen in equation \ref{Gamform}, we'll need
only three integrals to perform this task. They are:

\BQA \int_0^{p_T} dQ_T \delta (Q_T) &=& 1 \cr \int_0^{p_T} dQ_T
{\biggl[ {1\over Q_T} \biggr]}_+ &=& -{1\over 2} \ln {M_f^2 \over
p_T^2} + {1\over 2} \ln {M_f^2 \over p_j^2} \cr &=&  -{1\over 2}
\ln {Q^2 \over p_T^2} + {1\over 2} \ln {Q^2 \over p_j^2} \cr
\int_0^{p_T} dQ_T {\biggl[ {{\ln Q_T/Q}\over Q_T} \biggr]}_+ &=&
{1\over 8} \ln^2 {Q^2 \over p_T^2} - {1\over 8} \ln^2 {Q^2 \over
p_j^2} \;. \label{plusint} \EQA

\noindent Note that the coefficients of the plus-distributions
we've used are constant, and these distributions were defined
with an upper limit $p_j$. The integral from $0$ to $p_j$ is
therefore zero; the logs appearing in equations \ref{plusint}
arise solely from the integral from $p_j$ to $p_T$.

Applying these expressions to equation \ref{Gamform}, followed by
comparison with equation \ref{KTexpand}, yields the following
resummation parameters:

\BQA A^{(1)} &=& \sum_{mn} A_{mn} \cr B^{(1)} &=& \sum_{mn}
B_{mn} \cr C_{a/\tilde{a}}^{(1)} &=& \sum_{mn}  \biggl[ \delta_{a
\tilde{a}} \delta (1-z) \biggl( {{_\delta C_{mn}}\over 2} +
{A_{mn} \over 4} \ln^2 {Q^2\over p_j^2} + {B_{mn}\over 2} \ln
{Q^2\over p_j^2} \biggr) + {_{\tilde{a}} C_{mn}} \biggr] \cr &+&
\sum_{mn} P_{a\tilde{a}}^+ \biggl[ {_a D_{mn}}\ln {M_f^2 \over
p_j^2} +(1-{_a D_{mn}}) \ln {M_f^2 \over p_T^2} \biggr] \cr
C_{b/\tilde{b}}^{(1)} &=& \sum_{mn} \biggl[ \delta_{b \tilde{b}}
\delta (1-z) \biggl( {{_\delta C_{mn}}\over 2} + {A_{mn} \over 4}
\ln^2 {Q^2\over p_j^2} + {B_{mn}\over 2} \ln {Q^2\over p_j^2}
\biggr) + {_{\tilde{b}} C_{mn}} \biggr] \cr &+& \sum_{mn}
P_{b\tilde{b}}^+ \biggl[ {_b D_{mn}}\ln {M_f^2 \over p_j^2}
+(1-{_b D_{mn}}) \ln {M_f^2 \over p_T^2} \biggr] \;. \cr
&\phantom{\times}& \label{CSSparms} \EQA

Note that delegation of the $A_{mn}$, $B_{mn}$, and ${_\delta
C_{mn}}$ coefficients to one or the other leg ($a$ or $b$) is
arbitrary; we have chosen to assign half to each leg. We choose in
this dissertation to resum only initial-state ($IS$) pieces, and
for these, ${_a D_{mn}}={_b D_{mn}}=1$, so the $\ln M_f^2/p_T^2$
term drops in the above. Final-state ($FS$) pieces are left in
the NLO form shown in the last section. From here on out, we will
make a distinction among the following four types of
contributions:

\BQA d\sigma &=& d\sigma_{IS}^{sing} + d\sigma_{IS}^{finite} \;
\cr &+& d\sigma_{FS}^{sing} + d\sigma_{FS}^{finite} \;. \EQA

\noindent $d\sigma^{sing}$ are the (regulated) asymptotic
approximations. $d\sigma^{finite}$ are the finite corrections
(compare equation \ref{singreg}). Resummation will be performed
on only the first term, and afterwards we will call it
$d\sigma_{IS}^{Resum}$. The initial/final distinction cannot be
applied to virtual terms; we make the choice to include them in
$d\sigma_{IS}^{sing}$, where they show up in the CSS $C^{(1)}$
pieces.

Note also that in this transverse momentum space comparison, we
again obtain the logs of $Q/p_j$ mentioned above. As the ${_\delta
C_{mn}}$ pieces already contain logs of $p_j/M_f$, addition of
the new logarithms serves to produce $p_j$-independent
$C^{(1)}$-parameters, as expected. The new $D_{mn} P^+ \ln (M_f^2
/ p_j^2)$ terms also serve to cancel corresponding terms in ${_a
C_{mn}}$ and ${_b C_{mn}}$.

Finally, a word on free variables: references to $b$-space and
$k_T$-space resummed expressions usually assume the set of free
variables $\bigl\{ Q,y,Q_T,\phi_q,\cos\theta_{\rm CSS},\phi_{\rm
CSS} \bigr\}$, in which $Q^2$ and $\phi_q$ are the mass and
azimuthal angle of the observed particle pair, and $\{
\theta_{\rm CSS}, \phi_{\rm CSS} \}$ the direction of one of the
particles in a specific 2-body rest frame. We, of course, are
using a different set of variables, namely $\bigl\{
y_\gamma,p_j,y_j,\phi_j,Q_T,\phi_q \bigr\}$, but given the
existence of a nonsingular Jacobian between these two sets of
variables (as there must be), there is sufficient justification to
write the resummed result with either set of variables.
\footnote{Once the CSS coefficients are found, Monte-Carlo coding of 
the resummed piece proceeds as given by the CSS formalism, in which the
upper limit on $Q_T$ is the $\gamma j$ pair mass $Q$ (which 
$=\sqrt{2p_j^2(1+\cosh(y_\gamma-y_j))}$ in our variables),
 not the limit given by the theta function $\theta (p_j-Q_T)$ at NLO.
The question of whether this constitutes double-counting, or is physically 
justified given the {\bf multiple} emission being described, will be
discussed in a future work.}

\section{Matching Resummed and Fixed-Order Results}
\label{sec:matchme}

\begin{figure}[h]\centering
\begin{minipage}{0.75\linewidth}
\centering \epsfxsize = 0.75\linewidth
\leavevmode\epsffile{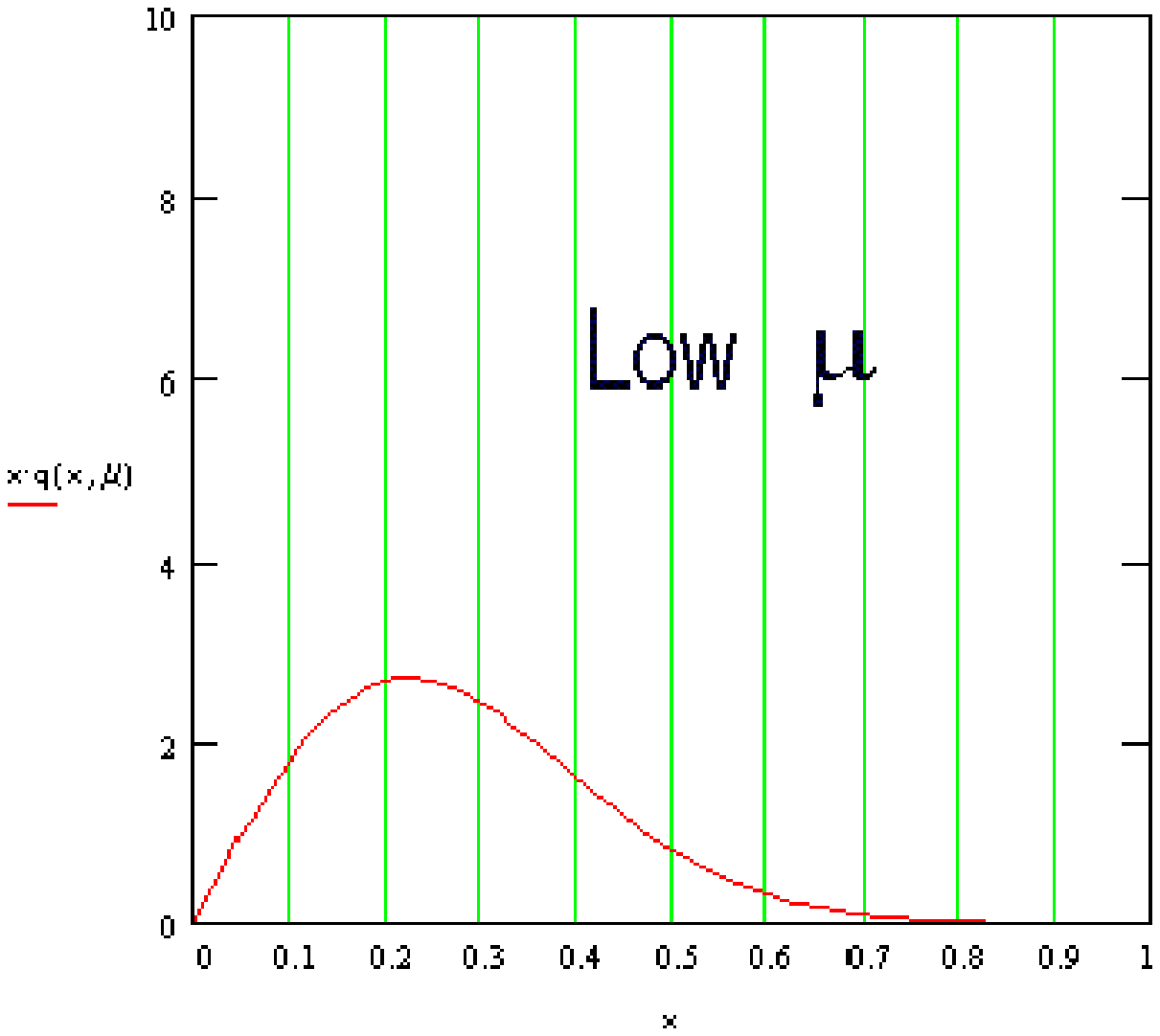}
\end{minipage} \begin{minipage}{0.75\linewidth}
\centering \epsfxsize = 0.75\linewidth
\leavevmode\epsffile{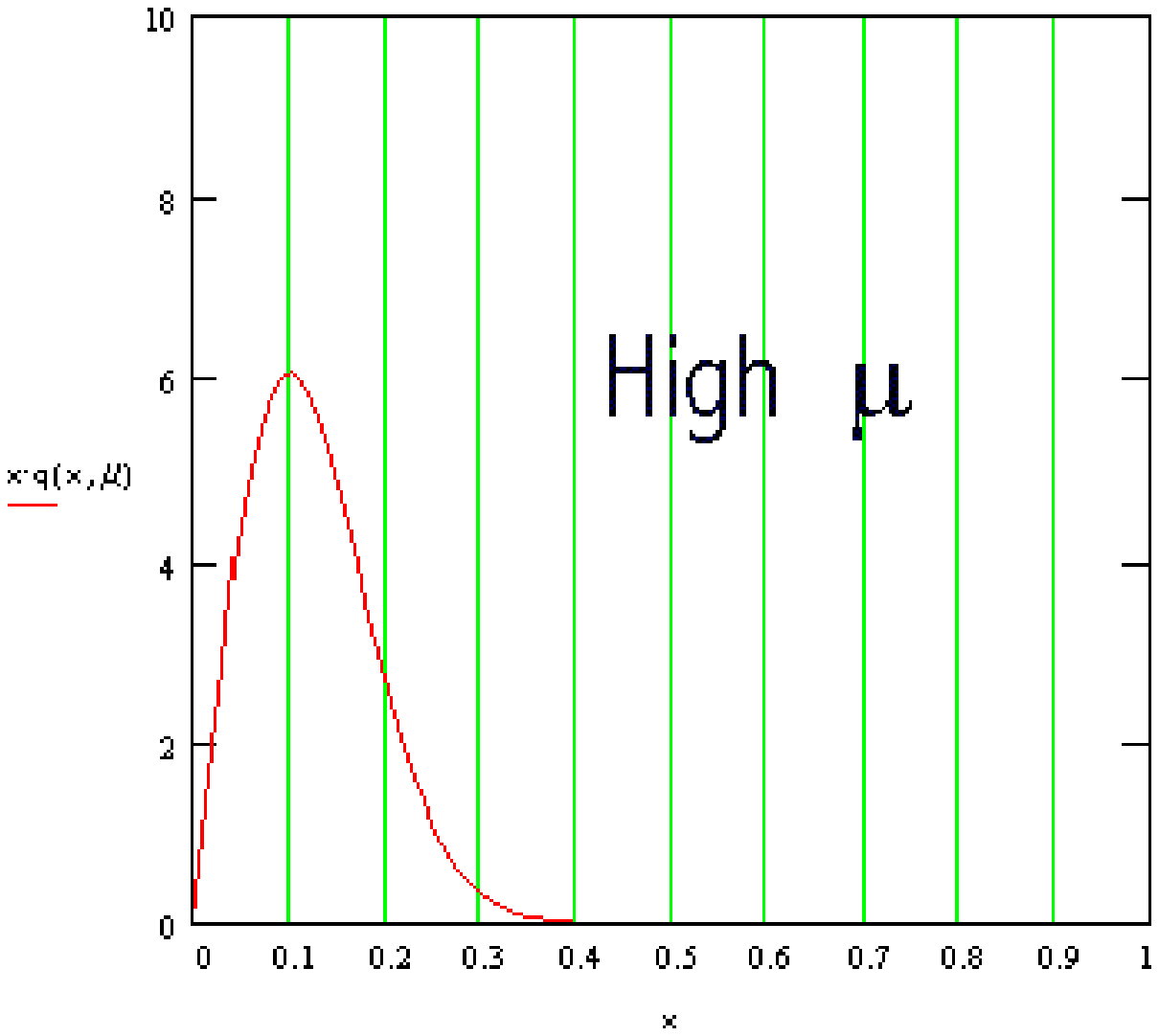}
\end{minipage}
\caption[Scale dependence of parton distributions.]{Scale
dependence of parton distributions.} \label{fig:lohimu}
\end{figure}

Depending on the quality of the approximations used in defining
the resummable asymptotic piece $d\sigma_{IS}^{sing}$, there is no
guarantee that it will continue to be a useful quantity at high
pair $Q_T$, either before or after resummation. In fact, it may
even become negative: looking at the resummed $k_T$-space
expression of equation \ref{EVfinal}, one can see that as $Q_T$
gets bigger, both the Sudakov and non-perturbative exponentials
tend to one, while their derivatives tend to zero. This leaves
only a term proportional to the derivative of the parton
distributions with respect to scale. According to the
Altarelli-Parisi equations, this contribution (and thus
$d\sigma_{IS}^{Resum}$) will be negative for all but the smallest
momentum fractions (see figure \ref{fig:lohimu}).

On the other hand, the 3-body perturbative piece (that is, the
unregulated NLO result) is fine in the high-$Q_T$ region. The act
of resumming its logs of $Q/Q_T$ should only make a difference in
the low-$Q_T$ region.

One is thus forced to make a switch from the resummed piece
$d\sigma_{IS}^{Resum}$ to the perturbative 3-body piece at some
value of $Q_T$ (see figure \ref{fig:match}). The best method is
still a matter of debate. Remember that in
$d\sigma_{IS}^{finite}$ there live both the perturbative 3-body
result and a 3-body asymptotic approximation thereof. The
resummed piece is a regulated and exponentiated version of the
latter, and if the resummation procedure leaves its high-$Q_T$
behavior significantly unchanged from the asymptotic
approximation on which it is based, there is hope that
cancellation will occur between these pieces, leaving only the
perturbative piece. If not, then this cancellation will require a
little help.

Kauffman ~\cite{Kauf91} proposed inclusion of a function $f(Q_T)$
to smoothly force this cancellation as $Q_T$ becomes greater than
some arbitrary {\it matching value} $Q_T^{\rm match}$. That is,

\BQN d\sigma \simeq f(Q_T,Q_T^{\rm match}) \bigl( {\rm Resummed}
- {\rm Asymptotic} \bigr) + {\rm Perturbative} \;, \EQN

\noindent where, it was suggested, an appropriate function might
be

\BQN f(Q_T,Q_T^{\rm match}) = {1\over {1+{\biggl({Q_T\over
Q_T^{\rm match}}\biggr)}^4}} \;. \EQN

\noindent This allows all contributions to do their job at low
$Q_T$, but shuts off the resummed and asymptotic contributions at
high-$Q_T$.

\figboxa{match}{Fixed-order perturbative result better at high
$Q_T$.}

Of course, it remains to determine $Q_T^{\rm match}$, and it is
by no means certain that one particular value will suffice for all
circumstances. If one's desired output is a single
$Q_T$-distribution at fixed pair mass and rapidity, it may be
easy to determine the optimum value by trial and error; better
would be to histogram each piece separately and, after the full
run, pick a $Q_T$-value at which the resummed and perturbative
pieces most closely agree, then drop all higher-$Q_T$ resummed and
asymptotic contributions.

The procedure we will use in this dissertation allows the
kinematics of the particular incoming state to determine the
proper cutoff (as opposed to a fixed $Q_T^{\rm match}$). However,
it does so only by allowing one distribution to be binned at a
time.

That is, we assume an output consisting of a single
one-dimensional histogram, the ordinate of which we will here
label $x$. Given a domain and step size in $x$, we have then a
defined number $N$ of bins in $x$.

1. For each $x$-bin, we construct a two dimensional array in $Q$
and $Q_T$ (up to $\sqrt{S}/2$ each). Thus in effect we have a
3-dimensional array in $x$, $Q$, and $Q_T$.

2. Proceeding with Monte-Carlo evaluation of $d\sigma^{resum}$, we
fill this 3-D array with weights.

3. We then collapse (integrate over $x$) this array into a 2-D
array in $Q$ and $Q_T$ only. For each $Q$-value, we start at
$Q_T=0$ and go up in $Q_T$ until we see the $x$-integrated weight
dip below zero. This will define a cutoff $Q_T^{\rm match}(Q)$
for each value of $Q$. That is, for each $Q$, all bins above this
$Q_T$-value are set to zero in the original 3-D array, for all
$x$.

4. The original 3-D array may now be integrated over $Q$ and
$Q_T$ to determine the proper $d\sigma_{IS}^{Resum}$ for each $x$.

5. The array of cutoffs $Q_T^{\rm match}(Q)$ is now used in the
subsequent evaluation of the 3-body weights ($d\sigma^{finite}$).
As each 3-body point is generated, $Q$ and $Q_T$ are calculated
for this point, and if $Q_T > Q_T^{\rm match}$ for that $Q$, the
asymptotic weight is not calculated, leaving only the
perturbative.

This procedure, while it has its limitations, nevertheless
respects the approximations in their regions of accuracy, while
not forcing a unique matching $Q_T$.

However, for the sake of relative simplicity, we're not using the
optimum cutoff criterion mentioned above, which would be to
compare the resummed cross-section to the perturbative, and
perform the cutoff at the $Q_T$ value at which they most closely
agree. This would require additional arrays and introduction of
some sort of tolerance parameter. It could be done, but the
results we've been able to achieve with the present system don't
reveal any need for the additional overhead.

In fact, at high enough hadronic-to-partonic energy ratios (low
momentum fraction), the resummed piece goes smoothly to zero
(without going negative) at high $Q_T$, and matching procedures
of any sort are not required. The above procedure alters nothing
in this case; by contrast, a fixed-$Q_T^{\rm match}$ procedure
would kick in anyway.

\chapter{Results and Conclusions}
\label{ch:Results}

\section{Photon $p_T$ Distributions}

The direct photon $p_T$ spectrum is probably the easiest test of
our formalism, as there are data, and no correlations with jets to
consider. We begin with results at collider energies, and end
with lower energy fixed-target results. We use the $k_T$-space
resummation formalism alone, at first, and use the
nonperturbative parameters $\tilde{a} = 0.3 {\rm GeV}^{-2}$ and
$k_{Tlim} = 4 {\rm GeV}$ for each subprocess. A discussion of
these parameters, and why we may want to use different fits for
different subprocesses, will follow in the next section.

In figure \ref{fig:d0before}, we show four curves: The LO
contribution, the NLO results, and finally resummed results using
the $k_T$ formalism, both with and without matching. Data are
from the Fermilab D0 experiment ~\cite{D099}. The process studied
is $p \bar{p} \rightarrow \gamma + X$ at $\sqrt{S}=1800{\rm
GeV}$, with the photon rapidity constrained at $|y_\gamma| \le
0.9$ and factorization scale $M_f = p_\gamma$, where $p_\gamma$
is the transverse momentum of the photon. There is also an
isolation cut on the photon, which rejects events with a jet of
energy $p_j > 2{\rm GeV}$ within $R \equiv
\sqrt{{(y_\gamma-y_j)}^2+{(\phi_\gamma-\phi_j)}^2} \leq 0.4$ of
the photon.

In figure \ref{fig:e800before}, we show the same four curves, this
time at the lower energy $\sqrt{S}=38.7 {\rm GeV}$, and for the
quantity $E_\gamma d^3\sigma / d^3 p_\gamma$, the invariant cross
section, as opposed to $d^2\sigma / dp_\gamma dy_\gamma$. Data
are from the Fermilab E706 experiment ~\cite{E70697}. The process
is $p Be \rightarrow \gamma + X$, with the photon rapidity
constrained at $-1.0 \leq y_\gamma \leq 0.5$.

Notable here is the effect of our matching prescription
(described in Section \ref{sec:matchme}), which cuts out the poor
high-$Q_T$ behavior of the resummed and asymptotic pieces. At
$p_\gamma \approx 3 {\rm GeV}$, the incoming partons' momentum
fractions are on the order $x \approx 2p_\gamma/\sqrt{S} \approx
0.2$, but as one approaches $p_\gamma \approx 6.5 \sim 12.0 {\rm
GeV}$, these fractions reach $x \approx 0.4 \sim 0.8$, and in
this range, the derivative of the parton distributions (which is
dominant in the resummed piece at high-$Q_T$) is negative. In
contrast, matching at the higher energy $\sqrt{S}=1800 {\rm GeV}$
is not an issue, as the momentum fractions concerned rarely go
above $x \approx 0.13$.

For completeness, we repeat the trials at the still lower energy
$\sqrt{S}=31.5{\rm GeV}$ (see figure \ref{fig:epv530}). Data here
are also from the Fermilab E706 experiment ~\cite{E70697}. The
process is $p Be \rightarrow \gamma + X$, with the photon
rapidity constrained at $|y_\gamma| \leq 0.75$. For all these
curves, the factorization scale $M_f = p_\gamma$ is used, and the
parton distributions are those of the CTEQ Collaboration,
specifically CTEQ5M ~\cite{CTEQ99}.

\figboxf{d0before}{$p \bar{p} \rightarrow \gamma + X$ at
$\sqrt{S}=1800{\rm GeV}$.}

\figboxf{e800before}{$p Be \rightarrow \gamma + X$ at
$\sqrt{S}=38.7 {\rm GeV}$.}

\figboxf{epv530}{$p Be \rightarrow \gamma + X$ at
$\sqrt{S}=31.5{\rm GeV}$.}

\clearpage

\subsection{Nonperturbative Parameter Choices}

As the lower-energy results are much more sensitive to changes in
the nonperturbative parameters, we focus exclusively on
comparison with the E706 results at $\sqrt{S}=31.5{\rm GeV}$.

In the plots discussed above, the $k_T$-space resummation
formalism was utilized exclusively. In accordance with the work
of Ellis and Veseli ~\cite{EV97}, this includes a nonperturbative
factor of the form

\BQN F_{NP} (Q_T) = 1-e^{-\tilde{a}Q_T^2} \;. \EQN

\noindent Besides $\tilde{a}$, there is also a second
nonperturbative parameter, $k_{Tlim}$, which controls the
boundary between perturbative and nonperturbative regions. We use
a value $k_{Tlim} = 4 {\rm GeV}$, but changes of up to 50\% in
this parameter have little effect, as can be seen in figure
\ref{fig:kcompt}. Thus, for the balance of this argument, we will
think of the $k_T$-space method as having effectively a
single-parameter nonperturbative form.

Starting with a value of $\tilde{a} = 0.3 {\rm GeV}^{-2}$, figure
\ref{fig:kcompa} shows the effect of raising and lowering
$\tilde{a}$ as $k_{Tlim}$ is held fixed. As can be seen here and
in the previous plots, our initial value works well with
available data, but the results are significantly dependent upon
the precise value at low $\sqrt{S}$.

\figboxf{kcompt}{$k_T$-space resummation : $k_{Tlim}$-dependence.}

\figboxf{kcompa}{$k_T$-space resummation :
$\tilde{a}$-dependence.}

\clearpage

This being the case, it is important to compare these $k_T$-space
results with the predictions given by the $b$-space formalism,
the parameters of which are, ostensibly, better known. Using the
$b$-space CSS formalism of Chapter \ref{ch:CSS}, and the Sudakov
parameters calculated in Chapter \ref{ch:photon}, figure
\ref{fig:scompref} shows the Born, NLO, and resummed theoretical
results compared with data. For the resummed curve, the
2-parameter nonperturbative model of Landry, Brock, Ladinsky, and
Yuan (LBLY) ~\cite{LBLY99} is used, with their best-fit
parameters $g_1 = 0.24 {\rm GeV}^2$ and $g_2 = 0.34 {\rm GeV}^2$.

Figures \ref{fig:scomp1} and \ref{fig:scomp2} show, respectively,
the result of altering $g_1$ and $g_2$ by 50\% each. As can be
seen from the parametrization

\BQN F_{NP} (b,Q) = \exp \Bigl[ -b^2 \biggl( g_2 \ln {Q\over
{2Q_0}} + g_1 \biggr) \Bigr] \;, \EQN

\noindent raising either $g_1$ or $g_2$ dampens out more of the
high-$b$ region, thereby broadening the $Q_T$ spectrum, and
consequently raising the average recoil momentum.

\figboxf{scompref}{$b$-space resummation : reference
distribution.} \figboxf{scomp1}{$b$-space resummation : $g_1$
dependence.} \figboxf{scomp2}{$b$-space resummation : $g_2$
dependence.}

\clearpage

Unfortunately, none of these curves agrees as well with the data
as does the single-parameter, $Q$-independent $k_T$-space result.
The first question to be answered, then, is whether the
difference lies entirely in the $Q$-dependence of the $b$-space
parametrization. To do this, we first look at the approximate
analytical relation between each parameter set and the average
$Q_T$ ``kick'' given to the system.

For example, in the $b$-space model, to first order in $\alpha_s$
we are resumming only the Born terms, and the $b$-dependence of
the Fourier integrand is simply that of the nonperturbative
exponential, which we write as simply $\exp(-gb^2)$. Then,

\BQA {{d\sigma}\over {dQ_T}} &\sim& {{2Q_T}\over {2\pi}} \int
d^2b e^{i\vec{b} \cdot \vec{Q_T}} e^{-gb^2} \; \cr &=& 2Q_T \int
db b J_0 (bQ_T) e^{-gb^2} \; \cr &=& {Q_T\over g} e^{-Q_T^2/4g}
\;, \EQA

\noindent which has a maximum at $<Q_T> = \sqrt{2g}$.

In the $k_T$-space model, we have

\BQA  {{d\sigma}\over {dQ_T}} &\sim& 2Q_T {d\over {dQ_T^2}}
\Bigl[ 1- e^{-\tilde{a}Q_T^2} \Bigr] \; \cr &=& 2Q_T \tilde{a}
e^{-\tilde{a}Q_T^2} \;, \EQA

\noindent which peaks at $<Q_T> = 1/\sqrt{2\tilde{a}}$. Equating
these two results gives a simple relation $g\tilde{a}=1/4$ which
we can use to test the effect of the $Q$-dependence in $g$.

Figure \ref{fig:qdepon} shows the result of a $k_T$-space run with
$\tilde{a} = 1/(4g)$ superimposed upon a $b$-space run of the
canonical two-parameter form $g(Q)= g_2 \ln {Q\over {2Q_0}} +
g_1$. One can see that adding the $Q$-dependence to $\tilde{a}$
reproduces the $b$-space result. Conversely, as shown in figure
\ref{fig:qdepoff}, {\bf removing} the $Q$-dependence by fixing
$g=1/(4\tilde{a})$ also results in agreement between the $b$ and
$k_T$-space formalisms.

\figboxf{qdepon}{Adding $Q$-dependence to $k_T$-space NP form.}
\figboxf{qdepoff}{Removing $Q$-dependence from $b$-space NP form.}

\clearpage

The next question is whether a $Q$-dependent form is truly
required by the existing data, and for that to be answered we
must look at the application of resummation methods to
intermediate vector boson production, {\it e.g.} Drell-Yan
processes, which comprise the bulk of the data upon which the
current nonperturbative fits have been based.

By simply swapping out our direct photon CSS parameters $\{A_i,
B_i, C_i \}$ for those calculated by Ellis, Ross, and Veseli
~\cite{ERV97} or Bal\'{a}zs and Yuan ~\cite{BY97}, we can begin
comparing resummed theory to data for dimuon production, as was
done in references ~\cite{LY94,LBLY99,DWS85} (see Section
\ref{sec:CSSNP}). Here we include only the resummed piece (which
gives the bulk of the cross section in this region), not the
finite corrections. Figure \ref{fig:e605ref} shows an invariant
cross section as a function of $Q_T$ for various $Q$-bins. The
data are from the E605 experiment ~\cite{E605}, and the theory
curves are the $b$-space resummed predictions using the
two-parameter $Q$-dependent nonperturbative form. The values
$g_1=0.24 {\rm GeV}^2$ and $g_2=0.34 {\rm GeV}^2$ are the result
of the global fit performed by Landry, Brock, Ladinsky and Yuan
(LBLY) ~\cite{LBLY99}, which considered data only out to $Q_T=1.4
{\rm GeV}$.

However, as shown in figure \ref{fig:e605f1}, we find that using a
single-parameter, $Q$-{\bf independent} form, with $g_1 \simeq 0.5
{\rm GeV}^2$ and $g_2 = 0.0 {\rm GeV}^2$, produces a better fit
to the shape of the cross section at every range of $Q$,
especially in light of the 15\% normalization uncertainty of the
data. This is true also for the dimuon data of the R209 experiment
~\cite{R209}, as shown in figure \ref{fig:r209comp}. These data
have a normalization uncertainty of 10\%, and were also used in
the LBLY fit.

\figboxf{e605ref}{E605 $b$-space predictions, standard 2-parameter
form.}

\figboxf{e605f1}{E605 $b$-space predictions, w/out
$Q$-dependence.}

\figboxf{r209comp}{R209 $b$-space predictions, with and w/out
$Q$-dependence.}

\clearpage

As seen in the previous section, $k_T$-space resummation did well
in describing direct photon data, as long as $\tilde{a}$ was not
too far from $0.3 {\rm GeV}^{-2}$. The value of $\tilde{a}$ which
corresponds to a $b$-space parameter $g=0.5 {\rm GeV}^2$,
however, is $\tilde{a} \simeq 1/(4g) \simeq 0.5 {\rm GeV}^{-2}$.
Drell-Yan dimuon production, however, has only quark annihilation
subprocesses as Born terms, while direct photon production has
gluon-induced Born-level terms, so it is possible that a
color-structure-dependent nonperturbative form is required.

To check this, we repeat the direct photon predictions of the
previous section, using single-parameter, $Q$-independent forms
for both $b$ and $k_T$-space resummations. For the $q\bar{q}$
contributions, we use $g=0.5 {\rm GeV}^2$ and $\tilde{a}=0.5 {\rm
GeV}^{-2}$, and for the $qg$ and $gq$ contributions,

\BQA g &= {{(C_F+N_C)}\over {2C_F}} 0.5 {\rm GeV}^2 &= 0.83 {\rm
GeV}^2 \cr {\rm and} \cr \tilde{a} &= 1/(4g) &= 0.3 {\rm
GeV}^{-2} \;. \EQA

\noindent The results, for E706 data at $\sqrt{S}=31.5 {\rm GeV}$
and $38.7 {\rm GeV}$, and for $D0$ data at $\sqrt{S}=1800 {\rm
GeV}$, are shown in figures \ref{fig:e530gw}, \ref{fig:e800gw},
and \ref{fig:d0gw}, respectively. The factorization scale is
$M_f=p_\gamma$ for all plots. All show improved agreement with
data for both $b$ and $k_T$-space resummation procedures.

\figboxf{e530gw}{E706 $\sqrt{S}=31.5{\rm GeV}$, with NP color
dependence.}

\figboxf{e800gw}{E706 $\sqrt{S}=38.7 {\rm GeV}$, with NP color
dependence.}

\figboxf{d0gw}{D0 $\sqrt{S}=1800{\rm GeV}$, with NP color
dependence.}

\clearpage

\section{Other Parameter Choices}

\subsection{Scale Dependence}
\label{subsec:ScaleDep}

In a leading-order calculation, the Born terms are combined with
energy-dependent couplings and distribution functions to produce
a Leading-Log (LL) approximation. As discussed in Chapter
\ref{ch:QCDpert}, when next-to-leading order contributions are
added, the dependence on renormalization and factorization scales
should diminish. We check this by looking at photon $p_T$
distributions in $p Be$ collisions at a fixed target energy of
$530 {\rm GeV}$. We set the renormalization and factorization
scales equal to each other, and run the LL and NLL calculations
for $3 \leq p_\gamma \leq 12{\rm GeV}$ at three values of this
scale. When we're done, for each $p_\gamma$ bin, we plot the
difference between the low and high scales in ratio with the
value at the middle scale. The NLL calculation should show
smaller percent differences, and does, as shown in figure
\ref{fig:scaledep}.

\figboxa{scaledep}{Scale Dependence.}

\subsection{Jet Cone Width}
\label{subsec:JetWidth}

The jet cone definition we have used counts two partons as
belonging to the same jet if they are within an angular radius
$R$ of each other, this radius defined by ~\cite{KS92}

\begin{equation}
r^2 =  \biggl[ {(\Delta y)}^2 + {(\Delta \phi)}^2 \biggr]
{\biggl({p_2 \over {p_2+p_3}} \biggr)}^2 \;.
\end{equation}.

However, our asymptotic pieces do not depend on this radius, as
we have taken a fixed jet definition for each (see Chapter
\ref{ch:photon}). The only possible dependence we have on this
parameter comes via the three-body perturbative contribution, and
then only through those observable quantities which are defined
differently depending on whether or not the cone condition is
met. For example, if we are looking at a photon $p_\gamma$
spectrum, with cuts on the photon rapidity $y_\gamma$, we should
see no dependence on the cone radius $R$, since $p_\gamma =
\sqrt{p_2^2+p_3^2 + 2p_2p_3 \cos (\phi_2 - \phi_3)}$ and
$y_\gamma = y_1$ everywhere in phase space. However, if we are
looking at a $Q_T$ distribution, say, then raising $R$ should
steepen the curve, since more weights are being binned as if the
photon were recoiling against a single jet, {\it i.e.} at $Q_T=0$.

\figboxf{eqtinit}{$Q_T$-distribution : Initial-state pieces.}

\figboxf{eqtfinal}{$Q_T$-distribution : Final-state pieces.}

Figures \ref{fig:eqtinit} and \ref{fig:eqtfinal} show the cross
section vs. $Q_T$ for those pieces associated with initial and
final-state singularities. In each, the perturbative, asymptotic,
2-body, finite (perturbative - asymptotic), and total are
displayed. Here $R=0.2$, $6.5 \leq Q \leq 7.5 {\rm GeV}$, $-0.75
\leq y_\gamma \leq 0.75$, $-0.3 \leq y_j \leq 0.3$, and the
reaction is $p Be \rightarrow \gamma + j + X$.

\clearpage

At present we know of no data for the $Q_T$ distribution of a
photon/jet system, which as just as well, as there are caveats
associated with predicting $Q_T$-distributions in our current
calculation.

As can be seen in Chapter \ref{ch:photon}, in order to extract the
logarithms of $Q/Q_T$ for resummation, we needed to analytically
integrate over the rapidity $y_3$ of the third final-state
particle. However, our jet definition depended upon $y_3$. We
were forced to approximate the kinematics as well as the matrix
elements, thus removing this $y_3$-dependence. For the
initial-state pieces, we took $Q_T=p_3$ everywhere, even though
radiation inside the jet cone would have resulted in $Q_T=0$.
This doesn't cause much of a problem for small cone width. For
the final-state pieces, we treated the soft terms differently
than the collinear, and this can cause a mismatch even outside
the cone. For example, if a 3-body event comes along for which
$p_3^\mu$ is soft and just outside the cone, we have the
following scenario:

\begin{table}[h]
\caption{Anomaly at low $Q_T$.} \vspace{.5cm}
\begin{center}
\thicklines
\begin {tabular} { lll }
\hline \hline \thinlines
Piece & ...looks like: & ...but gets binned at: \\
\hline \thicklines
3-body Pert. & SC & $Q_T=p_3$ \\
$m=2$ (Soft - SC) & 0 & $Q_T=p_3$  \\
$m=2$ Coll. & SC & $Q_T = 0$ \\
\hline \hline
\end{tabular}
\end{center}
\label{QTanomaly}
\end{table}

\noindent Here {\bf SC} stands for ``Soft-Collinear". The 3-body
perturbative piece and the $m=2$ asymptotic collinear piece both
give roughly the same weight, but the asymptotic collinear is
always binned at $Q_T=0$, while the perturbative piece isn't. If
the bin width in $Q_T$ is larger than this $p_3$, the
cancellation will take place. If not, it won't.

\figboxf{jdefprob}{Comparison of full and approximate jet
definition results.}

In figure \ref{fig:jdefprob}, we show a comparison of $Q_T$
distributions for the $m=2$ finite (perturbative - asymptotic)
piece, one curve corresponding to the approximated jet definition
we were forced to use, the other curve showing what would result
were we able to use the non-approximate version. The latter tends
nicely to zero as $Q_T \rightarrow 0$, while the former retains a
steep increase and is only counteracted in the first bin by the
large, subtracted collinear contribution.

The result is a non-continuous $Q_T$ distribution, and so further
work is required in order to deal properly with these final-state
terms. Eventually, of course, we'd like to resum these pieces,
and to do so in a way that is independent of the particular jet
definition in use. For the present, integration over $Q_T$, as in
the preceding inclusive $p_\gamma$-distributions, gives a good,
unambiguous result.

\section{Conclusions and Future Improvements}

We have seen the means by which soft radiative corrections can be
resummed to all orders, and the significant effect such
contributions have on the direct photon $p_T$ spectrum. In
principle, any steeply-falling spectrum is a candidate for this
procedure, including most single-particle inclusive transverse
momentum distributions. As contributions from all orders are
included, resummation improves agreement with data as compared to
fixed-order calculations. Thus there is the potential for a
significant impact on the prediction of a wide variety of
processes.

In this work, we have also seen that direct photon data can play
a role in constraining the nonperturbative function; in
particular we have seen evidence of a dependence on the color
structure of the subprocess, in the form of mass-independent
Gaussians for each structure. The nonperturbative parametrization
is thus far from pinned-down, and it is likely that further
exploration of perturbative methods can aid in probing the
nonperturbative region.

Proper matching of the resummed to fixed-order result has been
shown to be important, especially at fixed-target energies. The
matching procedure outlined in this work accomplishes this task
in a flexible and automatic manner, without influencing the result
in any way when not required.

The FORTRAN program written concurrently with this dissertation,
and from which the results of this chapter were produced, is
available upon request. It is capable of producing either NLO or
resummed output, as the user desires.

Without significantly altering the adopted procedure, it should
be possible to avoid problems associated with jet definitions by
looking at photon plus pion final states, for which there are a
good amount of data ~\cite{E70697}. Furthermore, $Q_T$ and
$p_{out}$ distributions exist for these data, providing a more
sensitive base upon which to examine nonperturbative
parametrizations.

Inclusion of threshhold effects is also possible, as indicated by
the recent work of Laenen, Sterman, and Vogelsang ~\cite{LSV00}.
This method involves both Fourier and Mellin transforms, and
looks, though complicated, to be quite powerful.

\appendix
\setcounter{footnote}{0}
\chapter{Diagrams and Matrix Elements for NLO Photon Hadroproduction}
\label{ap:Diagrams}

In this appendix, we include Feynman diagrams of the contributions
to photon production by hadrons, first to leading order, then to
next-to-leading order.

Figure \ref{fig:vertices} shows the basic couplings involved. The
first is a QED-like photon-quark coupling of strength $e_{\rm
QED}$. The second is the analogous QCD gluon-quark coupling, this
time of strength $g_s$. The third, a gluon-gluon coupling, has no
QED analogue. In general, all non-Abelian theories (such as QCD)
will have self-couplings among the gauge fields. Note that in QCD
there is another gluon coupling (with four attached propagators),
but as it is of order $g_s^2$, it will not be needed here.

\BFIG
\centerline{\psfig{figure=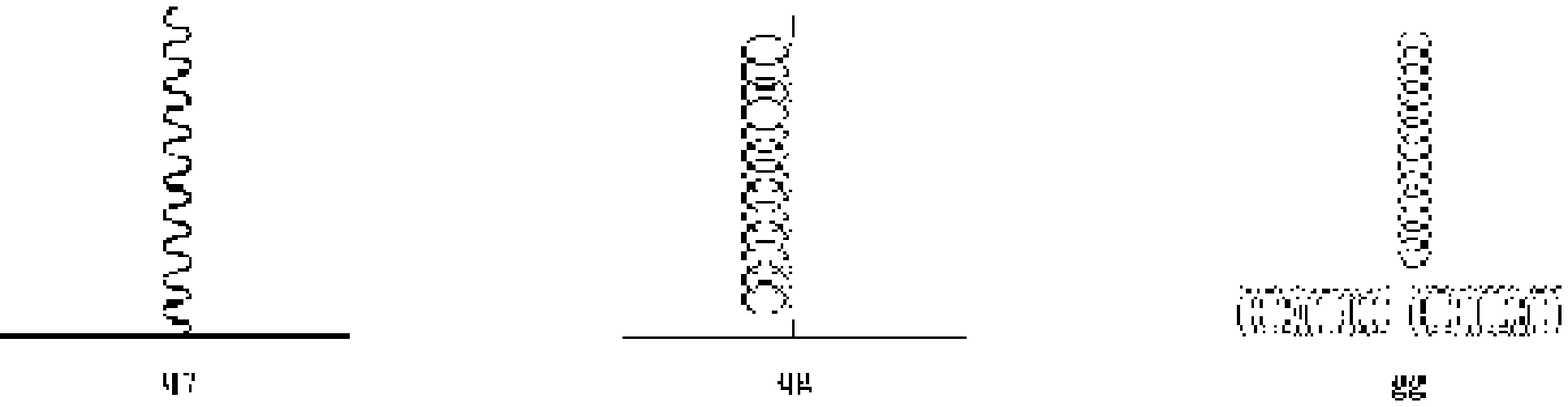,height=4.14cm,width=\linewidth}}
\caption[Basic couplings.]{$q\gamma$, $qg$, and $gg$ vertices.}
\label{fig:vertices} \EFIG

To find the leading-order contributions to photon
hadroproduction, we need to put these vertices together (along
with suitable propagators and external legs) in all possible
combinations consistent with the following conditions:

\begin{list}{}{}
\item 1. There must be one final-state photon.
\item 2. There must be two incoming particles, each either a quark
or gluon.
\item 3. Each diagram must be {\it simply-connected} and maintain momentum conservation.
\item 4. Each diagram must use as few vertices as possible.
\item 5. Each diagram must be topologically distinct.
\end{list}

These conditions constrain us to the diagrams shown in figure
\ref{fig:loqed}.


\BFIG
\centerline{\psfig{figure=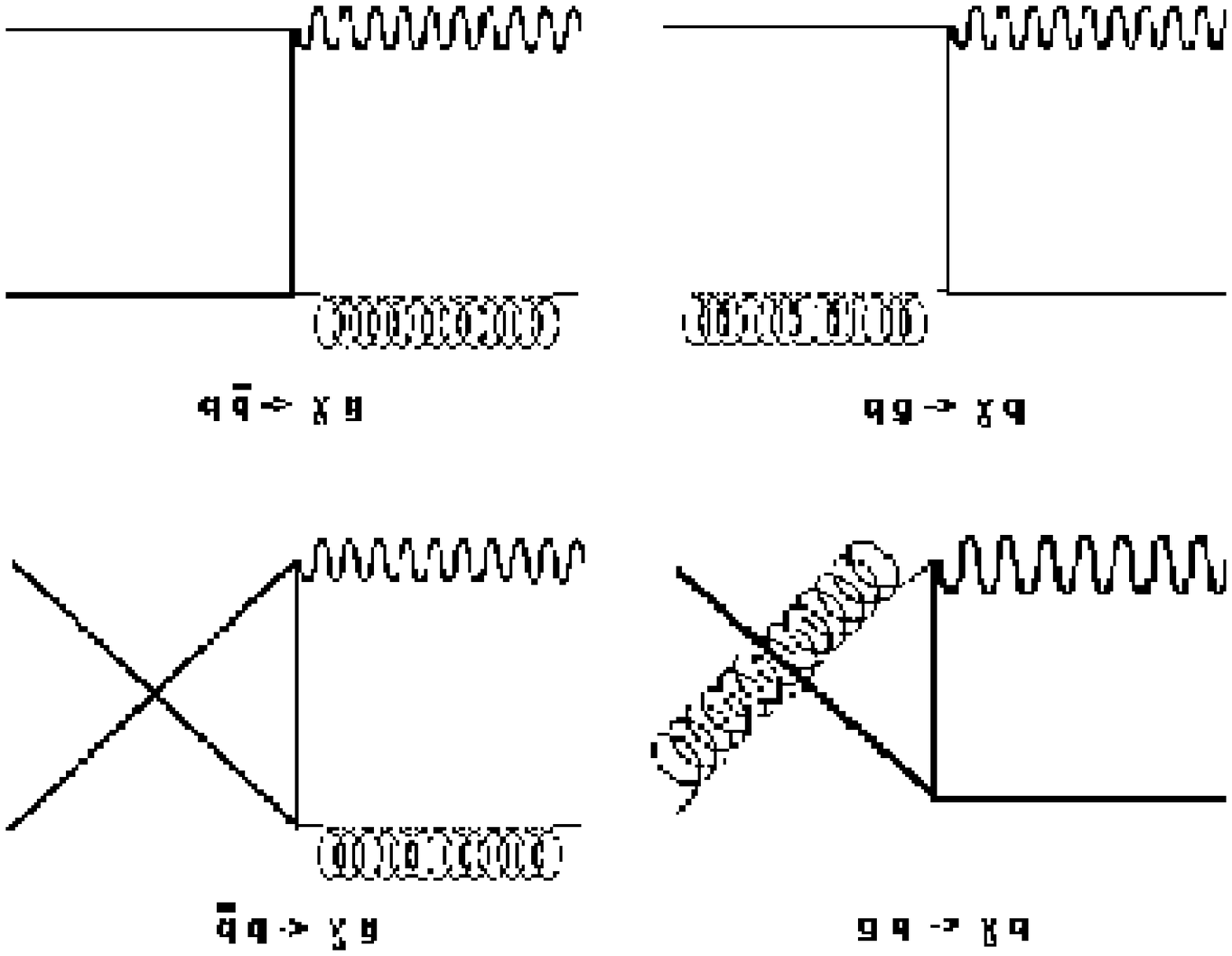,height=9.73cm,width=11.52cm}}
\caption[Leading order photon hadroproduction diagrams.]{Leading
order photon hadroproduction diagrams.} \label{fig:loqed} \EFIG

Now we go to next-to-leading-order (NLO). We first seek the basic
simply-connected Green's functions (that is, tree diagrams) which
include one photon-quark vertex and two vertices from the set $\{
qg, gg \}$. Given the ``connected" constraint, we rule out two
$gg$ vertices, and arrive at the following four diagrams (fig.
\ref{fig:Greens}).

\BFIG
\centerline{\psfig{figure=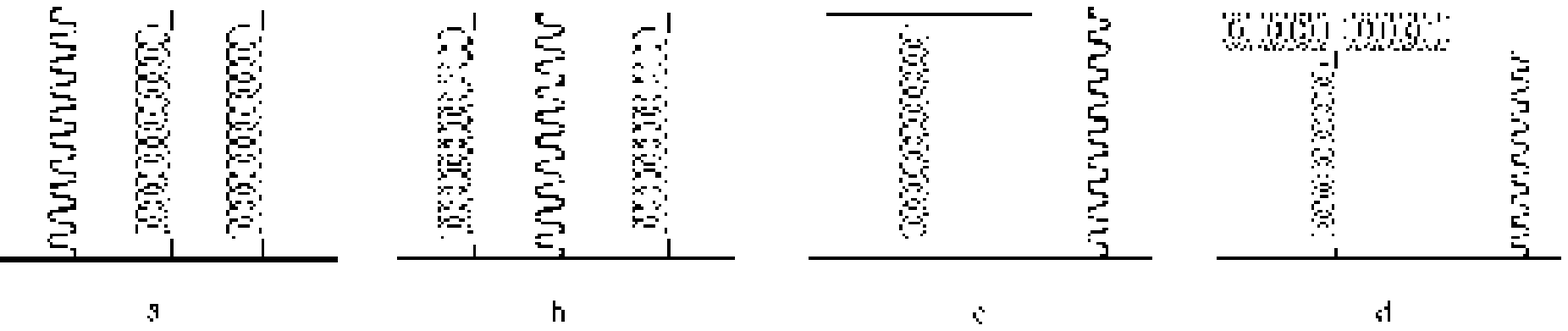,height=3.37cm,width=\linewidth}}
\caption[NLO direct photon Green's functions.]{$qq\gamma gg$ (a,
b and d), and $qq\gamma qq$ time-independent Feynman diagrams.}
\label{fig:Greens} \EFIG

Note that we have not yet shown any time direction. That is, we
have not assigned any of the external lines to "incoming" or
"outgoing" legs. Figures \ref{fig:abdiags} and \ref{fig:cddiags}
show the range of such associations, given that the incoming
particles (on the left) must be either quarks or gluons (this is
{\it hadroproduction}, after all), and the photon must be
outgoing. In addition, we will have next-to-leading order virtual
contributions, as shown in figure \ref{fig:virtual}.

\figboxf{abdiags}{Time-ordered Feynman diagrams from (a) and (b)}

\figboxf{cddiags}{Time-ordered Feynman diagrams from (c) and (d)}

\begin{figure}[h] \centering
$$
\begin{array}{cccc}

\begin{fmffile}{cfvirt01}
\fmfframe(5,5)(5,5){
\begin{fmfchar*}(25,25) \fmfset{curly_len}{2.0mm}
\fmfleft{i1,i2} \fmfright{o1,o2} \fmf{gluon,tension=3}{i1,v1}
\fmf{gluon,tension=3}{v2,v3} \fmf{quark,left,tension=2}{v1,v2,v1}
\fmf{quark}{i2,v3,v4} \fmf{quark}{v4,o1} \fmf{photon}{v4,o2}
\end{fmfchar*}
}
\end{fmffile}

&

\begin{fmffile}{cfvirt02}
\fmfframe(5,5)(5,5){
\begin{fmfchar*}(25,25) \fmfset{curly_len}{2.0mm}
\fmfleft{i1,i2} \fmfright{o1,o2} \fmf{gluon,tension=3}{i1,v1}
\fmf{gluon,tension=3}{v2,v3} \fmf{gluon,left,tension=1}{v1,v2,v1}
\fmf{quark}{i2,v3,v4} \fmf{quark}{v4,o1} \fmf{photon}{v4,o2}
\end{fmfchar*}
}
\end{fmffile}

&

\begin{fmffile}{cfvirt03}
\fmfframe(5,5)(5,5){
\begin{fmfchar*}(25,25) \fmfset{curly_len}{2.0mm} \fmfleft{i1,i2}
\fmfright{o1,o2} \fmf{gluon}{i1,v3} \fmf{quark,tension=3}{i2,v1}
\fmf{quark}{v1,v2} \fmf{quark,tension=3}{v2,v3}
\fmf{quark}{v3,v4,o1} \fmf{gluon,right}{v1,v2} \fmf{photon}{v4,o2}
\end{fmfchar*}
}
\end{fmffile}

&

\begin{fmffile}{cfvirt04}
\fmfframe(5,5)(5,5){
\begin{fmfchar*}(25,25) \fmfset{curly_len}{2.0mm} \fmfleft{i1,i2}
\fmfright{o1,o2} \fmf{gluon}{i1,v1} \fmf{quark,tension=2}{v1,v2}
\fmf{quark}{v2,v3} \fmf{quark,tension=2}{v3,v4} \fmf{quark}{i2,v1}
\fmf{quark}{v4,o1} \fmf{photon}{v4,o2} \fmffreeze
\fmf{gluon,right}{v2,v3}
\end{fmfchar*}
}
\end{fmffile}

\\

\begin{fmffile}{cfvirt05}
\fmfframe(5,5)(5,5){
\begin{fmfchar*}(25,25) \fmfset{curly_len}{2.0mm} \fmfleft{i1,i2}
\fmfright{o1,o2} \fmf{gluon}{i1,v1} \fmf{quark,tension=3}{v2,v3}
\fmf{quark}{v3,v4} \fmf{quark,tension=3}{v4,o1}
\fmf{quark}{i2,v1,v2} \fmf{photon}{v2,o2} \fmffreeze
\fmf{gluon,left,tension=0}{v3,v4}
\end{fmfchar*}
}
\end{fmffile}

&

\begin{fmffile}{cfvirt06}
\fmfframe(5,5)(5,5){
\begin{fmfchar*}(25,25) \fmfset{curly_len}{2.0mm}\fmfleft{i1,i2}
\fmfright{o1,o2} \fmf{gluon}{i1,v2}
\fmf{quark,tension=2}{i2,v1,v2} \fmf{quark}{v2,v3}
\fmf{quark,tension=2}{v3,v4,o1} \fmf{photon}{v3,o2} \fmffreeze
\fmf{gluon,tension=0}{v1,v4}
\end{fmfchar*}
}
\end{fmffile}

&

\begin{fmffile}{cfvirt07}
\fmfframe(5,5)(5,5){
\begin{fmfchar*}(25,25) \fmfset{curly_len}{2.0mm} \fmfleft{i1,i2}
\fmfright{o1,o2} \fmf{gluon,tension=2}{i1,v4,v1}
\fmf{quark}{i2,v1} \fmf{quark,tension=2}{v1,v2,v3}
\fmf{quark}{v3,o1} \fmf{photon}{v3,o2} \fmffreeze
\fmf{gluon,left,tension=0}{v2,v4}
\end{fmfchar*}
}
\end{fmffile}

&

\begin{fmffile}{cfvirt08}
\fmfframe(5,5)(5,5){
\begin{fmfchar*}(25,25) \fmfset{curly_len}{2.0mm} \fmfleft{i1,i2}
\fmfright{o1,o2} \fmf{gluon}{i1,v2}
\fmf{quark,tension=2}{i2,v1,v2} \fmf{quark,tension=2}{v2,v3,v4}
\fmf{quark}{v4,o1} \fmf{photon}{v4,o2} \fmffreeze
\fmf{gluon,left,tension=0}{v1,v3}
\end{fmfchar*}
}
\end{fmffile}

\\

\begin{fmffile}{cfvirt09}
\fmfframe(5,5)(5,5){
\begin{fmfchar*}(25,25) \fmfset{curly_len}{2.0mm} \fmfleft{i1,i2}
\fmfright{o1,o2} \fmf{gluon}{i1,v1} \fmf{quark}{i2,v1}
\fmf{quark,tension=2}{v1,v2,v3,v4,o1}
 \fmf{photon}{v3,o2} \fmffreeze
\fmf{gluon,right,tension=0}{v2,v4}
\end{fmfchar*}
}
\end{fmffile}

&

\begin{fmffile}{cfvirt10}
\fmfframe(5,5)(5,5){
\begin{fmfchar*}(25,15) \fmfset{curly_len}{2.0mm} \fmfleft{i1,i2}
\fmfright{o1,o2} \fmftop{t} \fmfbottom{b}
\fmf{gluon,tension=2}{i1,v1} \fmf{gluon,tension=2}{v2,b}
\fmf{quark,left}{v1,v2,v1} \fmf{quark}{i2,t}
\fmf{quark,tension=4}{t,b} \fmf{quark}{b,o1} \fmf{photon}{t,o2}
\end{fmfchar*}
}
\end{fmffile}

&

\begin{fmffile}{cfvirt11}
\fmfframe(5,5)(5,5){
\begin{fmfchar*}(25,15) \fmfset{curly_len}{2.0mm} \fmfleft{i1,i2}
\fmfright{o1,o2} \fmftop{t} \fmfbottom{b}
\fmf{gluon,tension=2}{i1,v1} \fmf{gluon,tension=2}{v2,b}
\fmf{gluon,left,tension=0.5}{v1,v2,v1} \fmf{quark}{i2,t}
\fmf{quark,tension=4}{t,b} \fmf{quark}{b,o1} \fmf{photon}{t,o2}
\end{fmfchar*}
}
\end{fmffile}

&

\begin{fmffile}{cfvirt12}
\fmfframe(5,5)(5,5){
\begin{fmfchar*}(25,15) \fmfset{curly_len}{2.0mm} \fmfleft{i1,i2}
\fmfright{o1,o2} \fmftop{t} \fmfbottom{b}
\fmf{quark,tension=1}{i2,v1} \fmf{quark}{v1,v2}
\fmf{quark,tension=1}{v2,t} \fmf{gluon}{i1,b}
\fmf{quark,tension=4}{t,b} \fmf{quark}{b,o1} \fmf{photon}{t,o2}
\fmffreeze \fmf{gluon,right,tension=0}{v1,v2}
\end{fmfchar*}
}
\end{fmffile}

\\

\begin{fmffile}{cfvirt13}
\fmfframe(5,5)(5,5){
\begin{fmfchar*}(25,15) \fmfset{curly_len}{2.0mm} \fmfleft{i1,i2}
\fmfright{o1,o2} \fmftop{t} \fmfbottom{b} \fmf{quark}{i2,t}
\fmf{quark,tension=3}{t,v1} \fmf{quark}{v1,v2}
\fmf{quark,tension=3}{v2,b} \fmf{quark}{b,o1} \fmf{gluon}{i1,b}
 \fmf{photon}{t,o2}
\fmffreeze \fmf{gluon,left,tension=0}{v1,v2}
\end{fmfchar*}
}
\end{fmffile}

&

\begin{fmffile}{cfvirt14}
\fmfframe(5,5)(5,5){
\begin{fmfchar*}(25,15) \fmfset{curly_len}{2.0mm} \fmfleft{i1,i2}
\fmfright{o1,o2} \fmftop{t} \fmfbottom{b} \fmf{quark}{i2,t,b}
\fmf{quark,tension=3}{b,v1} \fmf{quark}{v1,v2}
\fmf{quark,tension=3}{v2,o1} \fmf{gluon}{i1,b}
\fmf{photon,tension=1/3}{t,o2} \fmffreeze
\fmf{gluon,left,tension=0}{v1,v2}
\end{fmfchar*}
}
\end{fmffile}

&

\begin{fmffile}{cfvirt15}
\fmfframe(5,5)(5,5){
\begin{fmfchar*}(25,15) \fmfset{curly_len}{2.0mm} \fmfleft{i1,i2}
\fmfright{o1,o2} \fmftop{t} \fmfbottom{b}
\fmf{quark,tension=2}{i2,v1,t} \fmf{quark}{t,b}
\fmf{quark,tension=2}{b,v2,o1} \fmf{gluon}{i1,b}
\fmf{photon}{t,o2} \fmffreeze \fmf{gluon}{v1,v2}
\end{fmfchar*}
}
\end{fmffile}

&

\begin{fmffile}{cfvirt16}
\fmfframe(5,5)(5,5){
\begin{fmfchar*}(25,15) \fmfset{curly_len}{2.0mm} \fmfleft{i1,i2}
\fmfright{o1,o2} \fmftop{t} \fmfbottom{b} \fmf{quark}{i2,t}
\fmf{quark,tension=2}{t,v1,b} \fmf{quark}{b,o1}
\fmf{gluon,tension=2}{i1,v2,b} \fmf{photon}{t,o2} \fmffreeze
\fmf{gluon}{v1,v2}
\end{fmfchar*}
}
\end{fmffile}

\\

\begin{fmffile}{cfvirt17}
\fmfframe(5,5)(5,5){
\begin{fmfchar*}(25,15) \fmfset{curly_len}{2.0mm} \fmfleft{i1,i2}
\fmfright{o1,o2} \fmftop{t} \fmfbottom{b} \fmf{quark}{i2,t}
\fmf{quark,tension=2}{t,v1,b,v2,o1}
 \fmf{gluon}{i1,b} \fmf{photon}{t,o2} \fmffreeze \fmf{gluon}{v1,v2}
\end{fmfchar*}
}
\end{fmffile}

&

\begin{fmffile}{cfvirt18}
\fmfframe(5,5)(5,5){
\begin{fmfchar*}(25,15) \fmfset{curly_len}{2.0mm} \fmfleft{i1,i2}
\fmfright{o1,o2} \fmftop{t} \fmfbottom{b}
\fmf{quark,tension=2}{i2,v1,t,v2,b} \fmf{quark}{b,o1}
 \fmf{gluon}{i1,b} \fmf{photon}{t,o2} \fmffreeze \fmf{gluon}{v1,v2}
\end{fmfchar*}
}
\end{fmffile}

&

\begin{fmffile}{cfvirt19}
\fmfframe(5,5)(5,5){
\begin{fmfchar*}(25,15) \fmfset{curly_len}{2.0mm} \fmfleft{i1,i2}
\fmfright{o1,o2} \fmftop{t} \fmfbottom{b}
\fmf{quark,tension=2}{i2,v1,t} \fmf{quark}{t,b,o1}
 \fmf{gluon,tension=2}{i1,v2,b} \fmf{photon}{t,o2} \fmffreeze \fmf{gluon}{v1,v2}
\end{fmfchar*}
}
\end{fmffile}

&

$${{+ (1 \leftrightarrow 2)}\atop \hbox{\rm + ghost loops}}$$

\\

\end{array}
$$
\caption{Virtual Contributions.} \label{fig:virtual} \end{figure}

Finally, there will be Bremsstrahlung contributions, which arise
when one final-state parton in a purely QCD subprocess fragments
and a photon is produced. The signature for such an event is a
non-isolated photon (one with hadronic energy deposited in a
small cone about the photon axis) and a jet in the opposite
direction. The purely QCD 2-body subprocesses are as follows;
diagrams for these, as well as the derivation of their matrix
elements, can be found in Ellis and Sexton (1986) ~\cite{ES86}.

\BQA
 q \bar{q} \rightarrow g g &
 q g \rightarrow q g \cr
 g q \rightarrow q g &
 g g \rightarrow q \bar{q} \cr
 g g \rightarrow g g &
 q q \rightarrow q q \cr
 q \bar{q} \rightarrow q \bar{q} &
 q \bar{q} \rightarrow \bar{q} q \cr
 q q' \rightarrow q q' &
 q q' \rightarrow q' q \cr
 q \bar{q} \rightarrow q' \bar{q}' & \phantom{q \bar{q} \rightarrow q' \bar{q}'} \;
\EQA

\section{Two-Body Matrix Elements}

For dimension $D=4-2\epsilon$, and with the definitions

\begin{eqnarray}
\hat{s} &\equiv&  {(p_1^\mu + p_2^\mu)}^2 \; \cr \hat{t}
&\equiv&  {(p_1^\mu - p_3^\mu)}^2 \; \cr \hat{u} &\equiv&
{(p_1^\mu - p_4^\mu)}^2 \; \cr {K'}_{q \bar{q}} &\equiv& {1\over
3} \; \cr {K'}_{qg} = {K'}_{gq} &\equiv& {1\over {8(1-\epsilon)}}
\; \cr \omega_{2\gamma} &\equiv& 2 {(4\pi)}^2 C_F \alpha \alpha_s
\mu^{4\epsilon} \; \cr \omega_{2Q} &\equiv& 2 {(4\pi)}^2 C_F
\alpha_s^2 \mu^{4\epsilon} \;
\end{eqnarray}

\noindent we can write the two-body matrix elements as
~\cite{ES86,Owens87,ABDFS86}:

\subsection{$q(p_1) \bar{q}(p_2) \rightarrow \gamma(p_3) g(p_4)$}

\begin{equation}
\bar{\Sigma} {|M|}_{q \bar{q} \rightarrow \gamma g}^2 = {K'}_{q
\bar{q}} Q_q^2 \omega_{2\gamma} (1-\epsilon) \Bigl[ (1-\epsilon)
\Bigl( {\hat{t} \over \hat{u}} + {\hat{u} \over \hat{t}} \Bigr)
-2\epsilon \Bigr] \;
\end{equation}

\subsection{$q(p_1) g(p_2) \rightarrow \gamma(p_3) q(p_4)$}

\begin{equation}
\bar{\Sigma} {|M|}_{q g \rightarrow \gamma q}^2 = - {K'}_{qg}
Q_q^2 \omega_{2\gamma} (1-\epsilon) \Bigl[ (1-\epsilon) \Bigl(
{\hat{t} \over \hat{s}} + {\hat{s} \over \hat{t}} \Bigr)
-2\epsilon \Bigr] \;
\end{equation}

\subsection{$g(p_1) q(p_2) \rightarrow \gamma(p_3) q(p_4)$}

\begin{equation}
\bar{\Sigma} {|M|}_{g q \rightarrow \gamma q}^2 = - {K'}_{gq}
Q_q^2 \omega_{2\gamma} (1-\epsilon) \Bigl[ (1-\epsilon) \Bigl(
{\hat{u} \over \hat{s}} + {\hat{s} \over \hat{u}} \Bigr)
-2\epsilon \Bigr] \;
\end{equation}

\subsection{$q(p_1) \bar{q}(p_2) \rightarrow g(p_3) g(p_4)$}

\begin{equation}
\bar{\Sigma} {|M|}_{q \bar{q} \rightarrow g g}^2 = {{K'}_{q
\bar{q}}\over 2} \omega_{2Q} (1-\epsilon) \Bigl[ {{\hat{u}^2
+\hat{t}^2}\over \hat{s}^2} -\epsilon \Bigr] \Bigl[ C_F
{\hat{s}^2 \over {\hat{t} \hat{u}}} - N_C \Bigr] \;
\end{equation}

\subsection{$q(p_1) g(p_2) \rightarrow q(p_3) g(p_4)$}

\begin{equation}
\bar{\Sigma} {|M|}_{q g \rightarrow q g}^2 = -{K'}_{qg}
\omega_{2Q} (1-\epsilon) \Bigl[ {{\hat{u}^2 +\hat{s}^2}\over
\hat{t}^2} -\epsilon \Bigr] \Bigl[ C_F {\hat{t}^2 \over {\hat{s}
\hat{u}}} - N_C \Bigr] \;
\end{equation}

\subsection{$g(p_1) q(p_2) \rightarrow q(p_3) g(p_4)$}

\begin{equation}
\bar{\Sigma} {|M|}_{g q \rightarrow q g}^2 = -{K'}_{qg}
\omega_{2Q} (1-\epsilon) \Bigl[ {{\hat{t}^2 +\hat{s}^2}\over
\hat{u}^2} -\epsilon \Bigr] \Bigl[ C_F {\hat{u}^2 \over {\hat{s}
\hat{t}}} - N_C \Bigr] \;
\end{equation}

\subsection{$g(p_1) g(p_2) \rightarrow q(p_3) \bar{q}(p_4)$}

\begin{equation}
\bar{\Sigma} {|M|}_{g g \rightarrow q \bar{q}}^2 =
{{K'}_{qg}\over {2C_F}} \omega_{2Q} \Bigl[ {{\hat{t}^2
+\hat{u}^2}\over \hat{s}^2} -\epsilon \Bigr] \Bigl[ C_F {\hat{s}^2
\over {\hat{u} \hat{t}}} - N_C \Bigr] \;
\end{equation}

\subsection{$g(p_1) g(p_2) \rightarrow g(p_3) g(p_4)$}

\begin{equation}
\bar{\Sigma} {|M|}_{g g \rightarrow g g}^2 = {{K'}_{qg}\over 2}
9N_C \omega_{2Q} \Bigl[ 3 - {{\hat{t} \hat{u}}\over \hat{s}^2} -
{{\hat{s} \hat{u}}\over \hat{t}^2} - {{\hat{s} \hat{t}}\over
\hat{u}^2} \Bigr] \;
\end{equation}

\subsection{$q(p_1) q(p_2) \rightarrow q(p_3) q(p_4)$}

\begin{equation}
\bar{\Sigma} {|M|}_{q q \rightarrow q q}^2 = {{K'}_{q
\bar{q}}\over 4} \omega_{2Q}  \Bigl[ {{\hat{s}^2 + \hat{u}^2}\over
\hat{t}^2} + {{\hat{s}^2 + \hat{t}^2}\over \hat{u}^2} -2\epsilon
- {{2(1-\epsilon)}\over N_C} \Bigl( {\hat{s}^2\over {\hat{t}
\hat{u}}} + \epsilon \Bigr) \Bigr] \;
\end{equation}

\subsection{$q(p_1) \bar{q}(p_2) \rightarrow q(p_3) \bar{q}(p_4)$}

\begin{equation}
\bar{\Sigma} {|M|}_{q \bar{q} \rightarrow q \bar{q}}^2 = {{K'}_{q
\bar{q}}\over 2} \omega_{2Q}  \Bigl[ {{\hat{s}^2 + \hat{u}^2}\over
\hat{t}^2} + {{\hat{u}^2 + \hat{t}^2}\over \hat{s}^2} -2\epsilon -
{{2(1-\epsilon)}\over N_C} \Bigl( {\hat{u}^2\over {\hat{t}
\hat{s}}} + \epsilon \Bigr) \Bigr] \;
\end{equation}

\subsection{$q(p_1) \bar{q}(p_2) \rightarrow \bar{q}(p_3) q(p_4)$}

\begin{equation}
\bar{\Sigma} {|M|}_{q \bar{q} \rightarrow \bar{q} q}^2 = {{K'}_{q
\bar{q}}\over 2} \omega_{2Q}  \Bigl[ {{\hat{s}^2 + \hat{t}^2}\over
\hat{u}^2} + {{\hat{t}^2 + \hat{u}^2}\over \hat{s}^2} -2\epsilon -
{{2(1-\epsilon)}\over N_C} \Bigl( {\hat{t}^2\over {\hat{u}
\hat{s}}} + \epsilon \Bigr) \Bigr] \;
\end{equation}

\subsection{$q(p_1) q'(p_2) \rightarrow q(p_3) q'(p_4)$}

\begin{equation}
\bar{\Sigma} {|M|}_{q q' \rightarrow q q'}^2 = {{K'}_{q
\bar{q}}\over 2} \omega_{2Q}  \Bigl[ (1-\epsilon) - 2 {{\hat{u}
\hat{s}}\over \hat{t}^2} \Bigr] \;
\end{equation}

\subsection{$q(p_1) q'(p_2) \rightarrow q'(p_3) q(p_4)$}

\begin{equation}
\bar{\Sigma} {|M|}_{q q' \rightarrow q' q}^2 = {{K'}_{q
\bar{q}}\over 2} \omega_{2Q}  \Bigl[ (1-\epsilon) - 2 {{\hat{t}
\hat{s}}\over \hat{u}^2} \Bigr] \;
\end{equation}

\subsection{$q(p_1) \bar{q}(p_2) \rightarrow q'(p_3) \bar{q}'(p_4)$}

\begin{equation}
\bar{\Sigma} {|M|}_{q \bar{q} \rightarrow q' \bar{q}'}^2 =
{{K'}_{q \bar{q}}\over 2} \omega_{2Q}  \Bigl[ (1-\epsilon) - 2
{{\hat{t} \hat{u}}\over \hat{s}^2} \Bigr] \;
\end{equation}

\noindent Note that in the text, we will often use the definition

\BQN K'T_{f\tilde{f}}' (v) \equiv {{\overline{\Sigma}{|{\cal
M}|}^2}\over \omega_{2Q}} \;, \EQN

\noindent in which $v \equiv -\hat{u}/\hat{s}$. Use of either $K$
or $T_{f\tilde{f}}$ {\bf without} the prime refers to the
$\epsilon \rightarrow 0$ limit thereof. Here and in the text,
$C_F=4/3$ and $N_C=3$.

\section{Three-Body Matrix Elements}

With the definitions

\begin{eqnarray}
s_{12} &\equiv&  {(p_1^\mu + p_2^\mu)}^2 \; \cr t_{13} &\equiv&
{(p_1^\mu - p_3^\mu)}^2 \; \cr t_{14} &\equiv&  {(p_1^\mu -
p_4^\mu)}^2 \; \cr t_{15} &\equiv&  {(p_1^\mu - p_5^\mu)}^2 \; \cr
t_{23} &\equiv&  {(p_2^\mu - p_3^\mu)}^2 \; \cr t_{24} &\equiv&
{(p_2^\mu - p_4^\mu)}^2 \; \cr t_{25} &\equiv&  {(p_2^\mu -
p_5^\mu)}^2 \; \cr s_{34} &\equiv&  {(p_3^\mu + p_4^\mu)}^2 \; \cr
s_{35} &\equiv&  {(p_3^\mu + p_5^\mu)}^2 \; \cr s_{45} &\equiv&
{(p_4^\mu + p_5^\mu)}^2 \; \cr {K'}_{q \bar{q}} &\equiv& {1\over
3} \; \cr {K'}_{qg} = {K'}_{gq} &\equiv& {1\over {8(1-\epsilon)}}
\; \cr \omega_3 &\equiv& 4 {(4\pi)}^3 C_F \alpha \alpha_s^2
\mu^{6\epsilon} \;
\end{eqnarray}

\noindent we can write the 3-body matrix elements as follows
~\cite{ABDFS86}:

\subsection{$q(p_1) \bar{q}(p_2) \rightarrow \gamma(p_3) g(p_4) g(p_5)$}

\noindent We need:

\begin{eqnarray}
&M_{qqgg}&
(s_{12},t_{13},t_{14},t_{15},t_{23},t_{24},t_{25},s_{34},s_{35},s_{45})
\cr &\equiv& \Biggl[ \biggl( C_F - {N_C \over 2} \biggr) {s_{12}
\over {t_{15} t_{25}}} + {N_C \over 2} {t_{14} \over {t_{15}
s_{45}}} + {N_C \over 2} {t_{24} \over {t_{25} s_{45}}} \Biggl]
\cr &\times& \Biggl\{ {(1-\epsilon)}^2 \Bigl[ {{t_{13}^2 +
t_{23}^2}\over {t_{14} t_{24}}} + {{t_{14}^2 + t_{24}^2}\over
{t_{13} t_{23}}} + {{t_{15} t_{25} (t_{15}^2+t_{25}^2)}\over
{t_{13} t_{23} t_{14} t_{24}}} \Bigr] \cr &+& \epsilon
(1-2\epsilon) \Bigl[ {{(t_{14}-t_{24})(t_{15}-t_{25})}\over
{t_{14} t_{24}}} + {{(t_{13}-t_{23})(t_{15}-t_{25})}\over {t_{13}
t_{23}}} \; \cr &\phantom{+}& \phantom{\epsilon (1-2\epsilon)
\Bigl[ {{(t_{14}-t_{24})(t_{15}-t_{25})}\over {t_{14} t_{24}}}} +
{{t_{15} t_{25} (t_{13}-t_{23})(t_{14}-t_{24})}\over {t_{13}
t_{23} t_{14} t_{24}}} \Bigr] \cr &+& \epsilon (2+\epsilon) \Bigl[
{{t_{13}t_{23}}\over {t_{14}t_{24}}} + {{t_{14}t_{24}}\over
{t_{13}t_{23}}} + {{t_{15}^2 t_{25}^2}\over {t_{13} t_{23} t_{14}
t_{24}}} \Bigr] \cr &-& 2\epsilon (4-\epsilon) \Bigl[ 1+
{{t_{15}t_{25}}\over {t_{13}t_{23}}} + {{t_{15}t_{25}}\over
{t_{14}t_{24}}} \Bigr] \Biggr\} \;,
\end{eqnarray}

\noindent from which we obtain

\begin{equation}
\bar{\Sigma} {|M|}_{q \bar{q} \rightarrow \gamma g g}^2 = {{K'}_{q
\bar{q}}\over 2} Q_q^2 \omega_3 M_{qqgg}
(s_{12},t_{13},t_{14},t_{15},t_{23},t_{24},t_{25},s_{34},s_{35},s_{45})
\;.
\end{equation}

\subsection{$q(p_1) g(p_2) \rightarrow \gamma(p_3) q(p_4) g(p_5)$}

\begin{equation}
\bar{\Sigma} {|M|}_{q g \rightarrow \gamma q g}^2 = -{K'}_{qg}
Q_q^2 \omega_3 M_{qqgg}
(t_{14},t_{13},s_{12},t_{15},s_{34},t_{24},s_{45},t_{23},s_{35},t_{25})
\;.
\end{equation}

\subsection{$g(p_1) q(p_2) \rightarrow \gamma(p_3) q(p_4) g(p_5)$}

\begin{equation}
\bar{\Sigma} {|M|}_{g q \rightarrow \gamma q g}^2 = -{K'}_{gq}
Q_q^2 \omega_3 M_{qqgg}
(t_{24},t_{23},s_{12},t_{25},s_{34},t_{14},s_{45},t_{13},s_{35},t_{15})
\;.
\end{equation}

\subsection{$g(p_1) g(p_2) \rightarrow \gamma(p_3) q(p_4) \bar{q}(p_5)$}

\begin{equation}
\bar{\Sigma} {|M|}_{g g \rightarrow \gamma q \bar{q}}^2 =
{{K'}_{gq}\over {2C_F (1-\epsilon)}} Q_q^2 \omega_3 M_{qqgg}
(s_{45},s_{35},t_{25},t_{15},s_{34},t_{24},t_{14},t_{23},t_{13},s_{12})
\;.
\end{equation}

\subsection{$q(p_1) q(p_2) \rightarrow \gamma(p_3) q(p_4) q(p_5)$}

\noindent With

\begin{eqnarray} M_0 &=& {{-s_{12}}\over {
t_{13}t_{23}}}+{t_{25}\over { t_{23}s_{35}}}+{t_{24}\over
{t_{23}s_{34}}}+{t_{14}\over {t_{13}s_{34}}}-{s_{45}\over
{s_{35}s_{34}}}+{t_{15}\over { t_{13}s_{35}}} \; \cr M_1 &=&
{{s_{12}^2+t_{25}^2+t_{14}^2+s_{45}^2}\over {t_{24}t_{15}}} \;
\cr M_2 &=& {{s_{12}^2+t_{24}^2+t_{15}^2+s_{45}^2} \over
{t_{25}t_{14}}} \; \cr M_3 &=& (s_{12}+s_{45})
{{(t_{25}+t_{14})}\over {t_{24}t_{15}}} \; \cr M_4 &=&
(s_{12}+s_{45}) {{(t_{24}+t_{15})}\over {t_{25}t_{14}}} \; \cr M_5
&=& (s_{12}^2+s_{45}^2)(2s_{12}s_{45}-2t_{25}t_{14}-2t_{24}t_{15}
-(t_{25}+t_{14})(t_{24}+t_{15})) \cr &+&
(s_{12}-s_{45})^2(2(t_{25}+t_{14})(t_{24}+t_{15})+t_{25}t_{14}+t_{24}t_{15})
\cr &+&
((t_{25}+t_{24})^2-(t_{14}+t_{15})^2)(t_{15}t_{14}-t_{25}t_{24})
\cr &-& t_{15}t_{24}(t_{24}-t_{15})^2
-t_{14}t_{25}(t_{14}-t_{25})^2-t_{15}t_{25}(t_{15}+t_{25})^2
-t_{24}t_{14}(t_{14}+t_{24})^2 \cr &+&
8t_{25}t_{24}t_{14}t_{15}+(t_{24}-t_{25})(t_{15}-t_{14})(t_{24}t_{25}+t_{14}t_{15})
\;
\end{eqnarray}

\noindent and the further definition
\begin{eqnarray}
&M_{qqqq}&(s_{12},t_{13},t_{14},t_{15},t_{23},t_{24},t_{25},s_{34},s_{35},s_{45})
\cr &\equiv& \Biggl\{ 2M_0 (M_1+M_2-2\epsilon) \cr &-& 4\epsilon
\Bigl[ \biggl( {t_{24}\over {t_{23}s_{34}}}+{t_{15}\over
{t_{13}s_{35}}} \biggr) (M_1+M_3-2) \cr &+& \biggl( {t_{14}\over
{t_{13}s_{34}}}+{t_{25}\over {t_{23}s_{35}}} \biggr) (M_2+M_4-2)
\Bigr] \cr &-&2\epsilon \Biggl[ \biggl( {{-s_{12}}\over {
t_{13}t_{23}}}+{t_{25}\over { t_{23}s_{35}}}+{t_{14}\over
{t_{13}s_{34}}}-{s_{45}\over {s_{35}s_{34}}} \biggr) \cr &\times&
\bigl( {{{1\over 2} {(s_{12}-s_{45})}^2+{1\over
2}{(t_{14}-t_{25})}^2}\over { t_{24}t_{15}}}+M_1+M_3-2 \bigr) \cr
&+& \bigl( {{-s_{12}}\over {t_{13}t_{23}}}+{t_{24}\over {
t_{23}s_{34}}}-{s_{45}\over {s_{35}s_{34}}}+{t_{15}\over
{t_{13}s_{35}}} \bigr) \cr &\times& \bigl( {{{1\over 2}
{(s_{12}-s_{45})}^2+{1\over 2} {(t_{15}-t_{24})}^2}\over
{t_{25}t_{14}}}+M_2+M_4-2 \bigr) \Biggr] \cr &-&
2(1-\epsilon){M_0\over N_C} \Biggl[
{{(s_{12}^2+s_{45}^2)(s_{12}s_{45}-t_{14}t_{25}-t_{15}t_{24})}
\over {t_{25}t_{24}t_{14}t_{15}}}+4\epsilon \Biggr] \cr &+&
{{2\epsilon}\over N_C} \Bigl[
{{s_{35}(s_{12}s_{34}-t_{23}t_{14}-t_{24}t_{13})}\over
{t_{25}t_{23}t_{13}t_{15}}} + {{
s_{34}(s_{12}s_{35}-t_{25}t_{13}-t_{23}t_{15})}\over
{t_{23}t_{24}t_{14}t_{13}}} \cr &-&
{{t_{23}(-s_{45}t_{13}+t_{14}s_{35}+s_{34}t_{15})}\over
{t_{25}t_{24}s_{34}s_{35}}}
-{{t_{13}(-t_{23}s_{45}+t_{25}s_{34}+t_{24}s_{35})}\over
{t_{14}s_{34}s_{35}t_{15}}} \Bigr] \cr &+& {\epsilon\over N_C}
{{M_0 M_5}\over {t_{25}t_{24}t_{15}t_{14}}} \Biggr\} \;,
\end{eqnarray}

\noindent we can write

\begin{equation}
\bar{\Sigma} {|M|}_{q q \rightarrow \gamma q q}^2 = {{K'}_{q
g}\over 4} C_F Q_q^2 \omega_3 M_{qqqq}
(s_{12},t_{13},t_{14},t_{15},t_{23},t_{24},t_{25},s_{34},s_{35},s_{45})
\;.
\end{equation}

\subsection{$q(p_1) \bar{q}(p_2) \rightarrow \gamma(p_3) q(p_4) \bar{q}(p_5)$}

\begin{equation}
\bar{\Sigma} {|M|}_{q \bar{q} \rightarrow \gamma q \bar{q}}^2 =
{{K'}_{q g}\over 2} C_F Q_q^2 \omega_3 M_{qqqq}
(t_{15},t_{13},t_{14},s_{12},s_{35},s_{45},t_{25},s_{34},t_{23},t_{24})
\;.
\end{equation}

\subsection{$q(p_1) q'(p_2) \rightarrow \gamma(p_3) q(p_4) q'(p_5)$}

\noindent We begin with the definitions

\begin{eqnarray}
M_0 &=& {{-s_{12}}\over {t_{13}t_{23}}}+{t_{24}\over
{t_{23}s_{34}}}-{s_{45}\over {s_{35}s_{34}}}+{t_{15}\over
{t_{13}s_{35}}} \cr M_1 &=& {{{1\over 2} {(s_{12}-s_{45})}^2 +
{1\over 2}{(t_{15}-t_{24})}^2}\over {t_{25}t_{14}}} \cr M_2 &=&
{{s_{12}^2+t_{24}^2+t_{15}^2+s_{45}^2}\over {t_{25}t_{14}}} \cr
M_3 &=& Q_q^2 {t_{14}\over {t_{13}s_{34}}}+Q_{q'}^2 {t_{25}\over
{t_{23}s_{35}}} \cr M_4 &=&
{{(s_{12}+s_{45})(t_{24}+t_{15})}\over {t_{25}t_{14}}} \;
\end{eqnarray}

\begin{eqnarray}
&M_{qpqp}&
(s_{12},t_{13},t_{14},t_{15},t_{23},t_{24},t_{25},s_{34},s_{35},s_{45})
\cr &=& 2(M_2-2\epsilon)(Q_qQ_{q'}M_0+M_3) -
4\epsilon(M_2+M_4-2)M_3 \cr &-& 2\epsilon
(M_1+M_2+M_4-2)Q_qQ_{q'}M_0 \;.
\end{eqnarray}

\noindent and arrive at

\begin{equation}
\bar{\Sigma} {|M|}_{q q' \rightarrow \gamma q q'}^2 = {{K'}_{q
g}\over 2} C_F \omega_3 M_{qpqp}
(s_{12},t_{13},t_{14},t_{15},t_{23},t_{24},t_{25},s_{34},s_{35},s_{45})
\;.
\end{equation}

\subsection{$q(p_1) \bar{q}'(p_2) \rightarrow \gamma(p_3) q(p_4) \bar{q}'(p_5)$}

\begin{equation}
\bar{\Sigma} {|M|}_{q \bar{q}' \rightarrow \gamma q \bar{q}'}^2 =
{{K'}_{q g}\over 2} C_F \omega_3 M_{qpqp}
(t_{15},t_{13},t_{14},s_{12},s_{35},s_{45},t_{25},s_{34},t_{23},t_{24})
\;.
\end{equation}

\subsection{$q(p_1) \bar{q}(p_2) \rightarrow \gamma(p_3) q'(p_4) \bar{q}'(p_5)$}

\begin{equation}
\bar{\Sigma} {|M|}_{q \bar{q} \rightarrow \gamma q' \bar{q}'}^2 =
{{K'}_{q g}\over 2} C_F \omega_3 M_{qpqp}
(t_{14},t_{13},s_{12},t_{15},s_{34},t_{24},s_{45},t_{23},s_{35},t_{25})
\;.
\end{equation}

\setcounter{footnote}{0}

\chapter{Useful Integrals}
\label{ap:integrals}

Here we collect the details of the $y_3$ and $\phi_3$ integrals
used in Chapter \ref{ch:photon}. Certain sub-integrals have been
found in the tables of Gradshteyn and Ryzhik (G.R.) ~\cite{GR65}.

\section{$I_{ab}$}

\BQA I_{ab} (x) &\equiv& \int_{-y_{3m}^b}^{y_{3m}^a} dy_3
{e^{y_3}\over {e^{y_3}+xe^{-y_3}}} \; \cr &=&
\int_{-y_{3m}^b}^{y_{3m}^a} dy_3 {1\over {1+xe^{-2y_3}}} \; \cr
&=&  {\Bigl[ y_3 +{1\over 2} \ln (1+xe^{-2y_3})
\Bigr]}_{-y_{3m}^b}^{y_{3m}^a} \quad \quad \hbox{\rm (G.R.
2.313.1)} \; \cr &=& y_{3m}^a + y_{3m}^b + {1\over 2} \ln
{{1+xe^{-2y_{3m}^a}}\over {1+xe^{2y_{3m}^b}}} \;. \EQA

\noindent Similarly,

\BQA I_{ba} (x) &\equiv& \int_{-y_{3m}^b}^{y_{3m}^a} dy_3 {1\over
{1+xe^{2y_3}}} \; \cr &=&  y_{3m}^a + y_{3m}^b + {1\over 2} \ln
{{1+xe^{-2y_{3m}^b}}\over {1+xe^{2y_{3m}^a}}} \;. \EQA

\section{$I_{a2}$, $I_{b2}$}

\BQA I_{a2} (x) &\equiv& \int_{-y_{3m}^b}^{y_{3m}^a} dy_3
{e^{\Delta y}\over {xe^{-\Delta y} + \cosh \Delta y - \cos \Delta
\phi}} \; \cr &=& \int_{-y_{3m}^b-y_2}^{y_{3m}^a-y_2} d\Delta y
{e^{\Delta y}\over {xe^{-\Delta y} + \cosh \Delta y - \cos \Delta
\phi}} \;, \label{Ia2} \EQA

\noindent with $\Delta y \equiv y_3-y_2$, $\Delta \phi \equiv
\phi_3 - \phi_2$. With the change of variable $\eta \equiv
e^{\Delta y}$ ($\eta_{LO}=e^{-y_{3m}^b-y_2}$,
$\eta_{HI}=e^{y_{3m}^a-y_2}$) we can write this as

\BQA \int_{\eta_{LO}}^{\eta_{HI}} {{2\eta d\eta}\over {\eta^2
-2\eta \cos \Delta\phi +(1+2x)}} &=& \int_{\eta_{LO}}^{\eta_{HI}}
{{2\eta d\eta}\over {\eta^2+(1+2x)}} \cr + \quad \int_0^\infty
d\eta &\phantom{\times}& \Bigl[ {{2\eta}\over {\eta^2 -2\eta \cos
\Delta \phi +(1+2x)}} - {{2\eta}\over {\eta^2+(1+2x)}} \Bigr] \;,
\cr &\phantom{+}& \; \EQA

\noindent where the new limits are justified because in either
the soft or collinear regions, $\{ y_{3m}^a, y_{3m}^b, \eta \}$
all go to infinity, and the second integrand above goes to zero.

The first term is easy:

\BQN \int_{\eta_{LO}^2}^{\eta_{HI}^2} {{d\eta^2}\over {\eta^2
+(1+2x)}} = \ln {{e^{2(y_{3m}^a-y_2)} +(1+2x)}\over
{e^{-2(y_{3m}^b+y_2)} +(1+2x)}} \;, \EQN

\noindent while the second term, after defining $\cos \alpha
\equiv \cos\Delta\phi / \sqrt{1+2x}$, can be partial-fractioned as

\BQA \int_0^\infty d\eta \Biggl[ {{2\eta}\over { 2i
\sin\alpha\sqrt{1+2x}}} &\phantom{-}& \Bigl[ {1\over
{\eta-e^{i\alpha} \sqrt{1+2x}}} - {1\over {\eta-e^{-i\alpha}
\sqrt{1+2x}}} \Bigr] \; \cr &-& \Bigl[ {1\over {\eta- i
\sqrt{1+2x}}} + {1\over {\eta + i \sqrt{1+2x}}} \Bigr] \Biggr]
\;. \EQA

The first set of fractions has the form $z-z^* = 2i {\cal{IM}}
(z)$ for complex $z$, while the second set looks like $z+z^* =
2{\cal{RE}}(z)$. In this way, we can rewrite the integral as

\BQN \int_0^\infty d\eta {2 \over {\sqrt{1+2x} \sin\alpha}}
{\cal{IM}} \biggl[ {\eta\over {\eta-e^{i\alpha}\sqrt{1+2x}}}
\biggr] - \int_0^\infty d\eta 2 {\cal{RE}} \biggl[ {1\over
{\eta-i\sqrt{1+2x}}} \biggr] \;, \EQN

\noindent where

\BQN {\cal{IM}} \biggl[ {\eta\over {\eta-e^{i\alpha}\sqrt{1+2x}}}
\biggr] = {\cal{IM}} \biggl[ 1+ {{e^{i\alpha}\sqrt{1+2x}}\over
{\eta-e^{i\alpha}\sqrt{1+2x}}} \biggr] = {\cal{IM}} \biggl[
{{e^{i\alpha}\sqrt{1+2x}}\over {\eta-e^{i\alpha}\sqrt{1+2x}}}
\biggr] \;, \EQN

\noindent and so

\BQA \int_0^\infty d\eta {2 \over {\sqrt{1+2x} \sin\alpha}}
{\cal{IM}} \biggl[ {\eta\over {\eta-e^{i\alpha}\sqrt{1+2x}}}
\biggr] &=& 2 {\cal{IM}} \biggl[ {e^{i\alpha}\over {\sin\alpha}}
\int_0^\infty {{d\eta}\over {\eta - e^{i\alpha}\sqrt{1+2x}}}
\biggr] \; \cr &=& 2 {\cal{IM}} \biggl[ {e^{i\alpha}\over
{\sin\alpha}} \ln {\bigl( \eta - e^{i\alpha} \sqrt{1+2x} \bigr)}
\biggr] {\Bigr|}_0^\infty \;. \cr &\phantom{+}& \; \EQA

\noindent Meanwhile,

\BQN \int_0^\infty d\eta 2 {\cal{RE}} \biggl[ {1\over {\eta -
i\sqrt{1+2x}}} \biggr] = 2 {\cal{RE}} \biggl[ \ln {\bigl( \eta -
i\sqrt{1+2x} \bigr)} {\biggr]}_0^\infty \;, \EQN

\noindent So, since

\BQA e^{i\alpha} &=& \cos\alpha +i \sin\alpha \; \cr \ln {\bigl(
\eta - e^{i\alpha} \sqrt{1+2x} \bigr)} &=& \ln \sqrt{ \bigl( \eta
- e^{i\alpha}\sqrt{1+2x} \bigr) \bigl( \eta - e^{-i\alpha}
\sqrt{1+2x} \bigr) } \; \cr &+& i {\tan}^{-1} \biggl[
{{-\sqrt{1+2x} \sin\alpha}\over {\eta - \sqrt{1+2x} \cos\alpha}}
\biggr] \;, \EQA

\noindent we have

\BQA {\cal{IM}} \biggl[ e^{i\alpha} \ln \bigl( \eta - e^{i\alpha}
\sqrt{1+2x} \bigr) \biggr] &=& \sin\alpha \ln \sqrt{\eta^2 +
(1+2x) -2\eta \sqrt{1+2x} \cos\alpha} \; \cr &+& \cos\alpha
{\tan}^{-1} \biggl[ {{-\sqrt{1+2x} \sin\alpha}\over {\eta -
\sqrt{1+2x} \cos\alpha}} \biggr] \; \cr 2 {\cal{RE}} \ln ( \eta -
i \sqrt{1+2x} ) &=& 2 \ln \sqrt{(\eta -i\sqrt{1+2x})(\eta
+i\sqrt{1+2x})} \; \cr &=& \ln \bigl( \eta^2 + (1+2x) \bigr) \;.
\EQA

\noindent Putting it all together, we obtain

\BQA \biggl[ \ln {{\eta^2 + (1+2x)
-2\eta\sqrt{1+2x}\cos\alpha}\over {\eta^2 + (1+2x)}}
{\biggr]}_0^\infty &+& 2 \biggl[ \cot\alpha {\tan}^{-1}
{{-\sqrt{1+2x} \sin\alpha}\over {\eta - \sqrt{1+2x} \cos\alpha}}
{\biggr]}_0^\infty \; \cr &=& 2 \cot\alpha [\pi-\alpha] \;, \EQA

\noindent and so

\BQN I_{a2} (x) = \ln \Bigl[ {{e^{2(y_{3m}^a - y_2)}+(1+2x)}\over
{e^{2(-y_{3m}^b - y_2)} +(1+2x)}} \Bigr] + 2\cot \alpha [\pi -
\alpha ] \;. \EQN

\bigskip

$I_{b2}(x)$ will be similar. We here have

\BQN I_{b2}(x) \equiv \int_{y_2+y_{3m}^b}^{y_2-y_{3m}^a} d\Delta y
{e^{-\Delta y}\over {xe^{\Delta y} + \cosh \Delta y -\cos \Delta
\phi}} \;, \EQN

\noindent which, by comparison with equation \ref{Ia2}, can be
obtained via the negative of $I_{a2}$ with the signs of the
limits reversed. We then get

\BQN I_{b2}(x) = \ln \Bigl[ {{e^{2(y_{3m}^b + y_2)}+(1+2x)}\over
{e^{2(-y_{3m}^a + y_2)} +(1+2x)}} \Bigr] + 2\cot \alpha [\pi -
\alpha ] \;. \EQN

\noindent Again, $\cos \alpha \equiv \cos\Delta\phi /
\sqrt{1+2x}$.

\section{$I_{2a,2b}$}

\BQN I_{2a}(x) \equiv \int_{-\infty}^\infty d\Delta y {1\over
{\cosh\Delta y -\cos\Delta\phi}} \Bigl[ {x\over {xe^{-\Delta y}+
\cosh\Delta y - \cos\Delta\phi}} -1 \Bigr] \;. \label{I2a} \EQN

To start, consider the first term, which can be partial
fractioned as follows:

\BQA {\tilde{I}}_{2a}(x) &\equiv& \int_{-\infty}^\infty d\Delta y
\Bigl[ {e^{\Delta y}\over {\cosh \Delta y - \cos\Delta\phi}} -
{e^{\Delta y}\over {xe^{-\Delta y} + \cosh\Delta y
-\cos\Delta\phi}} \Bigr] \; \cr &=& 2 \int_0^\infty d\eta \Bigl[
{\eta\over {\eta^2 -2\eta\cos\Delta\phi +1}} - {\eta\over {\eta^2
-2\eta\cos\Delta\phi + (1+2x)}} \Bigr] \;, \cr &\phantom{+}& \;
\EQA

\noindent in which we've defined $\eta \equiv e^{\Delta y}$. Each
of these terms can itself be partial-fractioned and simplified as
follows:

\BQA {\eta\over {\eta^2 -2\eta\cos\Delta\phi +1}} &=& {\eta\over
{\bigl( \eta - e^{i\Delta\phi} \bigr) \bigl( \eta -
e^{-i\Delta\phi} \bigr)}} \; \cr &=& {1\over {2i\sin\Delta\phi}}
\Bigl[ {\eta\over {\eta - e^{i\Delta\phi}}} - {\eta\over {\eta -
e^{-i\Delta\phi}}} \Bigr] \; \cr &=& {1\over {\sin\Delta\phi}}
{\cal{IM}} \Bigl[ {\eta\over {\eta-e^{i\Delta\phi}}} \Bigr] \;;
\label{I2a1} \EQA

\noindent ... while, with the designation $\cos \alpha \equiv
\cos\Delta\phi / \sqrt{1+2x}$, we also have

\BQA {\eta\over {\eta^2 -2\eta\cos\Delta\phi + (1+2x)}} &=&
{\eta\over {\bigl( \eta - \sqrt{1+2x} e^{i\alpha} \bigr) \bigl(
\eta - \sqrt{1+2x} e^{-i\alpha} \bigr) }} \; \cr &=& {1\over
{2i\sin\alpha \sqrt{1+2x}}} \Bigl[ {\eta\over {\eta - \sqrt{1+2x}
e^{i\alpha}}} - {\eta\over {\eta - \sqrt{1+2x} e^{-i\alpha}}}
\Bigr] \; \cr &=& {1\over {\sin\alpha \sqrt{1+2x}}} {\cal{IM}}
\Bigl[ {\eta\over {\eta - \sqrt{1+2x} e^{i\alpha}}} \Bigr] \;.
\EQA

\noindent We're now left with the simple integrals

\BQA \int_0^\infty d\eta {\eta\over {\bigl( \eta-e^{i\Delta\phi}
\bigr) }} &=& \biggl[ \bigl( \eta-e^{i\Delta\phi} \bigr) +
e^{i\Delta\phi} \ln \bigl( \eta-e^{i\Delta\phi} \bigr)
{\biggr]}_0^\infty \; \cr \int_0^\infty d\eta {\eta\over {\bigl(
\eta-\sqrt{1+2x} e^{i\Delta\phi} \bigr) }} &=& \phantom{stuff} \cr
\biggl[ \bigl( \eta-\sqrt{1+2x} e^{i\Delta\phi} \bigr) &+&
\sqrt{1+2x} e^{i\Delta\phi} \ln \bigl( \eta-\sqrt{1+2x}
e^{i\Delta\phi} \bigr) {\biggr]}_0^\infty \;, \cr &\phantom{+}&
\; \EQA

\noindent the imaginary parts of which we now evaluate (before
taking limits). Of use will be the identities

\BQA e^{\pm i\phi} &=& \cos\phi \pm i\sin\phi \; \cr \ln z &=&
{1\over 2} \ln z z^* + i {\tan}^{-1} \Bigl[ {{{\cal{IM}} (z)}\over
{{\cal{RE}} (z)}} \Bigr] \;, \EQA

\noindent which lead to

\BQA {\cal{IM}} \biggl[ \bigl( \eta-e^{i\Delta\phi} \bigr) &+&
e^{i\Delta\phi} \ln \bigl( \eta-e^{i\Delta\phi} \bigr) \biggr] \;
\cr &=& -\sin\Delta\phi + {1\over 2} \sin\Delta\phi \ln (\eta^2
-2\eta\cos\Delta\phi +1) \; \cr &+& \cos\Delta\phi {\tan}^{-1}
\Bigl[ {{-\sin\Delta\phi}\over {\eta - \cos\Delta\phi}} \Bigr] \;
\EQA

\BQA {\cal{IM}} \biggl[ \bigl( \eta-\sqrt{1+2x} e^{i\Delta\phi}
\bigr) &+& \sqrt{1+2x} e^{i\Delta\phi} \ln \bigl( \eta-\sqrt{1+2x}
e^{i\Delta\phi} \bigr) \biggr] \; \cr &=&
{{\sqrt{1+2x}\sin\alpha}\over 2} \ln (\eta^2
-2\eta\sqrt{1+2x}\cos\alpha +(1+2x)) \; \cr &-&
\sqrt{1+2x}\sin\alpha \cr &+& \sqrt{1+2x}\cos\alpha {\tan}^{-1}
\Bigl[ {{-\sqrt{1+2x}\sin\alpha}\over {\eta -
\sqrt{1+2x}\cos\alpha}} \Bigr] \;.  \cr  &\phantom{+}& \EQA

\noindent Adding these two results, with the proper coefficients
from equation \ref{I2a1}, and taking the indicated limits results
in:

\BQN {\tilde{I}}_{2a} (x) = \ln (1+2x) + 2 \cot\Delta\phi
(\pi-\Delta\phi) - 2\cot\alpha (\pi-\alpha) \;. \EQN

\noindent From equation \ref{I2a}, we see that we still need the
integral

\BQA {\hat{I}}_{2a} (x) &\equiv& \int_{-\infty}^\infty {{d\Delta y
}\over {\cosh\Delta y - \cos\Delta\phi}} \; \cr &=& 2
\int_0^\infty {{d\eta}\over {\eta^2 -2\eta\cos\Delta\phi +1}} \;
\cr &=& 2 \int_0^\infty {{d\eta}\over {2i\sin\Delta\phi}} \Bigl[
{1\over {\eta -e^{i\Delta\phi}}} - {1\over {\eta -
e^{-i\Delta\phi}}} \Bigr] \; \cr  &=& {2\over {\sin\Delta\phi}}
{\cal{IM}} \int_0^\infty {{d\eta}\over {\eta-e^{i\Delta\phi}}} \;
\cr &=& {2\over {\sin\Delta\phi}} {\cal{IM}} \ln \bigl( \eta -
e^{i\Delta\phi} \bigr) {\Bigr|}_0^\infty \; \cr &=& {2\over
{\sin\Delta\phi}} {\tan}^{-1} \Bigl[ {{-\sin\Delta\phi}\over
{\eta - \cos\Delta\phi}} \Bigr] {\Bigr|}_0^\infty \; \cr &=&
{2\over {\sin\Delta\phi}} (\pi - \Delta\phi) \;. \EQA

\noindent This then makes

\BQA I_{2a}(x) &=& {\tilde{I}}_{2a}(x) - {\hat{I}}_{2a}(x) \; \cr
&=& \ln (1+2x) + 2 {{(\cos\Delta\phi-1)}\over {\sin\Delta\phi}}
(\pi-\Delta\phi) - 2\cot\alpha (\pi-\alpha) \;. \EQA

\noindent Since $I_{2b}$ is the same as $I_{2a}$ with $\Delta y
\rightarrow - \Delta y$, but both are symmetric with respect to
this transformation, we have $I_{2b}(x)=I_{2a}(x)$.

\section{$I_2^c$}

\BQA \int d^{1-2\epsilon} \phi_3 I_2^c (\phi_3) &\equiv& \int
d^{1-2\epsilon} \phi_3 {{2 (\pi-\Delta \phi)}\over {\sin \Delta
\phi}} \; \cr &=&  \int d^{1-2\epsilon} \Delta \phi \Bigl[ {{2
(\pi-\Delta \phi) (1-\cos \Delta \phi)}\over {\sin \Delta \phi}}
+ {{2 (\pi-\Delta \phi) \cos \Delta \phi}\over {\sin \Delta \phi}}
\Bigr] \;, \cr &\phantom{\times}& \; \label{intI2ca} \EQA

\noindent where the last term is

\BQN  \int d^{1-2\epsilon} \Delta \phi {{2 (\pi-\Delta \phi) \cos
\Delta \phi}\over {\sin \Delta \phi}} = 2 \int d
\Omega_{1-2\epsilon} \int_0^\pi d\Delta\phi \sin^{-1-2\epsilon}
\Delta \phi \cos \Delta \phi (\pi-\Delta \phi) \;. \EQN

\noindent Now, since $\cos \Delta \phi$ is odd around $\pi/2$, we
have

\BQN \pi \int_0^\pi d \Delta \phi \sin^{-1-2\epsilon} \Delta \phi
\cos \Delta \phi =0 \;, \EQN

\noindent while

$$\int_0^\pi d \Delta \phi
\sin^{-1-2\epsilon} \Delta \phi \cos \Delta \phi \Delta \phi$$

\noindent can be integrated by parts to give

\BQN {{\Delta \phi}\over {-2\epsilon}} \sin^{-2\epsilon} \Delta
\phi {\Bigr|}_0^\pi - \int_0^\pi d\Delta\phi
{\sin^{-2\epsilon}\over {-2\epsilon}} = {{\sqrt{\pi} \Gamma
({1\over 2}-\epsilon)}\over {2\epsilon \Gamma (1-\epsilon)}} \;.
\EQN

\noindent Since

\BQA \int d \Omega_{1-2\epsilon} &=& {{2\pi^{1/2-\epsilon}}\over
{\Gamma ({1\over 2}-\epsilon)}} \; \cr  \int d^{1-2\epsilon} \phi
\equiv \int d\Omega_{2-2\epsilon} &=& {{2\pi^{1-\epsilon}}\over
{\Gamma (1-\epsilon)}} \;, \EQA

\noindent it follows that

\BQN \int d^{1-2\epsilon} \Delta \phi {{2 (\pi-\Delta \phi) \cos
\Delta \phi}\over {\sin \Delta \phi}} = -{1\over \epsilon} \int
d^{1-2\epsilon} \phi_3 \;, \label{intI2cb} \EQN

\noindent and thus, in the sense of a distribution,

\BQN  \int d^{1-2\epsilon} \phi_3 I_2^c (\phi_3) =  \int
d^{1-2\epsilon} \phi_3 \Bigl[ {{2 (\pi-\Delta \phi) (1-\cos
\Delta \phi)}\over {\sin \Delta \phi}} - {1\over \epsilon} \Bigr]
\;. \EQN

It will be useful to perform this integral fully as well.
Concentrating on the first term, we can choose a change of
variable $w \equiv (1+\cos\Delta\phi)/2$, which leads to

\BQA \sin^2 \Delta\phi &=& 4w(1-w) \; \cr dw &=& -{1\over 2}
\sin\Delta\phi d\Delta\phi \; \cr \Delta\phi &=& -2i
\ln(\sqrt{w}+i\sqrt{1-w}) \;, \EQA

\noindent and the integral

\BQA \int d^{1-2\epsilon} \phi_3 {{2 (\pi-\Delta \phi) (1-\cos
\Delta \phi)}\over {\sin \Delta \phi}} &=&
 \int_0^1 dw  4^{-\epsilon} w^{-1-\epsilon}
{(1-w)}^{-\epsilon} \cr &\times& \biggl( \pi + 2i
\ln(\sqrt{w}+i\sqrt{1-w}) \biggr)  \;. \EQA

\noindent The log term can be integrated by parts. With

\BQA u &\equiv& \ln(\sqrt{w}+i\sqrt{1-w}) \; \cr dv &\equiv& dw
w^{-1-\epsilon} {(1-w)}^{-\epsilon} \;, \EQA

\noindent we find both $du = - {{idw}\over
{2\sqrt{w}\sqrt{1-w}}}$ and

\BQN v = \int dw w^{-1-\epsilon} {(1-w)}^{-\epsilon} =
-{{w^{-\epsilon}{(1-w)}^{-\epsilon}}\over \epsilon} - \int dw
w^{-\epsilon} {(1-w)}^{-1-\epsilon} \;, \EQN

\noindent the latter term of which is $\int dw w^{-1-\epsilon}
{(1-w)}^{-\epsilon} + {1\over \epsilon} w^{-\epsilon}
{(1-w)}^{-\epsilon}$ by symmetry. We then have

\BQN v=-{{w^{-\epsilon} {(1-w)}^{-\epsilon}}\over \epsilon} \;,
\EQN

\noindent so that

\BQA \int_0^1 dw w^{-1-\epsilon} {(1-w)}^{-\epsilon}
\ln(\sqrt{w}+i\sqrt{1-w}) &=& \phantom{stuff} \cr -{{w^{-\epsilon}
{(1-w)}^{-\epsilon}}\over \epsilon} \ln(\sqrt{w}+i\sqrt{1-w})
{\Bigr|}_0^1 &-& {i\over {2\epsilon}} \int_0^1 dw w^{-{1\over 2}
- \epsilon} {(1-w)}^{-{1\over 2} - \epsilon} \; \cr &=&
-{{i\pi}\over {2\epsilon}} {16}^\epsilon
{{\Gamma(1-2\epsilon)}\over {\Gamma^2 (1-\epsilon)}} \;. \EQA

\noindent This gives, finally,

\BQA \int d^{1-2\epsilon} \phi_3 {{2 (\pi-\Delta \phi) (1-\cos
\Delta \phi)}\over {\sin \Delta \phi}} &=& {1\over \epsilon}
 \Bigl[ 1-{{{16}^{-\epsilon} \Gamma^4 (1-\epsilon)}\over {\Gamma^2
(1-2\epsilon)}} \Bigr] {{2\pi^{1-\epsilon}}\over {\Gamma
(1-\epsilon)}} \; \cr &=& {1\over \epsilon} \Bigl[
1-{{{16}^{-\epsilon} \Gamma^4 (1-\epsilon)}\over {\Gamma^2
(1-2\epsilon)}} \Bigr] \int d^{1-2\epsilon} \phi_3 \;, \EQA

\noindent from which

\BQN  \int d^{1-2\epsilon} \phi_3 I_2^c (\phi_3) = -
{{16}^{-\epsilon}\over \epsilon} {{\Gamma^4 (1-\epsilon)}\over
{\Gamma^2 (1-2\epsilon)}} \int d^{1-2\epsilon} \phi_3 \; \EQN

\noindent follows by addition of equation \ref{intI2cb}.

\bigskip
\section{$I_{{2a}\atop {2b}} (x) - I_2^c$}

This integral is not used in the text, but may be useful to those
interested in treating all final-state pieces with a common
$Q_T=0$ definition. The crucial integral here is

\BQA \int d^{1-2\epsilon} \Delta \phi (\pi-\alpha) \cot \alpha
&=& \int d\Omega_{1-2\epsilon} \int_0^\pi d \Delta \phi
(\pi-\alpha) \cot \alpha \sin^{-2\epsilon} \Delta\phi \; \cr
&\equiv& I_1  (I_2 -2\epsilon I_3) \; \EQA

\noindent to ${\cal O}(\epsilon)$, where

\BQA I_1 &\equiv& \int d\Omega_{1-2\epsilon} = {2\over
{(4\pi)}^\epsilon} {{\Gamma (1-\epsilon)}\over {\Gamma
(1-2\epsilon)}} = {4^{-\epsilon}\over \pi} {{\Gamma^2
(1-\epsilon)}\over {\Gamma (1-2\epsilon)}} \int d^{1-2\epsilon}
\phi \; \cr I_2 &\equiv& \int_0^\pi d\Delta\phi (\pi-\alpha) \cot
\alpha \; \cr  I_3 &\equiv& {\pi\over 2} h(x) \;, \quad \quad {\rm
where} \; \cr h(x) &\equiv& {2\over \pi} \int_0^\pi d\Delta \phi
(\pi-\alpha) \cot \alpha \ln \sin \Delta\phi \;. \EQA

\noindent $I_2$ can be evaluated as follows:

\BQA d\cos\alpha &=& -\sin\alpha d\alpha = -{{\sin \Delta\phi
d\Delta\phi}\over \sqrt{1+2x}} \; \cr \sin\Delta\phi &=&
\sqrt{1-\cos^2 \Delta\phi} = \sqrt{1+2x} \sqrt{\sin^2 \alpha -
{{2x}\over {1+2x}}} \; \cr \alpha (\Delta\phi=0) &=& \cos^{-1}
\biggl( {1\over \sqrt{1+2x}} \biggr) \; \cr \alpha
(\Delta\phi=\pi) &=& \cos^{-1} \biggl( {{-1}\over \sqrt{1+2x}}
\biggr) \;, \EQA

\noindent and so

\BQN I_2 =
\int_{\cos^{-1}(1/\sqrt{1+2x})}^{\cos^{-1}(-1/\sqrt{1+2x})}
{{d\alpha (\pi-\alpha) \cos\alpha}\over \sqrt{\sin^2 \alpha -
{{2x}\over {1+2x}}}} \;. \EQN

\noindent Now, with $\pi-\alpha = {\pi\over 2} + ({\pi\over 2} -
\alpha)$, this can be broken up into two terms, the first of
which is odd about $\pi/2$ and the second even. Since our limits
of integration straddle $\pi/2$ evenly, integration over the
first term gives zero, while the second term gives the same
integral from $\cos^{-1}({1\over \sqrt{1+2x}}) \leq \alpha \leq
{\pi\over 2}$ as it does from ${\pi\over 2} \leq \alpha \leq
\cos^{-1}({1\over \sqrt{1+2x}})$. Thus

\BQN I_2 = 2 \int_{\cos^{-1}(1/\sqrt{1+2x})}^{\pi/2} d\alpha
{{({\pi\over 2}-\alpha) \cos\alpha}\over \sqrt{\sin^2 \alpha
-{{2x}\over {1+2x}}}} \;. \EQN

\noindent The first term here can be evaluated using the change
of variable $y \equiv \sin\alpha$, which yields

\BQA dy &=& \cos\alpha d\alpha \; \cr y_0 \equiv y(\alpha_0) &=&
\sin \cos^{-1} ({1\over \sqrt{1+2x}}) = \sqrt{{2x}\over {1+2x}}
\; \EQA

\noindent and thus

\BQA \pi \int_{y_0}^1 {{dy}\over \sqrt{y^2 - y_0^2}} &=& \pi \ln 2
\biggl( y+ \sqrt{y^2 - y_0^2} \biggr) {\Biggr|}_{y_0}^1 = \pi \ln
{{1+\sqrt{1+2x}}\over \sqrt{2x}} \; \EQA

\noindent by G.R. 2.261. The second term can be obtained via G.R.
3.842.2 with $u\equiv \cos^{-1} (1/\sqrt{1+2x})$:

\BQN 2 \int_u^{\pi / 2} {{d\alpha \alpha \cos\alpha}\over
\sqrt{\sin^2 \alpha - {{2x}\over {1+2x}}}} = \pi \ln (1+\cos u) =
\pi \ln {{1+\sqrt{1+2x}}\over \sqrt{1+2x}} \;. \EQN

\noindent Thus

\BQN I_2 = \pi \ln {{1+\sqrt{1+2x}}\over \sqrt{2x}} - \pi \ln
{{1+\sqrt{1+2x}}\over \sqrt{1+2x}} = {\pi\over 2} \ln
{{1+2x}\over {2x}} \;, \EQN

\noindent and so, to ${\cal O} (\epsilon)$

\BQN \int d^{1-2\epsilon} \Delta \phi (\pi-\alpha) \cot \alpha =
4^{-\epsilon} {{\Gamma^2 (1-\epsilon)}\over {\Gamma
(1-2\epsilon)}} \Bigl[ {1\over 2} \ln {{1+2x}\over {2x}}
-\epsilon h(x) \Bigr] \int d^{1-2\epsilon} \phi \;. \EQN

Now, from equations \ref{i2a2b} and \ref{i2c}, we find that

\BQA \phantom{\cdot} &\int& d^{1-2\epsilon} \phi_3 (I_{{2a}\atop
{2b}} (x)-I_2^c) \; \cr &=& \int d^{1-2\epsilon} \Delta\phi
\Bigl[ \ln {(1+2x)} -2 {{(\pi-\Delta\phi) (1-\cos\Delta\phi)}\over
{\sin\Delta\phi}} -2 \bigl( \pi - \alpha(x) \bigr) \cot \alpha
(x) \Bigr] \; \cr &=& \Bigl[ \ln {(1+2x)} -{1\over \epsilon} +
{{16}^{-\epsilon}\over \epsilon} {{\Gamma^4 (1-\epsilon)}\over
{\Gamma^2 (1-2\epsilon)}} - 4^{-\epsilon} {{\Gamma^2
(1-\epsilon)}\over {\Gamma (1-2\epsilon)}} \Bigl( \ln
{{1+2x}\over {2x}} -2\epsilon h(x) \Bigr) \Bigr]  \cr &\times&
\int d^{1-2\epsilon} \phi \; \cr &\simeq& \Bigl[ \Bigl( \ln {2x}
- \ln {16} \Bigr) + \epsilon \Bigl( 2h(x) + \ln 4 \ln
{{1+2x}\over {2x}} + {1\over 2} \ln^2 {16} - {\pi^2 \over 3}
\Bigr) \Bigr] \int d^{1-2\epsilon} \phi \;. \EQA

\setcounter{footnote}{0}
\chapter{Detailed NLO Results}
\label{ap:Contribs}

If we write out our $\Gamma_{mn}$ factors from Chapter
\ref{ch:photon} grouped by $Q_T$-dependence (and including poles):

\begin{eqnarray}
\Gamma_{mn} &\equiv & \delta(Q_T) \biggl( {{_2 \Gamma_{mn}}\over
\epsilon^2} + {{_1\Gamma_{mn}}\over \epsilon} \biggr) +
\delta(Q_T) {_\delta C_{mn}} \cr &+& \delta(Q_T) \biggl[
\sum_{\tilde{a}} \int_{x_a}^1 {dz\over z} {{H(x_a/z,x_b)}\over
H(x_a,x_b)} {_{\tilde{a}} C_{mn}} (z) \cr &+& \sum_{\tilde{b}}
\int_{x_b}^1 {dz\over z} {{H(x_a,x_b/z)}\over H(x_a,x_b)}
{_{\tilde{b}} C_{mn}} (z) \biggr] \cr &-&4 A_{mn} {\biggl[ {{\ln
Q_T/Q}\over Q_T} \biggr]}_+ + 2B_{mn} {\biggl[ {1\over Q_T}
\biggr] }_+ \cr &+& {\biggl[ {2\over Q_T} \biggr]}_+ \biggl[ {_a
D_{mn}} \sum_{\tilde{a}} \int_{x_a}^1 {dz\over z}
{{H(x_a/z,x_b)}\over H(x_a,x_b)} P_{a\tilde{a}}^+ (z) \cr &+& {_b
D_{mn}} \sum_{\tilde{b}} \int_{x_b}^1 {dz\over z}
{{H(x_a,x_b/z)}\over H(x_a,x_b)} P_{b\tilde{b}}^+ (z) \biggr] \;,
\end{eqnarray}

\noindent then each $\Gamma_{mn}$ is specified by writing out $\{
{_2 \Gamma}, {_1 \Gamma}, {_\delta C}, {_{\tilde{a}} C},
{_{\tilde{b}} C}, A,B,{_a D}, {_b D} \}$ for each $\{ mn \}$. On
the next few pages, we'll lay out our findings for these
parameters. As usual, $C_F=4/3$, $N_C=3$, and $N_f$ is the number
of participating flavors at the appropriate renormalization scale
$\mu$. In general, each $\Gamma_{mn}$ will depend on the
subprocess under consideration, labeled in the text by the index
$i$, or more specifically by indices $\{a,b,d\}$ over the types
of partons participating in the subprocess $ab \rightarrow \gamma
d$, for instance $qg \rightarrow \gamma q$. For the virtual
contributions, each $\Gamma_{(i)}$ is presented independently. The
rest are expressed in terms of constants ${\rm C}(f)$ and $\gamma
(f)$, where

\BQA {\rm C}(q) &\equiv& C_F \cr {\rm C}(g) &\equiv& N_C \cr
\gamma (q) &\equiv& {3\over 2} C_F \cr \gamma (g) &\equiv& {{11N_C
- 2N_f}\over 6} \; \EQA

\noindent For example, ${\rm C}(a) + {\rm C}(b) - {\rm C}(d)$ for
$q\bar{q} \rightarrow \gamma g$ is $2C_F-N_C$.

\section{$\Gamma_{Virt}$}

\begin{eqnarray}
{_2 \Gamma_{Virt}^{(q \bar{q} \rightarrow \gamma g)}} &=& -(2{\rm
C_F}+{\rm N_C}) \; \\ {_1 \Gamma_{Virt}^{(q \bar{q} \rightarrow
\gamma g)}} &=& -3{\rm C_F} -{{(11{\rm N_C}-2{\rm N_f})}\over 6}
+ (2{\rm C_F}-{\rm N_C}) \ln {\hat{s} \over M_f^2} + {\rm N_C}
\ln {{\hat{t} \hat{u}}\over M_f^4} \; \\ A_{Virt}^{(q \bar{q}
\rightarrow \gamma g)} &=&0 \; \\ B_{Virt}^{(q \bar{q}
\rightarrow \gamma g)} &=&0 \; \\ {_\delta C_{Virt}^{(q \bar{q}
\rightarrow \gamma g)}} &=& {\biggl[ A_{0\nu}+{B_\nu \over T_0}
\biggr]}_{(q \bar{q} \rightarrow \gamma g)} -(2{\rm C_F}+{\rm
N_C}) {\pi^2\over 6} \; \\ {_{\tilde{a}} C_{Virt}^{(q \bar{q}
\rightarrow \gamma g)}} &=&0 \; \\ {_{\tilde{b}} C_{Virt}^{(q
\bar{q} \rightarrow \gamma g)}} &=&0 \; \\ {_a D_{Virt}^{(q
\bar{q} \rightarrow \gamma g)}} &=&0 \; \\ {_b D_{Virt}^{(q
\bar{q} \rightarrow \gamma g)}} &=&0 \;
\end{eqnarray}

\noindent where

\begin{eqnarray}
A_{0\nu}^{(q \bar{q} \rightarrow \gamma g)} &=& -2\biggl( {\rm
C_F}+ {{\rm N_C}\over 2} \biggr) \biggl[ - {\pi^2 \over 6}
+{1\over 2} \ln^2 {M_f^2 \over \hat{s}} \biggr] -{{(11{\rm
N_C}-2{\rm N_f})}\over 6} \ln {\hat{s}\over \mu^2} \cr &-& \biggl[
3{\rm C_F} + {{(11{\rm N_C}-2{\rm N_f})}\over 6} + {\rm N_C} \ln
\bigl| {\hat{s}^2 \over {\hat{t} \hat{u}}} \bigr| \biggr] \ln
{M_f^2 \over \hat{s}} \cr &+& \biggl( 2{\rm C_F} - {{\rm
N_C}\over 2} \biggr) {\pi^2 \over 3} -7{\rm C_F} + {\rm C_F}
\biggl( \ln^2 \bigl| {\hat{t}\over \hat{s}} \bigr| + \ln^2 \bigl|
{\hat{u}\over \hat{s}} \bigr| \biggr) \cr &-& {{\rm N_C}\over 2}
\ln^2 {\hat{s}^2 \over {\hat{t} \hat{u}}} \; \\ B_\nu^{(q \bar{q}
\rightarrow \gamma g)} &=& 3{\rm C_F} \bigl| {\hat{t}\over
\hat{u}} \bigr| \ln \bigl| {\hat{u}\over \hat{s}} \bigr| \cr &+&
\biggl( {\rm C_F}-{{\rm N_C}\over 2} \biggr) \biggl[ \biggl( 2+
\bigl| {\hat{u}\over \hat{t}} \bigr| \biggr) \ln^2 \bigl|
{\hat{s}\over \hat{u}} \bigr| + \biggl( 2+ \bigl| {\hat{t}\over
\hat{u}} \bigr| \biggr) \ln^2 \bigl| {\hat{s}\over \hat{t}}
\bigr| \biggr] \cr &+& 3{\rm C_F} \bigl| {\hat{u}\over \hat{t}}
\bigr| \ln \bigl| {\hat{t}\over \hat{s}} \bigr| + 2\biggl( {\rm
C_F}-{{\rm N_C}\over 2} \biggr) \ln \bigl| {{\hat{t} \hat{u}}\over
\hat{s}^2} \bigr| \; \\ T_0^{(q \bar{q} \rightarrow \gamma g)}
&=&{\hat{t}\over \hat{u}}+{\hat{u}\over \hat{t}} \;
\end{eqnarray}

\noindent and

\begin{eqnarray}
\Gamma_{Virt}^{(q g \rightarrow \gamma q)} &=& \Gamma_{Virt}^{(q
\bar{q} \rightarrow \gamma g)} (\hat{s} \leftrightarrow \hat{u})
\\ \Gamma_{Virt}^{(g q \rightarrow \gamma q)} &=& \Gamma_{Virt}^{(q
\bar{q} \rightarrow \gamma g)} (\hat{s} \leftrightarrow \hat{t})
\;. \end{eqnarray}

\section{$\Gamma_{ab}^{\rm soft}$}

\begin{eqnarray}
{_2 \Gamma_{ab}} &=&{{{\rm C}(a)+{\rm C}(b)-{\rm C}(d)}\over 2}
\; \\ {_1 \Gamma_{ab}} &=&{{{\rm C}(a)+{\rm C}(b)-{\rm
C}(d)}\over 2} \biggl[ \ln {M_f^2\over
p_j^2} - \ln {{{(1-x_a)}^2 S x_b}\over {x_a p_j^2}} \biggr] \; \\
A_{ab} &=& {{{\rm C}(a)+{\rm C}(b)-{\rm C}(d)}\over 2} \; \\
B_{ab} &=& {{{\rm C}(a)+{\rm C}(b)-{\rm C}(d)}\over 2} \ln
{{{(1-x_a)}^2 S x_b}\over {x_a Q^2}} \; \\ {_\delta C_{ab}} &=&
{{{\rm C}(a)+{\rm C}(b)-{\rm C}(d)}\over 2} \biggl[ {1\over 2}
\ln^2 {M_f^2\over p_j^2} - \ln {M_f^2\over p_j^2} \ln
{{{(1-x_a)}^2 S x_b}\over {x_a p_j^2}} +{\pi^2\over 6} \biggr] \;
\\ {_{\tilde{a}} C_{ab}} &=&0 \;
\\ {_{\tilde{b}} C_{ab}} &=&0 \; \\ {_a D_{ab}} &=&0 \; \\ {_b D_{ab}}
&=&0 \; \end{eqnarray}

\section{$\Gamma_{ba}^{\rm soft}$}

\begin{eqnarray}
{_2 \Gamma_{ba}} &=&{{{\rm C}(a)+{\rm C}(b)-{\rm C}(d)}\over 2}
\; \\ {_1 \Gamma_{ba}} &=&{{{\rm C}(a)+{\rm C}(b)-{\rm
C}(d)}\over 2} \biggl[ \ln {M_f^2\over p_j^2} - \ln {{{(1-x_b)}^2
S x_a}\over {x_b p_j^2}} \biggr] \; \\ A_{ba} &=& {{{\rm
C}(a)+{\rm C}(b)-{\rm C}(d)}\over 2} \; \\ B_{ba} &=& {{{\rm
C}(a)+{\rm C}(b)-{\rm C}(d)}\over 2} \ln {{{(1-x_b)}^2 S
x_a}\over {x_b Q^2}} \; \\ {_\delta C_{ba}} &=& {{{\rm C}(a)+{\rm
C}(b)-{\rm C}(d)}\over 2} \biggl[ {1\over 2} \ln^2 {M_f^2\over
p_j^2} - \ln {M_f^2\over p_j^2} \ln {{{(1-x_b)}^2 S x_a}\over
{x_b p_j^2}} +{\pi^2\over 6} \biggr] \;
\\ {_{\tilde{a}} C_{ba}} &=&0 \; \\ {_{\tilde{b}} C_{ba}} &=&0 \; \\ {_a D_{ba}}
&=&0 \; \\ {_b D_{ba}} &=&0 \; \end{eqnarray}

\section{$\Gamma_{a2}^{\rm soft}$}

\begin{eqnarray}
{_2 \Gamma_{a2}} &=&{{{\rm C}(a)-{\rm C}(b)+{\rm C}(d)}\over 2}
\; \\ {_1 \Gamma_{a2}} &=&{{{\rm C}(a)-{\rm C}(b)+{\rm
C}(d)}\over 2} \cr &\times& \biggl[ \ln {M_f^2\over p_j^2} - \ln
{{{(1-x_a)}^2 S (1-v)}\over {(2-v) p_j^2}} -2(\pi-\alpha_a) \cot
\alpha_a+2y_j  \biggr] \; \\ A_{a2} &=& {{{\rm C}(a)-{\rm
C}(b)+{\rm C}(d)}\over 2} \; \\ B_{a2} &=& {{{\rm C}(a)-{\rm
C}(b)+{\rm C}(d)}\over 2} \cr &\times& \biggl[ \ln {{{(1-x_a)}^2 S
(1-v)}\over {(2-v) Q^2}} +2(\pi-\alpha_a) \cot \alpha_a-2y_j \biggr] \; \\
{_\delta C_{a2}} &=& {{{\rm C}(a)-{\rm C}(b)+{\rm C}(d)}\over 2}
\Biggl[ {1\over 2} \ln^2 {M_f^2\over p_j^2} \cr &-& \ln
{M_f^2\over p_j^2} \biggl[ \ln {{{(1-x_a)}^2 S (1-v)}\over {(2-v)
p_j^2}} +2(\pi-\alpha_a) \cot \alpha_a-2y_j
\biggr] +{\pi^2\over 6} \Biggr] \; \\ {_{\tilde{a}} C_{a2}} &=&0 \; \\
{_{\tilde{b}} C_{a2}} &=&0 \; \\ {_a D_{a2}} &=&0 \; \\ {_b
D_{a2}} &=&0 \;
\end{eqnarray}

\noindent where $v \equiv -\hat{u}/\hat{s}$, $\alpha_a \equiv
\alpha ({1\over {2(1-v)}})$ and $2y_j = \ln {{x_a (1-v)}\over
{x_b v}}$.

\section{$\Gamma_{b2}^{\rm soft}$}

\begin{eqnarray}
{_2 \Gamma_{b2}} &=&{{-{\rm C}(a)+{\rm C}(b)+{\rm C}(d)}\over 2}
\; \\ {_1 \Gamma_{b2}} &=&{{-{\rm C}(a)+{\rm C}(b)+{\rm
C}(d)}\over 2} \cr &\times& \biggl[ \ln {M_f^2\over p_j^2} - \ln
{{{(1-x_b)}^2 S v}\over {(1+v) p_j^2}} -2(\pi-\alpha_b) \cot
\alpha_b-2y_j  \biggr] \; \\ A_{b2} &=& {{-{\rm C}(a)+{\rm
C}(b)+{\rm C}(d)}\over 2} \; \\ B_{b2} &=& {{-{\rm C}(a)+{\rm
C}(b)+{\rm C}(d)}\over 2} \biggl[ \ln {{{(1-x_b)}^2 S v}\over
{(1+v) Q^2}} +2(\pi-\alpha_b) \cot \alpha_b+2y_j \biggr] \; \\
{_\delta C_{b2}} &=& {{-{\rm C}(a)+{\rm C}(b)+{\rm C}(d)}\over 2}
\Biggl[ {1\over 2} \ln^2 {M_f^2\over p_j^2} \cr &-& \ln
{M_f^2\over p_j^2} \biggl[ \ln {{{(1-x_b)}^2 S v}\over {(1+v)
p_j^2}} +2(\pi-\alpha_b) \cot \alpha_b+2y_j \biggr]
+{\pi^2\over 6} \Biggr] \; \\ {_{\tilde{a}} C_{b2}} &=&0 \; \\ {_{\tilde{b}} C_{b2}} &=&0 \; \\
{_a D_{b2}} &=&0 \; \\ {_b D_{b2}} &=&0 \;
\end{eqnarray}

\noindent where $\alpha_b \equiv \alpha ({1\over {2v}})$ and
$2y_j = \ln {{x_a (1-v)}\over {x_b v}}$.

\section{$\Gamma_{2a}^{\rm soft}$}

\begin{eqnarray}
{_2 \Gamma_{2a}} &=&0 \; \\ {_1 \Gamma_{2a}} &=&{{{\rm C}(a)-{\rm
C}(b)+{\rm C}(d)}\over 2} \cr &\times& \biggl[ \ln {{1-v}\over
{2-v}} +2{{(\pi-\Delta\phi)}\over \sin\Delta\phi}
(1-\cos\Delta\phi) +2(\pi-\alpha_a) \cot \alpha_a  \biggr] \; \\
A_{2a} &=& 0 \; \\ B_{2a} &=& {{{\rm C}(a)-{\rm C}(b)+{\rm
C}(d)}\over 2} \cr &\times& \biggl[ \ln {{2-v}\over {1-v}}
-2{{(\pi-\Delta\phi)}\over \sin\Delta\phi} (1-\cos\Delta\phi)
-2(\pi-\alpha_a) \cot \alpha_a  \biggr] \; \\ {_\delta C_{2a}}
&=& {{{\rm C}(a)-{\rm C}(b)+{\rm C}(d)}\over 2} \ln {p_j^2\over
M_f^2} \cr &\times& \biggl[ \ln {{2-v}\over {1-v}}
-2{{(\pi-\Delta\phi)}\over \sin\Delta\phi} (1-\cos\Delta\phi)
-2(\pi-\alpha_a) \cot \alpha_a  \biggr] \; \\ {_{\tilde{a}}
C_{2a}} &=&0 \;
\\ {_{\tilde{b}} C_{2a}} &=&0 \; \\ {_a D_{2a}} &=&0 \; \\ {_b D_{2a}}
&=&0 \; \end{eqnarray}

\noindent where $\alpha_a \equiv \alpha ({1\over {2(1-v)}})$ and
$\Delta\phi \equiv {\rm MOD} (\| \phi_q-\phi_j-\pi \| /\pi)$.

\section{$\Gamma_{2b}^{\rm soft}$}

\begin{eqnarray}
{_2 \Gamma_{2b}} &=&0 \; \\ {_1 \Gamma_{2b}} &=&{{-{\rm
C}(a)+{\rm C}(b)+{\rm C}(d)}\over 2} \cr &\times& \biggl[ \ln
{{v}\over {1+v}} +2{{(\pi-\Delta\phi)}\over \sin\Delta\phi}
(1-\cos\Delta\phi) +2(\pi-\alpha_b) \cot \alpha_b  \biggr] \; \\
A_{2b} &=& 0 \; \\ B_{2b} &=& {{-{\rm C}(a)+{\rm C}(b)+{\rm
C}(d)}\over 2} \cr &\times& \biggl[ \ln {{1+v}\over {v}}
-2{{(\pi-\Delta\phi)}\over \sin\Delta\phi}
(1-\cos\Delta\phi) -2(\pi-\alpha_b) \cot \alpha_b  \biggr] \; \\
{_\delta C_{2b}} &=& {{-{\rm C}(a)+{\rm C}(b)+{\rm C}(d)}\over 2}
\ln {p_j^2\over M_f^2} \cr &\times& \biggl[ \ln {{1+v}\over {v}}
-2{{(\pi-\Delta\phi)}\over \sin\Delta\phi} (1-\cos\Delta\phi)
-2(\pi-\alpha_b) \cot \alpha_b  \biggr] \; \\ {_{\tilde{a}}
C_{2b}} &=&0 \;
\\ {_{\tilde{b}} C_{2b}} &=&0 \; \\ {_a D_{2b}} &=&0 \; \\ {_b D_{2b}}
&=&0 \; \end{eqnarray}

\noindent where $\alpha_b \equiv \alpha ({1\over {2v}})$ and
$\Delta\phi \equiv {\rm MOD} (\| \phi_q-\phi_j-\pi \| /\pi)$.

\section{$\Gamma_a^{\rm coll}$}

\begin{eqnarray}
{_2 \Gamma_a} &=&0 \; \\ {_1 \Gamma_a} &=& \gamma (a) + 2{\rm
C}(a) \ln {{1-x_a}\over x_a} - \sum_{\tilde{a}} \int_{x_a}^1
{dz\over z} {{H(x_a/z,x_b)}\over H(x_a,x_b)} P_{a\tilde{a}}^+ (z)
\; \\ A_a &=& 0 \;
\\ B_a &=& -\gamma (a) - 2{\rm C}(a) \ln {{1-x_a}\over
x_a} \; \\ {_\delta C_a} &=& \ln {M_f^2\over p_j^2} \biggl[
\gamma (a) + 2{\rm C}(a) \ln {{1-x_a}\over x_a} \biggr] \; \\
{_{\tilde{a}} C_a} &=& P_{a\tilde{a}}^+ (z) \ln {p_j^2\over
M_f^2} -
P_{a\tilde{a}}^1 \; \\ {_{\tilde{b}} C_a} &=&0 \; \\ {_a D_a} &=&1 \; \\
{_b D_a} &=&0 \; \end{eqnarray}

\section{$\Gamma_b^{\rm coll}$}

\begin{eqnarray}
{_2 \Gamma_b} &=&0 \; \\ {_1 \Gamma_b} &=& \gamma (b) + 2{\rm
C}(b) \ln {{1-x_b}\over x_b} - \sum_{\tilde{b}} \int_{x_b}^1
{dz\over z} {{H(x_a,x_b/z)}\over H(x_a,x_b)} P_{b\tilde{b}}^+ (z)
\; \\ A_b &=& 0 \;
\\ B_b &=& -\gamma (b) - 2{\rm C}(b) \ln {{1-x_b}\over
x_b} \; \\ {_\delta C_b} &=& \ln {M_f^2\over p_j^2} \biggl[
\gamma (b) + 2{\rm C}(b) \ln {{1-x_b}\over x_b} \biggr] \; \\
{_{\tilde{a}} C_b} &=&0 \; \\ {_{\tilde{b}} C_b}
&=&P_{b\tilde{b}}^+ (z) \ln {p_j^2\over
M_f^2} - P_{b\tilde{b}}^1 \; \\ {_a D_b} &=&0 \; \\
{_b D_b} &=&1 \; \end{eqnarray}

\section{$\Gamma_a^{CT}$}

\begin{eqnarray}
{_2 \Gamma_a^{CT}} &=&0 \; \\ {_1 \Gamma_a^{CT}} &=&
\sum_{\tilde{a}} \int_{x_a}^1 {dz\over z} {{H(x_a/z,x_b)}\over
H(x_a,x_b)} P_{a\tilde{a}}^+ (z) \; \\ A_a^{CT} &=& 0 \; \\
B_a^{CT} &=& 0 \; \\ {_\delta C_a^{CT}} &=& 0 \; \\ {_{\tilde{a}}
C_a^{CT}} &=& 0 \; \\ {_{\tilde{b}} C_a^{CT}} &=&0 \; \\ {_a
D_a^{CT}} &=&0 \; \\ {_b D_a^{CT}} &=&0 \;
\end{eqnarray}

\section{$\Gamma_b^{CT}$}

\begin{eqnarray}
{_2 \Gamma_b^{CT}} &=&0 \; \\ {_1 \Gamma_b^{CT}} &=&
\sum_{\tilde{b}} \int_{x_b}^1 {dz\over z} {{H(x_a,x_b/z)}\over
H(x_a,x_b)} P_{b\tilde{b}}^+ (z) \; \\ A_b^{CT} &=& 0 \; \\
B_b^{CT} &=& 0 \; \\ {_\delta C_b^{CT}} &=& 0 \; \\ {_{\tilde{a}}
C_b^{CT}} &=&0 \; \\ {_{\tilde{b}} C_b^{CT}} &=& 0 \; \\ {_a
D_b^{CT}} &=&0 \; \\ {_b D_b^{CT}} &=&0 \; \end{eqnarray}

\section{$\Gamma_2^{\rm coll}$}

\begin{eqnarray}
{_2 \Gamma_2} &=&{\rm C}(d) \; \\ {_1 \Gamma_2} &=& \gamma
(d)-{\rm C} (d) \biggl[ 2{{(\pi-\Delta\phi)}\over \sin\Delta\phi}
(1-\cos\Delta\phi) -\ln {M_f^2\over p_j^2} \biggr] \;
\\ A_2 &=&0 \; \\ B_2 &=&0 \; \\ {_\delta
C_2} &=& - \gamma (d) \biggl[ 2{{(\pi-\Delta\phi)}\over
\sin\Delta\phi} (1-\cos\Delta\phi) -\ln {M_f^2\over p_j^2}
\biggr] + \delta_{dg} {{(17{\rm N_C}-23{\rm N_f})}\over 18} \cr
&-&{\rm C}(d) \biggl[ 2{{(\pi-\Delta\phi)}\over \sin\Delta\phi}
(1-\cos\Delta\phi) \ln {M_f^2\over p_j^2} \cr  &-& {1\over 2}
\ln^2 {M_f^2\over p_j^2} + \biggl( {{2\pi^2}\over 3} - {13\over
2} \biggr) -{\pi^2\over 6} \biggr] \; \\ {_{\tilde{a}} C_2} &=&0 \; \\
{_{\tilde{b}} C_2} &=&0 \; \\ {_a D_2} &=&0 \; \\ {_b D_2} &=&0 \;
\end{eqnarray}

\noindent where $\Delta\phi \equiv {\rm MOD} (\|
\phi_q-\phi_j-\pi \| /\pi)$.

One can easily verify that the poles all cancel; that is, the
sums $\sum_{mn} {_2 \Gamma_{mn}}$ and $\sum_{mn} {_1
\Gamma_{mn}}$ both equal $0$. Additional cancellations will occur
upon summing the remaining pieces, and further simplification
arises upon integration over $\phi_q$. In particular,

\begin{eqnarray}
\int_0^{2\pi} d\phi_q 2 {{(\pi-\Delta\phi)}\over
\sin\Delta\phi} (1-\cos\Delta\phi) &=& 2\pi\ln 16 \;, \\
\int_0^{2\pi} d\phi_q 2(\pi-\alpha(x_i)) \cot\alpha(x_i) &=& 2\pi
\ln {{1+2x_i}\over {2x_i}} \;. \end{eqnarray}

\noindent Thus, for example, terms involving $\cot\alpha({1\over
{2(1-v)}})$ will cancel corresponding $\ln (2-v)$ terms, and
$\cot\alpha({1\over {2v}})$ terms will get rid of $\ln (1+v)$
pieces.
\medskip

As it is, we have the initial-state and final-state sums:

\BQA A_{init} &\equiv & A_{Virt}+A_{ab}+A_{ba}+A_{a2}+A_{b2}  +
A_a^{\rm coll} +A_b^{\rm coll}+A_a^{CT}+A_b^{CT} \cr &=& {\rm C}(a)+{\rm C}(b) \; \\
A_{final} &\equiv & A_{2a}+A_{2b}+A_2^{\rm coll} \cr &=&0 \; \EQA

\BQA B_{init} &\equiv & B_{Virt}+B_{ab}+B_{ba}+B_{a2}+B_{b2} +
B_a^{\rm coll}+B_b^{\rm coll} +B_a^{CT}+B_b^{CT} \cr &=& {\rm
C}(a) \biggl[ {1\over 2} \ln {{1+v}\over {2-v}} {v\over {1-v}} +
\ln {\hat{s} \over Q^2} + (\pi-\alpha_a) \cot\alpha_a -
(\pi-\alpha_b) \cot\alpha_b \biggr] \cr &+& {\rm C}(b) \biggl[
{1\over 2} \ln {{2-v}\over {1+v}} {{1-v}\over v} + \ln
{\hat{s}\over Q^2} - (\pi-\alpha_a) \cot\alpha_a + (\pi-\alpha_b)
\cot\alpha_b \biggr] \cr &+& {\rm C}(d) \biggl[ {1\over 2} \ln
{{v(1-v)}\over {(2-v)(1+v)}} + (\pi-\alpha_a) \cot\alpha_a +
(\pi-\alpha_b) \cot\alpha_b \biggr] \cr &-& \bigl( \gamma(a) +
\gamma (b) \bigr) \; \\ B_{final} &\equiv &
B_{2a}+B_{2b}+B_2^{\rm coll} \cr &=& {\rm C}(a) \biggl[ {1\over 2}
\ln {{2-v}\over {1+v}} {v\over {1-v}} - (\pi-\alpha_a)
\cot\alpha_a + (\pi-\alpha_b) \cot\alpha_b \biggr] \cr &+& {\rm
C}(b) \biggl[ {1\over 2} \ln {{1+v}\over {2-v}} {{1-v}\over v} +
(\pi-\alpha_a) \cot\alpha_a - (\pi-\alpha_b) \cot\alpha_b \biggr]
\cr &+& {\rm C}(d) \biggl[ {1\over 2} \ln {{(2-v)(1+v)}\over
{v(1-v)}} - (\pi-\alpha_a) \cot\alpha_a - (\pi-\alpha_b)
\cot\alpha_b \cr &-& 2{{(\pi-\Delta\phi)} \over {\sin\Delta\phi}}
(1-\cos\Delta\phi) \biggr]  \; \EQA

\BQA  {_\delta C_{init}} &=& {\rm C}(a) \Biggl[ {1\over 2} \ln^2
{M_f^2\over p_j^2} - {1\over 2} \ln {M_f^2\over p_j^2} \biggl[\ln
{{1+v}\over {2-v}} {v\over {1-v}} +2\ln {\hat{s}\over p_j^2} \cr
&+& 2(\pi-\alpha_a) \cot\alpha_a -2 (\pi-\alpha_b) \cot\alpha_b
\biggr] \Biggr] \cr &+& {\rm C}(b) \Biggl[ {1\over 2} \ln^2
{M_f^2\over p_j^2} - {1\over 2} \ln {M_f^2\over p_j^2} \biggl[
\ln {{2-v}\over {1+v}} {{1-v}\over v} + 2\ln {\hat{s}\over p_j^2}
\cr &-& 2 (\pi-\alpha_a) \cot\alpha_a +2 (\pi-\alpha_b)
\cot\alpha_b \biggr] \Biggr] \cr &-& {{{\rm C}(d)}\over 2} \ln
{M_f^2\over p_j^2} \biggl[ \ln {{v(1-v)}\over {(2-v)(1+v)}} +2
(\pi-\alpha_a) \cot\alpha_a +2 (\pi-\alpha_b) \cot\alpha_b
\biggr] \cr &+& \bigl(\gamma(a) + \gamma (b) \bigr) \ln
{M_f^2\over p_j^2} +
A_{0\nu} +{B_\nu \over T_0} - {\rm C}(d) {\pi^2 \over 6} \; \\
{_\delta C_{final}} &=& {{{\rm C}(a)}\over 2} \ln {M_f^2\over
p_j^2} \biggl[\ln {{1+v}\over {2-v}} {{1-v}\over v}  +
2(\pi-\alpha_a) \cot\alpha_a -2 (\pi-\alpha_b) \cot\alpha_b
\biggr] \cr &+& {{{\rm C}(b)}\over 2} \ln {M_f^2\over p_j^2}
\biggl[ \ln {{2-v}\over {1+v}} {v\over {1-v}} -2(\pi-\alpha_a)
\cot\alpha_a +2 (\pi-\alpha_b) \cot\alpha_b \biggr] \cr &+& {\rm
C}(d) \Biggl[ {1\over 2} \ln {M_f^2\over p_j^2} \biggl[ \ln
{{v(1-v)}\over {(2-v)(1+v)}} \cr &+& 2 (\pi-\alpha_a)
\cot\alpha_a +2 (\pi-\alpha_b) \cot\alpha_b
+4{{(\pi-\Delta\phi)}\over {\sin\Delta\phi}}(1-\cos\Delta\phi)
\biggr] \cr &-& {1\over 2}\ln {M_f^2\over p_j^2}  \biggl[
4{{(\pi-\Delta\phi)}\over {\sin\Delta\phi}}(1-\cos\Delta\phi)
\biggr] + {1\over 2}\ln^2 {M_f^2\over p_j^2} -\biggl(
{{2\pi^2}\over 3}-{13\over 2}\biggr) +{\pi^2\over 6} \Biggr] \cr
&-& \gamma(d) \biggl[ 2{{(\pi-\Delta\phi)}\over
{\sin\Delta\phi}}(1-\cos\Delta\phi) - \ln {M_f^2\over p_j^2}
\biggr] + \delta_{dg} {{17{\rm N_C}-23{\rm N_f}}\over 18} \; \EQA

\BQA {_{\tilde{a}} C_{init}}+{_{\tilde{b}} C_{init}} &=& -\biggl[
P_{a \tilde{a}}^1 + P_{b \tilde{b}}^1 \biggr] - \ln {M_f^2 \over
p_j^2} \biggl( P_{a \tilde{a}}^+ + P_{b \tilde{b}}^+ \biggr) \;
\\ {_{\tilde{a}} C_{final}}+{_{\tilde{b}} C_{final}} &=&0 \; \EQA

\BQA {_a D_{init}} &=&1 \;
\\ {_b D_{init}} &=&1 \; \\ {_a D_{final}} &=&0 \; \\ {_b
D_{final}} &=&0 \; \EQA

\noindent Calculation of the CSS $\{A,B,C\}$ parameters from these
results is outlined in Section \ref{sec:ABC}.

For the $m=1$ Bremsstrahlung contributions, we derive (to order
$1$):

\BQA p_j {\Gamma}_1^{\rm coll} &=& -{1\over \epsilon} \sum_q
P_{\gamma q}^+ + \sum_q \Biggl[ \Biggl( {{2(\pi-\Delta\phi)}\over
{\sin\Delta\phi}} (1-\cos\Delta\phi) + 2 \ln
{{\tilde{z}(1-\tilde{z})p_j}\over M_f} \Biggr) \hat{P}_{\gamma q}
(\hat{z}) - P_{\gamma q}^1 (\hat{z}) \Biggr] \cr p_j
{\Gamma}_1^{\rm CT} &=& {1\over \epsilon} \sum_q P_{\gamma q}^+
(\tilde{z}) \;, \EQA

\noindent where $\tilde{z} = (p_j-Q_T)/p_j$, $\Delta\phi = \phi_q
-\phi_j-\pi$. The poles cancel.

\setcounter{footnote}{0}
\chapter{THE MONTE CARLO METHOD IN HIGH ENERGY PHYSICS}
\label{ap:MonteCarlo}

\section{Basic Theory}
\label{sec:MCTheory}

Monte Carlo is, at its essence, a numerical method of performing
integrations. Compared with other methods, it has certain
features which make it particularly well-suited to the types of
problems encountered in high energy physics. The fundamental
theorem is based upon the recognition that the average of a
function over a domain $V$ can be expressed as the average of a
set of discrete samples of the integrand, in the limit that the
number of samples goes to infinity ~\cite{BP87}:

\begin{equation}
\langle f \rangle = {\int_V d^nx f(\vec x) \over {V=\int d^nx}} =
\lim_{N \rightarrow \infty} {1\over N} \sum_{j=1}^N f(\vec x_j)
\;. \\
\end{equation}

\noindent Here $\vec x_j$ is a particular vector in the domain
$V$ of dimension $n$, and $N$ is the number of samples, each
generated randomly by computer.

In practice, the generally available random number generators
choose values on a $0 \to 1$ scale, and the function to be
integrated is then re-expressed in terms of variables $y_i$ which
have been thus scaled. Now (for each variable),

\begin{equation}
\int_{x_{min}}^{x_{max}} f(x) dx \quad {\rm becomes} \quad
\int_0^1 f(y) dy, \qquad {\rm where} \; \\
\end{equation}

\begin{equation}
y={(x-x_{min})\over {(x_{max}-x_{min})}} \quad \hbox{\rm and the
volume} \quad V=\prod_{i=1}^n \bigl[\int_0^1 dy_i \bigr]=1^n=1
\;. \\
\end{equation}

\noindent The theorem is then expressed as

\begin{equation}
\int d^ny f(\vec y)  = \lim_{N \rightarrow \infty} {1\over N}
\sum_{j=1}^N f(\vec y_j) \;. \\
\end{equation}

Often there will be variables $\vec z$ in $f$ that are not
integrated over, but are not fixed. They define a desired
distribution $\int d{\vec y} f(\vec y, \vec z)$. Instead of
stepping through values of ${\vec z}_j$ and calculating $\int
d{\vec y} f(\vec y,{\vec z}_j)$ at each, we can define {\it bins}
in $\vec z$ and expand our definition of the volume $V$ to
include the domain of $\vec z$. Each evaluation $f({\vec
y}_j,{\vec z}_j)$ is then added to the appropriate bin in $\vec
z$, and the result $\int d\vec y f(\vec y,{\{\vec z\}}_k)$ for
bin $k$ is calculated by dividing by the number of evaluations
$N$.

Further integrations can be done from here. For instance, if
$\vec z$ consists of two variables $z_1$ and $z_2$, we have a two
dimensional grid of bins in $\vec z{\rm -space}$:

\def\bigstrut{\vrule height12pt depth6pt width0pt}
\def\nvskip#1pt{\noalign{\vskip#1pt}}
$$ \vbox { \offinterlineskip
\halign{ #\hfil\hskip1.0em
    & #\vrule&\enskip\hfil#\hfil\enskip
    & #\vrule&\enskip\hfil#\hfil\enskip
    & #\vrule&\enskip\hfil#\hfil\enskip
    & #\vrule&\enskip\hfil#\hfil\enskip
    & #\vrule&\enskip\hfil#\hfil\enskip
    & #\vrule&\enskip\hfil#\hfil\enskip
    & #\vrule&\enskip\hfil#\hfil\enskip
    & #\vrule&\hskip1.0em \hfil#\cr
  &\multispan{15}\hrulefill\cr
      && && && && && && && && \bigstrut\cr
  &\multispan{15}\hrulefill\cr
   $\uparrow$ && && && && && && && && \bigstrut\cr
  &\multispan{15}\hrulefill\cr
   $z_1$   && && && && && && && && \bigstrut\cr
  &\multispan{15}\hrulefill\cr
      && && && && && && && && \bigstrut\cr
  &\multispan{15}\hrulefill\cr }} $$
\smallskip
\centerline{$\qquad \qquad z_2 \longrightarrow$}

\noindent Integration over $z_1$ then consists of summing the
contents of each column of bins into a new one-dimensional set of
bins in $z_2$:

$$ \vbox { \offinterlineskip
\halign{ #\hfil\hskip1.0em
    & #\vrule&\enskip\hfil#\hfil\enskip
    & #\vrule&\enskip\hfil#\hfil\enskip
    & #\vrule&\enskip\hfil#\hfil\enskip
    & #\vrule&\enskip\hfil#\hfil\enskip
    & #\vrule&\enskip\hfil#\hfil\enskip
    & #\vrule&\enskip\hfil#\hfil\enskip
    & #\vrule&\enskip\hfil#\hfil\enskip
    & #\vrule&\hskip1.0em \hfil#\cr
  &\multispan{15}\hrulefill\cr
      && && && && && && && && \bigstrut\cr
  &\multispan{15}\hrulefill\cr }} $$
\smallskip
\centerline{$\qquad \qquad z_2 \longrightarrow$}

While evaluating an integral, Monte Carlo can also be used to
compute derivatives of this integral with respect to a variable.
If $\triangle x$ is a region in the domain $V$, then the
derivative with respect to $x$ is found by binning in $x$ and
dividing by the width of the bin ~\cite{Berg79}:

\begin{equation}
{dI \over {dx}} = \lim_{\triangle x \rightarrow 0}
\bigl({\triangle I \over {\triangle x}}\bigr) \;, \\
\end{equation}

\noindent where $\triangle I$ is the sum of those weights $Vf/N$
which fall in the range $\triangle x$. In practice, we find the
approximate derivative $\triangle I \over {\triangle x}$, as we
can't take the limit. In fact, decreasing $\triangle x$ increases
the uncertainty since fewer points fall into each bin. Note that
we can differentiate with respect to any variables, not just the
ones being used to evaluate the integral, as long as the former
are expressible in terms of the latter.

\section{Advantages of Monte Carlo for High Energy Physics}
\label{sec:MCHEP}

Among the many quantities of interest in high energy physics,
surely the one most frequently evaluated through numerical
integration is the cross section $\sigma$ for a particular
reaction. The integrand is a {\it differential} cross section
$d\sigma \over { d\{p_i\} }$ which, depending on the order of the
calculation, may be a function of many four-momenta $\{p_i\}$, and
of course each four-vector has four components. Integrating over
all of these variables would yield one number -- the total cross
section; but often we will want to leave one or more
uninintegrated so as to see the distribution in these variables.
Also, some variables may be fixed by the associated experiment
(like the center-of-mass energy), or there may be complicated
relations among the variables which fix one of them with respect
to others. These things would all tend to reduce the number of
integrations required, but in practice it is not always clear at
the outset which variables will be of interest. Furthermore, some
of the integrations may be impossible analytically, either due to
the iterative nature of the integrand (as in the case of parton
distributions) or to the complicated, discontinuous domains of
integration that can arise when experimental constraints are
applied. It is of necessity, then, to seek the fastest, simplest,
and most flexible tool for the job.

The fundamental theorem of Monte Carlo integration allows us to think of the
situation in the following way: We have a space $V$ of randomly selected points
$\vec x$, each point carrying a contribution $Vf(\vec x)/N$ to the total
integral, and each section of our program, from selection of the points, through
application of the cuts, evaluation of the weight, and finally binning, can carve
up this space in its own way, with its own variables. The following
considerations should make this advantage clear.

{\bf Speed.} For integrals over many variables, Monte Carlo converges
faster than other methods. For example, to do an $n$ dimensional integral with
$N$-point Gaussian Quadrature, the integrand would need to be evaluated $N^n$
times. Using $N$ randomly distributed vectors, by contrast, requires only $N$
evaluations.

{\bf Variable Transformations.} Analytically, if one has calculated a
distribution (say $d\sigma \over {dx_1dx_2}$), and one now wants this as a
distribution in the related set $dyd\tau$, one must calculate the Jacobian
function for the transformation:

\begin{equation}
{d\sigma \over {dyd\tau}} = {d\sigma \over {dx_1dx_2}}
\left|\matrix{ {\partial y \over {\partial {x_1}}} & {\partial y
\over {\partial {x_2}}} \cr {\partial \tau \over {\partial
{x_1}}} & {\partial \tau \over {\partial {x_2}}} \cr} \right| \;.
\\
\end{equation}

\noindent In Monte Carlo, one simply bins in the new variables,
modifying the bin sizes if desired. That is, the weights $\sigma$
may still be evaluated using random $\{x_1,x_2\}$ values; one need
now only insert the calculation of $\{y,\tau\}$, and use these
values instead to decide on the bin.

{\bf Discontinuous Domains.} There are times when an integral we wish
to evaluate has a domain that is not analytically expressible, is
discontinuous, or is more easily expressed in variables other than the ones
which provide the simplest form for the integrand. In a typical high energy
cross section, it may be that each Lorentz-invariant piece is simplest in a
different frame, and experimental cuts may be most easily expressed in yet
another frame or coordinate system. There may be regions of the
detector which are blind by design or damage, and which therefore should be
similarly removed from the theoretical calculation. Creating these types of
``holes'' in the domain would violate the principles upon which Gaussian
quadrature and Simpson's Rule are based; in Monte Carlo the evaluation of the
integrand at any two points is divorced from continuity constraints between
these points, and the integral becomes possible.

The trick is to use a larger volume $\acute V$ which contains
$V$, and simply ask, for each random vector $\vec x$ in $\acute
V$, whether or not it is in $V$. This prescription allows us to
perform the random selections $x_i$ within a conveniently
continuous space, within one subroutine, and leave until later (a
different subroutine) the question of whether a particular point
$\vec x$ is in the desired domain $V$. The condition (or {\it
cut}) can be described using any quantities calculable from $\vec
x$, with the result being a decision on inclusion of the point's
weight in the integral:

\centerline{If $\vec x \in V$ then ${\rm WEIGHT} = f(\vec x) \acute V / N$}
\smallskip\noindent
\centerline{If $\vec x \notin V$ then ${\rm WEIGHT} = 0$}

\noindent To minimize the uncertainty, the volume $\acute V$
should be as close to $V$ as possible.

{\bf Estimate of Precision.} In contrast with Gaussian
Quadrature, Monte Carlo integration directly produces an
uncertainty estimate. For large $N$ (the number of sampled
points), the variance is given by ~\cite{Berg79}:

\begin{equation}
\sigma=\sqrt{\langle w^2 \rangle - {\langle w \rangle}^2 \over
{N-1}} \;, \\
\end{equation}

\begin{equation}
{\rm where}\qquad \langle w \rangle ={1 \over N}\sum_{i=1}^N
f({\vec x}_i), \quad \langle w^2 \rangle = {1 \over
N}\sum_{i=1}^N f^2({\vec x}_i) \;. \\
\end{equation}

\noindent Having a numerical value for the statistical variance of
an integral is a first step toward trying to decrease it, as we
will see.

{\bf Event Generation.} Monte Carlo can be used as an event
simulator in particle physics. The fact that a particular
randomly chosen event ${\vec x}_i$ yields an evaluation $f({\vec
x}_i)$ proportional to its probability of occurrence can be
turned around -- once we know which regions of phase space give
the greatest weights, we can generate sets of four-vectors
(events ${\vec x}_i$) with frequency $f({\vec x}_i) \triangle V$
~\cite{Sey97}. Such simulators (e.g. PYTHIA, HERWIG, ISAJET) are
helpful in determining background rates for unwanted processes.
This ``cross-referencing'' of the weights to the points usually
turns out to be necessary anyway, for reasons discussed below.

\section{Importance Sampling}
\label{sec:MCsampling}

Monte Carlo with uniformly chosen ordinates has one major
drawback: the frequency with which a region $\triangle V$ is
sampled depends only on the size of $\triangle V$, not on the
size of the integrand there. Thus just as much computing time is
spent generating contributions in regions of low weight as in
high-weight regions. Avoiding this inefficiency is especially
important for most particle physics cross sections, which are
rapidly falling functions of their arguments. If the function $f$
were simple enough, certain changes of variable might
sufficiently flatten it ~\cite{PFTV89}, but particle physics cross
sections are rarely simple, and for good results some means of
{\it importance sampling} is usually required.

If we select our points $\vec x$ with a different density $\rho
(\vec x)$ in each region of space, we will be more efficient; but
if we expect to get the same answer for the integral, then those
regions with more points should count less. In other words,
normalization will require that the size of the volume element
$\triangle V$ about point $\vec x$ go as $1/\rho (\vec x)$, and
in the above formul\ae $f(\vec x)$ is replaced with $f(\vec
x)/\rho (\vec x)$. It can be shown that the uncertainty is
minimized when the density has the same shape as the absolute
value of the integrand:

\begin{equation}
{\rm ``Best"}\quad \rho(\vec x)= {|f(\vec x)| \over {\int_V
|f(\vec x)| dV}} \\
\end{equation}

Of course, if we could calculate this, all our work would be done
already; we can't, but there are certain means of approximating
the best density on an iterative basis. If one divides the
integration volume into smaller volumes, then as implied above
there are two ways of modifying the density of selected points --
either have equal subvolumes with a different number of points in
each, or have varying subvolume sizes each with the same number
of points. The latter method, called {\it stratified sampling},
is used in most of the common algorithms, such as VEGAS and SHEP
~\cite{LePage78}. Initially, the subvolumes are of equal size. At
each iteration, a two-point Monte Carlo integration is performed
in each subvolume, generating a contribution to the total
integral and to the variance. This information is then used to
redefine the subvolume sizes for the next iteration, such that
concentrations of small subvolumes are built up in regions where
the variance was initially largest. The overall variance is thus
decreased at each iteration until no significant change occurs.
The better algorithms are able to use the integral information
from each step, not just the last.

\section{Program Structure and Example}
\label{sec:MCexample}

Here we will outline the main sections of a Monte Carlo program,
using hadronic two-photon production as our example. At Born
level, a quark from one hadron annihilates with an antiquark from
the other, to emit two photons whose momenta we detect. At higher
orders, we may have additional particles in the final state, for
example gluons that are radiated from the incoming quark legs.
We'll stick with one-gluon emission here, and ask the question,
``How much transverse momentum will this extra gluon impart to
the two-photon system?'' Without the gluon, the photon system has
zero transverse momentum, as there is none coming in (the quarks
are assumed collinear). With the additional kick, the question
makes sense, but so do many others, so we want to keep as many of
the photon variables unintegrated as possible. In other words, we
look at:

\BQA d\sigma(AB \rightarrow 2\gamma + X) &=& \sum_f \int_0^1 dx_A
\int_0^1 dx_B \bigl[ q_f(x_A)q_{\bar f}(x_B) + (f \leftrightarrow
\bar f) \bigr] d\hat \sigma(f\bar f \rightarrow 2\gamma) \;, \cr
&\phantom{\cdot}& \EQA

\noindent where

\BQA d\hat \sigma(f\bar f \rightarrow 2\gamma) &\sim& \int_g
d^4p_g d^4p_{\gamma_1} d^4p_{\gamma_2} \cr &\times&
\delta^{(4)}(p_f^\mu + p_{\bar f}^\mu - p_{\gamma_1}^\mu -
p_{\gamma_2}^\mu - p_g^\mu) \sum {|{\cal M}(f\bar f \rightarrow
\gamma\gamma g)|}^2 \;. \EQA

\noindent Note that we are integrating only over the gluon, as we
want all photon variables to be free.

Let's see what we can do to get this ready for Monte Carlo
integration. We know Monte Carlo doesn't like delta functions, so
we get rid of these first. Out of the 20 variables we started with
(five 4-vectors, one for each of the two incoming quarks, the
outgoing photons, and the gluon), we now are left with 16.
Additionally, we are assuming the partons and photons to be
massless, which means that one component of each 4-vector is
fixed relative to the others. This brings our variable count down
to 12. Fixing our $\hat z$-axis along the beam line also helps,
as the incoming partons now have no transverse momentum, which
gets rid of two variables apiece. Finally, we work at fixed total
hadronic center-of-mass energy $\sqrt{S}$, so we are left with
seven variables. Since it takes six to specify the two photons,
and these we wish to leave free, we will actually need only one
``forced'' integration (over one or the other momentum fraction
$x$). Integration over the others (the photons) will be left for
the user to decide upon through binning. Whatever is left
determines the desired distribution. That is, our total
integration volume $V$ is 7-dimensional, as we add the free
variables to the volume as discussed in Section
\ref{sec:MCTheory}. For the purposes of generating the random
points in this space, and their weights, we need to choose a set
for which we know the kinematic bounds. For binning of these
weights, the user may choose any set he or she wishes.

With a bit of analytical work, then, we obtain a simplified expression, each
piece of which we will write as a function of the simplest parameters for that
piece:

\begin{eqnarray}
d\sigma(AB \rightarrow 2\gamma + X) &\sim& \sum_f \int_{{Q^2 \over
S}e^y}^1 dx \bigl[ q_f(x,Q^2)q_{\bar f}(\tilde x,Q^2) + (f
\leftrightarrow \bar f) \bigr] P(Q^2,{p_T}^2,y,\phi_{2\gamma}) \;
\cr &\times& \sum {|{\cal M}({\hat s}_{12},{\hat t}_{13}, {\hat
t}_{14},{\hat t}_{15},{\hat t}_{23},{\hat t}_{24},{\hat
t}_{25})|}^2 \;.
\end{eqnarray}

\noindent Here $\tilde x$ is a known function of $x$, given by the
massless gluon constraint. $Q^2$, ${p_T}^2$, $y$, and
$\phi_{2\gamma}$ are, respectively, the (squared) energy,
transverse momentum, rapidity, and azimuthal angle of the
two-photon system in the hadronic cm frame (which we take to be
the lab frame -- at a collider). $P$ is a function which depends
purely on phase space.

The matrix element for the subprocess is most easily written in
terms of Mandelstam variables $\hat{s}_{ij}$ and $\hat{t}_{ij}$.
The astute reader will notice that there are seven of these, so
they cannot all be independent of the four we've already
mentioned. With a little extra work, we could choose a frame and
rewrite the matrix element in terms of $Q^2$, ${p_T}^2$, $y$,
$\phi_{2\gamma}$, $x$, and two more variables, but the result
would not be pretty. And indeed there is no need to. All that
matters is that we be able to calculate the 4-vectors of each
particle in whatever frame is easiest for that particle, boost
them all to the same frame, and calculate the Mandelstam
variables.

The best frame for the photons is the rest frame of the two-photon system, in
which their momenta are equal and opposite, and their energies are simply $Q/2$
each. The direction of their ``momentum line'' in this frame can be anywhere
from $0$ to $\pi$ in theta, and $0$ to $2\pi$ in phi. Those three variables
completely specify the photon 4-vectors in that frame, and now we're done
seeking variables. We have seven, and know their limits:

\begin{eqnarray}
0 &\le  Q^2 &\le  S \;\cr 0 &\le  {p_T}^2 &\le  {{(S+Q^2)^2}
\over {4S\cosh^2 y}}-Q^2 \;\cr y_{min} &\le  y &\le  y_{max} \;\cr
0 &\le  \phi_{2\gamma} &\le  2\pi \;\cr 0 &\le  \theta_\gamma
&\le  \pi \;\cr 0 &\le  \phi_\gamma &\le  2\pi \;\cr {Q^2 \over
S}e^y &\le  x &\le  1 \;
\end{eqnarray}

\noindent Note that in reality $Q^2$ will never approach $S$
(unless the hadrons disappear altogether!). We are taking a
larger volume (as discussed in Section \ref{sec:MCHEP}) for lack
of a better limit. Here $y_{min}$ and $y_{max}$ are assumed to be
given by the user, based on the viable range of the detector. We
are now ready to begin writing the code.

In the following, variables entering a subroutine will be
denoted by lowercase names, outgoing variables will be in uppercase, and
variables which are both inputs and outputs will be italicized. To avoid
confusion, we'll show the passed variables only in subroutine definitions, not
in calls.

\begin{list} {} {} \singlespace
\item {\bf Program MAIN}
\item Initialize fixed values.
\item Initialize bins
\item Loop ($i=1...N$)
\item $\phantom{xxxx}$   Call EVENT
\item $\phantom{xxxx}$   Call WEIGHT
\item $\phantom{xxxx}$   Call BINIT
\item $\phantom{xxxx}$   Write LABVECS,WEIGHT (optional)
\item end Loop
\end{list}

Here we read our fixed values (e.g. $S$, $y_{min}$, $y_{max}$,
etc.), make sure our bins are empty, and then process $N$
randomly selected events. For each event, we find the
four-vectors (EVENT), calculate this event's contribution
(WEIGHT), put the weight in the appropriate bin (BINIT), and
optionally, send the event out to a file.

\begin{list} {} {} \singlespace
\item {\bf Subroutine EVENT} (fixed, LABVECS[5,0:3], UNSCALED, VOLUME)
\item Random selection of scaled vars (e.g. {\tt XQ2=RAND(SEED)},{\tt XPT=RAND(SEED)}, etc.)
\item Scaled vars $\rightarrow$ Unscaled vars (e.g. {\tt Q2=S*XQ2}) \& VOLUME
\item Find photon cm 4-vecs (using $Q^2$, $\theta_\gamma$, $\phi_\gamma$)
\item Find boost parameters (using $Q^2$, ${p_T}^2$, $y$, $\phi_{2\gamma}$)
\item Boost photons to lab frame
\item Find parton 4-vecs in lab (from $S$, $x$, $\tilde x$)
\item Find gluon 4-vec ($= p_f^\mu + p_{\bar{f}}^\mu - p_{\gamma 1}^\mu - p_{\gamma 2}^\mu$)
\item return
\end{list}

Remember that we can't randomly choose (what I call) the ``unscaled'' variables
[$Q^2$, ${p_T}^2$, $y$, $\phi_{2\gamma}$, $\theta_\gamma$, $\phi_\gamma$, $x$]
directly; our random number generator picks between the limits 0 and 1. So we
define a ``scaled'' variable (e.g. {\tt XQ2} in the code) for
each, choose it randomly, and then calculate the unscaled variables from them.
The volume $V$ is the product of the unscaled variables' domains, and this will
be multiplied by the integrand in evaluation of the weight for this event. Once
we find the four-vectors for all the particles of the subprocess, we'll be able
to calculate the matrix element; we also send out the unscaled variables, as
they may be more directly useful for the phase space factor, cuts, and binning.

\begin{list} {} {} \singlespace
\item {\bf Subroutine WEIGHT} (fixed, labvecs, unscaled, volume, WEIGHT)
\item Implement CUTS (using labvecs, unscaled), e.g. {\tt IF(P5(0).LT.0.75) GOTO REJECT}
\item Calculate MANDELSTAM vars $t_{ij}$ (using labvecs)
\item Calculate MATRIX ELEMENT (using $t_{ij}$)
\item Calculate PARTON DISTRIBUTIONS $q_f (x,Q^2)$
\item Calculate PHASE SPACE FACTOR $P$ (using unscaled)
\item INTEGRAND = (partons)$\times$(phase space)$\times$(matrix element)
\item WEIGHT = (volume)$\times$(integrand)/($N$)
\item return
\item REJECT:
\item WEIGHT = 0
\item return
\end{list}

The first section here (CUTS) is open to user modification. The user has access to all
of the variables that describe the event, and may invent new variables as
needed, all in order to define {\tt IF} statements which either accept or deny
this event as valid. If denied, control is passed to the REJECT line, where the
weight is set to zero. This CUTS section may be put in the binning routine
instead, but this just wastes time, as the weight will be unnecessarily
calculated.

\begin{list} {} {} \singlespace
\item {\bf Subroutine BINIT} (fixed, labvecs, unscaled, {\it BIN}[1:100], WEIGHT)
\item Decide upon and calculate binning variable(s) (e.g. the pair PT in our case)
\item BIN\# = INT((PT-LO)/STEP) + 1
\item BIN(BIN\#)=BIN(BIN\#) + WEIGHT/STEP
\item return
\end{list}

Again, the first section here is user-accessible, for the purpose
of defining whatever variables are of interest. The user may wish
to look at the transverse momentum of the photon pair (as we do),
rapidity differences between the photons, or single photon
spectra (in which case each photon might contribute to a
different bin). Here we assume a one-dimensional bin array,
corresponding to $d\sigma \over {dp_T}$, but the generalization
to higher dimensions is trivial, and affects no other subroutine.
Note that our WEIGHT corresponds to $d\sigma$, so if we want
$d\sigma \over {dp_T}$, we must divide by the bin width (STEP) --
see the discussion on differentiation in Section
\ref{sec:MCTheory}.

In figure \ref{fig:monte1} we present results for two-photon
production in $\pi^-p$ collisions, to first order in $\alpha_s$.
The fixed quantities are: $S$=526.33 ${\rm GeV}^2$,
$\Lambda_{QCD}$=0.2305 GeV, STEP=0.08 GeV. The cuts that were used
are the following:

\begin{list} {} {}
\item 1. The photon of higher $p_T$ must have $p_T \ge 3.00{\rm GeV}$.
\item 2. The photon of lower  $p_T$ must have $p_T \ge 2.75{\rm GeV}$.
\item 3. The quantity $zz = -(\overrightarrow {p_{T_{\gamma 1}}} \cdot \overrightarrow {p_{T_{\gamma 2}}}) / \max{(p_{T_{\gamma 1}},p_{T_{\gamma 2}})}$ must exceed 2.75.
\end{list}

\noindent and the code that took care of these cuts:

\begin{list} {} {} \singlespace
\item {\tt C     --- CUTS ON LAB VARS ---}
\item {\tt PT3=DSQRT(P3(1)**2.D0+P3(2)**2.D0)}
\item {\tt PT4=DSQRT(P4(1)**2.D0+P4(2)**2.D0)}
\item {\tt ZZ=-1.D0*(P3(1)*P4(1)+P3(2)*P4(2))/DMAX1(PT3,PT4)}
\item {\tt IF (DMAX1(PT3,PT4).LT.3.D0) GOTO REJECT}
\item {\tt IF (DMIN1(PT3,PT4).LT.2.75D0) GOTO REJECT}
\item {\tt IF (ZZ.LE.2.75D0) GOTO REJECT}
\end{list}

\noindent On the graph, the $p_T$ distribution of the photon pair
is shown. The dashed line is the perturbative 1-gluon result; the
solid line is the result of resumming the soft gluons.

\figboxf{monte1}{$\pi^-p \rightarrow \gamma \gamma X$ pair $p_T$
distribution.}

\setcounter{footnote}{0}

\bibliography{thesis}
\biosketch{

\renewcommand{\baselinestretch}{1} \small
\bigskip
\bigskip
\bigskip

\noindent
{\large {\underline{\bf Personal}}}\\*[.3cm]
\begin{tabular}{p{2.7cm}p{.01cm}p{11.2cm}}
Date of Birth: & &{\bf December 25, 1964}. \\
Place of Birth: & &{\bf Milwaukee, WI., U.S.A.} \\
Marital Status: & &{\bf Single}. \\
Languages: & &{\bf English and Spanish}. \\
Nationality: & &{\bf U.S.A.} \\

\end{tabular}\\*[.75cm]

\noindent
{\large {\underline{\bf Education}}}\\*[.3cm]
\begin{tabular}{p{2.7cm}p{.01cm}p{11.2cm}}
2001& &{\bf Ph.D. in physics}, specialization in theoretical high
energy physics, Florida State University.
\newline
Dissertation: ``Transverse Resummation for Direct Photon Production''. \\
1995 & &{\bf M.S. in physics}, Florida State
University. \\
1987& &{\bf B.S., physics and philosophy}, University of Wisconsin, Madison.  \\
\end{tabular}\\*[.75cm]

\noindent
{\large {\underline{\bf Experience}}}\\*[.3cm]
\begin{tabular}{p{2.7cm}p{.01cm}p{11.2cm}}
1995 to present & &{\bf Research Assistant}, High Energy Physics Dept., Florida State University. \\
1994 to 1995 & &{\bf Teaching Assistant}, Department of Physics,
Florida State University. \\
1987 to 1994 & &{\bf Information Systems Analyst}, Temperature
Control Specialties, Middleton, WI. \\
\end{tabular}\\*[.75cm]

\newpage

\noindent {\large {\underline{\bf Research Interests}}}\\*[.3cm]
Theoretical high energy physics, particularly precision tests of
QCD and application of renormalization group methods to extend the
region of applicability of perturbation theory. Additional
interests include grand unification, physics education, quantum
coherence and computing, symbolic logic, and computational
methods.
\\*[.75cm]

\noindent
{\large {\underline{\bf Ph.D. Dissertation Advisor}}}\\*[.3cm]
{\bf Joseph F. Owens} at Florida State University.\\*[.75cm]

\noindent {\large {\underline{\bf Additional Research
Advisors}}}\\*[.3cm] {\bf Howard Baer}, {\bf Vasken Hagopian},
{\bf Elbio Dagotto}, {\bf Jack Quine}, all at Florida State
University.\\*[.75cm]

}

\end{document}